\documentclass[12pt,a4paper]{article}
\usepackage{amsmath,amssymb}
\usepackage{tocloft}     
\usepackage{titletoc}    
\usepackage{etoc}  
\usepackage[utf8]{inputenc}
\usepackage{natbib}
\usepackage{placeins}
\usepackage{graphicx}
\usepackage{ragged2e}
\usepackage{amsthm}
\usepackage{amsmath}
\usepackage{mathtools}
\usepackage{amsfonts}
\usepackage{amssymb}
\usepackage{float}
\usepackage{hyperref}
\hypersetup{
    colorlinks=true,
    linkcolor=red,
    citecolor=blue,
    filecolor=magenta,      
    urlcolor=cyan,
}
\usepackage{bbm}
\usepackage{bm}
\usepackage{tikz}
\usepackage{algpseudocode}
\usepackage{breqn}

\usepackage{nicefrac}
\usepackage[multiple]{footmisc}
\usepackage{enumitem}
\setenumerate{wide}
\usepackage{thmtools}
\usepackage{thm-restate}
\usepackage[nameinlink, capitalize]{cleveref}

\usepackage{cprotect}
\usepackage{etoc}
\usepackage{etaremune}

\usepackage[T1]{fontenc}    
\usepackage{booktabs}       
\usepackage{microtype}      
\usepackage{makecell}
\floatstyle{ruled}
\usepackage{xcolor}

\usepackage{caption}
\usepackage{comment}
\usepackage{subcaption}

\usetikzlibrary{arrows,automata, calc, positioning}
\usepackage{adjustbox}
\usepackage{wrapfig}
\DeclareCaptionLabelSeparator{custom}{---}

\usepackage{multirow}
\usepackage{accents}

\usepackage[subtle]{savetrees}
\usepackage{apptools}

\usepackage{threeparttable}
\newcommand{\indep}{\raisebox{0.05em}{\rotatebox[origin=c]{90}{$\models$}}}

\DeclareRobustCommand{\E}{\mathbb{E}}
\DeclareRobustCommand{\V}{\mathbb{V}}
\DeclareRobustCommand{\R}{\mathbb{R}}

\DeclareRobustCommand{\tr}{\textsc{train}}
\DeclareRobustCommand{\te}{\textsc{test}}

\algrenewcommand\algorithmicreturn{\textbf{Return:}}

\DeclareMathOperator{\cov}{\operatorname{Cov}}
\DeclareMathOperator{\var}{\operatorname{Var}}

\theoremstyle{definition}

\newtheorem{theorem}{Theorem}
\newtheorem{assumption}{Assumption}
\newtheorem{algorithm}{Algorithm}
\newtheorem{proposition}{Proposition}
\newtheorem{definition}{Definition}
\newtheorem{remark}{Remark}
\newtheorem{corollary}{Corollary}
\newtheorem{lemma}{Lemma}
\newtheorem{example}{Example}

\AtAppendix{\counterwithin{theorem}{section}}
\AtAppendix{\counterwithin{assumption}{section}}
\AtAppendix{\counterwithin{algorithm}{section}}
\AtAppendix{\counterwithin{proposition}{section}}
\AtAppendix{\counterwithin{definition}{section}}
\AtAppendix{\counterwithin{remark}{section}}
\AtAppendix{\counterwithin{corollary}{section}}
\AtAppendix{\counterwithin{lemma}{section}}
\AtAppendix{\counterwithin{example}{section}}

\DeclareRobustCommand{\VAN}[2]{#1}  
\usepackage{algorithm}
\usepackage{algpseudocode}

\begin{document}

\def\spacingset#1{\renewcommand{\baselinestretch}
{#1}\small\normalsize} \spacingset{1}

\title{Program Evaluation with Remotely Sensed Outcomes}
\author{
  Ashesh Rambachan \\
  MIT
  \and
  Rahul Singh \\ Harvard
  \and
  Davide Viviano\thanks{Email: asheshr@mit.edu, rahul\_singh@fas.harvard.edu, dviviano@fas.harvard.edu.
  We thank Isaiah Andrews, Josh Angrist, Arun Chandrasekhar, Raj Chetty, Kevin Chen, Ben Deaner,  Melissa Dell, Cristophe Gaillac, Paul Goldsmith-Pinkham, Seema Jayachandran, Namrata Kala, Sylvia Klosin, Ben Olken, Jonathan Roth, Jesse Shapiro, Matthieu Stigler, and Alex Tetenov, as well as audiences at Berkeley, Bocconi, CEMFI, CREST/Sciences Po, Harvard/MIT, the NBER Summer Institute, the Online Causal Inference Seminar,  Opportunity Insights, Princeton, Stanford, University of Bologna, University of Chicago, University of Southern California, University of Toronto, University of Geneva, and Yale for helpful discussions. Haya Alsharif, Peter Chen, Marvin Lob, Leonard Mushunje, Miriam Nelson, Kevin Wang, and Sammi Zhu provided excellent research assistance. Davide Viviano gratefully acknowledges funding from the Harvard Griffin Fund in Economics and NSF Grant SES 2447088. Replication code is available \href{https://github.com/asheshrambachan/rsv_replication}{here}. We provide the \href{https://github.com/asheshrambachan/remoteoutcome}{\texttt{remoteoutcome}} R package to implement our method.
  } \\ Harvard
}
\date{First draft: November 2024 \\ This draft: June 2026}
\maketitle

\begin{abstract}
%
We study causal inference in experiments and quasi-experiments, where the economic outcome is imperfectly measured by a remotely sensed variable. Remotely sensed variables are low-cost, scalable, and predictive of economic outcomes; examples include satellite imagery and mobile phone activity. We model the remotely sensed variable as post-outcome: variation in the economic outcome causes variation in the remotely sensed variable. For example, changes in environmental quality cause changes in satellite imagery, not vice versa. Under this assumption, we nonparametrically identify the causal parameter by combining experimental and observational data, and develop a method for robust $n^{-1/2}$ inference.
\vspace{1em}

\end{abstract}

\newpage

\spacingset{1.5} 


\section{Introduction}

Traditional program evaluation relies on surveys to measure outcomes, but many economic outcomes are costly or infeasible to collect at scale. 
Researchers increasingly rely on remotely sensed variables (Appendix Figure~\ref{fig:trend_papers}). 
Examples include night lights as measures of local economic activity, roofing materials as measures of housing quality, mobile phone transactions as measures of poverty or consumption, and satellite images as measures of pollution, deforestation, fires, flooding, and local poverty.\footnote{
Night lights \citep{chen2011using, henderson2012measuring, asher2021development}. Roofing material \citep{marx2019there, michaels2021planning, huang2021using}. Mobile phone transactions \citep{blumenstock2015predicting, aiken2025estimating}. Satellite images of pollution \citep{currie2023caused}; deforestation \citep{jayachandran2017cash, assunccao2023optimal}; fires \citep{jack2022money, balboni2024origins}; flooding \citep{chen2017validating, patel2024floods}; local poverty \citep{jean2016combining}. See \cite{donaldson2016view}, \cite{burke2021using}, and \cite{jack2023remotesensing} for reviews.}
These variables are cheap, scalable, and predictive of economic outcomes in observational data, but imperfect.\footnote{Recent work documents that remotely sensed measures of economic outcomes can be noisy \citep[see, e.g.,][]{bluhm2022nightlights, fowlie2019satellite, corral2025povertymapping, Josephson2026}.} 
We study how to identify treatment effects in experiments and quasi-experiments where the outcome is unobserved but remotely sensed variables are available.

To formalize this setting, we consider two samples, each containing the remotely sensed variable.
The experimental sample contains a randomly assigned treatment and the remotely sensed variable, but the outcome is missing due to logistical or cost constraints.
The observational sample links the outcome to the remotely sensed variable, but comes from a different, non-experimental context. 
For this reason, the treatment variable is missing or confounded in the observational sample, and treatment effects may differ across the two populations.

What the two samples share is the remotely sensed variable, which we view as a ``post-outcome'' measurement: variation in the outcome causes variation in the remotely sensed variable. 
For example, while evaluating financial incentives to reduce crop burning, the experimental sample contains randomized contracts and satellite images of fields, while the observational sample links satellite images to surveyor measurements of whether fields were burned.
In this example, crop burning causes changes in satellite images, not vice versa. 

Our main assumption is \textit{measurement stability}: the conditional distribution of the remotely sensed variable, given the outcome and treatment, is the same across samples.
This assumption is plausible when the same measurement technology is used across samples; for example, the same satellite generates the images in both datasets.
Therefore, we study a setting where the treatment mechanism and outcome mechanism vary across samples, but the \textit{sensing} mechanism is the same.

Our primary contribution is an identification formula for the treatment effect that requires no outcome data in the experimental sample. 
It exploits stability to combine the experimental and observational samples.
Intuitively, the experimental sample identifies how the treatment affects the remotely sensed variable, while the observational sample identifies how the outcome affects the remotely sensed variable. 
The treatment effect is identified by a conditional moment restriction, in which the remotely sensed variable relates experimental treatment variation to observational outcome variation (Theorems~\ref{theorem:disc_outcomes_incomplete}--\ref{theorem:disc_outcomes_complete}). 

By characterizing a set of conditional moment restrictions, we also provide diagnostic tests for the key identifying assumptions. 
One diagnostic test evaluates whether the remotely sensed variable is sufficiently informative about outcome variation, similar to testing for weak instruments.
Another diagnostic test evaluates whether different representations of the remotely sensed variable yield the same treatment effect estimate, similar to an overidentifying restrictions test.

As a secondary contribution, we develop a method for $n^{-1/2}$ inference on the treatment effect that is robust to misspecification and that accommodates complex machine learning. 
We explain how to use predicted outcomes for valid inference. 
Moreover, we explain how to use predicted outcomes, treatments, and sample indicators for more efficient inference through a connection to optimal instruments \citep{chamberlain1987asymptotic, newey1993efficient}.
Unlike the standard inference guarantees for double machine learning \citep{chernozhukov2018double}, $n^{-1/2}$ inference remains valid even if all of these predictors are misspecified and lack rate conditions.
From a practical perspective, this result allows complex neural networks to compress high-dimensional, unstructured data.

Our framework extends in many directions (Section~\ref{sec:generalizations}).
One extension (Section~\ref{sec:limited_experimental_outcomes}) considers not only an experimental sample and an observational sample but also a validation sample: units for which we observe the randomized treatment, the outcome, and the remotely sensed variable.
The validation sample may be non-randomly selected, with different treatment effects than the experimental sample.
Another extension (Section~\ref{sec:main_quasi_exp}) replaces the experimental sample with a quasi-experimental sample: instead of randomized treatments, we may have instrumental variables or a difference-in-differences design.

In the setting we study, we show that some existing methods are biased. 
Fifty percent of papers in general interest economics journals from 2015--2024 that use remotely sensed outcomes employ a two-step method: (i) train a predictor of the outcome from the remotely sensed variable in the observational sample; (ii) use it to predict outcomes in the experimental sample, and report a difference of predicted outcome means.\footnote{We survey American Economic Association (AEA) journals, \textit{Econometrica}, \textit{Journal of Political Economy}, \textit{Quarterly Journal of Economics}, and \textit{Review of Economic Studies}. The remaining fifty percent use a similar logic, without an explicit formula.}  
Without covariates, this method suffers from attenuation bias (Proposition~\ref{prop:bias_common_practice}), even if samples are randomly selected (Proposition~\ref{prop:bias_common_practice_random}). With covariates and heterogeneity, this method may flip the sign of the average treatment effect (Footnote~\ref{footnote:flip}).
Another important method is called prediction-powered inference (PPI) \citep{angelopoulos2023prediction}, which adds bias correction terms estimated from an appropriate auxiliary sample. 
In our setting, it can be biased when the difference in outcome means in the experimental sample does not equal the difference in outcome means in the observational sample (Proposition~\ref{prop:bias_PPI} and Remark~\ref{remark:ppi_simulations}). 
Such a distribution shift is plausible when the two samples come from different populations, time periods, or regions.

We illustrate our method in calibrated simulations and in real-world applications, pertaining to: (i) forest cover in Uganda \citep{hansen2013high,jayachandran2017cash}; (ii) household poverty in India \citep{muralidharan2016building, muralidharan2023general}, and (iii) crop burning in India \citep{jack2022money}.
Across settings, we assess our identifying assumptions using our diagnostic tests.
In simulations, our method is approximately unbiased with nominal coverage. 
By contrast, we find the two-step method and PPI are generally biased.
In the household poverty application, our method yields estimates close to the ``oracle'' difference-in-means estimate obtained using experimental outcomes.
In the crop burning application, our analysis suggests that the two-step method underestimates the effectiveness of the intervention by 47 percent.

\subsection{Related Work}\label{sec:related_work}

Our causal model differs from existing models along three dimensions: (i) the causal direction between the auxiliary variable and the outcome, (ii) robustness to misspecification, and (iii) data requirements. We provide formal comparisons in Section~\ref{section:main_comparisons}.

In the surrogacy model \citep{prentice1989surrogate, athey2024surrogate, kallus2024role}, the auxiliary variable---called a surrogate---is pre-outcome: it is a ``short-term'' measurement that mediates between the treatment and ``long-term'' outcome.
By contrast, in our model, the remotely sensed variable is post-outcome: the outcome mediates between the treatment and our measurement. 
We prove that misusing a remotely sensed variable as a surrogate leads to attenuation bias (Proposition~\ref{prop:bias_common_practice}). 
The negative control literature extends the surrogacy model to address unobserved confounding \citep{ghassami2022combining, imbens2024long}. In these extensions, misusing a remotely sensed variable incurs bias.

Several recent papers propose alternative methods for inference with remotely sensed outcomes. \cite{proctor2023parameter} propose a multiple imputation method that models the outcome as a function of the remotely sensed variable and imputes it across samples. 
It is akin to the surrogacy methods discussed above and in Section~\ref{sec:common_practice_main_text}.
\citet{alix2023remotely, torchiana2025improving} instead model the misclassification process for binary and panel outcomes, requiring correct specification of a model relating the remotely sensed variable to the outcome. 
By contrast, our method does not require correct specification of any model relating the remotely sensed variable to the outcome. This distinction is important because the remotely sensed variable may be high-dimensional or unstructured.

One major difference between our method and PPI methods \citep{angelopoulos2023prediction, angelopoulos2024ppiefficientpredictionpoweredinference, kluger2025prediction} is the data requirement. PPI methods require joint observations of the treatment, outcome, and remotely sensed variable, in both treatment arms; these methods are infeasible without such observations.
\cite{sanford2025remote, lu2025regressioncoefficientestimationremote,PELLETIER2026103655,fong2021machine, allon2023machine, egami2023using, carlson2025unifying, ji2025predictions} share this data requirement. 
By contrast, our method remains feasible when the experimental sample excludes outcomes, and when the observational sample excludes treatments (Theorem~\ref{theorem:disc_outcomes_incomplete}). 

Another difference is that our method requires stability of the sensing mechanism, whereas we show that PPI applied with an observational sample implicitly requires stability of the outcome mechanism (Proposition~\ref{prop:bias_PPI}).
By contrast, our method allows for the observational sample to come from a different population, time period, or region, with different baseline outcomes and treatment effects. 
See Remark~\ref{remark:ppi_simulations} for comparisons in other data environments.

Compared to the vast literature on data combination \citep[e.g.,][]{cross2002regressions, RidderMoffitt07, bareinboim2016causal, d2024partially} and nonclassical measurement error \citep[e.g.,][]{ChenHongNekipelov(11), schennach2020mismeasured}, we place a different main assumption. 
Several influential works handle measurement error in moment condition models by assuming that the conditional distribution of the variable of interest, given the imperfect measurement, is stable across samples \citep{chen2005measurement, chen2008semiparametric, GrahamPintoEgel2016}, akin to the surrogacy model.
Our main assumption is the opposite: the conditional distribution of the imperfect measurement, given the variable of interest, is stable. 
Our assumption is natural when we believe the sensing mechanism, e.g. the satellite technology, is stable across samples rather than the outcome mechanism.
This distinction leads to a distinct identifying formula.

Our framework is closer to nonclassical measurement error models with instruments or repeated measurements \citep{hu2008instrumental,hu2008identification}. 
However, their setting has no observational sample in which the outcome is observed, and therefore requires an additional normalization to pin down how the imperfect measurement is distributed around the true outcome. 
In our setting, the outcome is observed in the observational sample, and so we do not require an additional assumption beyond stability of the sensing mechanism.

Our framework is also related to the ``shadow-variable'' model for recovering the outcome distribution in a single sample with nonignorable missingness \citep[e.g.,][]{miao_tchetgen_2016_shadow_dr, miao_liu_li_tchetgen_geng_2024_shadow}. 
A shadow variable is an auxiliary variable that is relevant to the missing outcome yet excluded from the missingness mechanism.
The main difference is that we study a two-sample causal inference problem, rather than a single-sample missing-data problem studied in this literature.
Our framework includes scenarios where the treatment is only observed in one sample while the outcome is only observed in the other sample, so data combination is necessary for identification of the treatment effect; the single-sample shadow-variable approach would not, by itself, identify it.

Finally, a rich literature studies mismeasured or missing covariates \citep[e.g.,][]{fan2014identifying, d2024linear}, possibly imputed with machine learning \citep{battaglia2024inference}. 
By contrast, in our setting, only the outcome is missing yet a post-outcome variable is present. 
\section{Model and Identifying Assumptions}\label{sec:model}

\subsection{Setting and Causal Parameter}
The researcher observes units in two samples, indicated by the variable $S \in \{e, o\}$: an experimental sample ($S = e$) and an observational sample ($S = o$). 

In the experimental sample ($S = e$), we observe pre-treatment covariates $X \in \mathcal{X}$ and a binary treatment $D \in \{0, 1\}$. 
The outcome $Y \in \mathcal{Y}$, however, is entirely missing.
In its place, we observe a remotely sensed variable $R \in \mathcal{R}$. 
(In Section~\ref{sec:limited_experimental_outcomes}, we allow the outcome to be only partially missing in the experimental sample.)
We place no restriction on the form of the remotely sensed variable: examples include unstructured data such as satellite images or digital traces, a pretrained embedding of unstructured data, or the output of a pretrained predictor.
The researcher would like to use the remotely sensed variable $R$ as an imperfect measurement of the outcome $Y$ in the experimental sample. 

The parameter of interest is the effect of the treatment $D$ on the outcome $Y$ in the experimental sample. 
Though the outcome $Y$ is unobserved in the experimental sample, we may still define its potential outcomes $Y(d)$ and the causal parameter.\footnote{This definition rules out spillovers. We extend our framework to allow for spillovers in Section~\ref{sec:main_spillover}.}  

\begin{definition}[Causal parameter]\label{defn: causal param}
The average treatment effect (ATE) in the experimental sample is $\tau := \mu(1) - \mu(0)$, where $\mu(d) := \E\{Y(d) \mid S = e\}$. 
\end{definition}

\noindent Without further assumptions, when $R$ is an imperfect measurement of $Y$, it is generally impossible to point identify this causal parameter \citep[][]{horowitz1995identification}. 

Motivated by empirical work in environmental and development economics, we leverage an auxiliary sample that we refer to as the observational sample ($S=o$).
For these units, we observe the covariates $X$, outcome $Y$, and remotely sensed variable $R$. 
We may or may not observe the treatment $D$. 
If we do observe the treatment and there is variation in the treatment (though it may be confounded), we refer to this scenario as having ``complete'' observational cases. 
If we do not observe the treatment or if the treatment is deterministic in the observational sample, we set $D = 0$ for all units in the observational sample,
and we refer to this scenario as having ``incomplete'' observational cases.
Incomplete observational cases will require stronger assumptions for point identification.

\begin{table}[ht]
\centering
\caption{Summary of the Data Environment.}
\scalebox{0.8}{\begin{tabular}{r | c c c}
\toprule
& \multicolumn{3}{c}{Sample Indicator $S$} \\
& Experimental & Observational: Complete & Observational: Incomplete \\
\hline 
Covariates $X$ & \checkmark & \checkmark & \checkmark \\
Treatment $D$ & \checkmark & \checkmark & Missing or Deterministic \\
Outcome $Y$ & Missing & \checkmark & \checkmark \\
Remotely Sensed Variable $R$ & \checkmark & \checkmark & \checkmark \\
\bottomrule
\end{tabular}
}
\medskip

{\raggedright 
\footnotesize{\textit{Notes}: $\checkmark$ denotes that the variable is observed. When treatment is missing or deterministic in the observational sample, we encode it as $D = 0$ in the observational sample.}
\par}
\label{table:data}
\end{table}

Table~\ref{table:data} summarizes the data environment. 
Each unit is characterized by the vector $(S, X, D, Y(0), Y(1), R)$, which we assume to be independent and identically distributed across units. 
For units in the experimental sample ($S = e$), we observe $(X, D, R)$. 
For units in the observational sample ($S = o$), we observe $(X, D, Y, R)$ if there are complete cases or $(X, Y, R)$ if there are incomplete cases. 

\begin{example}[Environmental programs]\label{ex:environmental}
Consider a randomized experiment evaluating whether incentives for environmental conservation $D$, such as payments for ecosystem services (PES), reduce harmful environmental behaviors $Y$, such as deforestation or crop burning.
Directly measuring these outcomes is costly: monitoring tree cover requires on-site surveyors \citep{jayachandran2017cash}, and recording crop management practices requires field visits during the narrow post-harvest/pre-tilling window when crop burning occurs \citep{jack2022money}.
By contrast, satellite imagery $R$ is cheap and predictive of these environmental outcomes \citep[e.g.,][]{hansen2013high, burke2021using, walker2022detecting}.
In summary, in the experimental sample, only the treatment assignment $D$ and remotely sensed variable $R$ are easily observed.

Therefore, the researcher turns to an observational sample linking the remotely sensed variable $R$ to the outcome $Y$. 
For example, \citet[][]{jayachandran2017cash} collect ground-based forest measurements linked to satellite images, for a sample of locations distinct from the experimental units, collected at different dates than the images used in their experimental analysis.
\citet[][]{jack2022money} collect ground-based, random spot checks for crop burning, but the spot checks could not be synchronized with the narrow post-harvest/pre-tilling window (see Section~\ref{sec:empirical_applications} for discussion).
Neither is well-suited to serve as the primary basis for treatment effect estimation in the experiments. 
In our framework, both qualify as valid observational samples.
$\blacktriangle$
\end{example}

\begin{example}[Household poverty]\label{ex:poverty} 
Consider a randomized experiment evaluating whether an anti-poverty program $D$, such as unconditional cash transfers \citep[e.g.,][]{egger2022general, aiken2025estimating} or biometrically authenticated payments \citep{muralidharan2023general}, reduces village-level poverty $Y$.
Directly measuring poverty requires detailed household surveys on consumption, assets, and income in dispersed rural villages.
By contrast, remotely sensed variables $R$, such as satellite imagery \citep{jean2016combining, huang2021using, rolf2021generalizable} and digital trace data from mobile phones \citep[][]{blumenstock2015predicting, aiken2022machine, aiken2025estimating}, are widely available and predictive of poverty.
The researcher has access to an observational sample linking $R$ to poverty measurements $Y$ (e.g., census statistics or existing household surveys in other regions). 
Our framework combines these samples to identify the causal effect of the anti-poverty program. $\blacktriangle$
\end{example}

\subsection{Identifying Assumptions}

We present three assumptions under which the experimental and observational samples can be combined to identify the causal parameter. 

Our identifying assumptions allow $(X,Y,R)$ to be discrete or continuous. 
For exposition, we slightly abuse notation: for a random variable $W$, we use $f_W(\cdot \mid ...)$ to denote its (conditional) probability mass function if $W$ is discrete, or its (conditional) density if $W$ is continuous. Formally, $f_W(\cdot \mid ...)$ is the Radon-Nikodym derivative.

\subsubsection{Experimental Unconfoundedness}

\begin{assumption}[Experimental unconfoundedness]\label{assumption:experimental}
Suppose the following:
\begin{enumerate}[label=\roman*.]
\item Stable unit treatment value (SUTVA): $Y = D Y(1) + (1 - D) Y(0)$ almost surely.
\item Randomization: $D \indep \left\{ Y(0), Y(1) \right\} \mid X, S = e$.
\item Overlap: $\Pr(D = 1 \mid X, S = e) $ is bounded away from zero and one almost surely.
\end{enumerate}
\end{assumption}

Assumption~\ref{assumption:experimental} imposes standard conditions for causal inference in randomized experiments.
In many applications involving remotely sensed variables, such as Examples~\ref{ex:environmental} and~\ref{ex:poverty}, the assumption is satisfied by design: the treatment is randomly assigned to well-separated, aggregate units.
In Section \ref{sec:main_quasi_exp}, we extend our analysis to allow quasi-experimental designs in the experimental sample, such as instrumental variables and difference-in-differences. 
In Section~\ref{sec:main_spillover}, we extend our analysis to allow some spillovers across units in the experimental sample.

\subsubsection{Stability of the Remotely Sensed Variable}

Because the outcome is not observed in the experiment, our aim is to learn the relationship between the remotely sensed variable and the outcome in the observational sample and then to ``transport'' it to the experimental sample. 
Our next assumption justifies this intuition.

\begin{assumption}[Stability of the remotely sensed variable]\label{assumption:stability}
Suppose the following:
\begin{enumerate}[label=\roman*.]
\item  Stability: $S \indep R \mid X, D, Y$. 
\item  Common support: for some outcome support $\mathcal{Y}$, $\Pr(Y \in \mathcal{Y} \mid S = e, X) = 1$ almost surely, and $f_Y(y \mid S = o, X)$ is bounded away from zero almost surely for all $y \in \mathcal{Y}$.
\item  Coverage: $f_R( r \mid S, X, D, Y)$ is bounded away from zero almost surely for all $r \in \mathcal{R}$. 
\item  Two samples: $\Pr(S = e \mid X)$ is bounded away from zero and one almost surely. 
\end{enumerate}
\end{assumption}

Assumption~\ref{assumption:stability}(i) is the main assumption of our framework: conditional on $(X, D, Y)$, the distribution of the remotely sensed variable $R$ is \textit{stable} across the experimental and observational samples. 
It formalizes the idea that the sensing mechanism (i.e., how the outcome generates the remotely sensed variable) is invariant across samples: $f_{R}(r \mid S=e,X,D,Y)=f_{R}(r \mid S=o,X,D,Y)$. 
This assumption is most plausible when the sensing mechanism is a shared technology across the samples (e.g., the same satellite) and when factors that affect the measurement process are similar across the samples (e.g., the same season, soil conditions, and land-use patterns). 
In Sections~\ref{sec:simulations}-\ref{sec:empirical_applications}, we provide evidence illustrating that Assumption~\ref{assumption:stability}(i) is plausible across three applications in environmental and development economics. 

Assumption~\ref{assumption:stability}(i) does not require treatment effects to be stable across samples. The treatment mechanism $\Pr(D=1 \mid S,X)$ and the outcome mechanism $f_{Y(d)}(y \mid S,X,D)$ may depend on the sample $S$. 

As mentioned earlier, our framework is agnostic to whether $R$ is (i) unstructured data, (ii) a pretrained embedding of unstructured data, or (iii) the output of a pretrained predictor applied to unstructured data. Stability of (i) implies stability of (ii) and (iii).\footnote{Let $U$ be some unstructured data and $f(\cdot)$ be a fixed measurable function with $R = f(U)$. Stability of the unstructured data $S \indep U | X, D, Y$ implies stability of the pretrained output $S \indep R | X, D, Y$.}

The remaining conditions of Assumption~\ref{assumption:stability} are mild regularity conditions. 
Assumption~\ref{assumption:stability}(ii) requires that the outcome in the observational sample has a common (or larger) support than the outcome in the experiment. 
Assumption~\ref{assumption:stability}(iii) ensures that the distribution of the remotely sensed variable does not degenerate for any stratum.
Assumption~\ref{assumption:stability}(iv) requires that both samples are observed with positive probability across all covariate values.

\vspace{-1em}
\noindent \paragraph{Example~\ref{ex:environmental} (continued):} Returning to Example~\ref{ex:environmental}, satellite images detect environmental outcomes through physical processes: burned cropland alters soil reflectance, and deforestation changes color saturation.
Assumption~\ref{assumption:stability}(i) requires that these processes are stable across samples.
It does not require that PES contracts have the same effects on environmental outcomes across samples. $\blacktriangle$ 

\vspace{-1em}
\noindent \paragraph{Example~\ref{ex:poverty} (continued):}
In Example~\ref{ex:poverty}, satellite images detect poverty through observable markers such as roofing materials or the footprint of the house, while mobile phone data detect poverty through activity patterns.
Assumption~\ref{assumption:stability}(i) requires that these sensing mechanisms are stable across samples.
Again, it does not require that anti-poverty programs have the same effects, outcome mechanisms, or treatment mechanisms across samples. $\blacktriangle$ 

\subsubsection{Complete versus Incomplete Observational Cases}

Assumptions~\ref{assumption:experimental}-\ref{assumption:stability} suffice to identify the causal parameter when there are complete observational cases: when the (possibly confounded) treatment is observed and varies in the observational sample. 
However, when observational cases are incomplete---i.e., when the treatment is missing or deterministic in the observational sample---an additional assumption is needed for point identification.

\begin{assumption}[Observational completeness]\label{assumption:observational}
Suppose that \textit{either} condition holds:
\begin{enumerate}[label=\roman*.]
\item Complete observational cases: $\Pr(D = 1 \mid S = o, X, Y)$ is bounded away from zero and one almost surely; \textit{or}
\item No direct effect: $D \indep R \mid X, Y$. 
\end{enumerate}
\end{assumption}

\noindent Assumption~\ref{assumption:observational} requires only one of these two conditions to hold.

\vspace{-1em}
\paragraph{Complete observational cases:} Assumption~\ref{assumption:observational}(i) requires access to complete cases: joint observations of $(X, D, Y, R)$ in which $D$ varies in the observational sample. 
The treatment need not be randomized; it may suffer from unobserved confounding in the observational sample. 
The treatment $D$ may also have a direct effect on the remotely sensed variable $R$. 
Under Assumption~\ref{assumption:observational}(i), no further causal assumptions are needed. 

\vspace{-1em}
\paragraph{Incomplete observational cases:} If Assumption~\ref{assumption:observational}(i) fails, we have incomplete cases: no observations of $(X,D,Y,R)$ where $D$ varies (for at least some values of $(X,Y)$).
Without joint observations of the outcome and treatment, an additional causal restriction is necessary.
Assumption~\ref{assumption:observational}(ii) fills this gap, requiring that the treatment $D$ affects the remotely sensed variable $R$ only via its effect on the outcome $Y$. 
In Example~\ref{ex:environmental}, it requires that the PES contract affects the satellite image only via its effect on crop burning; the contract has no direct effect on the specific infrared bands used to measure charred soil. 
The assumption becomes more plausible when $Y$ is a vector of outcomes, which may capture all mechanisms through which the treatment affects the remotely sensed variable.\footnote{It is possible to have Assumption~\ref{assumption:observational}(i) hold for some covariate values $X = x$ and Assumption~\ref{assumption:observational}(ii) hold for other covariate values $X = x'$. We leave this extension to future research.}

Assumption~\ref{assumption:stability}(i) and Assumption~\ref{assumption:observational}(ii) imply that $\left( S, D \right) \indep R \mid X, Y$. 
These assumptions are jointly testable, as we discuss in Section \ref{sec:estimation}. Section~\ref{sec:main_partial_id} extends our framework to the setting with both incomplete observational cases and direct effects.

\begin{figure}
\centering
\begin{tikzpicture}[node distance=3cm]
    \tikzstyle{process} = [circle, minimum width=1cm, minimum height=1cm, text centered, draw=black, fill=white]
    \node (Sample) [process] {$S$};
    \node (Treatment) [process, above left of=Sample] {$D$};
    \node (Outcome) [process, above right of=Sample] {$Y$};
    \node (Trace) [process, right of=Outcome] {$R$};
    \draw[->, line width = 1.5pt] (Treatment) -- (Outcome);
    \draw[->, line width = 1.5pt] (Outcome) -- (Trace);
    \draw[->, line width = 1.5pt] (Sample) -- (Treatment);
    \draw[->, line width = 1.5pt] (Sample) -- (Outcome);
    \draw[->, line width = 1.5pt, dotted]  (Treatment) edge[bend right=-30] node [left] {} (Trace);
\end{tikzpicture}
\caption{Causal graph for remotely sensed variables under Assumptions~\ref{assumption:observational}(i) versus~\ref{assumption:observational}(ii). Complete cases allow the dotted line. Assumption~\ref{assumption:stability} rules out the line from $S$ to $R$.}
\label{fig:dag_rsv}
\end{figure}

Figure~\ref{fig:dag_rsv} illustrates our identifying assumptions as a causal graph. 
The treatment affects the outcome, which in turn affects the remotely sensed variable. 
Under our identifying assumptions, the remotely sensed variable $R$ is \textit{post-outcome}: $R$ is caused by $Y$. 
By Assumption~\ref{assumption:stability}, the sample indicator does not change the conditional distribution of the remotely sensed variable given the treatment and outcome.
Depending on which version of Assumption~\ref{assumption:observational} is imposed, the treatment may have a direct effect on the remotely sensed variable, as illustrated by the dotted line. 
See also Appendix Table~\ref{tab:implications}.

\section{Identification via Stability of Remotely Sensed Variables}\label{sec:identification}

Our primary contribution is to nonparametrically identify the treatment effect by combining the experimental and observational samples, appealing to stability of the sensing mechanism.  
Under our assumptions, the treatment effect is identified by a conditional moment restriction, in which the remotely sensed variable relates experimental treatment variation to observational outcome variation. 
We compare our identification strategy to alternative approaches, clarifying how our assumptions and data requirements differ. In our setting, several commonly used methods are infeasible, biased, or inefficient.

\subsection{Main Result}

We derive a formula for data combination, without restricting either the distribution or the dimension of the remotely sensed variable.

In this section, we impose that the outcome is discrete with finite support $\mathcal{Y} = \{y_1, \ldots, y_K\}$ for $K \geq 2$. 
A discrete outcome simplifies our discussion of identification, estimation, and inference.
It is also realistic in some empirical applications, where the outcome is often binary (e.g., crop burning versus no crop burning).
Section~\ref{sec:main_cts_outcomes} extends our results to continuous outcomes. 

Throughout, we assume that the support of $R$ is at least as large as the support of $Y$. Specifically, $R$ may be discrete or continuous, and we are agnostic about its complexity.
We separately consider incomplete and complete observational cases, under Assumption~\ref{assumption:observational}(ii) and~\ref{assumption:observational}(i), respectively. 

\subsubsection{Incomplete Observational Cases: No Direct Effects}

We express the distribution of the remotely sensed variable in the experimental sample as a mixture (i.e., linear combination) of distributions in the observational sample. 
The mixture weights are the potential outcome probabilities we wish to identify.

\begin{lemma}[Identification as conditional densities]\label{lemma:discrete_mixture_incomplete}
Suppose Assumptions~\ref{assumption:experimental},~\ref{assumption:stability}, and~\ref{assumption:observational}(ii) hold.
Then, for any $d \in \{0, 1\}$,
$$
f_R( R \mid S = e, X, D = d) = \sum_{k=1}^{K} f_R( R \mid S = o, X, Y = y_k) \mu_k(d, X),\quad \mu_k(d, X) := \Pr\{Y(d) = y_k \mid S = e, X\}.
$$
\end{lemma}

\begin{proof} 
See Appendix~\ref{proof:lemma:discrete_mixture_incomplete}. 
\end{proof} 

By Lemma~\ref{lemma:discrete_mixture_incomplete}, we recover the treatment effect by combining (i) how the remotely sensed variable reflects the treatment in the experimental sample, and (ii) how the remotely sensed variable reflects the outcome in the observational sample.

While Lemma~\ref{lemma:discrete_mixture_incomplete} identifies the treatment effect, it involves conditional densities of the remotely sensed variable. 
Estimating these densities could be challenging, especially when the remotely sensed variable is high-dimensional, as in contexts with unstructured data or pretrained embeddings.
Therefore, we use Bayes' rule to rewrite Lemma~\ref{lemma:discrete_mixture_incomplete} as a conditional moment equation. 
This transformation avoids estimation of the conditional densities.

Let $\theta(X) := \left( \mu_1(1, X) - \mu_1(0, X), \ldots, \mu_{K-1}(1, X) - \mu_{K-1}(0, X) \right)^\top$ denote the vector of conditional treatment effects on outcome probabilities. 
Identification of $\theta(X)$ implies identification of the conditional average effect $\tau(X) := \E\{ Y(1) - Y(0) \mid S = e, X \}$ since $\tau(X) = \sum_{k=1}^{K-1} (y_k - y_K) \theta_k(X)$.

\begin{theorem}[Identification as conditional moment]\label{theorem:disc_outcomes_incomplete}
Under the conditions of Lemma~\ref{lemma:discrete_mixture_incomplete}, 
$$
    \E\{\Delta^e(X) - \Delta^o(X)^\top \theta(X) \mid X, R \} = 0 \text{ almost surely, where}
$$
$$  
\small 
\begin{aligned}
    \Delta^e(X)&:= \frac{1\{ D = 1, S = e \}}{\Pr(D = 1, S = e \mid X)} - \frac{1\{ D = 0, S = e \}}{\Pr(D = 0, S = e \mid X)}, \\ 
    \Delta_k^o(X) &:=\frac{1\{Y = y_k, S = o \}}{\Pr(Y = y_k, S = o \mid X)} - \frac{1\{Y = y_K, S = o \}}{\Pr(Y = y_K, S = o \mid X)} \text{ with } \Delta^o(X) = \left( \Delta_1^o(X), \ldots, \Delta_{K-1}^o(X) \right)^\top.
\end{aligned}
$$ 
The scalar $\Delta^e(X)$ summarizes the treatment variation in the experimental sample, while the vector $\Delta^o(X)$ summarizes the outcome variation in the observational sample.
\end{theorem}

\begin{proof} 
See Appendix~\ref{proof:theorem:disc_outcomes_incomplete}. 
\end{proof}

\begin{example}[Binary outcome, no covariates]\label{ex:binary_y}
For intuition, consider the setting where the outcome is binary and there are no covariates: $K = 2$, $y_1 = 1$, $y_2 = 0$, and $\mathcal{X} = \varnothing$. 
Treatment and outcome variation simplify to scalars $\Delta^e$ and $\Delta^o$. Theorem~\ref{theorem:disc_outcomes_incomplete} reduces to a ratio:
$$
\tau=\frac{\E[\Delta^e\mid R]}{\E[\Delta^o \mid R]},\quad \Delta^e= \frac{1\{D = 1, S = e\}}{\Pr(D = 1, S = e)} - \frac{1\{ D = 0, S = e \} }{\Pr(D = 0, S = e)}, \quad  \Delta^o = \frac{1\{ Y = 1, S = o \}}{ \Pr(Y = 1, S = o)} - \frac{1\{Y = 0, S = o\}}{ \Pr(Y = 0, S = o) }.
$$
This ratio appears to be a new formula for data combination, where the numerator and denominator are from different samples. 
The numerator captures the effect of both $D$ and $Y$ on $R$, so we must divide it by the effect of $Y$ on $R$ in order to isolate the effect of $D$ on $Y$. 
Crucially, there is no need to specify the conditional distribution of $R$ given $Y$. $\blacktriangle$
\end{example}

More generally, Theorem~\ref{theorem:disc_outcomes_incomplete} identifies the conditional average treatment effect, for each covariate stratum, through a system of conditional moments. Each conditional moment contrasts one possible outcome value against a baseline outcome value. These conditional moments imply unconditional moments.

\begin{corollary}[Identification as representation]\label{corr:disc_outcomes_incomplete_representation}
Suppose the conditions of Lemma~\ref{lemma:discrete_mixture_incomplete} hold, and consider any measurable function $H \colon \mathcal{X} \times \mathcal{R} \rightarrow \mathbb{R}^J$ with $J \geq K - 1$. Then, 
$$
    \E[ H(X, R) \Delta^e(X) \mid X ] = \E[H(X, R) \Delta^o(X)^\top \mid X ] \theta(X).
$$
If $\E[ H(X, R) \Delta^o(X)^\top \mid X ]$ has full column rank, then $\theta(X)$ and hence $\tau$ are identified.
\end{corollary}

\vspace{-1em}
\noindent \paragraph{Example~\ref{ex:binary_y} (continued):} Return to the setting of binary $Y$ and no $X$ for intuition. Consider any measurable function $H:\mathcal{R}\rightarrow\mathbb{R}$. Corollary~\ref{corr:disc_outcomes_incomplete_representation} simplifies to the ratio
$$
\tau=\frac{\E[H(R)\Delta^e]}{\E[H(R)\Delta^o]} \text{ when } \E[H(R)\Delta^o ] \neq 0.
$$
The ratio resembles the Wald estimand from instrumental variable analysis, with many parallel insights. 
Any representation $H(R)$ of the remotely sensed variable $R$ is valid, as long as the representation is relevant to outcome variation: $\E[ H(R)\Delta^o]\neq 0$. A weak remotely sensed variable is one where $\E[H(R) \Delta^o ] \approx 0$. A testable implication of our framework is that two representations $H(R)$ and $H'(R)$ should give similar causal estimates.  
$\blacktriangle$ 

Corollary~\ref{corr:disc_outcomes_incomplete_representation} shows that we may introduce a representation $H(X,R)$ of the remotely sensed variable $R$. The representation may be arbitrary, as long as it satisfies a rank condition: $H(X, R)$ must generate sufficient variation to distinguish each outcome category. 
The representation must predict outcome variation in the observational sample well, which may be viewed as an analogue of an instrument relevance condition.

By Corollary~\ref{corr:disc_outcomes_incomplete_representation}, many representations suffice for identification. This has two important consequences. 
First, naive choices of the representation may be inefficient. 
To find an efficient representation, we therefore connect the literature on ``representation learning'' \cite[e.g.,][]{johannemann2019sufficient, vafa2024estimatingwagedisparitiesusing} to classical results for conditional moment equalities \citep{chamberlain1987asymptotic, newey1993efficient} in Section~\ref{sec:estimation}.
Second, there are testable implications of our assumptions through overidentifying restrictions.

\begin{remark}[Overidentifying restrictions and a specification test]\label{rem:overid_and_spec_test}
Our framework generates testable overidentifying restrictions. 
With incomplete observational cases, Corollary~\ref{corr:disc_outcomes_incomplete_representation} shows that any representation $H(X, R)$ identifies the treatment effect as long as $\E[H(X, R) \Delta^o(X)^\top \mid X ]$ has full column rank. 
If Assumptions~\ref{assumption:stability} and~\ref{assumption:observational}(ii) hold, then different representations should yield the same treatment effect. Disagreement across representations constitutes evidence against our identifying assumptions. 
This can be formalized as a test of overidentifying restrictions using a $J$-statistic \citep[e.g.,][]{Sargan(58), Hansen(82)}.
An analogous test can be constructed with complete observational cases, using Corollary~\ref{corr:disc_outcomes_complete_representation} below.
\end{remark}

\begin{remark}[Weak remotely sensed variables]\label{rem:weak_rsvs}
Just as instrumental variable methods require a sufficiently strong first stage, our method requires that the chosen representation is a sufficiently strong predictor of outcome variation in the observational sample. 
In Example~\ref{ex:binary_y}, when $\E[H(R)\Delta^o]$ is close to zero, the ratio $\E[H(R)\Delta^e] / \E[H(R) \Delta^o]$ is weakly identified and conventional inference can be unreliable. 
Researchers can check for weak remotely sensed variables by assessing whether $\E[H(R)\Delta^o]$ is bounded away from zero, using existing frameworks that test for weak instruments.
\end{remark}

\subsubsection{Complete Observational Cases: Allowing for Direct Effects}\label{sec:main_identification_complete}
Our identification result also holds for the setting of Assumption~\ref{assumption:observational}(i), which requires complete observational cases yet allows for direct effects.
Complete observational cases do not require randomization of treatment $D$ in the observational sample nor transportability of treatment effects between the observational and experimental sample.
Let $\mu(d,X) := \left( \mu_1(d, X), \ldots \mu_{K-1}(d, X) \right)^\top$.

\begin{theorem}[Identification as conditional moment]\label{theorem:disc_outcomes_complete}
Under Assumptions~\ref{assumption:experimental},~ \ref{assumption:stability}, and~\ref{assumption:observational}(i),
$$
\E[\widetilde{\Delta}^e(d,X) - \widetilde{\Delta}^o(d, X)^\top \mu(d, X) \mid X, R ] = 0 \text{ almost surely, for any } d\in\{0,1\}, \text{ where}
$$
\begin{align*}
    \widetilde{\Delta}^e(d, X)&:= \frac{1\{D = d, S = e\}}{\Pr(D = d, S = e \mid X )} - \frac{1\{ Y = y_K, D = d, S = o \}}{\Pr( Y = y_K, D = d, S = o \mid X )}, \\
    \widetilde{\Delta}_k^o(d, X) &:= \frac{1\{Y = y_k, D = d, S = o \}}{\Pr(Y = y_k, D = d, S = o \mid X)} - \frac{1\{Y = y_K, D = d, S = o \}}{\Pr(Y = y_K, D = d, S = o \mid X )}.
\end{align*}
\end{theorem}

\begin{proof}
See Appendix~\ref{proof:theorem:disc_outcomes_complete}.  
\end{proof} 

\begin{corollary}[Identification as representation]\label{corr:disc_outcomes_complete_representation}
Suppose the conditions of Theorem \ref{theorem:disc_outcomes_complete} hold, and consider any measurable function $H_d \colon \mathcal{X} \times \mathcal{R} \rightarrow \mathbb{R}^{J}$ with $J \geq K - 1$.
Then,
$$
    \E[ H_d(X, R) \widetilde{\Delta}^e(d, X) \mid X] = \E[ H_d(X, R) \widetilde{\Delta}^o(d, X)^\top \mid X ] \mu(d, X) \text{ for any } d\in\{0,1\}.
$$
If $\E[H_d(X, R) \widetilde{\Delta}^o(d, X)^\top \mid X]$ has full column rank, then $\mu(d, X)$ and hence $\tau$ are identified.
\end{corollary}

\subsection{Comparison with Alternative Approaches}\label{section:main_comparisons}

We compare our identification strategy to alternative approaches and clarify how our identifying assumptions and data requirements differ. 
Applying some existing methods to post-outcome variables may be infeasible, biased, or inefficient. 
To simplify the discussion, we will focus on Example~\ref{ex:binary_y}, with a binary outcome $Y\in \{0,1\}$ and no covariates $X=\varnothing$, although our discussion extends to the more general setting. 

\subsubsection{A Common Practice: Surrogacy Methods with Post-Outcome Variables}\label{sec:common_practice_main_text}

For exposition, suppose Assumption~\ref{assumption:observational}(ii) holds (with $D = 0$ almost surely in the observational sample): we have incomplete observational cases and no direct effects.

An intuitive method to combine the experimental and observational samples proceeds in two steps: (i) train a predictor $m(R)$ of the outcome $Y$ from the remotely sensed variable $R$ in the observational sample; (ii) apply the predictor $m(R)$ to the experimental sample and calculate treatment effects on the predicted outcomes. 
This method appears in roughly half the papers we surveyed in general interest economics journals from 2015--2024 that use remotely sensed outcomes.
While intuitive, it is generally biased for the causal parameter $\tau$.

This two-step method implicitly targets the estimand 
$$
\widetilde{\tau} = \widetilde{\mu}(1) - \widetilde{\mu}(0)
=\E[m(R) \mid D=1,S=e]-\E[m(R) \mid D=0,S=e],\quad m(R):=\E[Y \mid R, S = o].
$$
The first step estimates the function $m(R)$. The second step evaluates and averages $m(R)$ over the treated and untreated groups in the experimental sample.

As a starting point, suppose the remotely sensed variable fails to predict the outcome. The two-step method would return a precise estimate of zero, even though the remotely sensed variable provides no information about treatment effects in this case. Formally, if $\E[Y \mid R,S=o]=\E[Y \mid S=o]$, then $m(R)$ is constant and $\widetilde{\tau}=0$ regardless of the true treatment effect.
    
More generally, even if the remotely sensed variable is predictive of the outcome, the implicit target $\widetilde{\tau}$ is biased away from the experimental treatment effect $\tau$.
 
\begin{proposition}[Bias of a common practice]\label{prop:bias_common_practice}
If Assumptions~\ref{assumption:experimental},~\ref{assumption:stability}, and~\ref{assumption:observational}(ii) hold in Example~\ref{ex:binary_y}:
\begin{enumerate}
\item[i.] $\widetilde{\mu}(d)=\mu(d)+ \E[ Y \left\{ w_d(R,Y) - 1 \right\} \mid D = d, S = e]$, where $w_d(r, y) := \frac{\Pr(Y = y | S = o)}{\Pr(Y= y|D=d, S = e)} \frac{f_{R}(r\mid D = d, S = e)}{f_{R}(r\mid S = o)}$;
\item[ii.] there exists $\kappa \in [0, 1)$ such that $\widetilde{\tau} = \kappa \tau$, as long as $R$ is an imperfect predictor of $Y$ (i.e., $\var(Y \mid R, S = o)>0$ almost surely).
\end{enumerate}
\end{proposition}

\begin{proof} 
See Appendix~\ref{proof:bias_practice} for the proof and its generalization to discrete $Y$.
\end{proof} 

Proposition~\ref{prop:bias_common_practice}(i) characterizes the implicit estimand of the common practice.
Under Assumptions~\ref{assumption:stability} and~\ref{assumption:observational}(ii), the conditional distribution of the remotely sensed variable $f_R(r\mid Y,S=o)$ is stable across the experimental and observational samples.
However, the common practice attempts to transport the reverse relationship---the predictions $m(R)=\E[Y\mid R,S=o]$---into the experimental sample. 
By Bayes' rule, this reversal induces bias due to differences in the marginal distributions of outcomes between the two samples.

Proposition~\ref{prop:bias_common_practice}(ii) further shows that the common practice has attenuation bias: it is biased towards zero, whenever the remotely sensed variable is an imperfect predictor of the outcome. 
Empirical papers using this method may systematically under-report the impacts of environmental and anti-poverty interventions. 
The attenuation factor $\kappa$ reflects the strength of the correlation between the remotely sensed variable and the outcome.\footnote{
Two related results appear in the literature. \citet{PELLETIER2026103655} show that one component of the bias from using predicted outcomes in difference-in-differences is attenuation bias due to variance compression, though their remaining bias terms are application-specific and generally unsigned. \citet{crossley2022regression} show that, in a linear model, imputing a dependent variable across two samples induces a Berkson measurement error that attenuates the regression coefficient. Proposition~\ref{prop:bias_common_practice} fully characterizes the estimand and bias of the two-step method when the remotely sensed variable is post-outcome and without assuming a linear model.}  

In the presence of covariates, each conditional average treatment effect is attenuated, implying arbitrary sign flipping for the average treatment effect.\footnote{Denote the conditional average treatment effect by $\tau(x)=\E\{Y(1)-Y(0)|S=e,X=x\}$ and the density by $p(x)=f_{X}(x|S=e)$.  By the proof of Proposition~\ref{prop:bias_common_practice}, $\widetilde{\tau}(x)=\kappa(x)\tau(x)$ for $\kappa(x)=\frac{\var\{m(R)|S=o,X=x\}}{\var(Y|S=o,X=x)}\in[0,1)$. There exist $\tau(x)$ and $\kappa(x)$ such that the signs of $\tau=\int \tau(x) p(x) \mathrm{d}x$ and $\widetilde{\tau} = \int \kappa(x) \tau(x) p(x) \mathrm{d}x$ are reversed. For example, with a binary covariate $X$, if $\tau(1)>0$ and $\tau(2)<0$ with $\tau=\tau(1) p(1)+\tau(2) p(2) > 0$, then there exist small $\kappa(1)$ and large $\kappa(2)$ such that $\widetilde{\tau}=p(1) \kappa(1) \tau(1) + p(2) \kappa(2) \tau(2) < 0$.\label{footnote:flip}} 
The two-step method transports outcome predictions within each covariate cell. However, under our identifying assumptions, the stable object is the sensing mechanism within each covariate cell.

We generalize Proposition~\ref{prop:bias_common_practice} for several empirically relevant settings.
Appendix~\ref{sec:bias_common_practice_complete} provides an analogous result with complete observational cases (allowing for direct effects) under Assumption~\ref{assumption:observational}(i). 
Appendix~\ref{sec:bias_random_samples} shows that the attenuation bias arises even if units are randomly allocated into the experimental and observational samples.

\begin{remark}[Surrogacy and negative control methods] 
Proposition~\ref{prop:bias_common_practice} provides a direct comparison to the surrogacy model. 
The implicit target $\widetilde{\tau}$ of the two-step method is the surrogate formula \citep[][]{prentice1989surrogate, athey2024surrogate}, so the two-step method recovers the causal parameter $\tau$ under the standard surrogacy assumptions $(D, S) \indep Y \mid R$.
These surrogacy assumptions state that $R$ fully mediates the effect of the treatment on the outcome; that is, the surrogate $R$ is \textit{pre-outcome}. 
By contrast, Assumptions~\ref{assumption:stability} and~\ref{assumption:observational} imply the opposite: the outcome mediates the effect of the treatment on the remotely sensed variable, either partially (Assumption~\ref{assumption:observational}(i)) or fully (Assumption~\ref{assumption:observational}(ii)); the remotely sensed variable $R$ is \textit{post-outcome}.
Appendix Figure~\ref{fig:dag_surrogates} illustrates the difference via a causal graph.

The surrogacy method, and related methods that lead to the estimand $\widetilde{\tau}$, have been extended to include negative controls \cite[e.g.,][]{ghassami2022combining,imbens2024long}. The surrogacy method and its negative control generalizations would all incur the bias formalized in Proposition~\ref{prop:bias_common_practice} when applied to post-outcome variables.

Similar to our method, the single negative control method \citep{park2024single} uses one auxiliary variable. Our method has two key differences. 
Unlike the single negative control method, we allow the treatment $D$ to affect both the outcome $Y$ and the auxiliary variable $R$ when Assumption~\ref{assumption:observational}(i) holds. 
When no complete cases are available (i.e., the setting covered by Assumption~\ref{assumption:observational}(ii)), the single negative control method 
becomes infeasible.
\end{remark} 

\begin{remark}[Imputation methods]
A recent approach to inference with remotely sensed variables is multiple imputation \citep[][]{rubin1987multiple, proctor2023parameter}, which fits a model for the missing outcome, draws repeated imputations of the missing outcome, and combines the resulting estimates.
Even with a correctly specified imputation model, multiple imputation targets the same estimand as the two-step method. 
For Example~\ref{ex:binary_y} with incomplete observational cases, the imputation model is $\Pr(Y=1\mid R,S=o)$, so the multiple imputation estimand coincides with the two-step estimand $\widetilde{\tau}$. 
With complete observational cases, the imputation model additionally conditions on the treatment $D$, and the multiple imputation estimand coincides with the two-step estimand discussed in Appendix~\ref{sec:bias_common_practice_complete}. 
In our framework, bias can arise because multiple imputation implicitly transports the outcome model across samples, whereas identification instead relies on transporting the stable sensing mechanism.
\end{remark}

\subsubsection{Prediction-Powered Inference}\label{section:main_text_PPI}

We compare our approach with prediction-powered inference (PPI) \citep[][]{angelopoulos2023prediction, PELLETIER2026103655}. 
The PPI method takes as input a pretrained machine learning predictor of the outcome variable $Y$. 
Let $m_1(R)$ and $m_0(R)$ denote arbitrary predictors for treated and untreated units, respectively.

Like the common practice described above, PPI computes a difference of means using predicted outcomes for the observations of $(D,R)$. 
It also adds bias correction terms, called ``rectifiers,'' using complete observations of $(D,Y,R)$. 
Within Example~\ref{ex:binary_y}, the PPI estimand is 
{\small
\begin{align*}
 \widetilde{\tau}^{\mathrm{PPI}} = & \E[m_1(R) \mid  D = 1,S = e] - \E[m_0(R) \mid  D = 0,S = e]  
  + \E[Y - m_1(R) \mid  D=1,S=o]-\E[Y - m_0(R) \mid D=0,S=o].
\end{align*}
}

First, suppose that we have incomplete observational cases. By definition, we have no joint observations of $(D,Y,R)$ with variation in $D$. Consequently, it is not possible to observe $Y$ for some treated units and also some untreated units. At least one PPI rectifier (i.e., $\E[Y - m_d(R) \mid D=d,S=o]$) cannot be computed, and so PPI cannot be implemented.

Second, suppose that we have complete observational cases, and we use the observational sample to construct the rectifiers. 
In this case, the relevant question is whether PPI rectifiers estimated in \(S=o\) can be transported to \(S=e\).
PPI may be biased in several plausible scenarios: (i) the treatment is confounded in the observational sample; (ii) the experimental and observational samples have different treatment effects; (iii) the experimental and observational samples have different ``baselines'' (i.e., different untreated outcome means).\footnote{PPI is unbiased if and only if $\E\{Y-m_d(R)|D=d,S=e\}=\E\{Y-m_d(R)|D=d,S=o\}$ for $d\in\{0,1\}$. Each scenario we describe is a way to violate this condition. See Section~\ref{sec:limited_experimental_outcomes} for further discussion.}
There may be bias even when complete observational cases are available and $D$ has no direct effect on $R$.

\begin{proposition}[Bias of transporting PPI rectifiers]\label{prop:bias_PPI}
Suppose Assumptions~\ref{assumption:experimental},~\ref{assumption:stability},~\ref{assumption:observational}(i) and~\ref{assumption:observational}(ii) hold in Example~\ref{ex:binary_y}. Let $m_1(R)$ and $m_0(R)$ be any bounded and measurable functions. Let $\widetilde{\tau}^o := \E[Y \mid D = 1,S = o] - \E[Y |D = 0, S = o]$ denote the difference in means in the observational sample. Then 
\begin{enumerate} 
\item $\widetilde{\tau}^{\mathrm{PPI}}=\tau+ (1 - \kappa_1)(\widetilde{\tau}^o-\tau) + (\kappa_1-\kappa_0)\delta$ where $\kappa_d = \E[m_d(R) \mid Y = 1] - \E[m_d(R) \mid Y = 0]$ and $\delta = \E[Y \mid D=0, S = e] - \E[Y \mid D = 0, S = o]$;
\item for any predictors $m_1(R)$, $m_0(R)$ and any values $\widetilde{\tau}^o$ and $\tau$ (with possibly $\widetilde{\tau}^o \neq \tau$), there exists a data-generating process with a relevant $R$ (i.e., $R \not \indep Y \mid D, S$) such that  $\widetilde{\tau}^{\mathrm{PPI}} = \widetilde{\tau}^o$.
\end{enumerate}  
\end{proposition}
\begin{proof} 
See Appendix~\ref{proof:prop:bias_PPI}.  
\end{proof} 

Proposition~\ref{prop:bias_PPI}(i) characterizes the PPI estimand as a mixture of the desired estimand $\tau$, the difference in means in the observational sample $\widetilde{\tau}^o$, and the difference in baselines across samples $\delta$, with weights depending on the predictors $m_1(R)$ and $m_0(R)$. PPI is therefore biased whenever the difference in means in the observational sample differs from the treatment effect in the experimental sample (either due to heterogeneity of treatment effects or because of confounding) or whenever there are differences in baselines across samples.
As a consequence, Proposition~\ref{prop:bias_PPI}(ii) further shows that PPI can have arbitrary bias when it uses the observational sample: it may be targeting $\widetilde{\tau}^o>\tau$, $\widetilde{\tau}^o<\tau$, or something else entirely.

By contrast, our method relies on stability of the sensing mechanism.
This distinction matters in applications like Example~\ref{ex:environmental}, where the observational sample is collected on different units and at different times than the experimental sample.
Concretely, \citet[][]{jayachandran2017cash} use ground-based forest measurements at locations distinct from the experimental units and measured at different dates, while  \citet[][]{jack2022money} use randomized spot checks that cannot be synchronized with the narrow post-harvest/pre-tilling window.
These mismatches make it plausible that baselines differ ($\delta \neq 0$), the treated–untreated contrasts differ across samples ($\widetilde{\tau}^o \neq \tau$), or both---precisely the conditions under which Proposition~\ref{prop:bias_PPI}(i) shows PPI can be biased.

\subsubsection{Measurement Error and Data Fusion}

A large literature studies identification using noisy measurements of a latent variable as well as auxiliary data. Works such as \citet{chen2005measurement, chen2008semiparametric}, \citet{GrahamPintoEgel2016}, \citet{RidderMoffitt07}, \citet{fan2014identifying}, and \citet{d2024linear, d2024partially} establish identification by assuming stability of the distribution of the latent variable given the noisy measurement. In our notation, they require outcome stability: $f_{Y}(y\mid R, D,S=e)=f_{Y}(y\mid R, D, S=o)$. 

We study a different regime, where the noisy measurement is post-outcome: the remotely sensed variable is caused by the outcome of interest. 
We instead assume that the sensing mechanism is stable: $f_{R}(r\mid Y, D, S = e) = f_{R}(r\mid Y,D, S = o)$. Meanwhile, we explicitly allow the treatment effects, treatment mechanism, and outcome mechanism to differ across samples.
This distinction---in what information can be transported across samples---leads to a different identifying formula.

Finally, a closer point of contact is work by  \citet{hu2008instrumental,hu2008identification}.
Those authors also impose invariance of the noisy measurement distribution conditional on the latent variable. However, their data environment is different: the true variable is
never directly observed, so identification requires a normalization that pins down the latent scale, e.g. conditionally mean zero measurement error \cite[Assumption 5]{hu2008instrumental}. 
In our setting, the economic outcome is observed in the observational sample, eliminating the latent-scale normalization problem and yielding a different identification strategy based on conditional moment restrictions.

\section{Estimation and Inference}\label{sec:estimation}

As our secondary contribution, we show how our identification result guides the choice of representation for the remotely sensed variable, enabling precise inference on the treatment effect. 
We also prove that our confidence intervals are robust to mis-specification.
We derive valid $n^{-1/2}$ inference without rate conditions or complexity restrictions on the researcher's remotely sensed variable-based predictions.
In particular, we justify the use of complex deep learning algorithms that may be misspecified.

For exposition, throughout this section we focus on incomplete observational cases (Assumption~\ref{assumption:observational}(ii)) and no covariates ($X=\varnothing$).
Estimation and inference for complete observational cases are similar, using the identification results in Section~\ref{sec:main_identification_complete}; see Appendix~\ref{sec:est_inf_direct_effects} for details.
We generalize our estimation algorithm to include covariates in Section~\ref{sec:main_covariates}.

\subsection{Choice of the Representation}\label{sec:optimal}

Consider the binary outcome setting (Example~\ref{ex:binary_y}), where the treatment effect is identified as $\tau=\frac{\E\{H(R)\Delta^e \}}{\E\{H(R)\Delta^o \}}$ for any representation $H(R)$ satisfying $\E\{ H(R) \Delta^o\}\neq0$.
Different choices of representation have different efficiency properties. We now ask, which representation is a reasonable choice in practice?

To answer this question, we view $\tau$ as the coefficient in the regression model
\begin{equation} \label{eq:regression}
 \Delta^e = \tau \Delta^o + \epsilon, \quad \E(\epsilon \mid R) = 0,
\end{equation} 
where the residual $\epsilon:=\Delta^e- \tau \Delta^o$ is mean zero conditional on $R$ by Theorem~\ref{theorem:disc_outcomes_incomplete}. 
Results in \cite{chamberlain1987asymptotic} and \cite{newey1993efficient} imply that the semiparametrically efficient representation within a class of models satisfying~\eqref{eq:regression} is $H^*(R)=\frac{\E(\Delta^o \mid R)}{\sigma^2(\tau, R)}$, where $\sigma^2(\tau, R) := \E\{(\Delta^e -  \tau \Delta^o )^2 \mid R\}$. 
This ``optimal instrument'' $H^*(R)$ is a conditional expectation divided by a conditional variance.

The discrete case is similar. Applying results in \citet[][]{chamberlain1987asymptotic} and \citet[][]{newey1993efficient}, the semiparametrically efficient representation for models satisfying the conditional moment restriction in Theorem~\ref{theorem:disc_outcomes_incomplete} is
$$
H^*(R)=\frac{\E(\Delta^o \mid R)}{\sigma^2(\theta, R)},\quad \sigma^2(\theta, R) := \E[\{\Delta^e - (\Delta^o)^\top \theta \}^2 \mid R],
$$
where now $H^*(R)\in\R^{K-1}$, $\E(\Delta^o \mid R)\in \R^{K-1}$, and $\sigma^2(\theta, R)\in\R$.\footnote{Although $R$ can be high-dimensional, $H^*(R)$ is low dimensional because $K$ is fixed, so results from \cite{chamberlain1987asymptotic} directly apply.} 

This formula reveals that improving the efficiency of treatment effect estimates requires three types of predictions: (i) predictions of each outcome category from the remotely sensed variable, appearing in the numerator $\E(\Delta^o \mid R)$; (ii) prediction of the treatment from the remotely sensed variable, appearing in the denominator $\sigma^2(\theta,R)$; and (iii) prediction of the sample indicator from the remotely sensed variable, appearing in both the numerator and the denominator. That is, improving efficiency requires predicting not only the outcome $Y$ (using the observational sample) but also the treatment $D$ (using the experimental sample) and the sample indicator $S$ (using all data).

In Section~\ref{sec:inference_learned}, we propose an algorithm based on the choice $H^*(R)$. In Section~\ref{sec:inference_robust}, we prove robust inference guarantees for a family of algorithms, allowing a family of representations, including the optimal choice $H^*(R)$ and the simple choice $\Pr(Y=y_k \mid S=o,R)$.

\subsection{Estimation and Inference with Learned Representations}\label{sec:inference_learned}

Our estimation algorithm follows three steps: divide the sample into $\tr$ and $\te$ folds; learn the representation on $\tr$; and apply the learned representation to estimate the treatment effect on $\te$.
Algorithm~\ref{alg:discrete_nocovar_overview} provides an overview of our estimation algorithm using sample splitting with $\tr$ and $\te$ folds, as well as two-fold cross-fitting, though our proof in Appendix~\ref{sec:estimation_proof} allows for cross-fitting with any fixed number of folds. 
Algorithm \ref{algorithm:discrete_covar_details} provides a detailed implementation of the estimation algorithm with covariates. The algorithm for complete observational cases is similar. 
Sample splitting and cross-fitting may be eliminated under complexity restrictions that tolerate simple machine learning procedures. See, for example, \cite{chernozhukov2020adversarial} for a recent summary. 

\begin{algorithm}[t]
\caption{Estimation with Incomplete Observational Cases (overview)}\label{alg:discrete_nocovar_overview}
\begin{algorithmic}[1]
\Require Data $\{S_i,1\{S_i=e\}D_i,1\{S_i=o\}Y_i,R_i\}_{i=1}^n$ with $Y\in\mathcal Y=\{y_1,\ldots,y_K\}$. 

\State Randomly split indices into a $\tr$ fold  and a $\te$ fold. Define the empirical mean operators $\E_{\tr}(\cdot) = \frac{1}{ \left|\tr\right| }\sum_{i \in \tr} (\cdot)$ and $\E_{\te}(\cdot) = \frac{1}{ \left|  \te \right| } \sum_{i \in \te } (\cdot)$. 

\Statex \textbf{Step 1: Learn a representation on $\tr$.}
\State Using $\tr$, estimate
$\widehat{\Pr}(Y=y_k \mid S=o,R)$,
$\widehat{\Pr}(D=1\mid S=e,R)$, and
$\widehat{\Pr}(S=e\mid R)$.
\State Combine these estimates into a learned representation $\widehat H(R)\in\mathbb R^{K-1}$ that approximates the efficient representation $H^*(R)$ (see Algorithm~\ref{algorithm:discrete_covar_details} for the explicit construction).

\Statex \textbf{Step 2: Estimate $\widehat\theta$ on $\te$.}
\State On $\te$, compute the sample analogues $(\widehat\Delta^e,\widehat\Delta^o)$ of the treatment and outcome variation.
\State Compute
$
\widehat\theta
=
\left[\E_{\te}\left\{\widehat H(R)(\widehat\Delta^o)^\top\right\}\right]^{-1}
\E_{\te}\left\{\widehat H(R) \widehat\Delta^e\right\}.$

\Statex \textbf{Step 3: Compute the ATE and standard error.}
\State Compute $\widehat{\tau} = \sum_{j=1}^{K-1} (y_j - y_K) \widehat{\theta}_j$.
\State Bootstrap the standard error of $\widehat\tau$ on $\te$, holding $\widehat H$ fixed.

\State If \textbf{cross-fitting}: Reverse the role of $\tr$ and $\te$, then re-estimate the ATE and its standard error. Average the ATEs and standard errors across the two iterations.  

\State \Return $\widehat\tau$ and its confidence interval.

\end{algorithmic}
\end{algorithm}

Within Algorithm~\ref{alg:discrete_nocovar_overview}, we propose a learned representation $\widehat{H}(R)$ based on $H^*(R)$ above. However, our guarantees below allow any learned representation $\widehat{H}(R)$ that has some limit $\widetilde{H}(R)$. For example, a researcher may prefer to use a pre-trained representation $\widehat{\Pr}(Y=y_k \mid S=o,R)$, sacrificing some precision to reduce computation.

\subsection{Robustness to Misspecification}\label{sec:inference_robust}

Remotely sensed variables take many forms---from unstructured data to pretrained embeddings to outputs of complex machine learning predictors---and our identification results accommodate all of these cases. We therefore analyze our estimator under weak regularity conditions. 

For example, the prediction may be constructed from a complex machine learning algorithm, such as a deep convolutional neural network applied to satellite images or even a pre-trained model. In such settings, rates of convergence are often unknown. Even carefully crafted architectures positing a generative model would typically be misspecified.

For this reason, we do not require convergence rates for the machine learning algorithms used to construct predictions. Under cross-fitting, we only require that the predictions, and hence the representation estimator, have some probability limit; they may be misspecified. Under sample splitting, we can further relax this weak condition; see Remark~\ref{rem:sample_splitting} below.

\begin{assumption}[Limit]\label{assumption:limit}
The learned representation has some mean-square limit $\widetilde{H}(R)$: $\E_R\left\{ \| \widehat{H}(R) - \widetilde{H}(R) \|^2 \right\} = o_p(1)$, where $\E\{\| \widetilde{H}(R) \|^2\}$ is finite, and possibly $\widetilde{H}(R)\neq H^*(R)$.  
This limit is correlated with outcome variation: $\E\{\widetilde{H}(R) (\Delta^o)^\top\}$ is nonsingular, and its smallest singular value is bounded away from zero.
\end{assumption}

While the overall structure of Algorithm~\ref{alg:discrete_nocovar_overview} is familiar \citep{angrist1999jackknife,chernozhukov2018double}, valid $n^{-1/2}$-inference on $\tau$ requires no rate conditions on the learned representation. 
The moment restriction in~\eqref{eq:regression} is infinite-order Neyman orthogonal \citep{mackey2018orthogonal,chen2020mostly}, and therefore Assumption~\ref{assumption:limit} requires no complexity restriction \citep{chernozhukov2018double,chernozhukov2023simple} and no rate conditions \citep{chamberlain1987asymptotic,newey1993efficient}.

Concretely, within Algorithm~\ref{alg:discrete_nocovar_overview}, all three predictors $\widehat{\Pr}(Y=y_k \mid S=o,R)$,
$\widehat{\Pr}(D=1\mid S=e,R)$, and
$\widehat{\Pr}(S=e\mid R)$ may be misspecified. Any other choice of learned representation may be misspecified. Nonetheless, we have valid $n^{-1/2}$ inference, which is a strong form of robustness.

In the following statements, we refer to $\Pr(D = d, S = e)$ for $d\in\{0,1\}$ and $\Pr(Y = y_k, S = o)$ for $y_k\in\{y_1,\ldots,y_K\}$ as the marginal probabilities.

\begin{proposition}[Inference with known marginal probabilities]\label{prop:known}
Consider the cross-fitted estimator in Algorithm \ref{alg:discrete_nocovar_overview}. Suppose Theorem~\ref{theorem:disc_outcomes_incomplete}'s conditions and Assumption~\ref{assumption:limit} hold. 
Suppose the marginal probabilities are known and bounded away from zero. 
Then, for $\lambda = (y_1 - y_K, \ldots, y_{K-1} - y_K)^\top$, we have
$$\sqrt{n} \left( \widehat{\tau} - \tau \right) \rightsquigarrow \mathcal{N}(0, \lambda^\top A B A^\top \lambda ),\quad A = [\E\{\widetilde{H}(R) (\Delta^o)^\top\}]^{-1},\quad B = \E[\{\Delta^e - (\Delta^o)^\top \theta\}^2 \widetilde{H}(R) \widetilde{H}(R)^\top].$$
Moreover, if $\widetilde{H}(R) = H^*(R)$, then $\widehat{\tau}=\lambda^{\top}\widehat{\theta}$ is semiparametrically efficient for $\tau := \lambda^\top \theta$ satisfying the conditional moment restriction $\E\{\Delta^e - (\Delta^{o})^\top \theta \mid R\} = 0$,  with known marginal probabilities.
\end{proposition}

\begin{proof} 
See Appendix~\ref{proof:prop:known}.  
\end{proof} 

\begin{proposition}[Inference with estimated marginal probabilities]\label{prop:unknown}
Consider the cross-fitted estimator in Algorithm \ref{alg:discrete_nocovar_overview}. 
Suppose Theorem~\ref{theorem:disc_outcomes_incomplete}'s conditions and Assumption~\ref{assumption:limit} hold. 
If the marginal probabilities and their estimators are bounded away from zero, then $\sqrt{n} (\widehat{\tau} - \tau) \rightsquigarrow \mathcal{N}(0, \lambda^\top A V A^\top \lambda)$ for $V$ defined in Appendix Lemma~\ref{lemma:num2_unknown}. 
\end{proposition}

\begin{proof} 
See Appendix~\ref{proof:prop:unknown}.  
\end{proof} 

When marginal probabilities are known, the asymptotic variance is standard.
The asymptotic variance in Proposition~\ref{prop:unknown} differs from the one in Proposition~\ref{prop:known} because estimation of the marginal probabilities introduces an additional estimation error.

Standard errors may be calculated by either a bootstrap estimator or an analytic estimator. The former is described in  Algorithm~\ref{alg:discrete_nocovar_overview}, and may be easier to code. 
The latter is described in Appendix Lemma~\ref{lemma:num2_unknown}, and involves less computation.

Theorem~\ref{theorem:disc_outcomes_complete} allows for direct effects of the treatment on the remotely sensed variable, provided there are complete observational cases. Since we have already derived the conditional moment equations, estimation and inference proceed in a similar fashion.

\begin{remark}[Inference without a probability limit for $\widehat{H}$] \label{rem:sample_splitting}
In the absence of Assumption~\ref{assumption:limit}, valid conditional inference is still possible by using sample splitting, with some caveats. With sample splitting, $\widehat{H}$ is a deterministic function conditional on $\tr$. Therefore, Algorithm~\ref{alg:discrete_nocovar_overview} with sample splitting is asymptotically normal conditional on $\tr$: $|\te|^{1/2}\sigma^{-1}_{\widehat{H}}(\widehat   \tau-\tau) | \tr \rightsquigarrow \mathcal{N}(0,1)$, provided that $\widehat{H}(R)$ is uniformly bounded, $\E\{\widehat{H}(R)(\Delta^o)^{\top}|\tr\}$ has its smallest singular value bounded away from zero with probability approaching one, and the conditional asymptotic variance $\sigma^2_{\widehat{H}}$ is nondegenerate with probability approaching one. There are two caveats: 
(i) the convergence rate is $|\te|^{-1/2}$ rather than $n^{-1/2}$; (ii) the variance $\sigma_{\widehat{H}}^2$ is random rather than nonrandom because it depends on the realization of $\tr$. 
\end{remark} 
\section{Extensions and Generalizations}\label{sec:generalizations}

Our framework for identification and estimation extends in several directions that are important for practice: 
(i) a ``validation'' sample with randomized treatments and observed outcomes; 
(ii) a quasi-experimental sample rather than an experimental sample; 
(iii) some spillovers across units; 
(iv) variables collected at different times; 
(v) sensitivity analyses under relaxations of our identifying assumptions; 
(vi) continuous outcomes; 
(vii) low-dimensional covariates.
We provide a brief overview of each extension in this section, deferring formal details to the appendix.

\subsection{A Validation Sample}\label{sec:limited_experimental_outcomes}

In empirical applications, we may also observe outcomes for a (possibly small) subset of experimental units, which we call a ``validation'' sample.
This third sample fits naturally into our framework.
Define an extended sampling indicator $\widetilde{S} \in \{ e,o,v\}$. As before, $\widetilde{S} = o$ denotes an observational unit and $\widetilde{S} = e$ denotes an experimental unit with missing outcome. Now, $\widetilde{S} = v$ denotes an experimental unit for which the outcome is observed.

With this notation, our identification and estimation results hold by replacing $S = e$ with $\widetilde{S}\in\{e,v\}$, and $S = o$ with $\widetilde{S}\in\{o,v\}$, in the appropriate expressions. 
Intuitively, units with $\widetilde{S} = v$ can be reused: because they contain $(X,D,Y,R)$,  they contribute to both treatment variation and outcome variation, and therefore inform both sides of our moment conditions.

This extension does not require assumptions about which experimental units have observed outcomes: the subset with $\widetilde{S} = v$ need not be random (e.g., it may be selected based on logistics or targeting). Our key requirement is that the remotely sensed variable is stable across all three samples: $ \widetilde{S} \indep R \mid X,D,Y$.\footnote{In fact, by the proof of Lemma~\ref{lemma:discrete_mixture_incomplete}, it suffices that $f_R( r\mid \widetilde{S}\in \{e,v\}, X, D, Y) = f_R( r\mid \widetilde{S}\in\{o,v\}, X, D, Y)$.} Only the sensing mechanism must be invariant.

Remark~\ref{remark:ppi_simulations} below explains how a particular variation of PPI can be unbiased at the expense of (i) placing an extra assumption on the validation sample, and (ii) discarding the observational sample.\footnote{Only the validation sample, and not the observational sample, must be used for the rectifiers. This PPI variation is unbiased if and only if $\E\{Y-m_d(R) \mid D=d,\widetilde{S}=e\}=\E\{Y-m_d(R) \mid D=d,\widetilde{S}=v\}$ for $d\in\{0,1\}$. By contrast, our method is unbiased and also leverages the observational sample, so it is more efficient.} Simulations in Section~\ref{sec:simulations} show that, in the presence of a validation sample, our method is more efficient.
 
\subsection{A Quasi-Experimental Sample}\label{sec:main_quasi_exp} 

Many empirical applications with remotely sensed variables exploit quasi-experimental variation \citep[e.g.,][among many others]{BurgessEtAl(12), ratledge2022using, assunccao2023optimal, PELLETIER2026103655}.
Appendix~\ref{sec:other_id_strategies} extends our framework to these settings by modifying our assumption for the $S=e$ sample: we replace experimental unconfoundedness (Assumption~\ref{assumption:experimental}) with instrumental variable or difference-in-differences assumptions. 
Under stability of the remotely sensed variable (Assumption~\ref{assumption:stability}), we show how researchers can combine a quasi-experimental sample with an observational sample to identify familiar causal parameters: the local average treatment effect (LATE) using instruments, or the average treatment effect on the treated (ATT) using difference-in-differences. 
Overall, our main identification technique seamlessly integrates into identification arguments for quasi-experiments.

\subsection{Spillovers}\label{sec:main_spillover}

When evaluating environmental conservation and anti-poverty interventions, empirical researchers often discuss spillovers: how treatment of one unit may affect the outcome of another unit. In Appendix~\ref{sec:spillover}, we extend our framework to allow for some simple types of spillovers.
After expanding the potential outcome notation to reflect spillovers, we define the average direct treatment effect and the average global treatment effect.
We identify the average direct treatment effect using stability (Assumption~\ref{assumption:stability}). Then, under the assumption that spillovers are either (i) within but not across higher levels of geography, or (ii) within a known geographic radius, we identify the average global treatment effect.

\subsection{Time Variation}\label{sec:main_timing}

In practice, data are indexed by time. The remotely sensed variable and the outcome may be measured at different times. 
Moreover, the treatment variable may be defined as early adoption. 
In Appendix~\ref{sec:timing}, we confirm that our method accommodates such settings, and we clarify the interpretation of our reported estimate.

\subsection{Sensitivity Analysis}\label{sec:main_partial_id} 

Our identification result relies on two conditions beyond experimental unconfoundedness: stability of the remotely sensed variable (Assumption~\ref{assumption:stability}) and, with incomplete observational cases, no direct effects of the treatment on the remotely sensed variable (Assumption~\ref{assumption:observational}(ii)).  
While we provide evidence supporting these assumptions in our empirical applications (Sections~\ref{sec:simulations} and~\ref{sec:empirical_applications}), researchers may wish to assess how sensitive their conclusions are to departures from these assumptions.
Appendix~\ref{sec:partial_id} develops a partial identification framework for this purpose, which we summarize below.

First, we relax Assumption~\ref{assumption:observational}(ii): we maintain stability but allow the treatment to have a small direct effect on the remotely sensed variable. Suppose we are willing to bound the size of the direct effect and can calculate the covariance between the outcome variation and the representation $H(X,R)$. We show that the identified set for the average treatment effect is an interval whose size is increasing in the direct effect bound and decreasing in the calculated correlation.

Next, we relax Assumptions~\ref{assumption:stability} and~\ref{assumption:observational}(ii) simultaneously: the treatment may have a direct effect on the remotely sensed variable, and the remotely sensed variable's conditional distribution may vary across samples. We provide an outer bound on the identified set for the average treatment effect. This outer bound recovers the identified set described above when Assumption~\ref{assumption:stability} is imposed again.

We may report these identified sets for a range of plausible violation magnitudes to assess how robust causal findings are to departures from the identifying assumptions. For example, a researcher may report a ``breakdown'' analysis illustrating the smallest direct effect or stability violation required to overturn the study's conclusions \citep[e.g.,][]{MastenPoirier(20)}.

\subsection{Continuous Outcomes}\label{sec:main_cts_outcomes}

Appendix~\ref{sec:cts_outcomes_appendix} extends our results to settings with continuous outcomes. While the identification remains essentially the same, estimation is more complex because the statistical inverse problem of recovering the treatment effect becomes more difficult.

We propose a practical method: discretize the continuous outcome into finitely many bins and appeal to our results for discrete outcomes, while keeping track of the discretization error. 
Under a mild smoothness condition (that the conditional density of the remotely sensed variable $f_R(r \mid  S = o,X,D,Y)$ is uniformly continuous in $Y$), we control the discretization error in Corollary~\ref{cor:continuous1}. 
This yields an approximate version of Theorem~\ref{theorem:disc_outcomes_incomplete}, with a discretization error that we bound explicitly.
We can therefore apply our estimation algorithm to the discretized outcome and adjust inference for a non-vanishing discretization error, for example, by using bias-aware confidence intervals. 

We also discuss the nonparametric generalization. 
Allowing the bin width to vanish amounts to kernel smoothing, which eliminates the discretization error at the cost of slower convergence rates and more subtle regularity conditions. 
We clarify how our relevance condition $\E\{H(R)\Delta^o\}\neq 0$ generalizes to a completeness condition when outcomes are continuous.
Overall, continuous outcomes require stronger conditions for identification and estimation. 

\subsection{Pre-treatment Covariates}\label{sec:main_covariates}

Section~\ref{sec:identification} derived identification with unrestricted, pre-treatment covariates. For ease of exposition, Section~\ref{sec:estimation} derived inference with no covariates. We now describe how inference extends to low dimensional covariates. Future work may study inference with high dimensional covariates.

When covariates are discrete with finite support, inference is straightforward: apply Algorithm~\ref{alg:discrete_nocovar_overview} within each covariate stratum, then average over covariate strata. Appendix~\ref{sec:covariates_general} presents this algorithm and shows that our asymptotic analysis suitably generalizes.

When covariates are low dimensional and continuous, inference requires standard nonparametric techniques. A researcher must estimate the conditional probabilities $\Pr(D=d, S=e \mid X)$ and $\Pr(Y=y, S=o \mid X)$ as smooth functions of $X$ using kernels or series, then the conditional average treatment effects via analogous methods, before averaging over $X$; see Appendix~\ref{sec:est_inf_cts_x}.
\section{Calibrated Simulations}\label{sec:simulations}

We conduct a semi-synthetic exercise calibrated to two real-world settings. 

First, we study forest cover measurements from \citet{hansen2013high} for Uganda. 
These widely used measurements are derived from Landsat satellite imagery at 30-meter resolution, defining forest cover as the fraction of a grid cell containing vegetation taller than 5 meters. 
We define a binary outcome $Y \in \{0, 1\}$ indicating whether at least 80\% of a grid cell is forested, consistent with conservation policies that mandate forest cover retention.\footnote{Examples include Brazil's Forest Code \citep[][]{AzevedoEtAl(17)} as well as conservation programs that determine eligibility based on forest cover thresholds \citep[e.g.,][]{WunderEtAl(08)}.}

We define the experimental sample as grid cells within the area covered by the payments for ecosystem services experiment conducted by \citet{jayachandran2017cash}. 
The observational sample consists of grid cells in surrounding geographic bands. 
Appendix Figure~\ref{fig:map_uganda_forestcover} provides a map, and Appendix~\ref{section: appendix, forest cover, details} provides additional details.  

To illustrate how one might use the output of a pretrained predictor as the remotely sensed variable, we construct one for this setting: using grid cells from the rest of Uganda, we train a random forest to predict $Y$ from MOSAIKS satellite embeddings \citep{rolf2021generalizable}, which are low-cost pretrained embeddings of publicly available satellite images distinct from the Landsat images underlying the \citet{hansen2013high} data \citep[see also][]{ProctorEtAl(25)}.
The random forest predictor achieves an area under the curve (AUC) of 0.967 on held-out grid cells.
We define the remotely sensed variable $R$ as the scalar output of this predictor.

Second, we use data from \citet{muralidharan2016building, muralidharan2023general},  who conducted a randomized experiment to evaluate the introduction of biometrically authenticated payment infrastructure (Smartcards) in Andhra Pradesh, India. 
We define a binary outcome $Y \in \{0, 1\}$ indicating whether the village has only low- and middle-income households, using administrative data from 2012--2013; see Appendix~\ref{section: appendix, smartcards data, details} for details.

The experimental sample consists of treated, untreated, and buffer villages in the original evaluation. The observational sample consists of non-study villages; see Appendix Figure~\ref{fig:smartcards_sample_map}. 

We use this setting to illustrate our framework when the remotely sensed variable is high-dimensional.
We link each village's geographic coordinates to nighttime luminosity measures (a vector in $\mathbb{R}^{50}$) \citep{asher2021development} and MOSAIKS satellite embeddings (a vector in $\mathbb{R}^{4000}$) \citep{rolf2021generalizable}. 
Both have been validated as predictors of poverty \citep{hsw2009, jean2016combining, huang2021using}. 
We define the remotely sensed variable $R$ as their concatenation.

Several diagnostic tests in Appendix \ref{sec:main_simulation_stability} support the plausibility of our assumptions in both empirical settings.

\subsection{Simulation Design}\label{sec:main_simulation_design}

We first consider the most cost-effective data collection strategy: no experimental outcomes are collected, with incomplete observational cases (Assumption~\ref{assumption:observational}(ii)).

Let $\alpha_e := \Pr\{Y(0) = 1 \mid S = e\}$ and $\alpha_o := \Pr\{Y(0) = 1 \mid S = o\}$ denote the baseline outcome probabilities in the experimental and observational samples, respectively.
The simulation design allows $\alpha_e \neq \alpha_o$, so that baseline outcomes may shift across samples. 
In the forest cover setting, $\alpha_e$ and $\alpha_o$ are calibrated from grid cells in the experimental and observational samples, respectively.\footnote{In the main text, we report results calibrating $\alpha_o$ based on a two-kilometer band around the experimental sample. Appendix~\ref{section:additional_sims_without_outcomes} provides similar results in which $\alpha_o$ is calibrated using five- and ten-kilometer bands.} 
Similarly, in the Smartcards setting, $\alpha_e$ and $\alpha_o$ are calibrated from the experimental and observational samples, respectively. 
We vary the treatment effect across a range of values: $\tau \in \{0, 0.05, 0.10, 0.15, 0.20\}$. 
For each treatment effect value, we conduct $500$ simulations.

Each simulation is generated with $n_e$ experimental units and $n_o$ observational units, according to a data-generating process satisfying both stability and no direct effects.
For each experimental unit, we draw the potential outcomes $Y_i(0)\sim \text{Bernoulli}(\alpha_e)$, $Y_i(1) \sim \text{Bernoulli}(\alpha_e + \tau)$ and randomly assign treatment as $D_i \sim \text{Bernoulli}(0.5)$. The observed outcome is $Y_i = (1 - D_i) Y_i(0) + D_i Y_i(1)$.
We then draw the remotely sensed variable $R_i$ from the empirical distribution of $R_i \mid Y_i$ in the real data, and delete $Y_i$, recording only $(D_i, R_i)$. 
For each observational unit, we draw the outcome $Y_i \sim \text{Bernoulli}(\alpha_o)$. 
Then, we draw the remotely sensed variable $R_i$ from the same empirical distribution, and record $(Y_i, R_i)$. 
In the forest cover setting, we set $n_e = 1000$, roughly corresponding to the sample size in \citet{jayachandran2017cash}, and $n_o = 500$. 
In the Smartcards setting, we set $n_e = 3000$ and $n_o = 1000$, and we also truncate the remotely sensed variable to $\mathbb{R}^{1050}$ for computational tractability.
We report results for alternative sample sizes in Appendix~\ref{section:additional_sims_without_outcomes}. Appendix~\ref{section:additional_sims_with_limitedoutcomes} provides additional details on the simulation design. 

We compare two estimators: the commonly used two-step method (Section~\ref{sec:common_practice_main_text}), which trains a predictor of $Y$ from $R$ in the observational sample and regresses predicted outcomes on $D$ in the experimental sample; and our method (Algorithm~\ref{alg:discrete_nocovar_overview}).
PPI methods (Section~\ref{section:main_text_PPI}) are not feasible in this setting because there are no units with jointly observed $(D,Y,R)$ with variation in $D$.

\subsection{Results}\label{sec:main_simulation_results}

\begin{figure}[ht]
\begin{subfigure}[t]{0.45\textwidth}
  \centering
  {\includegraphics[width=\textwidth]{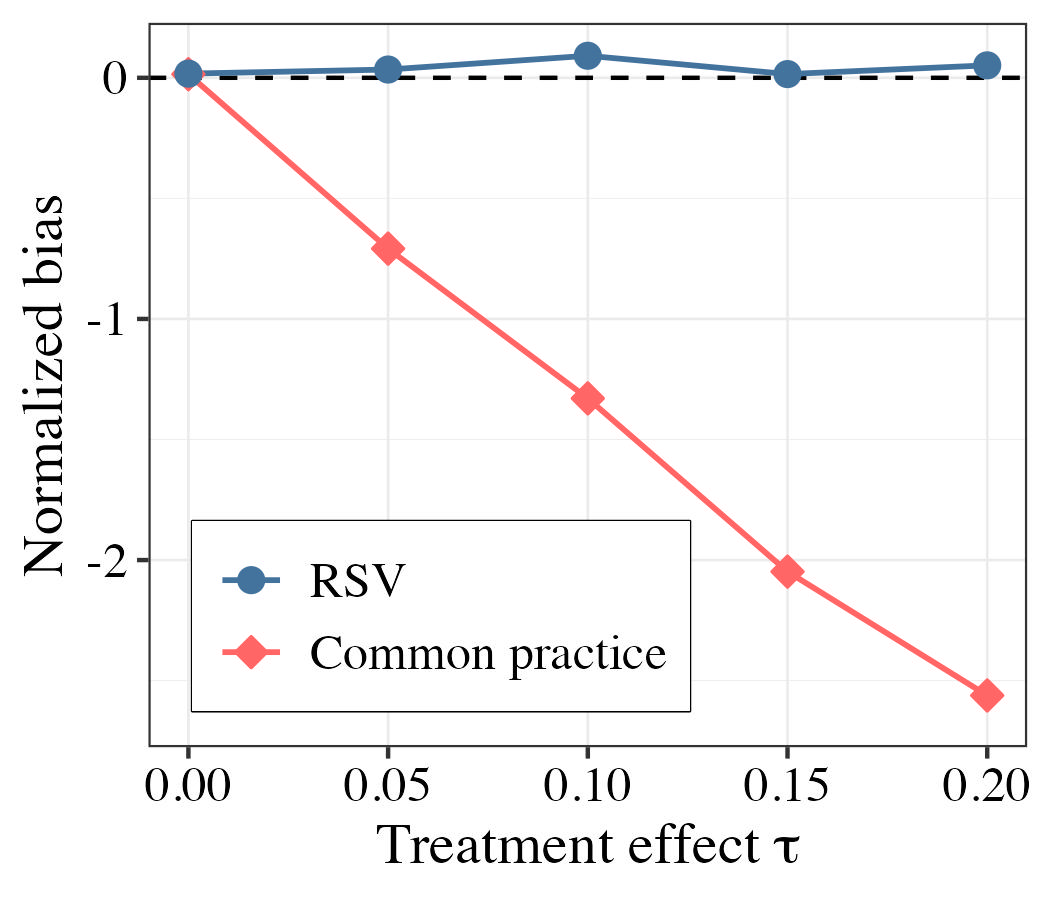}}
  \caption{Forest Cover in Uganda.}
\end{subfigure}\hfill
\begin{subfigure}[t]{0.45\textwidth}
  \centering
  {\includegraphics[width=\textwidth]{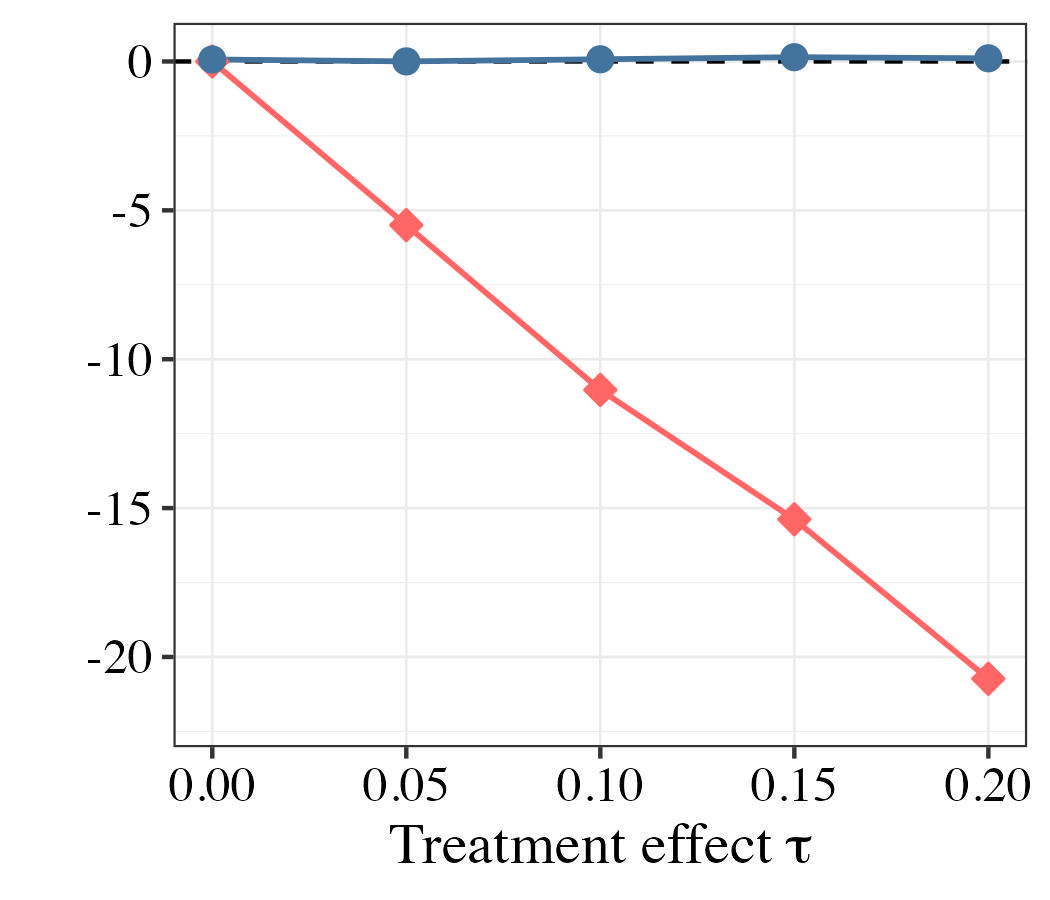}}
  \caption{Smartcards Experiment in India.}
\end{subfigure}
\caption{
Our estimator outperforms the two-step estimator in terms of normalized average bias, defined as the average bias of the estimator divided by its standard deviation across simulations. 
For each value of the synthetic treatment effect $\tau$, we conduct $500$ simulations.}
\label{fig:maintext_simulation_bias}
\end{figure}

First, we compare the quality of the point estimates by comparing their bias, across treatment effect values and empirical settings. 
Figure~\ref{fig:maintext_simulation_bias} visualizes the results. 
The commonly used two-step method exhibits large bias that is proportional to the treatment effect.
This pattern confirms Proposition~\ref{prop:bias_common_practice}: this method targets $\kappa \tau$ for some $\kappa \in [0,1)$. 
Though the remotely sensed variables are good predictors of the outcome in both settings, accurate prediction alone does not ensure valid causal inference when $R$ is post-outcome.
By contrast, our method exhibits negligible bias across all treatment effect values and empirical settings. 

\begin{figure}
\begin{subfigure}[t]{0.45\textwidth}
  \centering
  {\includegraphics[width=\textwidth]{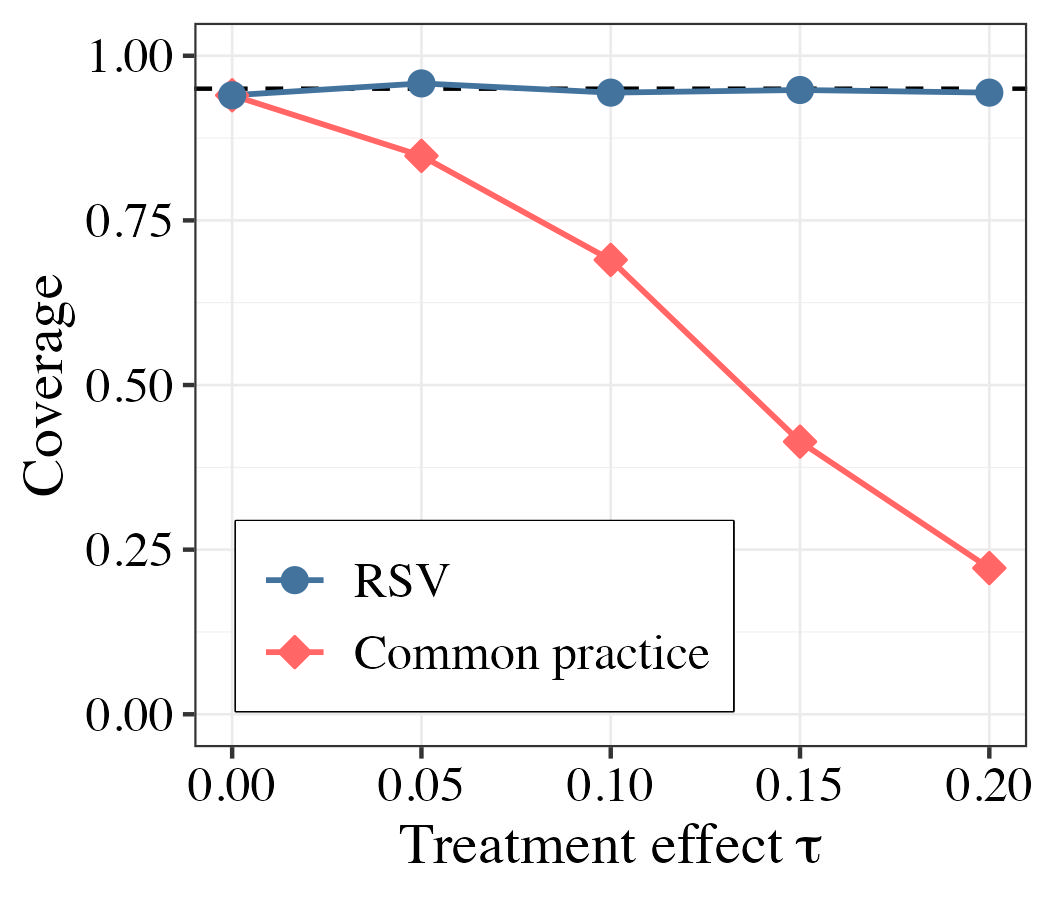}}
  \caption{Forest Cover in Uganda.}
\end{subfigure}\hfill
\begin{subfigure}[t]{0.45\textwidth}
  \centering
  \resizebox{\textwidth}{!}
  {\includegraphics[width=\textwidth]{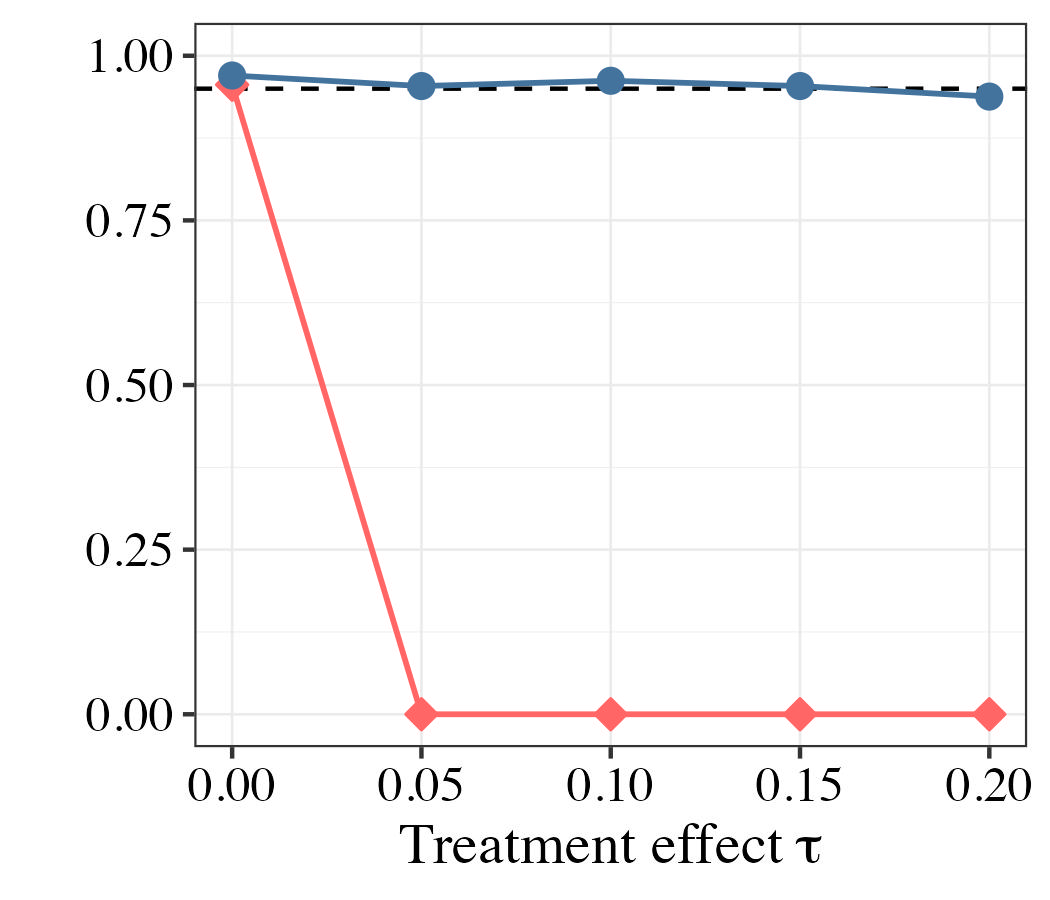}}
  \caption{Smartcards Experiment in India.}
\end{subfigure}
\caption{
Our confidence intervals achieve approximately nominal coverage across simulations.
For each value of the synthetic treatment effect $\tau$, we conduct $500$ simulations.}
\label{fig:maintext_simulation_coverage}
\end{figure}

Next, we compare the quality of the confidence intervals by comparing their coverage, across treatment effect values and empirical settings.
Figure~\ref{fig:maintext_simulation_coverage} visualizes the results. 
The 95\% confidence intervals based on the commonly used method can undercover significantly---as low as 25\% coverage in the forest cover setting and 0\% coverage in the Smartcards setting, when the treatment effects are nonzero.
By contrast, 95\% confidence intervals based on our method achieve close to nominal coverage across all treatment effect values and empirical settings, confirming that the asymptotic theory in Section~\ref{sec:estimation} provides reliable finite-sample guidance.

\begin{remark}[Comparison with prediction-powered inference (PPI)]\label{remark:ppi_simulations}
In the simulations above, PPI methods are infeasible because no experimental outcomes are observed. Appendix~\ref{section:additional_sims_with_limitedoutcomes} extends the simulations to include a small, randomly selected validation sample (Section~\ref{sec:limited_experimental_outcomes}), enabling comparison with PPI. 
Two findings emerge. 
First, a PPI method that uses both the validation sample and the observational sample in the rectifiers is biased whenever $\alpha_e \neq \alpha_o$, confirming Proposition~\ref{prop:bias_PPI} (see Appendix Figure~\ref{fig:normalizedbias_limitedoutcomes}).
Second, a PPI method that only uses the random validation sample in the rectifiers (disregarding the observational sample) is unbiased but inefficient; our estimator yields standard errors up to 40\% smaller than PPI (see Appendix Figure~\ref{fig:main_relse_limitedoutcomes}).\footnote{Only this variation of PPI is unbiased, since with a randomly selected validation sample $\E\{Y-m_d(R) \mid D=d,\widetilde{S}=e\}=\E\{Y-m_d(R) \mid D=d,\widetilde{S}=v\}\neq \E\{Y-m_d(R) \mid D=d,\widetilde{S}=o\}$. It is inefficient since it discards the sample $\widetilde{S}=o$. Even this variation of PPI would be biased if the validation sample were not randomly selected.}
Therefore, our method can improve efficiency, even in the presence of a random validation sample.
By exploiting the post-outcome structure of the remotely sensed variable, our method efficiently incorporates information from the observational sample.
\end{remark}

\section{Applications with Remotely Sensed Outcomes}\label{sec:empirical_applications}

Next, we apply our framework to two real-world settings. 

First, we revisit the Smartcards experiment of \citet[][]{muralidharan2016building, muralidharan2023general}, who study the effect of biometrically authenticated payments on village-level poverty in Andhra Pradesh, India.  
Experimental outcomes were directly measured, so we can benchmark our method (using satellite images in the experimental sample) against an ``oracle'' difference-in-means (using direct outcome measurements in the experimental sample). Our method closely matches the benchmark, despite limited access to outcomes.

Second, we revisit the experiment of \citet[][]{jack2022money}, who study whether payments for ecosystem services (PES) contracts reduce crop burning by farmers in Punjab, India. 
Unlike the Smartcards experiment, no experimental outcomes are available, so we cannot benchmark against an unbiased difference-in-means. 
Instead, we compare our method (which uses the authors' pretrained classifier as a remotely sensed variable) to the commonly used two-step method (which uses the authors' pretrained classifier as a surrogate outcome). 
We find that the commonly used two-step method may underestimate the treatment effect by 47\%.

\subsection{Smartcards and Village-Level Poverty}\label{section:smartcards_empirical_maintext}

\paragraph{Data.} 
We use data from \citet[][]{muralidharan2016building, muralidharan2023general}. The treatment $D \in\{0,1\}$ is the early introduction of Smartcards in 2010, and the outcome $Y\in\{0,1\}$ is a measure of village-level poverty constructed from administrative data in 2012--2013. 
To assess robustness across different poverty definitions, we consider three outcomes: (i) ``consumption'' indicates whether the village is in the bottom quartile of per capita consumption; (ii) ``low income'' indicates whether the village has only low-income households; and (iii) ``middle income'' indicates whether the village has only low- and middle-income households. While (i) captures average consumption, (ii) and (iii) describe the village income distribution using categories from Indian administrative data; see Appendix~\ref{section: appendix, smartcards data, details} for details. 
The remotely sensed variable $R \in\mathbb{R}^{4050}$ concatenates the nighttime luminosity and pretrained satellite embeddings described in Section~\ref{sec:simulations}.\footnote{Here, we use the full embedding rather than the truncated version used in the simulations.}

We analyze this experiment as having incomplete observational cases (Assumption~\ref{assumption:observational}(ii)) and a non-randomly selected validation sample (Section~\ref{sec:limited_experimental_outcomes}). The experimental sample $\widetilde{S}=e$ consists of treated and untreated villages, for which we observe $(D,R)$. The validation sample $\widetilde{S}=v$ consists of buffer villages, for which we observe $(D,Y,R)$. The observational sample $\widetilde{S}=o$ consists of non-study villages, for which we observe $(Y,R)$. Appendix Figure~\ref{fig:smartcards_sample_map2} provides a map. Further details are in Appendix~\ref{section:smartcards_empirical_additionaldetails}.

Two aspects of this exercise are noteworthy. 
First, the experimental sample has 155 mandals while the validation sample has 136 mandals. Whereas the benchmark has ``oracle'' data access, using outcomes for all 291 mandals, our method has ``imperfect'' data access, using outcomes for only the 136 mandals in the validation sample. 
Second, the validation sample is non-randomly selected: rather than containing treated and untreated villages, it only contains untreated villages. As a consequence, there are no joint observations of $(D=1,Y,R)$. The PPI rectifier for the treated group cannot be implemented, rendering PPI infeasible in this setting.\footnote{Formally, the term $\E(Y \mid D=1, \widetilde{S} \in\{v,o\})$ cannot be computed.}

\paragraph{Plausibility of identifying assumptions.}
The treatment was randomized in this experiment (Assumption~\ref{assumption:experimental}). 
Appendix~\ref{section:smartcards_empirical_additionaldetails} provides evidence that stability is plausible (Assumption~\ref{assumption:stability}) for each of the three poverty outcomes.

\begin{figure}[htbp!]
\captionsetup[subfigure]{justification=centering}
\begin{subfigure}[t]{0.45\textwidth}
    \centering
    \includegraphics[width=\textwidth]{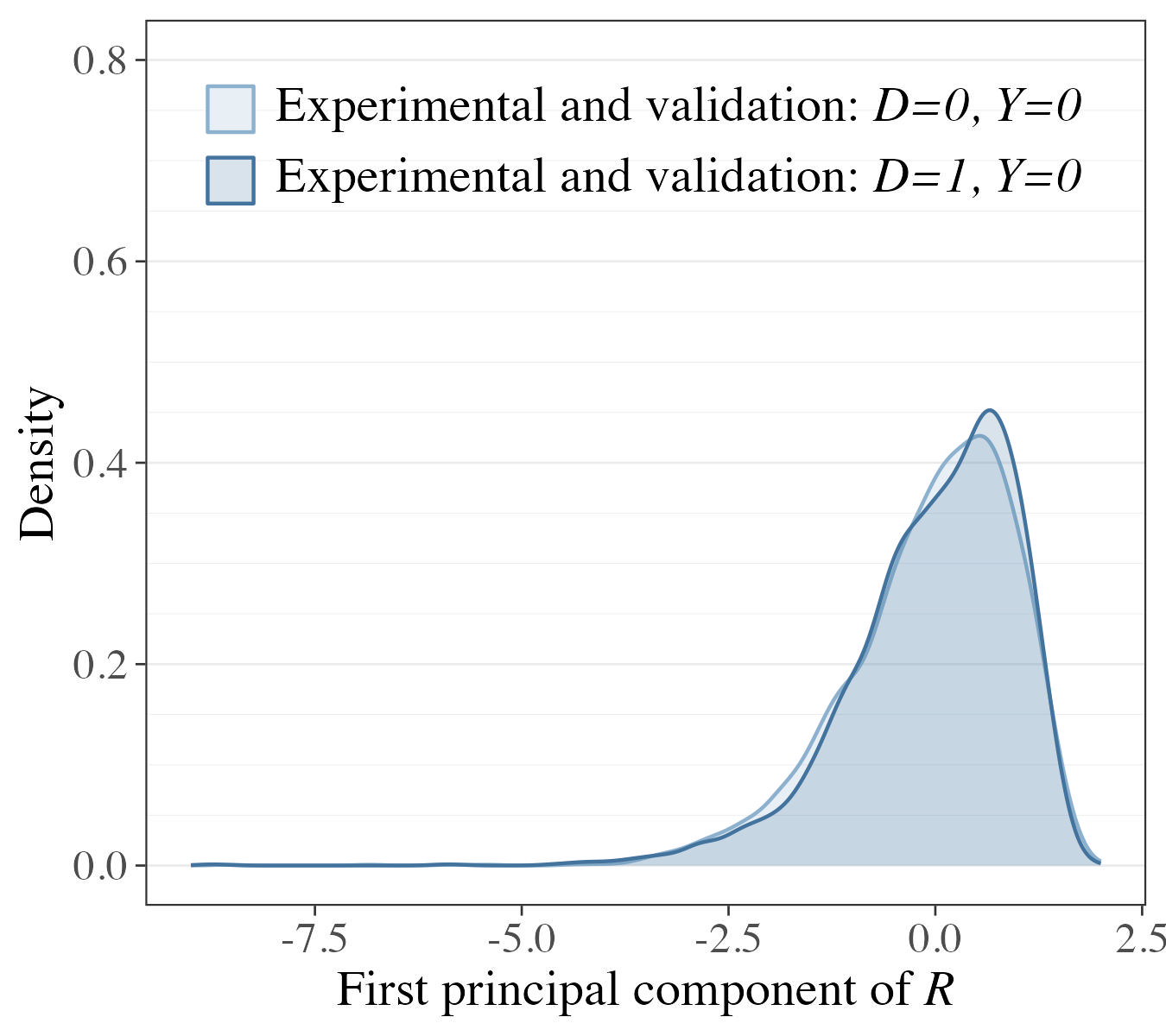}
    \caption{\footnotesize Densities of $R \mid \widetilde{S}\in \{e,v\}$, $D$, $Y=0$.}
\end{subfigure}
\hfill 
\begin{subfigure}[t]{0.45\textwidth}
    \centering
    \includegraphics[width=\textwidth]{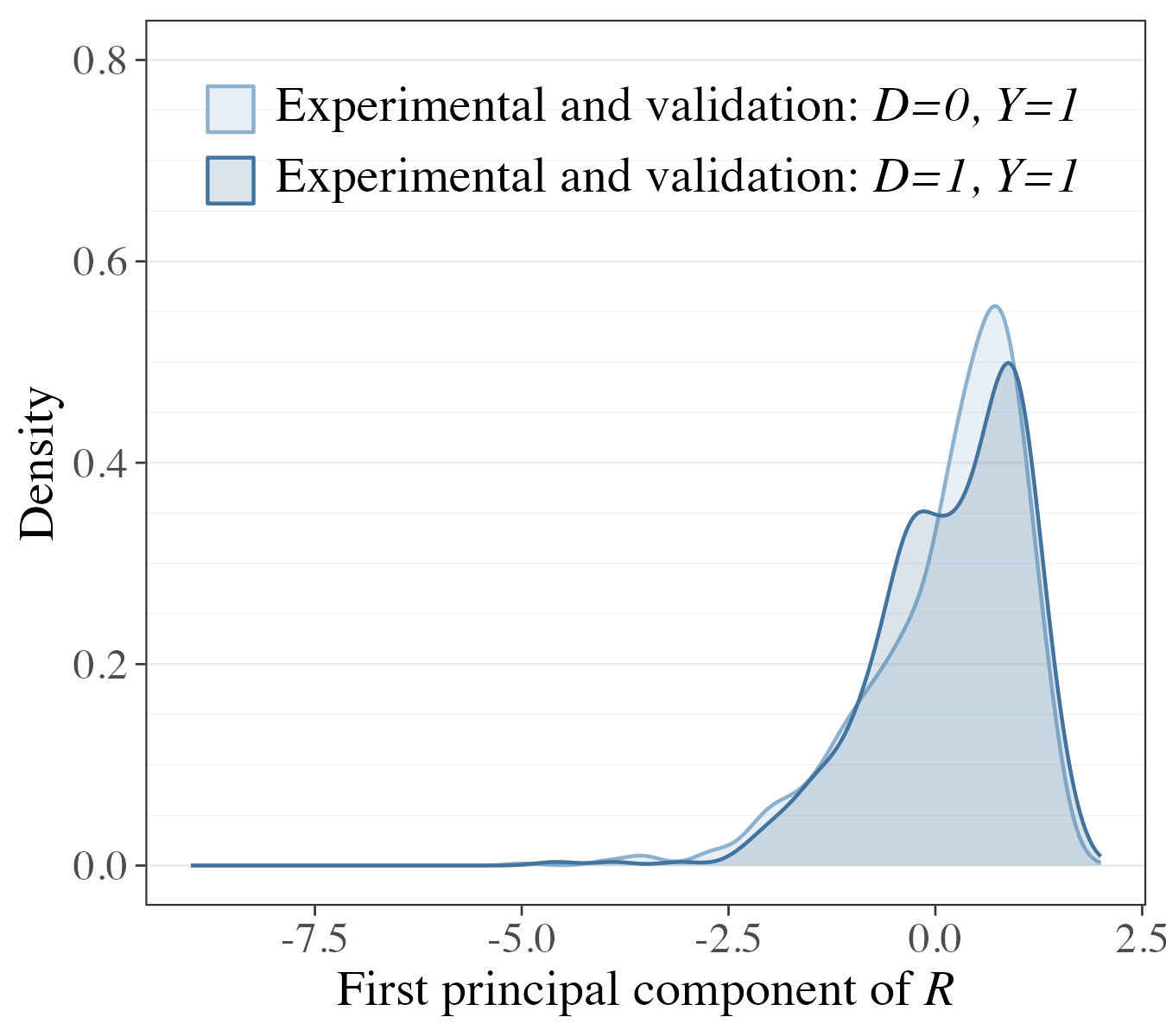}
    \caption{\footnotesize Densities of $R \mid \widetilde{S}\in \{e,v\}$, $D$, $Y=1$.}
\end{subfigure}
\caption{No direct effects (Assumption~\ref{assumption:observational}(ii)) is plausible in the Smartcards experiment.
We compare $f_R(R \mid \widetilde{S}\in \{e,v\}, D = 0, Y = y)$ with $f_{R}(R \mid \widetilde{S}\in \{e,v\}, D = 1, Y = y)$ for $y\in\{0,1\}$. Here, the outcome $Y$ is ``middle income.''
Because $R$ is high-dimensional, we visualize the density of its standardized first principal component. } 
\label{fig:no_direct_effects_maintext}
\end{figure}

With incomplete observational cases, we require no direct effects (Assumption~\ref{assumption:observational}(ii)): the treatment should affect the remotely sensed variable only through the outcome. 
Using ``oracle'' data access, we can visually assess this condition; the densities $f_R(r \mid \widetilde{S}\in \{e,v\}, D = 0, Y = y)$ and $f_R(r \mid \widetilde{S}\in \{e,v\}, D = 1, Y = y)$ should coincide for $y\in\{0,1\}$.
In Figure~\ref{fig:no_direct_effects_maintext}, the conditional densities closely align for one of the poverty outcomes (``middle income''), and Kolmogorov–Smirnov tests cannot reject equality at the 5\% level. 
Appendix \ref{section:smartcards_empirical_additionaldetails} reports the corresponding diagnostics for the other outcomes.\footnote{Tests for ``low income'' also do not reject, while tests for ``consumption'' marginally reject.}

The remotely sensed variable must be relevant: its representation $H(R)$ must be correlated with outcome variation in the observational sample (Corollary~\ref{corr:disc_outcomes_incomplete_representation}). 
For the learned representation in Algorithm~\ref{alg:discrete_nocovar_overview}, we test whether $\E_n\{\widehat{H}(R)\widehat{\Delta}^o\}$ is significantly different from zero. 
Table~\ref{tab:rsv_relevance_smartcards} confirms that the remotely sensed variable is relevant for each of the three poverty outcomes.

\begin{table}
\centering
\begin{threeparttable}
\begin{tabular}{l | ccc}
\toprule
 & \multicolumn{1}{c}{Consumption} & \multicolumn{1}{c}{Low income} & \multicolumn{1}{c}{Middle income} \\
\midrule
  Relevance, $\E_n\{\widehat{H}(R) \widehat{\Delta}^o\}$ & 0.5000 & 1.0102 & 0.4530 \\
           & (0.0601) & (0.1156) & (0.0493) \\
\bottomrule
\end{tabular}
\end{threeparttable}

\caption{Satellite images are relevant to poverty outcomes. 
For each poverty outcome, we report $\E_n\{\widehat{H}(R) \widehat{\Delta}^o\}$ and its standard error using our learned representation. 
Bootstrap standard errors, based on $1000$ replications, are clustered at the sub-district level.}
\label{tab:rsv_relevance_smartcards}
\end{table}
 
\begin{figure}[ht]
\centering
\begin{subfigure}[t]{0.32\textwidth}
    \centering
    \includegraphics[width=\textwidth]{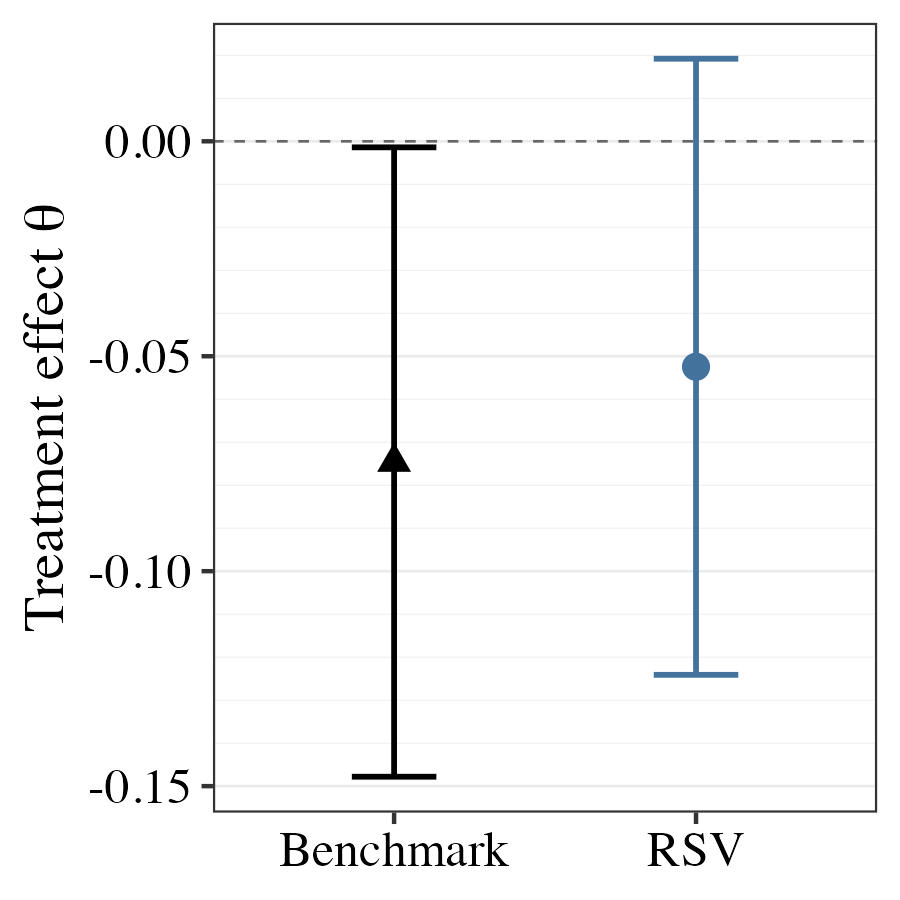}
    \caption{\footnotesize Consumption.}
\end{subfigure}
\hfill 
\begin{subfigure}[t]{0.32\textwidth}
    \centering
    \includegraphics[width=\textwidth]{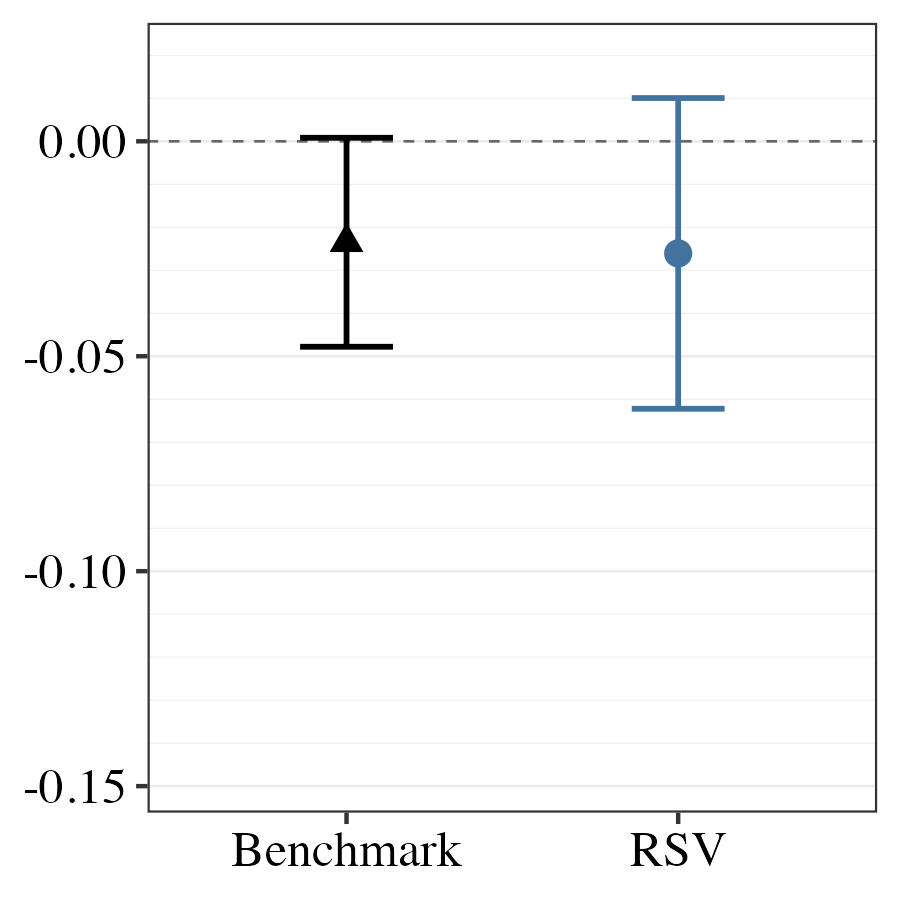}
    \caption{\footnotesize Low income.}
\end{subfigure}
\hfill 
\begin{subfigure}[t]{0.32\textwidth}
    \centering
    \includegraphics[width=\textwidth]{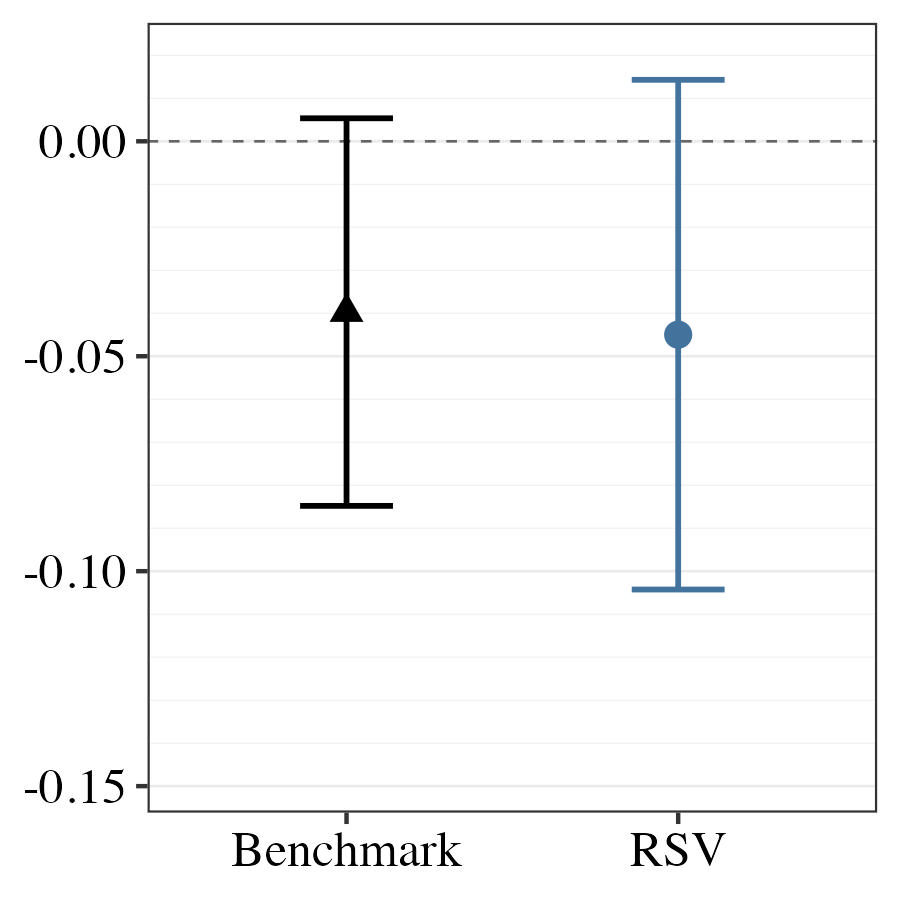}
    \caption{\footnotesize Middle income.}
\end{subfigure}
\caption{Our method's point estimates approximately recover the unbiased benchmark estimates.
For each poverty outcome, we report 90\% confidence intervals for the benchmark and for our method. 
Bootstrap standard errors, based on $1000$ replications, are clustered at the sub-district level.
}\label{fig:smartcards_main_results}
\end{figure}

\paragraph{Main results.} Figure~\ref{fig:smartcards_main_results} summarizes our findings. 
For each poverty outcome, we compare confidence intervals of (i) the benchmark, a difference-in-means using all 291 mandal outcomes; and (ii) our method, using only 136 mandal outcomes but leveraging an observational sample.

Our method yields estimates that are close to those obtained by the benchmark.
For each poverty outcome, our point estimates have the same sign and similar magnitude as the benchmark, i.e. as if the economist could observe all treatments and outcomes in the Smartcards experiment. 
Consistent with \citet[][]{muralidharan2023general}, we find that early adoption of Smartcards reduces poverty. 
Appendix Table~\ref{table:smartcards_main_results} gives details. 
For each outcome, we cannot reject the null hypothesis that our method and the benchmark yield the same point estimate.

Furthermore, our method's confidence intervals are nearly the same length as the benchmark's. 
For the consumption outcome, our interval is essentially of the same length.
For the middle-income outcome, our interval is 31\% longer.

This exercise illustrates the potential for survey-cost reductions when an appropriate observational sample is available.
An economist could forgo poverty surveys for a large fraction of villages in the Smartcards experiment, and instead rely on publicly available satellite images and an observational sample linking satellite images to existing poverty measurements---for example, from prior surveys or censuses.
In this illustration, if surveying each individual in a village costs \$0.50, this choice could save \$3 million.\footnote{Appendix Table~\ref{tab:summary_stats_smartcards} summarizes the number of villages and their populations, which we use to calculate savings. This calculation is conservative; \cite{viviano2022policy} find that phone surveys in Pakistan cost \$7 per individual, rather than \$0.50.}

\vspace{-1em}
\paragraph{Specification test and efficiency gains.}
Our model generates testable overidentifying restrictions (Remark~\ref{rem:overid_and_spec_test}): if stability and no direct effects hold, then different representations $H(R)$ and $H'(R)$ should yield the same treatment effect. 
We implement this test by comparing estimates from two representations: (i) the optimal representation $H(R)=H^*(R)$, combining three predictions; and (ii) the simple representation $H'(R)=\Pr(Y = 1 \mid R, \widetilde{S} \in \{o,v\})$, using one prediction.
Table~\ref{tab:smartcards_spectest} reports the results. 
For each poverty outcome, we cannot reject the null of equal treatment effects, supporting the plausibility of our model.

\begin{table}[ht]
    \centering
    \begin{threeparttable}
\begin{tabular}{l | ccc}
\toprule
 & \multicolumn{1}{c}{Consumption} & \multicolumn{1}{c}{Low income} & \multicolumn{1}{c}{Middle income} \\
\midrule
  Optimal representation & -0.0525 & -0.0261 & -0.0450 \\
           & (0.0436) & (0.0220) & (0.0360) \\
\addlinespace[0.3em]
  Simple representation & -0.0494 & -0.0242 & -0.0431 \\
           & (0.0733) & (0.0341) & (0.0542) \\
\midrule
  $J$ statistic & 0.0267 & 0.0600 & 0.0095 \\
  $p$ value     & 0.8702 & 0.8065 & 0.9223 \\
\bottomrule
\end{tabular}
\end{threeparttable}

    \caption{
    We cannot reject the overidentifying restrictions implied by our model, and the optimal representation yields more precise inference. 
    The optimal representation is $H^*(R)$ from Section~\ref{sec:optimal}. 
    The simple representation is $\Pr(Y = 1 \mid R, \widetilde{S} \in \{o,v\})$. 
    The $J$ statistic compares treatment effect estimates based on these two representations. 
    Bootstrap standard errors, based on $1000$ replications, are clustered at the sub-district level.
    }
    \label{tab:smartcards_spectest}
\end{table}

Table~\ref{tab:smartcards_spectest} also confirms that the standard error using the optimal representation is meaningfully smaller than the standard error using the simple representation. 
For example, for the consumption outcome, the optimal representation yields a standard error that is 40\% smaller.
Consistent with Proposition~\ref{prop:known}, our method is valid using various representations, but the optimal representation confers efficiency. 
Therefore, if it is computationally feasible, researchers should consider predicting not only the outcome but also the treatment and sample indicator.
The optimal representation allows for extrapolation (i.e., negative weights on villages), whereas the simple representation does not. We further interpret and compare the two representations in Appendix~\ref{section:smartcards_empirical_additionaldetails}.

\begin{remark}[Spillovers]\label{remark:spillovers}
We compare our method to a natural benchmark: the difference-in-means estimate an economist would obtain if they could observe all treatments and outcomes in the experiment. 
If there are no spillovers, then we are comparing our method to an unbiased estimate of the treatment effect in Definition~\ref{defn: causal param}. 
If there are spillovers, then we are comparing to an unbiased estimate of a reasonable estimand: a difference-in-means of expected outcomes, indexing by each unit's treatment status and averaging over all other units. 
As discussed in Section~\ref{sec:main_spillover}, our method can be adapted to estimate the global average treatment effect, which is often a parameter of interest in models with spillovers. 
We provide this extension and apply it to the Smartcards experiment in Appendix~\ref{section:smartcards_empirical_additionaldetails}.
\end{remark}

\begin{remark}[Timing]\label{remark:timing}
Outcomes were measured at the end of the experiment in 2013, so we study the effect on poverty in 2013. 
The remotely sensed variables were measured through 2020. Appendix~\ref{section:smartcards_empirical_additionaldetails} confirms that it is valid to use remotely sensed variables measured after the outcomes. 
The observational sample links outcomes from 2013 with remotely sensed variables through 2020, so the estimand and method remain unchanged despite differences in timing.
\end{remark}

\subsection{Payments for Ecosystem Services and Crop Burning}\label{sec:crop_burning}

\paragraph{Data.} We use data from~\citet{jack2022money}. The treatment $D \in \{0, 1\}$ is the offer of a PES contract, and the outcome $Y\in\{0,1\}$ indicates whether a field was \textit{not} burned, so a positive effect means an environmental benefit.\footnote{We modify the specification of \citet{jack2022money} in two ways. While the authors distinguish between standard and up-front PES contracts, we define the treatment as whether any PES contract was offered. The authors analyze effects at the farmer level, aggregating across fields, while we analyze effects at the field level.}

It is costly to measure whether the crop residue on a particular field has been burned, requiring a surveyor to make frequent visits to rural areas. 
Therefore, it is natural to turn to a remotely sensed variable $R$. 
To construct such a variable, \citet{jack2022money} link surveyor-collected measurements of crop burning to satellite-based spectral indices from PlanetScope and Sentinel-2 imagery, and then train a random forest to predict whether fields have been burned \citep{walker2022detecting}. 
The random forest outputs a continuous score at the pixel level, which is the proportion of decision trees that classify a pixel as burned; these pixel-level scores are aggregated to field-level scores. 
\citet{jack2022money} then construct binary classifiers at the field level by applying threshold rules to field-level scores.\footnote{\citet{jack2022money} construct two binary classifiers for crop burning by applying two different threshold rules to the field-level scores. We use their ``max accuracy'' classifier in the main text, and we report analogous results using their ``balanced accuracy'' classifier in Appendix~\ref{section:cropburning_empirical_additionaldetails}.} 

The authors conducted randomized spot checks, sending a surveyor to inspect certain fields, but the timing was imperfect. Because crop burning is a short-lived event, ``spot checks could not be synchronized with the farmer's residue management timing...[spot checks] indicate no burning in days immediately preceding the spot check visit but provide no information about burning outside that window'' \citep[p.~45]{jack2022money}. For this reason, the authors caution against interpreting the spot check sample as a randomly collected validation sample.

We consider two variations. Throughout, the experimental sample $(\widetilde{S}=e)$ consists of treated and untreated fields, for which we observe $(D,R)$. 
In one variation, we view the spot check sample as a validation sample $(\widetilde{S}=v)$ (Section~\ref{sec:limited_experimental_outcomes}) for which we observe $(D,Y,R)$. 
In another variation, we view the spot check sample as an observational sample $(\widetilde{S}=o)$, for which we observe $(D,Y,R)$. 
In the spot check sample, there are no joint observations of $(D,Y=1,R)$ in which $D$ varies, so Assumption~\ref{assumption:observational}(i) fails and we must appeal to Assumption~\ref{assumption:observational}(ii).

Because of the imperfect timing of the spot checks, the outcomes from the spot check sample do not provide unbiased estimates of potential outcomes, making PPI biased (Proposition~\ref{prop:bias_PPI}).\footnote{For example, $\tau\neq \widetilde{\tau}^{\mathrm{PPI}}$ when the correlation between crop-burning-ever and PES contracts in the experimental sample is different than the correlation between crop-burning-in-a-time-window and PES contracts in the spot check sample.} 
On the other hand, stability of the sensing mechanism is plausible, since it requires that the relationship between satellite images and crop burning is stable. 

\paragraph{Plausibility of identifying assumptions.} 
Since offers of PES contracts were randomized at the village level, experimental unconfoundedness (Assumption~\ref{assumption:experimental}) is satisfied by design.

Unlike the Smartcards application, ``oracle'' data access is unavailable in this application, so we appeal to domain knowledge to justify the remaining assumptions.
To justify stability (Assumption~\ref{assumption:stability}), we appeal to the fact that the same sensing technology was used across samples.  Conditional on burning status, how crop burning manifests in the satellite-based spectral indices is determined by physical properties of the soil surface, which are plausibly the same across samples.
To justify no direct effects (Assumption~\ref{assumption:observational}(ii)), we appeal to the specificity of the sensing mechanism: it detects particular infrared spectral bands. Other changes that PES contracts might induce, such as investments in farm equipment, should not affect the spectral indices used to detect burning. 

As before, the remotely sensed variable must be relevant. We confirm that the learned representation is correlated with outcome variation: $\E_n\{\widehat{H}(R)\widehat{\Delta}^o\}=0.262$, with standard error $0.048$.\footnote{The standard error is based on 5000 bootstrap replications clustered at the village level.}

\paragraph{Main results.} Table~\ref{table:crop} summarizes our findings. For each variation, we compare estimates of (i) the common practice, which regresses the predicted outcome on the treatment; (ii) our method using the optimal representation; and (iii) our method using a simple representation.

Our method yields a substantially larger treatment effect. We find that offering any PES contract reduces burning by 14.1\%. By contrast, the common practice estimates an effect of 7.5\%, which is 47\% smaller in magnitude.\footnote{\citet[Table 2]{jack2022money} report effects of 7.7\% for up-front PES and 2.0\% for standard PES, at the farmer level. Our estimate pools across the two types of PES, at the field level.} 
Interpreted through our framework, the common two-step method appears to attenuate the treatment effect by 47\% (Proposition~\ref{prop:bias_common_practice}).

\paragraph{Specification test and efficiency gains.}
As before, our model generates testable overidentifying restrictions (Remark~\ref{rem:overid_and_spec_test}). 
We implement this test by comparing estimates from two representations. 
Table~\ref{table:crop} reports the results.
For each variation, we cannot reject the null of equal treatment effects, supporting the plausibility of our model.

\begin{table}[ht]
    \centering
    \begin{threeparttable}
\begin{tabular}{l | cc}
\toprule
 & Spot checks as validation & Spot checks as observational \\
\midrule
  Optimal representation &  0.1210  &  0.1417  \\
                         & (0.0684) & (0.0782) \\
  Simple representation  &  0.1565  &  0.1929  \\
                         & (0.0803) & (0.0989) \\
  Common practice        &  0.0588  &  0.0752  \\
                         & (0.0320) & (0.0380) \\
\midrule
  $J$ statistic & 1.3242 & 1.7439 \\
  $p$ value     & 0.2498 & 0.1866 \\
\bottomrule
\end{tabular}
\end{threeparttable}

    \caption{
    Our method yields substantially larger treatment effects than the common practice. The common practice is from Section~\ref{sec:common_practice_main_text}. 
    Within our method, the optimal representation is $H^*(R)$ from Section~\ref{sec:optimal}, and the simple representation is $\Pr(Y = 1 \mid R, \widetilde{S} \in \{o,v\})$. 
    The $J$ statistic compares treatment effect estimates based on these two representations. 
    Bootstrap standard errors, based on $5000$ replications, are clustered at the village level. 
    }
    \label{table:crop}
\end{table}

Table~\ref{table:crop} also confirms that the optimal representation gives a meaningfully smaller standard error than the simple representation. For example, when viewing the spot check sample as an observational sample, the optimal representation yields a standard error that is 21\% smaller.
\section{Recommendations for Practice}\label{sec:discussion}

We study remotely sensed variables as \textit{post-outcome}: variation in the economic outcome causes variation in the remotely sensed variable, not vice versa. 
We formalize this causal structure and derive a new formula to combine an experimental sample (containing randomized treatments and remotely sensed variables) with an observational sample (linking remotely sensed variables to outcomes).

In this setting, some alternative methods are problematic. 
A common practice implicitly uses the remotely sensed variable as a surrogate variable, which mediates the effect of the treatment on the outcome. 
Because the remotely sensed variable is post-outcome, this method suffers from attenuation bias (Proposition~\ref{prop:bias_common_practice}). 
Prediction-powered inference (PPI) methods require an auxiliary sample that contains jointly observed treatments, outcomes, and remotely sensed variables (for both treatment values) to construct bias corrections.
When such an auxiliary sample is unavailable, PPI is infeasible.
Even when such an auxiliary sample is available, PPI can be biased whenever the experimental and auxiliary samples differ in baseline outcomes or treatment effects (Proposition~\ref{prop:bias_PPI}). 
Such a distribution shift is plausible when the two samples come from different populations, time periods, or regions. 

However, the core intuition underlying empirical practice is powerful: the conditional distribution of the remotely sensed variable given the outcome and treatment should be \emph{stable} across samples. 
Our stability assumption formalizes the idea that the sensing mechanism---how satellite images or mobile phone data reflect economic conditions---operates the same way in different settings. 
Under stability, we nonparametrically identify treatment effects and precisely estimate them using remotely sensed variables.

Based on our framework, we summarize concrete recommendations for researchers conducting program evaluation with remotely sensed outcomes.

\textbf{Choosing the remotely sensed variable and the observational sample.} 
The researcher faces two key decisions: which remotely sensed variable to use, and how to construct the observational sample linking it to outcomes.

Our framework is agnostic about what the remotely sensed variable is: it may be the output of a pretrained predictor, a pretrained embedding of unstructured remote sensing inputs (e.g., imagery), or the remote sensing inputs themselves.
Our main recommendation is that it should reflect the outcome through physical or technological channels.
For example, satellite imagery captures crop burning through changes in soil reflectance from charred fields \citep{walker2022detecting, jack2022money}, deforestation through changes in color saturation \citep{hansen2013high, jayachandran2017cash}, and poverty through changes in roofing materials and building footprints \citep{jean2016combining, huang2021using}.

Given this choice, the observational sample should be constructed to make stability (Assumption~\ref{assumption:stability}) credible. 
Researchers should select observational units that have the same sensing mechanism (e.g., the same satellite technology) as well as similar conditions that might affect the measurement process (e.g., similar soil type and urban density).
Differences in the outcome mechanism (perhaps due to timing of data collection) are permitted. The treatment need not be observed, let alone randomized, in the observational sample.

\textbf{Inference with remotely sensed outcomes.} 
Valid inference requires the remotely sensed variable to be predictive of the outcome.
More efficient inference leverages three predictions from the remotely sensed variable: predictions of the outcome $Y$, the treatment $D$, and the sample indicator $S$.
We aggregate these three predictions into a more efficient representation of the remotely sensed variable, reducing its dimension while preserving information relevant for causal inference (Algorithm~\ref{alg:discrete_nocovar_overview}). 
Crucially, our method is robust to misspecification. 
In practice, more precise predictions deliver more efficient inference. 

\textbf{Diagnostics for remotely sensed outcomes.} Our framework yields two diagnostics that should be reported when conducting program evaluation with remotely sensed outcomes. 

First, just as weak instruments threaten inference in instrumental variable models, weak remotely sensed variables threaten inference in our framework.  
Researchers should test for weak remotely sensed variables using existing tests for weak instruments. 

Second, our identifying assumptions generate testable overidentifying restrictions (Remark~\ref{rem:overid_and_spec_test}). 
For example, if stability and no direct effects hold, then different representations of the remotely sensed variable should yield the same treatment effect estimate. Disagreement across representations constitutes evidence against the identifying assumptions, via a $J$-test. 

\textbf{Future work.} While we focus on applications to environmental and development economics---where the sensing mechanism is a physical technology that is plausibly stable---other possible applications may be in applied microeconomics. 
For example, arrests may be post-outcome measurements of crime; test scores may be post-outcome measurements of understanding; and biomarkers may be post-outcome measurements of latent disease. 

Finally, our framework poses new questions. 
Future research may study how to simultaneously design treatment assignment, outcome collection, and sensor deployment to maximize statistical power, subject to budget constraints. 
The solution to this problem would inform not only the use of satellite images, but also the design of cheap, noisy surveys.

\spacingset{1}
\DeclareRobustCommand{\VAN}[2]{#2}  %
\DeclareRobustCommand{\VAN}[2]{#1} 

\bibliographystyle{chicago}
\bibliography{Bibliography}

\begin{thebibliography}{}

\bibitem[\protect\citeauthoryear{Ai and Chen}{Ai and Chen}{2003}]{ai2003efficient}
Ai, C. and X.~Chen (2003).
\newblock Efficient estimation of models with conditional moment restrictions containing unknown functions.
\newblock {\em Econometrica\/}~{\em 71\/}(6), 1795--1843.

\bibitem[\protect\citeauthoryear{Aiken, Bellue, Blumenstock, Karlan, and Udry}{Aiken et~al.}{2025}]{aiken2025estimating}
Aiken, E., S.~Bellue, J.~E. Blumenstock, D.~Karlan, and C.~Udry (2025).
\newblock Estimating impact with surveys versus digital traces: Evidence from randomized cash transfers in {T}ogo.
\newblock {\em Journal of Development Economics\/}~{\em 175}, 103477.

\bibitem[\protect\citeauthoryear{Aiken, Bellue, Karlan, Udry, and Blumenstock}{Aiken et~al.}{2022}]{aiken2022machine}
Aiken, E., S.~Bellue, D.~Karlan, C.~Udry, and J.~E. Blumenstock (2022).
\newblock Machine learning and phone data can improve targeting of humanitarian aid.
\newblock {\em Nature\/}~{\em 603\/}(7903), 864--870.

\bibitem[\protect\citeauthoryear{Alix-Garcia and Millimet}{Alix-Garcia and Millimet}{2023}]{alix2023remotely}
Alix-Garcia, J. and D.~L. Millimet (2023).
\newblock Remotely incorrect? {A}ccounting for nonclassical measurement error in satellite data on deforestation.
\newblock {\em Journal of the Association of Environmental and Resource Economists\/}~{\em 10\/}(5), 1335--1367.

\bibitem[\protect\citeauthoryear{Allon, Chen, Jiang, and Zhang}{Allon et~al.}{2023}]{allon2023machine}
Allon, G., D.~Chen, Z.~Jiang, and D.~Zhang (2023).
\newblock Machine learning and prediction errors in causal inference.
\newblock Technical report, The Wharton School.
\newblock SSRN ID 4480696.

\bibitem[\protect\citeauthoryear{Andrews}{Andrews}{2017}]{andrews2017examples}
Andrews, D.~W. (2017).
\newblock Examples of {L}2-complete and boundedly-complete distributions.
\newblock {\em Journal of Econometrics\/}~{\em 199\/}(2), 213--220.

\bibitem[\protect\citeauthoryear{Angelopoulos, Bates, Fannjiang, Jordan, and Zrnic}{Angelopoulos et~al.}{2023}]{angelopoulos2023prediction}
Angelopoulos, A.~N., S.~Bates, C.~Fannjiang, M.~I. Jordan, and T.~Zrnic (2023).
\newblock Prediction-powered inference.
\newblock {\em Science\/}~{\em 382\/}(6671), 669--674.

\bibitem[\protect\citeauthoryear{Angelopoulos, Duchi, and Zrnic}{Angelopoulos et~al.}{2024}]{angelopoulos2024ppiefficientpredictionpoweredinference}
Angelopoulos, A.~N., J.~C. Duchi, and T.~Zrnic (2024).
\newblock Ppi++: Efficient prediction-powered inference.

\bibitem[\protect\citeauthoryear{Angrist, Imbens, and Krueger}{Angrist et~al.}{1999}]{angrist1999jackknife}
Angrist, J.~D., G.~W. Imbens, and A.~B. Krueger (1999).
\newblock Jackknife instrumental variables estimation.
\newblock {\em Journal of Applied Econometrics\/}~{\em 14\/}(1), 57--67.

\bibitem[\protect\citeauthoryear{Asher, Lunt, Matsuura, and Novosad}{Asher et~al.}{2021}]{asher2021development}
Asher, S., T.~Lunt, R.~Matsuura, and P.~Novosad (2021).
\newblock Development research at high geographic resolution: {A}n analysis of night-lights, firms, and poverty in {I}ndia using the {SHRUG} open data platform.
\newblock {\em The World Bank Economic Review\/}~{\em 35\/}(4), 845--871.

\bibitem[\protect\citeauthoryear{Asher and Novosad}{Asher and Novosad}{2020}]{asher2020rural}
Asher, S. and P.~Novosad (2020).
\newblock Rural roads and local economic development.
\newblock {\em American Economic Review\/}~{\em 110\/}(3), 797--823.

\bibitem[\protect\citeauthoryear{Assuncao, McMillan, Murphy, and Souza-Rodrigues}{Assuncao et~al.}{2023}]{assunccao2023optimal}
Assuncao, J., R.~McMillan, J.~Murphy, and E.~Souza-Rodrigues (2023).
\newblock Optimal environmental targeting in the {A}mazon rainforest.
\newblock {\em The Review of Economic Studies\/}~{\em 90\/}(4), 1608--1641.

\bibitem[\protect\citeauthoryear{Athey, Chetty, Imbens, and Kang}{Athey et~al.}{2025}]{athey2024surrogate}
Athey, S., R.~Chetty, G.~W. Imbens, and H.~Kang (2025, 09).
\newblock The surrogate index: Combining short-term proxies to estimate long-term treatment effects more rapidly and precisely.
\newblock {\em The Review of Economic Studies\/}, rdaf087.

\bibitem[\protect\citeauthoryear{Azevedo, Rajão, Costa, Stabile, Macedo, dos Reis, Alencar, Soares-Filho, and Pacheco}{Azevedo et~al.}{2017}]{AzevedoEtAl(17)}
Azevedo, A.~A., R.~Rajão, M.~A. Costa, M.~C.~C. Stabile, M.~N. Macedo, T.~N.~P. dos Reis, A.~Alencar, B.~S. Soares-Filho, and R.~Pacheco (2017).
\newblock Limits of brazil’s forest code as a means to end illegal deforestation.
\newblock {\em Proceedings of the National Academy of Sciences\/}~{\em 114\/}(29), 7653--7658.

\bibitem[\protect\citeauthoryear{Balboni, Burgess, and Olken}{Balboni et~al.}{2025}]{balboni2024origins}
Balboni, C., R.~Burgess, and B.~A. Olken (2025).
\newblock The origins and control of forest fires in the tropics.
\newblock {\em The Review of Economic Studies\/}, rdaf088.

\bibitem[\protect\citeauthoryear{Bareinboim and Pearl}{Bareinboim and Pearl}{2016}]{bareinboim2016causal}
Bareinboim, E. and J.~Pearl (2016).
\newblock Causal inference and the data-fusion problem.
\newblock {\em Proceedings of the National Academy of Sciences\/}~{\em 113\/}(27), 7345--7352.

\bibitem[\protect\citeauthoryear{Battaglia, Christensen, Hansen, and Sacher}{Battaglia et~al.}{2024}]{battaglia2024inference}
Battaglia, L., T.~Christensen, S.~Hansen, and S.~Sacher (2024, February).
\newblock Inference for regression with variables generated by {AI} or machine learning.

\bibitem[\protect\citeauthoryear{Bluhm and Krause}{Bluhm and Krause}{2022}]{bluhm2022nightlights}
Bluhm, R. and M.~Krause (2022).
\newblock Top lights: Bright cities and their contribution to economic development.
\newblock {\em Journal of Development Economics\/}~{\em 157}, 102880.

\bibitem[\protect\citeauthoryear{Blumenstock, Cadamuro, and On}{Blumenstock et~al.}{2015}]{blumenstock2015predicting}
Blumenstock, J., G.~Cadamuro, and R.~On (2015).
\newblock Predicting poverty and wealth from mobile phone metadata.
\newblock {\em Science\/}~{\em 350\/}(6264), 1073--1076.

\bibitem[\protect\citeauthoryear{Burgess, Hansen, Olken, Potapov, and Sieber}{Burgess et~al.}{2012}]{BurgessEtAl(12)}
Burgess, R., M.~Hansen, B.~A. Olken, P.~Potapov, and S.~Sieber (2012, 11).
\newblock The political economy of deforestation in the tropics*.
\newblock {\em The Quarterly Journal of Economics\/}~{\em 127\/}(4), 1707--1754.

\bibitem[\protect\citeauthoryear{Burke, Driscoll, Lobell, and Ermon}{Burke et~al.}{2021}]{burke2021using}
Burke, M., A.~Driscoll, D.~B. Lobell, and S.~Ermon (2021).
\newblock Using satellite imagery to understand and promote sustainable development.
\newblock {\em Science\/}~{\em 371\/}(6535), eabe8628.

\bibitem[\protect\citeauthoryear{Callaway and Sant'Anna}{Callaway and Sant'Anna}{2021}]{CallawaySantanna}
Callaway, B. and P.~H. Sant'Anna (2021).
\newblock Difference-in-differences with multiple time periods.
\newblock {\em Journal of Econometrics\/}~{\em 225\/}(2), 200--230.

\bibitem[\protect\citeauthoryear{Canay, Santos, and Shaikh}{Canay et~al.}{2013}]{canay2013testability}
Canay, I.~A., A.~Santos, and A.~M. Shaikh (2013).
\newblock On the testability of identification in some nonparametric models with endogeneity.
\newblock {\em Econometrica\/}~{\em 81\/}(6), 2535--2559.

\bibitem[\protect\citeauthoryear{Carlson and Dell}{Carlson and Dell}{2025}]{carlson2025unifying}
Carlson, J. and M.~Dell (2025).
\newblock A unifying framework for robust and efficient inference with unstructured data.
\newblock {\em arXiv:2505.00282\/}.

\bibitem[\protect\citeauthoryear{Carrasco, Florens, and Renault}{Carrasco et~al.}{2007}]{carrasco2007linear}
Carrasco, M., J.-P. Florens, and E.~Renault (2007).
\newblock Linear inverse problems in structural econometrics estimation based on spectral decomposition and regularization.
\newblock {\em Handbook of Econometrics\/}~{\em 6}, 5633--5751.

\bibitem[\protect\citeauthoryear{Chamberlain}{Chamberlain}{1987}]{chamberlain1987asymptotic}
Chamberlain, G. (1987).
\newblock Asymptotic efficiency in estimation with conditional moment restrictions.
\newblock {\em Journal of Econometrics\/}~{\em 34\/}(3), 305--334.

\bibitem[\protect\citeauthoryear{Chen, Chen, and Lewis}{Chen et~al.}{2020}]{chen2020mostly}
Chen, J., D.~L. Chen, and G.~Lewis (2020).
\newblock Mostly harmless machine learning: Learning optimal instruments in linear {IV} models.
\newblock {\em arXiv preprint arXiv:2011.06158\/}.
\newblock Accepted at NeurIPS 2020 Workshop on Machine Learning for Economic Policy.

\bibitem[\protect\citeauthoryear{Chen, Mueller, Jia, and Tseng}{Chen et~al.}{2017}]{chen2017validating}
Chen, J.~J., V.~Mueller, Y.~Jia, and S.~K.-H. Tseng (2017).
\newblock Validating migration responses to flooding using satellite and vital registration data.
\newblock {\em American Economic Review\/}~{\em 107\/}(5), 441--45.

\bibitem[\protect\citeauthoryear{Chen}{Chen}{2007}]{Chen(07)}
Chen, X. (2007).
\newblock Chapter 76 large sample sieve estimation of semi-nonparametric models.
\newblock Volume~6 of {\em Handbook of Econometrics}, pp.\  5549--5632. Elsevier.

\bibitem[\protect\citeauthoryear{Chen, Hong, and Nekipelov}{Chen et~al.}{2011}]{ChenHongNekipelov(11)}
Chen, X., H.~Hong, and D.~Nekipelov (2011).
\newblock Nonlinear models of measurement errors.
\newblock {\em Journal of Economic Literature\/}~{\em 49\/}(4), 901–37.

\bibitem[\protect\citeauthoryear{Chen, Hong, and Tamer}{Chen et~al.}{2005}]{chen2005measurement}
Chen, X., H.~Hong, and E.~Tamer (2005).
\newblock Measurement error models with auxiliary data.
\newblock {\em The Review of Economic Studies\/}~{\em 72\/}(2), 343--366.

\bibitem[\protect\citeauthoryear{Chen, Hong, and Tarozzi}{Chen et~al.}{2008}]{chen2008semiparametric}
Chen, X., H.~Hong, and A.~Tarozzi (2008).
\newblock Semiparametric efficiency in {GMM} models with auxiliary data.
\newblock {\em The Annals of Statistics\/}~{\em 36\/}(2), 808--843.

\bibitem[\protect\citeauthoryear{Chen and Nordhaus}{Chen and Nordhaus}{2011}]{chen2011using}
Chen, X. and W.~D. Nordhaus (2011).
\newblock Using luminosity data as a proxy for economic statistics.
\newblock {\em Proceedings of the National Academy of Sciences\/}~{\em 108\/}(21), 8589--8594.

\bibitem[\protect\citeauthoryear{Chen and Reiss}{Chen and Reiss}{2011}]{chen2011rate}
Chen, X. and M.~Reiss (2011).
\newblock On rate optimality for ill-posed inverse problems in econometrics.
\newblock {\em Econometric Theory\/}~{\em 27\/}(3), 497--521.

\bibitem[\protect\citeauthoryear{Chernozhukov, Chetverikov, Demirer, Duflo, Hansen, Newey, and Robins}{Chernozhukov et~al.}{2018}]{chernozhukov2018double}
Chernozhukov, V., D.~Chetverikov, M.~Demirer, E.~Duflo, C.~Hansen, W.~K. Newey, and J.~Robins (2018).
\newblock Double/debiased machine learning for treatment and structural parameters.
\newblock {\em The Econometrics Journal\/}~{\em 21\/}(1), C1--C68.

\bibitem[\protect\citeauthoryear{Chernozhukov, Newey, Singh, and Syrgkanis}{Chernozhukov et~al.}{2025}]{chernozhukov2020adversarial}
Chernozhukov, V., W.~Newey, R.~Singh, and V.~Syrgkanis (2025).
\newblock Adversarial estimation of {R}iesz representers.
\newblock {\em Journal of the American Statistical Association\/}.

\bibitem[\protect\citeauthoryear{Chernozhukov, Newey, and Singh}{Chernozhukov et~al.}{2023}]{chernozhukov2023simple}
Chernozhukov, V., W.~K. Newey, and R.~Singh (2023).
\newblock A simple and general debiased machine learning theorem with finite-sample guarantees.
\newblock {\em Biometrika\/}~{\em 110\/}(1), 257--264.

\bibitem[\protect\citeauthoryear{Corral, Henderson, and Segovia}{Corral et~al.}{2025}]{corral2025povertymapping}
Corral, P., H.~Henderson, and S.~Segovia (2025).
\newblock Poverty mapping in the age of machine learning.
\newblock {\em Journal of Development Economics\/}~{\em 172}, 103377.

\bibitem[\protect\citeauthoryear{Cross and Manski}{Cross and Manski}{2002}]{cross2002regressions}
Cross, P.~J. and C.~F. Manski (2002).
\newblock Regressions, short and long.
\newblock {\em Econometrica\/}~{\em 70\/}(1), 357--368.

\bibitem[\protect\citeauthoryear{Crossley, Levell, and Poupakis}{Crossley et~al.}{2022}]{crossley2022regression}
Crossley, T.~F., P.~Levell, and S.~Poupakis (2022).
\newblock Regression with an imputed dependent variable.
\newblock {\em Journal of Applied Econometrics\/}~{\em 37\/}(7), 1277--1294.

\bibitem[\protect\citeauthoryear{Currie, Voorheis, and Walker}{Currie et~al.}{2023}]{currie2023caused}
Currie, J., J.~Voorheis, and R.~Walker (2023).
\newblock What caused racial disparities in particulate exposure to fall? new evidence from the clean air act and satellite-based measures of air quality.
\newblock {\em American Economic Review\/}~{\em 113\/}(1), 71--97.

\bibitem[\protect\citeauthoryear{de~Chaisemartin and D'Haultfœuille}{de~Chaisemartin and D'Haultfœuille}{2020}]{dCdH-DiD}
de~Chaisemartin, C. and X.~D'Haultfœuille (2020, September).
\newblock Two-way fixed effects estimators with heterogeneous treatment effects.
\newblock {\em American Economic Review\/}~{\em 110\/}(9), 2964–96.

\bibitem[\protect\citeauthoryear{D'Haultf{\oe}uille, Gaillac, and Maurel}{D'Haultf{\oe}uille et~al.}{2025}]{d2024partially}
D'Haultf{\oe}uille, X., C.~Gaillac, and A.~Maurel (2025).
\newblock Partially linear models under data combination.
\newblock {\em The Review of Economic Studies\/}~{\em 92\/}(1), 238--267.

\bibitem[\protect\citeauthoryear{Donaldson and Storeygard}{Donaldson and Storeygard}{2016}]{donaldson2016view}
Donaldson, D. and A.~Storeygard (2016).
\newblock The view from above: Applications of satellite data in economics.
\newblock {\em Journal of Economic Perspectives\/}~{\em 30\/}(4), 171--98.

\bibitem[\protect\citeauthoryear{D’Haultf{\oe}uille, Gaillac, and Maurel}{D’Haultf{\oe}uille et~al.}{2024}]{d2024linear}
D’Haultf{\oe}uille, X., C.~Gaillac, and A.~Maurel (2024).
\newblock Linear regressions with combined data.
\newblock {\em arXiv:2412.04816\/}.

\bibitem[\protect\citeauthoryear{Egami, Hinck, Stewart, and Wei}{Egami et~al.}{2023}]{egami2023using}
Egami, N., M.~Hinck, B.~Stewart, and H.~Wei (2023).
\newblock Using imperfect surrogates for downstream inference: Design-based supervised learning for social science applications of large language models.
\newblock {\em Advances in Neural Information Processing Systems\/}~{\em 36}, 68589--68601.

\bibitem[\protect\citeauthoryear{Egger, Haushofer, Miguel, Niehaus, and Walker}{Egger et~al.}{2022}]{egger2022general}
Egger, D., J.~Haushofer, E.~Miguel, P.~Niehaus, and M.~Walker (2022).
\newblock General equilibrium effects of cash transfers: Experimental evidence from {K}enya.
\newblock {\em Econometrica\/}~{\em 90\/}(6), 2603--2643.

\bibitem[\protect\citeauthoryear{Fan, Sherman, and Shum}{Fan et~al.}{2014}]{fan2014identifying}
Fan, Y., R.~Sherman, and M.~Shum (2014).
\newblock Identifying treatment effects under data combination.
\newblock {\em Econometrica\/}~{\em 82\/}(2), 811--822.

\bibitem[\protect\citeauthoryear{Fong and Tyler}{Fong and Tyler}{2021}]{fong2021machine}
Fong, C. and M.~Tyler (2021).
\newblock Machine learning predictions as regression covariates.
\newblock {\em Political Analysis\/}~{\em 29\/}(4), 467--484.

\bibitem[\protect\citeauthoryear{Fowlie, Rubin, and Walker}{Fowlie et~al.}{2019}]{fowlie2019satellite}
Fowlie, M., E.~Rubin, and R.~Walker (2019, May).
\newblock Bringing satellite-based air quality estimates down to earth.
\newblock {\em AEA Papers and Proceedings\/}~{\em 109}, 283–88.

\bibitem[\protect\citeauthoryear{Ghassami, Liu, Yang, Richardson, Shpitser, and Tchetgen}{Ghassami et~al.}{2022}]{ghassami2022combining}
Ghassami, A., C.~Liu, A.~Yang, D.~Richardson, I.~Shpitser, and E.~T. Tchetgen (2022).
\newblock Combining experimental and observational data for identification and estimation of long-term causal effects.
\newblock {\em arXiv:2201.10743\/}.

\bibitem[\protect\citeauthoryear{Graham, de~Xavier~Pinto, and Egel}{Graham et~al.}{2016}]{GrahamPintoEgel2016}
Graham, B.~S., C.~C. de~Xavier~Pinto, and D.~Egel (2016).
\newblock Efficient estimation of data combination models by the method of auxiliary-to-study tilting ({AST}).
\newblock {\em Journal of Business \& Economic Statistics\/}~{\em 34\/}(2), 288--301.

\bibitem[\protect\citeauthoryear{Hansen}{Hansen}{1982}]{Hansen(82)}
Hansen, L.~P. (1982).
\newblock Large sample properties of generalized method of moments estimators.
\newblock {\em Econometrica\/}~{\em 50\/}(4), 1029--1054.

\bibitem[\protect\citeauthoryear{Hansen, Potapov, Moore, Hancher, Turubanova, Tyukavina, Thau, Stehman, Goetz, Loveland, et~al.}{Hansen et~al.}{2013}]{hansen2013high}
Hansen, M.~C., P.~V. Potapov, R.~Moore, M.~Hancher, S.~A. Turubanova, A.~Tyukavina, D.~Thau, S.~V. Stehman, S.~J. Goetz, T.~R. Loveland, et~al. (2013).
\newblock High-resolution global maps of 21st-century forest cover change.
\newblock {\em Science\/}~{\em 342\/}(6160), 850--853.

\bibitem[\protect\citeauthoryear{Henderson, Storeygard, and Weil}{Henderson et~al.}{2011}]{hsw2009}
Henderson, V., A.~Storeygard, and D.~N. Weil (2011, May).
\newblock A bright idea for measuring economic growth.
\newblock {\em American Economic Review\/}~{\em 101\/}(3), 194–99.

\bibitem[\protect\citeauthoryear{Henderson, Storeygard, and Weil}{Henderson et~al.}{2012}]{henderson2012measuring}
Henderson, V., A.~Storeygard, and D.~N. Weil (2012).
\newblock Measuring economic growth from outer space.
\newblock {\em American Economic Review\/}~{\em 102\/}(2), 994--1028.

\bibitem[\protect\citeauthoryear{Horowitz and Manski}{Horowitz and Manski}{1995}]{horowitz1995identification}
Horowitz, J.~L. and C.~F. Manski (1995).
\newblock Identification and robustness with contaminated and corrupted data.
\newblock {\em Econometrica\/}~{\em 63\/}(2), 281--302.

\bibitem[\protect\citeauthoryear{Hu}{Hu}{2008}]{hu2008identification}
Hu, Y. (2008).
\newblock Identification and estimation of nonlinear models with misclassification error using instrumental variables: A general solution.
\newblock {\em Journal of Econometrics\/}~{\em 144\/}(1), 27--61.

\bibitem[\protect\citeauthoryear{Hu and Schennach}{Hu and Schennach}{2008}]{hu2008instrumental}
Hu, Y. and S.~M. Schennach (2008).
\newblock Instrumental variable treatment of nonclassical measurement error models.
\newblock {\em Econometrica\/}~{\em 76\/}(1), 195--216.

\bibitem[\protect\citeauthoryear{Huang, Hsiang, and Gonzalez-Navarro}{Huang et~al.}{2021}]{huang2021using}
Huang, L.~Y., S.~M. Hsiang, and M.~Gonzalez-Navarro (2021).
\newblock Using satellite imagery and deep learning to evaluate the impact of anti-poverty programs.
\newblock Working Paper w29105, National Bureau of Economic Research.

\bibitem[\protect\citeauthoryear{Imbens, Kallus, Mao, and Wang}{Imbens et~al.}{2025}]{imbens2024long}
Imbens, G., N.~Kallus, X.~Mao, and Y.~Wang (2025, 4).
\newblock Long-term causal inference under persistent confounding via data combination.
\newblock {\em Journal of the Royal Statistical Society Series B: Statistical Methodology\/}~{\em 87\/}(2), 362--388.

\bibitem[\protect\citeauthoryear{Imbens and Angrist}{Imbens and Angrist}{1994}]{AngristImbens(94)}
Imbens, G.~W. and J.~D. Angrist (1994).
\newblock Identification and estimation of local average treatment effects.
\newblock {\em Econometrica\/}~{\em 62\/}(2), 467--475.

\bibitem[\protect\citeauthoryear{Jack, Jayachandran, Kala, and Pande}{Jack et~al.}{2025}]{jack2022money}
Jack, B.~K., S.~Jayachandran, N.~Kala, and R.~Pande (2025).
\newblock Money (not) to burn: Payments for ecosystem services to reduce crop residue burning.
\newblock {\em American Economic Review: Insights\/}~{\em 7\/}(1), 39–55.

\bibitem[\protect\citeauthoryear{Jack and Walker}{Jack and Walker}{2023}]{jack2023remotesensing}
Jack, B.~K. and K.~Walker (2023).
\newblock Integrating remote sensing and randomized controlled trials: Challenges, opportunities, and practical guidance.
\newblock Mimeo, Bren School of Environmental Science \& Management, UC Santa Barbara.

\bibitem[\protect\citeauthoryear{Jayachandran, De~Laat, Lambin, Stanton, Audy, and Thomas}{Jayachandran et~al.}{2017}]{jayachandran2017cash}
Jayachandran, S., J.~De~Laat, E.~F. Lambin, C.~Y. Stanton, R.~Audy, and N.~E. Thomas (2017).
\newblock Cash for carbon: A randomized trial of payments for ecosystem services to reduce deforestation.
\newblock {\em Science\/}~{\em 357\/}(6348), 267--273.

\bibitem[\protect\citeauthoryear{Jean, Burke, Xie, Davis, Lobell, and Ermon}{Jean et~al.}{2016}]{jean2016combining}
Jean, N., M.~Burke, M.~Xie, W.~M. Davis, D.~B. Lobell, and S.~Ermon (2016).
\newblock Combining satellite imagery and machine learning to predict poverty.
\newblock {\em Science\/}~{\em 353\/}(6301), 790--794.

\bibitem[\protect\citeauthoryear{Ji, Lei, and Zrnic}{Ji et~al.}{2025}]{ji2025predictions}
Ji, W., L.~Lei, and T.~Zrnic (2025).
\newblock Predictions as surrogates: Revisiting surrogate outcomes in the age of {AI}.

\bibitem[\protect\citeauthoryear{Johannemann, Hadad, Athey, and Wager}{Johannemann et~al.}{2019}]{johannemann2019sufficient}
Johannemann, J., V.~Hadad, S.~Athey, and S.~Wager (2019).
\newblock Sufficient representations for categorical variables.
\newblock {\em arXiv preprint arXiv:1908.09874\/}.

\bibitem[\protect\citeauthoryear{Josephson, Michler, Kilic, and Murray}{Josephson et~al.}{2026}]{Josephson2026}
Josephson, A., J.~D. Michler, T.~Kilic, and S.~Murray (2026).
\newblock The mismeasure of weather: Using earth observation data for estimation of socioeconomic outcomes.
\newblock {\em Journal of Development Economics\/}~{\em 178}, 103553.

\bibitem[\protect\citeauthoryear{Kallus and Mao}{Kallus and Mao}{2024}]{kallus2024role}
Kallus, N. and X.~Mao (2024, 10).
\newblock On the role of surrogates in the efficient estimation of treatment effects with limited outcome data.
\newblock {\em Journal of the Royal Statistical Society Series B: Statistical Methodology\/}~{\em 87\/}(2), 480--509.

\bibitem[\protect\citeauthoryear{Kluger, Lu, Zrnic, Wang, and Bates}{Kluger et~al.}{2025}]{kluger2025prediction}
Kluger, D.~M., K.~Lu, T.~Zrnic, S.~Wang, and S.~Bates (2025).
\newblock Prediction-powered inference with imputed covariates and nonuniform sampling.
\newblock {\em arXiv preprint\/}.

\bibitem[\protect\citeauthoryear{Kress}{Kress}{1989}]{kress1989linear}
Kress, R. (1989).
\newblock {\em Linear Integral Equations}, Volume~82 of {\em Applied Mathematical Sciences}.
\newblock Springer.

\bibitem[\protect\citeauthoryear{Lu, Kluger, Bates, and Wang}{Lu et~al.}{2025}]{lu2025regressioncoefficientestimationremote}
Lu, K., D.~M. Kluger, S.~Bates, and S.~Wang (2025).
\newblock Regression coefficient estimation from remote sensing maps.
\newblock {\em Remote Sensing of Environment\/}~{\em 330}, 114949.

\bibitem[\protect\citeauthoryear{Mackey, Syrgkanis, and Zadik}{Mackey et~al.}{2018}]{mackey2018orthogonal}
Mackey, L., V.~Syrgkanis, and I.~Zadik (2018).
\newblock Orthogonal machine learning: Power and limitations.
\newblock In {\em International Conference on Machine Learning}, pp.\  3375--3383. PMLR.

\bibitem[\protect\citeauthoryear{Marx, Stoker, and Suri}{Marx et~al.}{2019}]{marx2019there}
Marx, B., T.~M. Stoker, and T.~Suri (2019).
\newblock There is no free house: Ethnic patronage in a {K}enyan slum.
\newblock {\em American Economic Journal: Applied Economics\/}~{\em 11\/}(4), 36--70.

\bibitem[\protect\citeauthoryear{Masten and Poirier}{Masten and Poirier}{2020}]{MastenPoirier(20)}
Masten, M.~A. and A.~Poirier (2020).
\newblock Inference on breakdown frontiers.
\newblock {\em Quantitative Economics\/}~{\em 11\/}(1), 41--111.

\bibitem[\protect\citeauthoryear{Miao, Liu, Li, Tchetgen~Tchetgen, and Geng}{Miao et~al.}{2024}]{miao_liu_li_tchetgen_geng_2024_shadow}
Miao, W., L.~Liu, Y.~Li, E.~J. Tchetgen~Tchetgen, and Z.~Geng (2024).
\newblock Identification and semiparametric efficiency theory of nonignorable missing data with a shadow variable.
\newblock {\em ACM/IMS Journal of Data Science\/}~{\em 1\/}(2), 1--23.

\bibitem[\protect\citeauthoryear{Miao and Tchetgen~Tchetgen}{Miao and Tchetgen~Tchetgen}{2016}]{miao_tchetgen_2016_shadow_dr}
Miao, W. and E.~J. Tchetgen~Tchetgen (2016).
\newblock On varieties of doubly robust estimators under missingness not at random with a shadow variable.
\newblock {\em Biometrika\/}~{\em 103\/}(2), 475--482.

\bibitem[\protect\citeauthoryear{Michaels, Nigmatulina, Rauch, Regan, Baruah, and Dahlstrand-Rudin}{Michaels et~al.}{2021}]{michaels2021planning}
Michaels, G., D.~Nigmatulina, F.~Rauch, T.~Regan, N.~Baruah, and A.~Dahlstrand-Rudin (2021).
\newblock Planning ahead for better neighborhoods: Long-run evidence from {T}anzania.
\newblock {\em Journal of Political Economy\/}~{\em 129\/}(7), 2112--2156.

\bibitem[\protect\citeauthoryear{Muralidharan, Niehaus, and Sukhtankar}{Muralidharan et~al.}{2016}]{muralidharan2016building}
Muralidharan, K., P.~Niehaus, and S.~Sukhtankar (2016).
\newblock Building state capacity: Evidence from biometric smartcards in {I}ndia.
\newblock {\em American Economic Review\/}~{\em 106\/}(10), 2895--2929.

\bibitem[\protect\citeauthoryear{Muralidharan, Niehaus, and Sukhtankar}{Muralidharan et~al.}{2023}]{muralidharan2023general}
Muralidharan, K., P.~Niehaus, and S.~Sukhtankar (2023).
\newblock General equilibrium effects of (improving) public employment programs: Experimental evidence from {I}ndia.
\newblock {\em Econometrica\/}~{\em 91\/}(4), 1261--1295.

\bibitem[\protect\citeauthoryear{Newey}{Newey}{1993}]{newey1993efficient}
Newey, W.~K. (1993).
\newblock Efficient estimation of models with conditional moment restrictions.
\newblock In G.~S. Maddala, C.~R. Rao, and H.~D. Vinod (Eds.), {\em Handbook of Statistics}, Volume~11, Chapter~16, pp.\  419--454. Elsevier.

\bibitem[\protect\citeauthoryear{Newey}{Newey}{1997}]{Newey(97)}
Newey, W.~K. (1997).
\newblock Convergence rates and asymptotic normality for series estimators.
\newblock {\em Journal of Econometrics\/}~{\em 79\/}(1), 147--168.

\bibitem[\protect\citeauthoryear{Newey and Powell}{Newey and Powell}{2003}]{newey2003instrumental}
Newey, W.~K. and J.~L. Powell (2003).
\newblock Instrumental variable estimation of nonparametric models.
\newblock {\em Econometrica\/}~{\em 71\/}(5), 1565--1578.

\bibitem[\protect\citeauthoryear{Park, Richardson, and Tchetgen~Tchetgen}{Park et~al.}{2024}]{park2024single}
Park, C., D.~B. Richardson, and E.~J. Tchetgen~Tchetgen (2024).
\newblock Single proxy control.
\newblock {\em Biometrics\/}~{\em 80\/}(2), ujae027.

\bibitem[\protect\citeauthoryear{Patel}{Patel}{2024}]{patel2024floods}
Patel, D. (2024).
\newblock Floods.
\newblock Technical report, Department of Economics, Harvard University.

\bibitem[\protect\citeauthoryear{Pelletier, Korb, Alemu, Yonis, Lybbert, and Stigler}{Pelletier et~al.}{2026}]{PELLETIER2026103655}
Pelletier, J., M.~Korb, S.~Alemu, M.~B. Yonis, T.~J. Lybbert, and M.~Stigler (2026).
\newblock Causal inference with predicted outcomes: Correcting prediction error bias in satellite-based impact evaluation.
\newblock {\em Journal of Development Economics\/}~{\em 179}, 103655.

\bibitem[\protect\citeauthoryear{Prentice}{Prentice}{1989}]{prentice1989surrogate}
Prentice, R.~L. (1989).
\newblock Surrogate endpoints in clinical trials: {D}efinition and operational criteria.
\newblock {\em Statistics in Medicine\/}~{\em 8\/}(4), 431--440.

\bibitem[\protect\citeauthoryear{Proctor, Carleton, Chong, Fransen, Greenhill, Katz, Murayama, Sherman, Tseng, Druckenmiller, and Hsiang}{Proctor et~al.}{2025}]{ProctorEtAl(25)}
Proctor, J., T.~Carleton, T.~Chong, T.~Fransen, S.~Greenhill, J.~Katz, H.~Murayama, L.~Sherman, J.~Tseng, H.~Druckenmiller, and S.~Hsiang (2025, October).
\newblock What can satellite imagery and machine learning measure?
\newblock Working Paper 34315, National Bureau of Economic Research.

\bibitem[\protect\citeauthoryear{Proctor, Carleton, and Sum}{Proctor et~al.}{2023}]{proctor2023parameter}
Proctor, J., T.~Carleton, and S.~Sum (2023).
\newblock Parameter recovery using remotely sensed variables.
\newblock Technical Report 30861, National Bureau of Economic Research.
\newblock NBER Working Paper.

\bibitem[\protect\citeauthoryear{Ratledge, Cadamuro, De~La~Cuesta, Stigler, and Burke}{Ratledge et~al.}{2022}]{ratledge2022using}
Ratledge, N., G.~Cadamuro, B.~De~La~Cuesta, M.~Stigler, and M.~Burke (2022).
\newblock Using machine learning to assess the livelihood impact of electricity access.
\newblock {\em Nature\/}~{\em 611\/}(7936), 491--495.

\bibitem[\protect\citeauthoryear{Ridder and Moffitt}{Ridder and Moffitt}{2007}]{RidderMoffitt07}
Ridder, G. and R.~Moffitt (2007).
\newblock The econometrics of data combination.
\newblock In J.~J. Heckman and E.~E. Leamer (Eds.), {\em Handbook of Econometrics}, Volume~6, Chapter~75, pp.\  5469--5547. Amsterdam: Elsevier.

\bibitem[\protect\citeauthoryear{Rolf, Proctor, Carleton, Bolliger, Shankar, Ishihara, Recht, and Hsiang}{Rolf et~al.}{2021}]{rolf2021generalizable}
Rolf, E., J.~Proctor, T.~Carleton, I.~Bolliger, V.~Shankar, M.~Ishihara, B.~Recht, and S.~Hsiang (2021).
\newblock A generalizable and accessible approach to machine learning with global satellite imagery.
\newblock {\em Nature Communications\/}~{\em 12}, 4392.

\bibitem[\protect\citeauthoryear{Roth and Sant'Anna}{Roth and Sant'Anna}{2023}]{RothSantAnna(23)}
Roth, J. and P.~H.~C. Sant'Anna (2023).
\newblock When is parallel trends sensitive to functional form?
\newblock {\em Econometrica\/}~{\em 91\/}(2), 737--747.

\bibitem[\protect\citeauthoryear{Rubin}{Rubin}{1987}]{rubin1987multiple}
Rubin, D.~B. (1987).
\newblock {\em Multiple Imputation for Nonresponse in Surveys}.
\newblock Wiley Series in Probability and Statistics. New York: John Wiley \& Sons.

\bibitem[\protect\citeauthoryear{Sanford, Ayers, Gordon, and Stone}{Sanford et~al.}{2025}]{sanford2025remote}
Sanford, L.~C., M.~Ayers, M.~Gordon, and E.~Stone (2025).
\newblock Adversarial debiasing for unbiased parameter recovery.

\bibitem[\protect\citeauthoryear{Sargan}{Sargan}{1958}]{Sargan(58)}
Sargan, J.~D. (1958).
\newblock The estimation of economic relationships using instrumental variables.
\newblock {\em Econometrica\/}~{\em 26\/}(3), 393--415.

\bibitem[\protect\citeauthoryear{Schennach}{Schennach}{2020}]{schennach2020mismeasured}
Schennach, S.~M. (2020).
\newblock Mismeasured and unobserved variables.
\newblock In S.~N. Durlauf, L.~P. Hansen, J.~J. Heckman, and R.~L. Matzkin (Eds.), {\em Handbook of Econometrics}, Volume~7, pp.\  487--565. Elsevier.

\bibitem[\protect\citeauthoryear{Torchiana, Rosenbaum, Scott, and Souza-Rodrigues}{Torchiana et~al.}{2025}]{torchiana2025improving}
Torchiana, A.~L., T.~Rosenbaum, P.~T. Scott, and E.~Souza-Rodrigues (2025).
\newblock Improving estimates of transitions from satellite data: A hidden markov model approach.
\newblock {\em The Review of Economics and Statistics\/}~{\em 107\/}(2), 426--441.

\bibitem[\protect\citeauthoryear{Vafa, Athey, and Blei}{Vafa et~al.}{2025}]{vafa2024estimatingwagedisparitiesusing}
Vafa, K., S.~Athey, and D.~M. Blei (2025).
\newblock Estimating wage disparities using foundation models.
\newblock {\em Proceedings of the National Academy of Sciences\/}~{\em 122\/}(22), e2427298122.

\bibitem[\protect\citeauthoryear{Viviano and Rudder}{Viviano and Rudder}{2020}]{viviano2022policy}
Viviano, D. and J.~Rudder (2020).
\newblock Policy design in experiments with unknown interference.
\newblock arXiv preprint 2011.08174.

\bibitem[\protect\citeauthoryear{Walker, Moscona, Jack, Jayachandran, Kala, Pande, Xue, and Burke}{Walker et~al.}{2022}]{walker2022detecting}
Walker, K., B.~Moscona, K.~Jack, S.~Jayachandran, N.~Kala, R.~Pande, J.~Xue, and M.~Burke (2022).
\newblock Detecting crop burning in {I}ndia using satellite data.
\newblock arXiv preprint arXiv:2209.10148.

\bibitem[\protect\citeauthoryear{Wunder, Engel, and Pagiola}{Wunder et~al.}{2008}]{WunderEtAl(08)}
Wunder, S., S.~Engel, and S.~Pagiola (2008).
\newblock Taking stock: A comparative analysis of payments for environmental services programs in developed and developing countries.
\newblock {\em Ecological Economics\/}~{\em 65\/}(4), 834--852.

\end{thebibliography}

\newpage 

\spacingset{1.5}
\appendix

\begin{center}
\vspace{-1.8cm}{\Large \textbf{Program Evaluation with Remotely Sensed Outcomes}} 
\bigskip \\
{\Large \textit{Appendix Materials}}
\bigskip \\
\large Ashesh Rambachan, Rahul Singh, Davide Viviano \medskip \\
\bigskip
\end{center}

\renewcommand\thefigure{\thesection.\arabic{figure}} 
\setcounter{figure}{0}   
\renewcommand\thetable{\thesection.\arabic{table}} 
\setcounter{table}{0} 
\setcounter{page}{1}
\startcontents[appendix]
\printcontents[appendix]{}{1}{\setcounter{tocdepth}{3}}
\numberwithin{equation}{section}
\numberwithin{table}{section}
\numberwithin{figure}{section}
\newpage 

\section{Additional Model Details}\label{sec:figures}

\begin{figure}[htbp!]
\centering 
\includegraphics[width=0.5\textwidth]{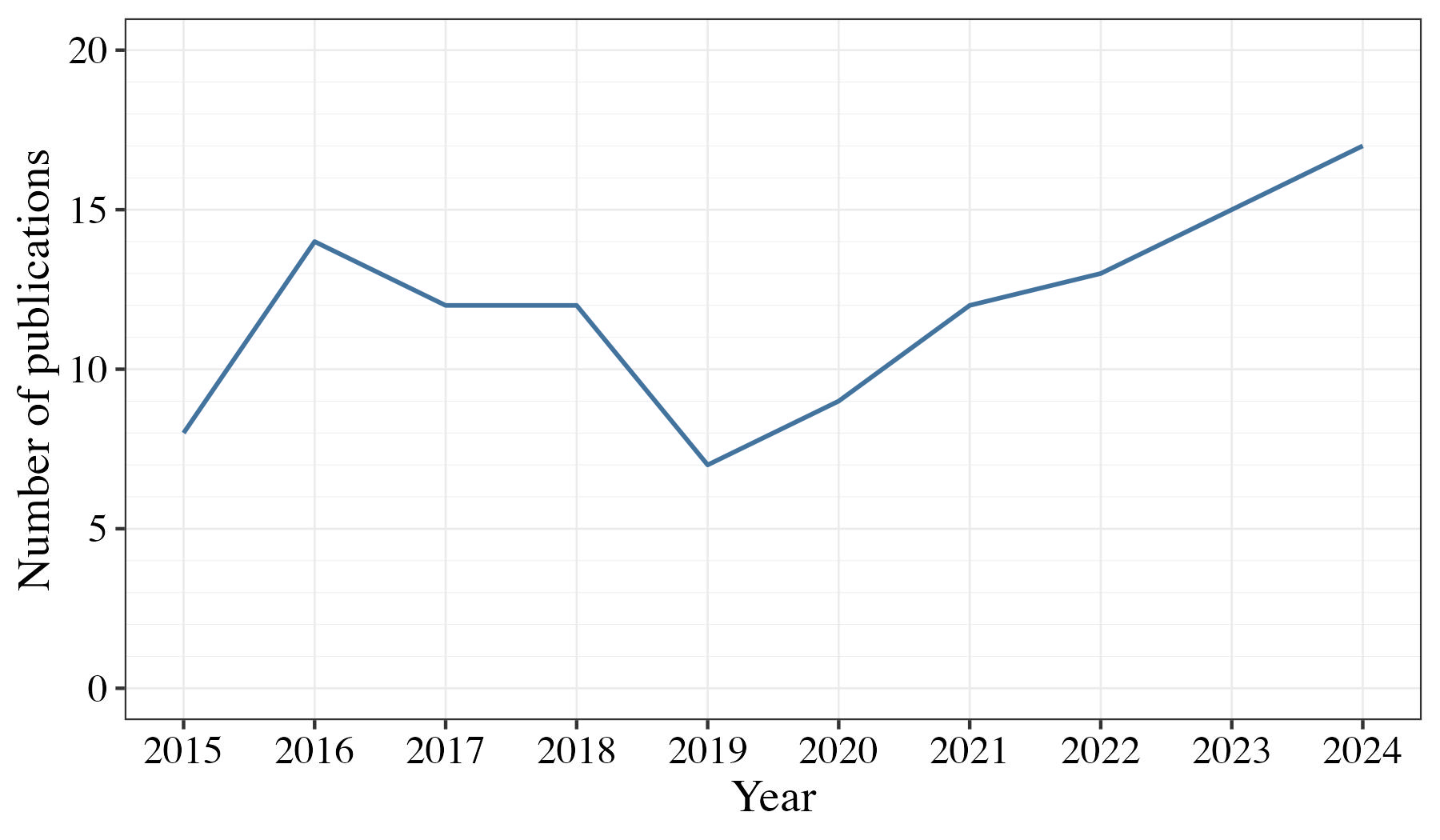}
\caption{Remotely sensed variables are increasingly popular in published papers.}
\label{fig:trend_papers}
\end{figure}

Figure~\ref{fig:trend_papers} illustrates the increasing popularity of remotely sensed variables in empirical research.
First, we collected papers published in the American Economic Association (AEA) journals, \textit{Econometrica}, \textit{Quarterly Journal of Economics}, \textit{Review of Economic Studies} and \textit{Journal of Political Economy} using a keyword search of ``remotely sensed variables'', ``mobile phone'', ``satellite'', ``machine learning'', and ``drones / aerial'' on their websites. 
Next, we collected  papers published in \textit{Nature} and \textit{Science} using a keyword search of ``remotely sensed variables'' on their websites. 
We subset to the papers with remotely sensed variables in their main empirical analysis.

\begin{figure}[htbp!]
\centering
\begin{tikzpicture}[node distance=3cm]
    \tikzstyle{process} = [circle, minimum width=1cm, minimum height=1cm, text centered, draw=black, fill=white]
    \node (Trt) [process] {$D$};
    \node (Trace) [process, right of=Trt] {$R$};
    \node (Out) [process, right of=Trace] {$Y$};
    \draw[->, line width = 1.5pt] (Trt) -- (Trace);
    \draw[->, line width = 1.5pt] (Trace) -- (Out);
\end{tikzpicture}
\caption{Causal graph of surrogacy model.}
\label{fig:dag_surrogates}
\end{figure}

Figure~\ref{fig:dag_surrogates} illustrates the causal graph associated with the surrogacy identifying assumptions  \citep{prentice1989surrogate,athey2024surrogate}: 
the surrogate $R$ fully mediates the effect of the treatment $D$ on the outcome $Y$. 
Our identifying assumptions are the opposite, as illustrated by Figure~\ref{fig:dag_rsv}.

\begin{table}[!ht]
\centering 
\caption{Implications of the identifying assumptions.}

\scalebox{0.7}{\begin{tabular}{c|cccl}
\toprule 
Assumption& Quantity & Experimental  & Observational & Description  \\
\hline 
\ref{assumption:experimental} & $ \Pr\left(Y(d) \mid S,X,D\right)$ & $ \Pr\left(Y(d) \mid S=e,X\right)$ & $ \Pr\left(Y(d) \mid  S=o,X,D\right)$ & Differs across samples \\
\ref{assumption:stability} & $ \Pr(R \mid S,X,D,Y)$ & $ \Pr(R \mid X,D,Y)$& $ \Pr(R|X,D,Y)$& Stable across samples \\
\ref{assumption:observational}(i) & $ \Pr(R \mid S,X,D,Y)$ & $ \Pr(R \mid X,D,Y)$& $ \Pr(R \mid X,D,Y)$& Differs across treatments if complete cases  \\
\ref{assumption:observational}(ii) & $ \Pr(R \mid S,X,D,Y)$ & $ \Pr(R \mid  X,Y)$& $ \Pr(R \mid X,Y)$ & Stable across treatments if incomplete cases \\
\bottomrule
\end{tabular}}
\label{tab:implications}
\end{table}

Table~\ref{tab:implications} summarizes the implications of our identifying assumptions. 
Our identifying assumptions allow the propensity score $\Pr(D=1 \mid S,X)$ to differ across samples. 
In summary, the outcome and treatment mechanisms may vary across samples, but the sensing mechanism must be stable across samples.

\section{Proofs for Theoretical Results in the Main Text}
\subsection{Proofs for Section \ref{sec:identification}}\label{sec:identification_proof}

\subsubsection{Proof of Lemma \ref{lemma:discrete_mixture_incomplete}} \label{proof:lemma:discrete_mixture_incomplete}

We consider the general case where $(X,Y,R)$ may be discrete or continuous. Then, we specialize the result to the case where $Y$ is discrete. Recall that $f_W(\cdot \mid \ldots)$ is the Radon-Nikodym derivative. 

By the law of total probability,
\begin{align*}
    \delta_{d}^{e}(X,R) &:=f_R( R\mid S = e, X, D = d) \\
    &= \int f_{R,Y}( y,R \mid S = e, X, D = d) \mathrm{d}y \\
    &=\int f_R( R\mid S = e, X, D = d, Y=y) f_Y(y \mid S=e,X,D=d) \mathrm{d}y.
\end{align*}
By Assumption \ref{assumption:experimental}, $f_Y(y \mid S=e,X,D=d)=f_{Y(d)}(y \mid S=e,X)$. 
Next, notice that 
\begin{align*}
    f_R( R\mid S = e, X, D = d, Y=y) &= f_R( R\mid X, Y=y) \\
    &= f_R( R\mid S = o, X, Y=y) =: \delta_y^o(X,R),
\end{align*}
where the equalities apply the contraction implied by Assumptions \ref{assumption:stability} and \ref{assumption:observational}(ii): $(S,D)\indep R|X,Y$. 

Combining the previous displays, we arrive at the general result 
$$
    \delta_{d}^{e}(X,R)=\int \delta_y^o(X,R) f_{Y(d)}(y \mid S=e,X) \mathrm{d}y.
$$
When $Y$ is discrete, $ f_{Y(d)}(y_k \mid S=e,X)=\mu_k(d,X)$ and the integral becomes a sum.
\qed

\subsubsection{Proof of Theorem \ref{theorem:disc_outcomes_incomplete}} \label{proof:theorem:disc_outcomes_incomplete}

We proceed in three steps. 

\begin{enumerate}
\item  Define the treatment weights in the experimental sample as 
$$
   \pi_d(X, R) := \frac{ \Pr\left( D = d, S = e \mid X, R \right) }{\Pr\left( D = d, S = e \mid X \right) },
$$
and the outcome weights in the observational sample as 
$$
\gamma_{y}(X, R) := \frac{\Pr(Y = y, S = o \mid X, R)}{ \Pr(Y = y, S = o \mid X)}.
$$
By Bayes' rule, 
\begin{align*}
f_R(R \mid Y = y, S = o, X) &= \frac{\Pr(Y = y, S = o \mid X,R) f_R(R \mid X)}{\Pr(Y = y, S = o \mid X)}
=\gamma_{y}(X, R) f_R(R|X), \\
f_R\left( R \mid D = d, S = e, X \right) &= \frac{ \Pr(D = d, S = e \mid X,R) f_R(R \mid X) }{ \Pr(D = d, S = e \mid X) }
=\pi_d(X, R) f_R(R \mid X).
\end{align*}
Substituting these expressions into Lemma~\ref{lemma:discrete_mixture_incomplete} and canceling $f_R(R \mid X)$ yields
$$
\pi_d(X, R) = \sum_{k=1}^{K} \gamma_{y_k}(X, R) \Pr\{Y(d) = y_k \mid S = e, X\}.
$$

\item Since $\sum_{k=1}^K \Pr\{Y(d) = y_k \mid S=e, X\} = 1$, we can rewrite the previous display as  
\begin{align*}
\pi_d(X,R) &= \sum_{k=1}^{K-1} \gamma_{y_k}(X,R) \Pr\{Y(d) = y_k \mid S=e, X\} + \gamma_{y_K}(X,R) \left[1 - \sum_{k=1}^{K-1} \Pr\{Y(d) = y_k \mid S=e, X\}\right].
\end{align*}
Re-arranging yields
\begin{equation*}
\pi_d(X,R) - \gamma_{y_K}(X,R) = \sum_{k=1}^{K-1} \left\{\gamma_{y_k}(X,R) - \gamma_{y_K}(X,R)\right\} \Pr\{Y(d) = y_k \mid S=e, X\}.
\end{equation*}
We then define 
$$
\tilde{\gamma}(X,R) := \begin{bmatrix}
\gamma_{y_1}(X,R) - \gamma_{y_K}(X,R) \\
\vdots \\
\gamma_{y_{K-1}}(X,R) - \gamma_{y_K}(X,R)
\end{bmatrix}, \quad \mu(d,X) := \begin{bmatrix}
\Pr\{Y(d)=y_1 \mid S=e, X\} \\
\vdots \\
\Pr\{Y(d)=y_{K-1} \mid S=e, X\}
\end{bmatrix}
$$
and rewrite the previous display as 
$$
\pi_d(X,R) - \gamma_{y_K}(X,R) = \tilde{\gamma}(X,R)^\top \mu(d,X).
$$
Applying iterated expectations, we have shown, for $d \in \{0, 1\}$,  
$$
\E\left\{ \Delta^{e}(d, X) - \Delta^o(X)^\top \mu(d,X) \mid X, R \right\} = 0 \mbox{ almost surely},
$$
where $\Delta^e(d, x) := \frac{1\{D=d, S=e\}}{\Pr(D=d, S=e \mid X)} - \frac{1\{Y=y_K, S=o\}}{\Pr(Y=y_K, S=o \mid X)}$ and 
$$
\Delta^o(x) := \begin{bmatrix}
\frac{1\{Y=y_1, S=o\}}{\Pr(Y=y_1, S=o \mid X)} - \frac{1\{Y=y_K, S=o\}}{\Pr(Y=y_K, S=o \mid X)} \\
\vdots \\
\frac{1\{Y=y_{K-1}, S=o\}}{\Pr(Y=y_{K-1}, S=o \mid X)} - \frac{1\{Y=y_K, S=o\}}{\Pr(Y=y_K, S=o \mid X)}
\end{bmatrix} \in \mathbb{R}^{K-1}.
$$

\item We take the difference of the treated and untreated moment restrictions, thereby proving the result. \qed

\end{enumerate}

\subsubsection{Proof of Theorem \ref{theorem:disc_outcomes_complete}} \label{proof:theorem:disc_outcomes_complete}

The proof follows the same steps as the proofs of Lemma~\ref{lemma:discrete_mixture_incomplete} and Theorem~\ref{theorem:disc_outcomes_incomplete}. 

\begin{enumerate}
\item As in Lemma~\ref{lemma:discrete_mixture_incomplete}, for general $(X,Y,R)$,
$$
    \delta_{d}^{e}(X,R):=f_R(R \mid S = e, X, D = d)=\int f_R(R \mid S = e, X, D = d, Y=y) f_Y(y \mid S=e,X,D=d) \mathrm{d}y.
$$
By Assumption \ref{assumption:experimental}, $f_Y(y \mid S=e,X,D=d)=f_{Y(d)}(y \mid S=e,X)$. 
Now, however,
\begin{align*}
    f_R( R\mid S = e, X, D = d, Y=y) &= f_R( R\mid S = o, X, D = d, Y=y)
    =:\delta_{y,d}^o(X,R),
\end{align*}
where the first equality applies Assumption \ref{assumption:stability} and Assumption \ref{assumption:observational}(i) so that the conditioning event has non zero probability. 

Combining the previous displays, we arrive at the general result 
$$
    \delta_{d}^{e}(X,R)= \int \delta_{y,d}^o(X,R) f_{Y(d)}(y \mid S=e,X) \mathrm{d}y.
$$
When $Y$ is discrete, $ f_{Y(d)}(y_k \mid S=e,X)=\mu_k(d,X)$ and the integral becomes a sum.

\item As in Theorem~\ref{theorem:disc_outcomes_incomplete}, by Bayes' rule,
\begin{align*}
     \delta_{y,d}^{o}(X,R) &= \frac{ \Pr(Y = y, D = d, S = o \mid X,R) f_R(R \mid X) }{ \Pr(Y = y, D = d, S = o \mid X) }
     =\gamma_{y, d}(X, R) f_R(R \mid X) \\
      \delta_{d}^{e}(X,R) &= \frac{ \Pr(D = d, S = e \mid X,R) f_R(R \mid X) }{ \Pr(D = d, S = e \mid X) }
      =\pi_d(X, R)f_R(R \mid X),
\end{align*}
where
$$
\pi_d(X, R) := \frac{ \Pr\left( D = d, S = e \mid X, R \right) }{\Pr\left( D = d, S = e \mid X \right) },\quad \gamma_{y, d}(X, R) = \frac{\Pr(Y = y, D = d, S = o \mid X, R)}{ \Pr(Y = y, D = d, S = o \mid X)}.
$$
Substituting these expressions into the previous step and canceling $f_R(R \mid X)$ yields
$$
\pi_d(X, R) = \sum_{k=1}^K \gamma_{y_k, d}(X, R) \Pr\{Y(d) = y_k \mid S = e, X\}.
$$

\item The remainder of the proof is identical to step 2 of Theorem~\ref{theorem:disc_outcomes_incomplete}, replacing $\gamma_{y_k}$ with $\gamma_{y_k,d}$. \qed
\end{enumerate}

\subsubsection{Proof of Proposition \ref{prop:bias_common_practice}} \label{proof:bias_practice}

Under Assumption \ref{assumption:experimental}, \ref{assumption:stability} and \ref{assumption:observational}(ii), $(S, D) \indep R \mid Y$.

\begin{enumerate}
\item We prove (i) with possibly discrete outcomes. Fix $d \in \{0, 1\}$. Let 
    $$m(R) = \E(Y \mid R, S = o) = \int y f_{Y}(y \mid R, S = o) \mathrm{d}y.$$
By Bayes' rule,
\begin{align*}
    f_{Y}(Y \mid R, S = o) &= \frac{f_{R}(R \mid Y, S = o) f_{Y }(Y \mid S = o)}{f_{R }(R\mid S = o)} \\
    f_{R}(R\mid Y, D = d, S = e) &= \frac{f_{Y}(Y \mid R, D = d, S = e) f_{R}(R \mid D = d, S = e)}{f_{Y }(Y \mid D = d, S =e )}.
\end{align*}
By $(S, D) \indep R \mid Y$, 
$$
 f_{R}(R\mid Y, S = o) = f_{R}(R \mid Y, D = d, S = e).
$$

We combine these expressions to rewrite
$$
f_{Y}(Y \mid R, S = o) = f_{Y}(Y \mid R, D = d, S = e) w_d(R,Y),\quad  w_d(R,Y) = \frac{f_{Y}(Y \mid S = o)}{f_{Y}(Y \mid D = d, S =e )} \frac{f_{R}(R\mid D = d, S = e)}{f_{R}(R\mid S = o)}.
$$
Consequently, 
$$
m(R) = \int y f_{Y }(y \mid R, D = d, S = e) w_d(R,y)  \mathrm{d}y = \E\{Y w_d(R,Y) \mid R, D = d, S = e\}.
$$
Since $\widetilde{\mu}(d) = \E\{m(R) \mid D = d, S = e\}$, the result follows by iterated expectations. 

\item We prove (ii) for binary outcomes, where $\mu(d) = \Pr(Y = 1 \mid D = d, S = e)$. Fix $d\in\{0,1\}$. By iterated expectations and  $(S, D) \indep R \mid Y$, 
\begin{align*}
     f_{R}(R\mid D = d, S = e) 
     &= \mu(d) f_{R}(R\mid Y = 1,D=d, S = e) + \{1 -  \mu(d)\} f_{R}(R \mid Y = 0, D=d,S = e)\\
     &= \mu(d) f_{R}(R\mid Y = 1, S = o) + \{1 -  \mu(d)\} f_{R }(R \mid Y = 0, S = o).
\end{align*}
As a consequence, 
\begin{align*}
     \widetilde{\mu}(d) 
     &= \E\{m(R) \mid D = d, S = e\} \\
     &= \mu(d) \E\{m(R) \mid Y = 1, S = o\} + \{1 - \mu(d)\} \E\{m(R) \mid Y = 0, S = o\}.
\end{align*}
Taking differences, 
\begin{align*}
    \widetilde{\tau} &:= \widetilde{\mu}(1)-\widetilde{\mu}(0) \\  
    &=\left[ \E\{m(R) \mid Y = 1, S = o\} - \E\{m(R) \mid Y = 0, S = o\} \right] \left\{ \mu(1) - \mu(0) \right\} \\
    &=\left[ \E\{m(R) \mid Y = 1, S = o\} - \E\{m(R) \mid Y = 0, S = o\} \right]\tau\\
    &=\frac{\cov\{m(R), Y \mid S = o\}}{\var(Y \mid S = o)} \tau \\
    &=\frac{\var\{m(R) \mid S = o\}}{\var(Y \mid S = o)} \tau,
\end{align*}
yielding $\kappa := \frac{\var\{m(R) \mid S = o\}}{\var(Y \mid S = o)}$. 

By the law of total variance,
$$
\var(Y \mid S = o)=\var\{m(R)\mid S=o\}+\E\{\var(Y\mid R,S=o)\mid S=o\}, 
$$
so $\kappa\in[0,1]$. Moreover, $\kappa = 1$ if and only if $\E\{\var(Y \mid R, S = o) \mid S = o\}= 0$, which is equivalent to $\var(Y \mid R, S = o) = 0$ almost surely. 
\end{enumerate}
\qed

\subsubsection{Proof of Proposition \ref{prop:bias_PPI}} \label{proof:prop:bias_PPI}

We prove each part of the proposition separately. Since $m_d(R)$ is a bounded and measurable function, the expectation is well defined. As before, the assumptions imply $(S,D)\indep R\mid Y$.

\begin{enumerate}
\item Fix $d\in\{0,1\}$ and $s\in\{e,o\}$.  By iterated expectations, the fact that $Y$ is binary, and $(S,D)\indep R\mid Y$,
\begin{align*}
    &\E\{m_d(R)\mid D=d,S=s\} \\
&=\E[\E\{m_d(R)\mid Y,D=d,S=s\}\mid D=d,S=s]\\
& =\E\{m_d(R)\mid Y=1,D=d,S=s\}\E(Y|D=d,S=s)+\E\{m_d(R)\mid Y=0,D=d,S=s\}\{1-\E(Y|D=d,S=s)\} \\
& =\E\{m_d(R)\mid Y=1 \}\E(Y|D=d,S=s)+\E\{m_d(R)\mid Y=0 \}\{1-\E(Y|D=d,S=s)\} \\
&=\E\{m_d(R)\mid Y=0\} + \kappa_d\E(Y\mid D=d,S=s),
\end{align*}
where $\kappa_d=\E\{m_d(R)\mid Y=1\}-\E\{m_d(R)\mid Y=0\}$. 

The first term does not depend on $s$. Therefore, substitution yields
\begin{align*}
 \tilde{\tau}^{\mathrm{PPI}}& =  \E\{m_1(R) \mid  D = 1,S = e\} - \E\{m_0(R) \mid  D = 0,S = e\}  \\
 & + \E\{Y - m_1(R) \mid  D=1,S=o\} - \E\{Y - m_0(R) \mid D=0,S=o\} \\
 &=\kappa_1\E(Y\mid D=1,S=e) - \kappa_0\E(Y\mid D=0,S=e)  \\
 & + \E(Y \mid D=1,S=o) - \kappa_1\E(Y\mid D=1,S=o) - \E(Y\mid D=0,S=o) + \kappa_0\E(Y\mid D=0,S=o)  \\
 &=\tilde{\tau}^o+\kappa_1\{\E(Y\mid D=1,S=e)-\E(Y\mid D=1,S=o)\} -\kappa_0\{\E(Y\mid D=0,S=e)-\E(Y\mid D=0,S=o)\}.
\end{align*}

By Assumption~\ref{assumption:experimental}(i, ii), $\E\{Y(d)\mid S=e\}=\E(Y\mid D=d,S=e)$, so
\begin{align*}
\tau&=\E(Y\mid D=1,S=e)-\E(Y\mid D=0,S=e) \\
\tilde{\tau}^o&=\E(Y\mid D=1,S=o)-\E(Y\mid D=0,S=o) \\
    \delta&=\E(Y\mid D=0,S=e)-\E(Y\mid D=0,S=o) \\
    \tau-\tilde{\tau}^o+\delta&=\E(Y\mid D=1,S=e)-\E(Y\mid D=1,S=o).
\end{align*}

Substituting these identities gives
\begin{align*}
\tilde{\tau}^{\mathrm{PPI}} &= \tilde{\tau}^o+\kappa_1(\tau-\tilde{\tau}^o+\delta ) -\kappa_0\delta \\
&= \tau+ (1-\kappa_1)(\tilde{\tau}^o-\tau)+(\kappa_1-\kappa_0) \delta.
\end{align*}

\item Fix arbitrary measurable functions $m_0,m_1$. We construct a relevant $R$ with $\kappa_0=\kappa_1=0$.
    
Let $R$ be discrete with support $\{r_1,r_2,r_3,r_4\}$. 
Construct a nonzero vector
$v=(v_1,\dots,v_4)$ such that
\[
\sum_{j=1}^4 v_j=0,\qquad \sum_{j=1}^4 m_0(r_j) v_j=0,\qquad \sum_{j=1}^4 m_1(r_j) v_j=0.
\]
Such vector always exists since we have three linear constraints and four degrees of freedom. 

Choose $\varepsilon>0$ small enough that $\frac{1}{4}+\varepsilon v_j$ and $\frac{1}{4}-\varepsilon v_j$ are bounded away from zero and one for all $j$. For $s \in \{e,o\}$ and $d \in \{0,1\}$, set
\[
\Pr(R=r_j\mid Y=1, D = d, S = s)=\frac{1}{4}+\varepsilon v_j,\quad
\Pr(R=r_j\mid Y=0, D = d, S = s)=\frac{1}{4}-\varepsilon v_j.
\]
Then $\Pr(R=r_j\mid Y=1, D = d, S = s)\neq\Pr(R=r_j\mid Y=0, D = d, S = s)$ (since $v\neq 0$), so $R \not \perp Y \mid D, S$.
Moreover, for each $d\in\{0,1\}$,
\begin{align*}
    \kappa_d&=\E\{m_d(R)\mid Y=1\}-\E\{m_d(R)\mid Y=0\} \\
&=\sum_{j=1}^4 m_d(r_j)\left\{\left(\frac{1}{4}+\varepsilon v_j\right)-\left(\frac{1}{4}-\varepsilon v_j\right)\right\} \\
&=2\varepsilon \sum_{j=1}^4 m_d(r_j)v_j \\
&=0
\end{align*}
where the last equality uses the construction of $v$.

By construction, this DGP satisfies stability, coverage, and no direct effects.
\end{enumerate}

\qed
\subsection{Proofs for Section \ref{sec:estimation}}\label{sec:estimation_proof}

\subsubsection{Proof of Proposition~\ref{prop:known}} \label{proof:prop:known}

To lighten notation, define $Y = \Delta^e$, $W = \Delta^o$, $U = Y - W^\top \theta$, $Z = \widetilde{H}(R)$, and $\widehat{Z} = \widehat{H}(R)$, so that
$$
Y=W^{\top}\theta+U,\quad \E(U|R)=0,\quad \E(UZ)=0.
$$

Note that $\|W\|_{\max}\leq \bar{W}$ almost surely for some constant $\bar{W}$ due to the assumption that the marginal probabilities are bounded away from zero. Moreover,   $\E(U^2|R)\leq \bar{\sigma}^2_U$ for some constant $\bar{\sigma}^2_U<\infty$ for the same reason.

While proving this result, we use standard notation for cross-fitting. 
Let there be $L$ folds, each denoted by $I_\ell$ with $\ell \in [L]$. 
Each fold contains $n_{\ell} = n/L$ observations.
The complement of $I_\ell$ is $I_{-\ell}$. 
If $i \in I_{\ell}$, then $\widehat{Z}_i = \widehat{H}_{\ell}(R_i)$ is constructed from $\widehat{H}_{\ell}$ estimated on the remaining folds $I_{-\ell}$.

With cross fitting, the estimator in Algorithm~\ref{alg:discrete_nocovar_overview} generalizes to  
$$\widehat{\theta} := \{\E_n(\widehat{Z} W^\top)\}^{-1} \E_n( \widehat{Z} Y )
=
\left(\frac{1}{L}\frac{1}{n_{\ell}}   \sum_{\ell = 1}^{L} \sum_{i \in I_{\ell}} \widehat{Z}_i W_i^\top\right)^{-1} \left(\frac{1}{L} \frac{1}{n_{\ell}}  \sum_{\ell = 1}^{L} \sum_{i \in I_{\ell}} \widehat{Z}_i Y_i\right).
$$ 

With known marginal probabilities, the argument uses standard techniques, similar to \cite{mackey2018orthogonal,chen2020mostly}.
In this lighter notation, 
$$
n^{1/2}(\widehat{\theta}-\theta)
=n^{1/2}\left[\{\E_n(\widehat{Z} W^\top)\}^{-1} \E_n( \widehat{Z} Y )- \{\E_n(\widehat{Z} W^\top)\}^{-1} \E_n(\widehat{Z} W^\top) \theta   \right]
= \{\E_n(\widehat{Z} W^\top)\}^{-1} n^{1/2}\E_n( \widehat{Z} U).
$$
We derive the probability limit of $\E_n(\widehat{Z} W^\top)$, then the Gaussian approximation of $n^{1/2} \E_n( \widehat{Z} U)$.

\begin{lemma}\label{lemma:denom1}
Under Proposition~\ref{prop:known}'s conditions, $\E_{n}(\widehat{Z} W^\top) \overset{p}{\rightarrow} \E_n(Z W^\top)$.
\end{lemma}

\begin{proof}
We express the difference as
    $$
    \E_n\{(\widehat{Z}-Z)W^{\top}\}
    =\frac{1}{L} \frac{1}{n_{\ell}} \sum_{\ell=1}^L \sum_{i\in I_{\ell}}  (\widehat{Z}_i-Z_i)W_i^{\top}
    =\frac{1}{L}\sum_{\ell=1}^L \frac{1}{n_{\ell}}  \sum_{i\in I_{\ell}}  (\widehat{Z}_i-Z_i)W_i^{\top}.
    $$
    Focusing on the foldwise quantity, we first control 
    $ \E \left\{\left|\frac{1}{n_{\ell}}  \sum_{i\in I_{\ell}}  (\widehat{Z}_{ij}-Z_{ij})W_{ik}\right| | I_{-\ell}\right\}.
    $
   We can write
    \begin{align*}
        &\E \left\{\frac{1}{n_{\ell}}  \sum_{i\in I_{\ell}}  \left|(\widehat{Z}_{ij}-Z_{ij})W_{ik}\right| | I_{-\ell}\right\} 
        \leq \E \left\{\frac{\bar{W}}{n_{\ell}} \cdot   \sum_{i\in I_{\ell}}  \left|\widehat{Z}_{ij}-Z_{ij}\right| | I_{-\ell}\right\} \\
        &= \bar{W}   \E \left\{\left|\widehat{Z}_{ij}-Z_{ij}\right| | I_{-\ell}\right\} 
        \leq \bar{W}   [\E \{(\widehat{Z}_{ij}-Z_{ij})^2 | I_{-\ell}\}]^{1/2} 
        \leq \bar{W}  \mathcal{R}(\widehat{Z})^{1/2}=o_p(1).
    \end{align*}
    We use $|W_{ij}|\leq \bar{W}$ almost surely,  since marginal probabilities are bounded away from zero. We also write $\mathcal{R}(\widehat{Z})=\E (\|\widehat{Z}_i-Z_i\|^2 \mid I_{-\ell})=o_p(1)$ for the mean square limit.

    Finally note that for any $\epsilon > 0$, $\Pr( |\frac{1}{n_{\ell}}  \sum_{i\in I_{\ell}}  (\widehat{Z}_{ij}-Z_{ij})W_{ik}| > \epsilon | I_{-\ell}) \le \bar{W} \mathcal{R}(\hat{Z})^{1/2}/\epsilon \rightarrow_p 0$. Therefore by the dominated convergence theorem (since $\Pr( |\frac{1}{n_{\ell}}  \sum_{i\in I_{\ell}}  (\widehat{Z}_{ij}-Z_{ij})W_{ik}| > \epsilon | I_{-\ell}) \in [0,1]$), it follows $\mathbb{E}[\Pr( |\frac{1}{n_{\ell}}  \sum_{i\in I_{\ell}}  (\widehat{Z}_{ij}-Z_{ij})W_{ik}| > \epsilon | I_{-\ell})] \rightarrow 0$. Applying this to each of the $\ell \in \{1,\ldots, L\}$ the result follows directly from the union bound (since $L$ is finite). 
\end{proof}

\begin{lemma}\label{lemma:denom2}
Under Proposition~\ref{prop:known}'s conditions, $\E_{n}(Z W^\top) \overset{p}{\rightarrow} \E(Z W^\top)$.
\end{lemma}

\begin{proof}
   By Chebyshev's inequality in Frobenius norm, it suffices to bound $\V(Z_{ij}W_{ik})\leq \E(Z_{ij}^2W_{ik}^2)\leq \bar{W}^2\E(\|Z_i\|^2)$. In summary, we use $\|W_i\|_{\max} \leq \bar{W}$ almost surely since marginal probabilities are bounded away from zero and $\E(\|Z_i\|^2)<\infty$ due to Assumption~\ref{assumption:limit}.
\end{proof}

\begin{lemma}\label{lemma:num1}
Under Proposition~\ref{prop:known}'s conditions, $n^{1/2} \E_{n}(\widehat{Z} U) \overset{p}{\rightarrow} n^{1/2} \E_{n}(Z U)$.
\end{lemma}

\begin{proof}
  We express the difference as 
\begin{align*}
n^{1/2} \E_n\{U(\widehat{Z}-Z)\} 
&=n^{1/2} \frac{1}{L} \frac{1}{n_{\ell}} \sum_{\ell=1}^L \sum_{i\in I_{\ell}} U_i(\widehat{Z}_i-Z_i)
=L^{1/2}\frac{1}{L}\sum_{\ell=1}^L  n_{\ell}^{1/2} \frac{1}{n_{\ell}}  \sum_{i\in I_{\ell}} U_i(\widehat{Z}_i-Z_i).
\end{align*}
Focusing on the foldwise quantity, it suffices to control
$$
\E \left[\left\{ n_{\ell}^{1/2} \frac{1}{n_{\ell}}  \sum_{i\in I_{\ell}} U_i(\widehat{Z}_{ik}-Z_{ik})\right\}^2\right]
=\E\left(\E \left[\left\{ n_{\ell}^{1/2} \frac{1}{n_{\ell}}  \sum_{i\in I_{\ell}} U_i(\widehat{Z}_{ik}-Z_{ik})\right\}^2\mid I_{-\ell}\right]\right).
$$
Due to cross-fitting, the inner expectation is
\begin{align*}
&\E \left[\left\{ n_{\ell}^{1/2} \frac{1}{n_{\ell}}  \sum_{i\in I_{\ell}} U_i(\widehat{Z}_{ik}-Z_{ik})\right\}^2\mid I_{-\ell}\right]
=
\frac{1}{n_{\ell}} \E \left\{ \sum_{i,j\in I_{\ell}} U_i(\widehat{Z}_{ik}-Z_{ik})U_j(\widehat{Z}_{jk}-Z_{jk}) \mid I_{-\ell}\right\} \\
&=\frac{1}{n_{\ell}} \E \left\{ \sum_{i\in I_{\ell}} U_i^2(\widehat{Z}_{ik}-Z_{ik})^2 \mid I_{-\ell}\right\} 
= \E \{U_i^2(\widehat{Z}_{ik}-Z_{ik})^2 \mid I_{-\ell}\} 
=\E \{\E(U_i^2|R_i,I_{-\ell})(\widehat{Z}_{ik}-Z_{ik})^2 \mid I_{-\ell}\}  \\
&\leq \bar{\sigma}^2_U \E \{(\widehat{Z}_{ik}-Z_{ik})^2 \mid I_{-\ell}\}
\leq \bar{\sigma}^2_U\mathcal{R}(\widehat{Z})=o_p(1).
\end{align*}
In the first inequality, we use $\E(U_i^2|R_i,I_{-\ell})=\E(U_i^2|R_i)\leq \bar{\sigma}^2_U$, where $\bar{\sigma}^2_U<\infty$ under our assumptions. In the second inequality, we write $\mathcal{R}(\widehat{Z})=\E (\|\widehat{Z}_i-Z_i\|^2 \mid I_{-\ell})=o_p(1)$ for the mean square limit.
\end{proof}

\begin{lemma}\label{lemma:num2}
Under Proposition~\ref{prop:known}'s conditions, $n^{1/2} \E_{n}(Z U ) \rightsquigarrow \mathcal{N}\{0, \E(U^2 Z Z^\top)\}$.
\end{lemma}

\begin{proof}
    By Theorem~\ref{theorem:disc_outcomes_incomplete}, $\E(U_iZ_i)=\E\{\E(U_i|R_i)Z_i\}=0$. Moreover, $\E(U_i^2Z_iZ_i^{\top})=\E\{\E(U_i^2|R_i)Z_iZ_i^{\top}\}\leq \bar{\sigma}^2_U \E(Z_iZ_i^{\top})$ since $\E(U_i^2|R_i)\leq \bar{\sigma}^2_U$ under our assumptions. Here, $\E(Z_iZ_i^{\top})$ is finite by Assumption~\ref{assumption:limit}, so we apply the Lindeberg-Levy central limit theorem.
\end{proof}

We are now ready to prove Proposition~\ref{prop:known}. Recall that 
$$
n^{1/2}(\widehat{\theta}-\theta)
= \{\E_n(\widehat{Z} W^\top)\}^{-1} n^{1/2}\E_n( \widehat{Z} U).
$$
By the continuous mapping theorem, Lemmas~\ref{lemma:denom1} and~\ref{lemma:denom2} imply $\E_n(\widehat{Z} W^\top)\overset{p}{\rightarrow}\E(Z W^\top)$. By Slutsky's theorem, Lemmas~\ref{lemma:num1} and~\ref{lemma:num2} imply $n^{1/2}\E_n( \widehat{Z} U)\rightsquigarrow \mathcal{N}\{0, \E(U^2 Z Z^\top)\}$. Overall, by Slutsky's theorem, we conclude that $n^{1/2}(\widehat{\theta}-\theta) \rightsquigarrow \mathcal{N}[0, \{\E(Z W^\top)\}^{-1} \E(U^2 Z Z^\top) \{\E(W Z^\top)\}^{-1}]$.

Finally, recall that for $\lambda = (y_1 - y_K, \hdots, y_{K-1} - y_K)^\top$, we have  $\widehat{\tau} =\lambda^{\top} \widehat{\theta}$ in Algorithm~\ref{alg:discrete_nocovar_overview} and $\tau =\lambda^{\top} \theta$  by definition. Therefore, $n^{1/2}(\widehat{\tau}-\tau)=n^{1/2}\lambda^{\top}(\widehat{\theta}-\theta)$ and we appeal to the delta method.

Efficiency follows directly from \cite{chamberlain1987asymptotic} and \cite{newey1993efficient}.  Specifically, $\widehat{\theta}$ is semiparametrically efficient for $\theta$ within the class of  estimators based upon the conditional moment $\E\{\Delta^e - (\Delta^o)^{\top}\theta|R\} = 0$. Therefore, $\widehat{\tau}$ is semiparametrically efficient for $\tau$, since they are known  linear transformations of $\widehat{\theta}$ and $\theta$, respectively. \qed

\subsubsection{Proof of Proposition~\ref{prop:unknown}} \label{proof:prop:unknown}

With unknown marginal probabilities, some extra care is required.
 
Extending the notation from the proof of Proposition~\ref{prop:known}, we let 
$
\widehat{Y}=\widehat{\Delta}^e$, $\widehat{W}=\widehat{\Delta}^o$, and $\widehat{U}=\widehat{Y}-\widehat{W}\theta.
$
If $i \in I_{\ell}$, then the marginal probabilities in $\widehat{Y}_i$ and  $\widehat{W}_i$ are estimated on the same fold $I_{\ell}$. Meanwhile, $\widehat{Z}_i = \widehat{H}_{\ell}(R_i)$ is constructed from $\widehat{H}_{\ell}$ estimated on the remaining folds $I_{-\ell}$.

The generalized decomposition is
$$
n^{1/2}(\widehat{\theta}-\theta)
=n^{1/2}\left[\{\E_n(\widehat{Z} \widehat{W}^\top)\}^{-1} \E_n( \widehat{Z} \widehat{Y})- \{\E_n(\widehat{Z} \widehat{W}^\top)\}^{-1} \E_n(\widehat{Z} \widehat{W}^\top) \theta   \right]
= \{\E_n(\widehat{Z} \widehat{W}^\top)\}^{-1} n^{1/2}\E_n( \widehat{Z} \widehat{U}).
$$
We derive the probability limit of $\E_n(\widehat{Z} \widehat{W}^\top)$, then the Gaussian approximation of $n^{1/2} \E_n( \widehat{Z} \widehat{U})$.

\begin{lemma}\label{lemma:denom1_unknown}
    Under Proposition~\ref{prop:unknown}'s conditions,  $\E_n(\widehat{Z}\widehat{W}^{\top})\overset{p}{\rightarrow}\E_n(Z\widehat{W}^{\top})$.
\end{lemma}

\begin{proof}
    The argument is similar to Lemma~\ref{lemma:denom1}, using $|\widehat{W}_{ij}|\leq \bar{W}'$ almost surely, which follows from our assumptions.
\end{proof}

\begin{lemma}\label{lemma:denom2_unknown}
    Under Proposition~\ref{prop:unknown}'s conditions,   $\E_n(Z\widehat{W}^{\top})\overset{p}{\rightarrow}\E_n(ZW^{\top})$.
\end{lemma}

\begin{proof}
We express the difference as 
$$\left\|\E_n\{Z(\widehat{W}-W)^{\top}\}\right\|_F \leq \left\{\E_n(\|\widehat{W}-W\|^2)\right\}^{1/2} \left\{\E_n(\|Z\|^2)\right\}^{1/2}, 
$$
where $||\cdot||_F$ indicates the Frobenius norm. 
Since $\E_n(\|Z\|^2)\overset{p}{\rightarrow}\E(\|Z\|^2)$ when $\E(\|Z\|^2)<\infty$ by the weak law of large numbers, it suffices to study 
$$\E_n(\|\widehat{W}-W\|^2)
=\frac{1}{L}\sum_{\ell=1}^L \frac{1}{n_{\ell}}\sum_{i\in I_{\ell}}\|\widehat{W}_i-W_i\|^2
=\frac{1}{L}\sum_{\ell=1}^L \E_{\ell}(\|\widehat{W}-W\|^2),\quad \E_{\ell}(\cdot)=\frac{1}{n_{\ell}}\sum_{i\in I_{\ell}}(\cdot).$$
Unpacking the notation of $\E_{\ell}(\|\widehat{W}-W\|^2)$,
\begin{align*}
    &\widehat{W}_k-W_k
    =\widehat{\Delta}^o_k-\Delta^o_k
    =\frac{1_{ Y=y_k, S = o}}{\E_{\ell}(1_{ Y=y_k, S = o})} - \frac{1_{ Y = y_K, S = o }}{\E_{\ell}(1_{ Y = y_K, S = o })}-\left\{\frac{1_{ Y=y_k, S = o}}{\E(1_{ Y=y_k, S = o})} - \frac{1_{ Y = y_K, S = o }}{\E(1_{ Y = y_K, S = o })}\right\} \\
    &= 1_{ Y=y_k, S = o} \left\{\frac{1}{\E_{\ell}(1_{ Y=y_k, S = o})}-\frac{1}{\E(1_{ Y=y_k, S = o})}\right\} -  1_{ Y = y_K, S = o } \left\{ \frac{1}{\E_{\ell}(1_{ Y = y_K, S = o })}-\frac{1}{\E(1_{ Y = y_K, S = o })}\right\} \\
    &=1_{ Y=y_k, S = o} \left\{\frac{\E(1_{ Y=y_k, S = o})-\E_{\ell}(1_{ Y=y_k, S = o})}{\E_{\ell}(1_{ Y=y_k, S = o})\E(1_{ Y=y_k, S = o})}\right\} -  1_{ Y = y_K, S = o } \left\{ \frac{\E(1_{ Y = y_K, S = o })-\E_{\ell}(1_{ Y = y_K, S = o })}{\E_{\ell}(1_{ Y = y_K, S = o })\E(1_{ Y = y_K, S = o })}\right\}.
\end{align*}
With population and empirical counts bounded away from zero,
$$
|\widehat{W}_k-W_k| \lesssim |\E(1_{ Y=y_k, S = o})-\E_{\ell}(1_{ Y=y_k, S = o})|+|\E(1_{ Y = y_K, S = o })-\E_{\ell}(1_{ Y = y_K, S = o })|.
$$
By Hoeffding's inequality and the union bound, with probability $1-\delta$,
$$
|\E(1_{ Y=y_k, S = o})-\E_{\ell}(1_{ Y=y_k, S = o})| \leq \left\{\frac{\ln(2/\delta)}{2n_{\ell}}\right\}^{1/2}.
$$
Therefore with probability $1-\delta$ for all $i\in[n]$ and $k\in[K]$ simultaneously,
$$
|\widehat{W}_{ik}-W_{ik}|\lesssim  2\left\{\frac{\ln(2K/\delta)}{2n_{\ell}}\right\}^{1/2} \lesssim \frac{\ln(2K/\delta)^{1/2}}{n_{\ell}^{1/2}}.
$$
We conclude that, under this union event that holds with probability $1-\delta$,
$$
\E_{\ell}(\|\widehat{W}-W\|^2) \lesssim \E_{\ell}\left\{K\frac{\ln(2K/\delta)}{n_{\ell}}\right\} = K\frac{\ln(2K/\delta)}{n_{\ell}}.
$$
For all $\delta>0$ and fixed $K$, this quantity vanishes for large $n_{\ell}=n/L$, so $\E_{\ell}(\|\widehat{W}-W\|^2)=o_p(1)$. The continuous mapping theorem implies the desired result.
\end{proof}

\begin{lemma}\label{lemma:denom3_unknown}
    Under Proposition~\ref{prop:unknown}'s conditions, $\E_n(ZW^{\top})\overset{p}{\rightarrow}\E(ZW^{\top})$.
\end{lemma}

\begin{proof}
    The argument is identical to Lemma~\ref{lemma:denom2}.
\end{proof}

\begin{lemma}\label{lemma:num1_unknown}
Under Proposition~\ref{prop:unknown}'s conditions, $n^{1/2} \E_n(\widehat{Z}\widehat{U}) \overset{p}{\rightarrow}n^{1/2} \E_n(Z\widehat{U})$.
\end{lemma}

\begin{proof}
    The argument is similar to Lemma~\ref{lemma:num1}, using up front that $|\widehat{U}_i|\leq \bar{U}'$ almost surely, which follows from our assumptions.
\end{proof}

\begin{lemma}\label{lemma:num2_unknown}
    Under Proposition~\ref{prop:unknown}'s conditions,  $n_{\ell}^{1/2} \E_{\ell}(Z\widehat{U})\rightsquigarrow \mathcal{N}(0,V)$ where $V=v^{\top}\Sigma v$, $\Sigma_{ij}=\cov(C_i,C_j)$, and
    $$
    C
    =
    \begin{pmatrix}
        C_1 \\ C_2  \\ C_3 \\ C_4 \\ C_5 \\ C_6 \\ \vdots \\ C_{2K+1} \\ C_{2K+2} \\ C_{2K+3} \\ C_{2K+4}
    \end{pmatrix}
 =\begin{pmatrix}
  1_{ D = 1, S = e }
   \\
   1_{ D = 1, S = e }Z
   \\
    1_{ D = 0, S = e } \\
    1_{ D = 0, S = e }Z \\
    1_{ Y=y_1, S = o } \\
    1_{ Y=y_1, S = o }Z \\
    \vdots \\
     1_{ Y=y_{K-1}, S = o } \\
    1_{ Y=y_{K-1}, S = o }Z \\
     1_{ Y = y_K, S = o }  \\
      1_{ Y = y_K, S = o }Z 
   \end{pmatrix},\quad 
v=
\begin{pmatrix}
- \E(C_2)^{\top}\E(C_1)^{-2}\\
\E(C_1)^{-1} I_{K-1} \\
\E(C_4)^{\top}\E(C_3)^{-2}\\
- \E(C_3)^{-1} I_{K-1} \\
\theta_1 \E(C_6)^{\top}\E(C_5)^{-2}\\
- \theta_1 \E(C_5)^{-1} I_{K-1} \\
\vdots\\
\theta_{K-1} \E(C_{2K+2})^{\top}\E(C_{2K+1})^{-2}\\
- \theta_{K-1} \E(C_{2K+1})^{-1} I_{K-1} \\
- \Big(\sum_{k=1}^{K-1}\theta_k\Big)\E(C_{2K+4})^{\top}\E(C_{2K+3})^{-2}\\
\Big(\sum_{k=1}^{K-1}\theta_k\Big)\E(C_{2K+3})^{-1} I_{K-1}
\end{pmatrix},
$$
where the odd entries are scalars and the even entries are vectors in $\mathbb{R}^{K-1}$.
\end{lemma}

\begin{proof}
  We unpack the definition of $\widehat{U}$:
\begin{align*}
    \widehat{U}
    &=\widehat{Y}-\widehat{W}^{\top}\theta \\
    &=\widehat{\Delta}^e-(\widehat{\Delta}^o)^{\top}\theta \\
    &=\frac{1_{ D = 1, S = e }}{\E_{\ell}(1_{ D = 1, S = e })} - \frac{1_{ D = 0, S = e }}{\E_{\ell}(1_{ D = 0, S = e })} 
    -\sum_{k=1}^{K-1}\left\{\frac{1_{ Y=y_k, S = o }}{\E_{\ell}(1_{ Y=y_k, S = o })} - \frac{1_{ Y = y_K, S = o }}{\E_{\ell}(1_{ Y = y_K, S = o })}\right\}\theta_k.
\end{align*}
Therefore, for $  h(C)=\frac{C_2}{C_1}-\frac{C_4}{C_3}-\sum_{k=1}^{K-1}\left(\frac{C_{2k+4}}{C_{2k+3}}-\frac{C_{2K+4}}{C_{{2K+3}}}\right)\theta_k$,
\begin{align*}
n_{\ell}^{1/2}\E_{\ell}(\widehat{U}Z)
&=n_{\ell}^{1/2}\left[\frac{\E_{\ell}(1_{ D = 1, S = e }Z)}{\E_{\ell}(1_{ D = 1, S = e })} - \frac{\E_{\ell}(1_{ D = 0, S = e }Z)}{\E_{\ell}(1_{ D = 0, S = e })} 
    -\sum_{k=1}^{K-1}\left\{\frac{\E_{\ell}(1_{ Y=y_k, S = o }Z)}{\E_{\ell}(1_{ Y=y_k, S = o })} - \frac{\E_{\ell}(1_{ Y = y_K, S = o }Z)}{\E_{\ell}(1_{ Y = y_K, S = o })}\right\}\theta_k\right] \\
    &=n_{\ell}^{1/2}h\{\E_{\ell}(C)\}.
\end{align*}

We make three observations. 
First, $\E(\|Z\|^2)<\infty$ by Assumption~\ref{assumption:limit}, so by the Lindeberg-Levy central limit theorem,
$
n_{\ell}^{1/2}\{\E_{\ell}(C)-\E(C)\}\rightsquigarrow \mathcal{N}(0,\Sigma).
$

Second, by the conditional moment equation,
\begin{align*}
    h\{\E(C)\}
    &=\frac{\E(1_{ D = 1, S = e }Z)}{\E(1_{ D = 1, S = e })}
    -\frac{\E(1_{ D = 0, S = e }Z)}{\E(1_{ D = 0, S = e })}
    -\sum_{k=1}^{K-1} \left\{\frac{\E(1_{ Y=y_k, S = o }Z)}{\E(1_{ Y=y_k, S = o })}-\frac{\E(1_{ Y = y_K, S = o }Z)}{\E(1_{ Y = y_K, S = o })}\right\}\theta_k \\
    &=\E [\{\Delta^e-(\Delta^o)^{\top}\theta\} Z]
    =\E [\E\{\Delta^e-(\Delta^o)^{\top}\theta|R\}Z ]
    =0.
\end{align*}
Third, the Jacobian is
$$
\nabla h(C)
=
\begin{pmatrix}
- C_2^{\top} C_1^{-2}\\
C_1^{-1} I_{K-1} \\
C_4^{\top} C_3^{-2}\\
- C_3^{-1} I_{K-1} \\
\theta_1 C_6^{\top} C_5^{-2}\\
- \theta_1 C_5^{-1} I_{K-1} \\
\vdots\\
\theta_{K-1} C_{2K+2}^{\top} C_{2K+1}^{-2}\\
- \theta_{K-1} C_{2K+1}^{-1} I_{K-1} \\
- \Big(\sum_{k=1}^{K-1}\theta_k\Big) C_{2K+4}^{\top} C_{2K+3}^{-2}\\
\Big(\sum_{k=1}^{K-1}\theta_k\Big) C_{2K+3}^{-1} I_{K-1}
\end{pmatrix}.
$$
Therefore, by the delta method,
$$
    n_{\ell}^{1/2}\E_{\ell}(Z\widehat{U})
    =n_{\ell}^{1/2}[h\{\E_{\ell}(C)\}- h\{\E(C)\}]
    \rightsquigarrow \mathcal{N} (0, [\nabla h\{\E(C)\}]^{\top}\Sigma [\nabla h\{\E(C)\}]).
$$
\end{proof}

We are now ready to prove Proposition~\ref{prop:unknown}. Recall that 
$$
n^{1/2}(\widehat{\theta}-\theta)
= \{\E_n(\widehat{Z} \widehat{W}^\top)\}^{-1} n^{1/2}\E_n( \widehat{Z} \widehat{U}).
$$
By the continuous mapping theorem, Lemmas~\ref{lemma:denom1_unknown},~\ref{lemma:denom2_unknown}, and~\ref{lemma:denom3_unknown} imply $\E_n(\widehat{Z} \widehat{W}^\top)\overset{p}{\rightarrow}\E(Z W^\top)$. By Slutsky's theorem, Lemmas~\ref{lemma:num1_unknown} and~\ref{lemma:num2_unknown} imply $n^{1/2}\E_n( \widehat{Z} \widehat{U})\rightsquigarrow \mathcal{N}(0, V)$. Overall, by Slutsky's theorem, we conclude that $n^{1/2}(\widehat{\theta}-\theta) \rightsquigarrow \mathcal{N}[0, \{\E(Z W^\top)\}^{-1} V \{\E(W Z^\top)\}^{-1}]$. The rest of the argument is identical to Proposition~\ref{prop:known}.

\qed

\section{Additional Theoretical Results and Discussion}
\subsection{Bias of Alternative Approaches: Extensions} \label{proofs:common_practice}

In Proposition \ref{prop:bias_common_practice}, we characterized the bias of a common two-step method when there are incomplete observational cases (Assumption \ref{assumption:observational}(ii)). 
In this appendix, we extend our characterization of the two-step method's bias in two ways: (i) complete observational cases, and (ii) settings where units are randomly assigned to the experimental or observational sample.

\subsubsection{Bias of Common Practice with Complete Observational Cases}\label{sec:bias_common_practice_complete}

Consider the setting of Assumption \ref{assumption:observational}(i) with complete observational cases. For simplicity, suppose there are no pretreatment covariates. 
The first step of the method is now to train a predictor $m_d(R)$ of the outcome $Y$ from the remotely sensed variable $R$ and treatment $D$ in the observational sample. The method implicitly targets the estimand 
$$\widetilde{\tau}^\prime = \widetilde{\mu}^\prime(1) - \widetilde{\mu}^\prime(0)=\E\{m_1(R) \mid D=1,S=e\}-\E\{m_0(R) \mid D=0,S=e\},\quad m_d(R):=\E(Y \mid R,D=d,S=o).$$

\begin{proposition}[Bias with complete observational cases]\label{proposition:common_practice_complete}
Suppose Assumptions \ref{assumption:experimental}, \ref{assumption:stability}, and \ref{assumption:observational}(i) hold,  with binary outcome and no covariates. 
Then, we have that 
\begin{enumerate}
\item[i.] $\widetilde{\mu}^\prime(d)=\mu(d)+ \E[Y \left\{ v_d(R,Y) - 1 \right\} \mid D = d, S = e]$, where $v_d(r,y) := \frac{\Pr(Y = y \mid D = d, S = o)}{\Pr(Y = y \mid D = d, S = e)} \frac{ f_{R }(r\mid D = d, S = e) }{ f_{R }(r\mid D = d, S = o) }$;

\item[ii.] there exists $\kappa_d \in [0, 1)$ such that $\widetilde{\mu}^\prime(d) = \kappa_d \mu(d)+(1 - \kappa_d) \E(Y \mid D = d, S = o)$, as long as $R$ is an imperfect predictor of $Y$ in the treatment arm $D=d$, i.e. as long as $\var(Y \mid R, D = d, S = o)>0$ almost surely.
\end{enumerate}
\end{proposition}

\begin{proof} See Appendix \ref{proof:proposition:common_practice_complete}. 
\end{proof} 

The common practice is biased towards the conditional expectation of the outcome in the observational sample.

\subsubsection{Bias of Common Practice with Random Sample Selection}\label{sec:bias_random_samples}

We now verify that the two-step method is biased, even when the researcher randomly assigns units to the experimental and observational samples. As in Proposition~\ref{prop:bias_common_practice}, consider the setting with binary outcome, no covariates, and incomplete observational cases (Assumption \ref{assumption:observational}(ii)). 

In this thought experiment, the researcher implements the following study. First, they randomize units into the experimental or observational sample. Then, they randomize experimental units to treatment or control. No observational units are treated.

\begin{proposition}[Bias with random sample selection]\label{prop:bias_common_practice_random}
Suppose the conditions of Proposition \ref{prop:bias_common_practice} hold. Further assume $S \indep \{Y(0), Y(1)\}$. 
Then, we have that:
\begin{enumerate}
\item[i.] $\widetilde{\mu}(0)= \mu(0)$ and $\widetilde{\mu}(1)= \mu(1)+ \E\left[Y \left\{w_1(R,Y) - 1 \right\}\mid D = 1, S = e \right]$, where $w_1(r,y):=\frac{\Pr\{Y(0) = y|S =e\}}{\Pr\{Y(1) = y|S =e\}} \frac{f_{R }(r\mid D = 1, S = e)}{f_{R }(r\mid D = 0, S = e)}$.

\item[ii.] there exists a scalar $\kappa \in [0, 1)$ such that $\widetilde{\mu}(1) = \kappa\mu(1)+(1-\kappa)\mu(0)$, as long as $R$ is an imperfect predictor of $Y$, i.e. as long as $\var(Y \mid R,S=o)>0$ almost surely.
\end{enumerate}
\end{proposition}

\begin{proof} See Appendix  \ref{proof:prop:bias_common_practice_random}. 
\end{proof} 

Even when units are randomly assigned to the experimental and observational samples, the two step method yields an estimate of the average treated potential outcome  that is biased towards the average untreated potential outcome. As such, it attenuates the treatment effect.

\subsection{Quasi-Experiments with Remotely Sensed Outcomes}\label{sec:other_id_strategies}

In this section, we describe how our main identifying assumption (Assumption \ref{assumption:stability}(i)) can be used to identify treatment effects in quasi-experimental samples via data combination with the observational sample.
We discuss two quasi-experimental strategies: instrumental variables and difference-in-differences.
To ease exposition, we discuss Example~\ref{ex:binary_y}; we consider a binary outcome, omit pre-treatment covariates, and focus on ``incomplete'' observational cases (Assumption~\ref{assumption:observational}(ii)). 
The generalization to a discrete outcome with discrete covariates, as in Section \ref{sec:identification} of the main text, is straightforward. 

\subsubsection{Instrumental Variables}

Each unit is now characterized by the random vector 
$$\{S, Z, D(0), D(1), Y(0, 0), Y(0, 1), Y(1, 0), Y(1, 1), R\},$$ where $Z \in \{0, 1\}$ is a binary instrument, $D(z)$ are potential treatments and $Y(d, z)$ are potential outcomes, following \citet[][]{AngristImbens(94)}. 
For units in the quasi-experimental sample $(S = e)$, we observe $(Z, D, R)$. 
For units in the observational sample $(S = o)$, we observe $(Y, R)$ since we focus on incomplete observational cases.

We modify our previous assumption about the experimental sample to accommodate the instrument. This sample may be viewed as a quasi-experiment, or as an experiment with imperfect compliance.

\begin{assumption}[Instrumental variable]\label{assumption: IV assumptions}
Suppose the following:
\begin{enumerate}[label=\roman*.]
\item Instrument exclusion: for all $z \in \{0, 1\}$ and $d \in \{0, 1\}$, $Y(d, z) := Y(d)$ almost surely.
\item SUTVA: $D = Z D(1) + (1 - Z) D(0)$ and $Y = D Y(1) + (1 - D) Y(0)$ almost surely.
\item Instrument randomization: $Z \indep \{ D(0), D(1), Y(0), Y(1) \} \mid S = e$. 
\item Overlap: $\Pr(Z = 1 \mid S = e)$ and $P(Y = 1\mid S = o)$  are bounded away from zero and one almost surely.
\item Monotonicity: $\Pr\{D(1)\geq D(0)\mid S=e\}=1$ and $\Pr\{D(1)>D(0) \mid S=e\}>0$.
\end{enumerate}
\end{assumption}

\noindent Under Assumption \ref{assumption: IV assumptions}, if we were to observe the outcome in the quasi-experimental sample, then the local average treatment effect (LATE) $\theta^{LATE} := \E\{Y(1) - Y(0) \mid D(1) > D(0), S=e\}$ could be identified using standard arguments. 
Since the outcome is unobserved in the quasi-experimental sample, we will use the observational sample as in the main text.

We next modify our stability and observational completeness assumptions to accommodate the instrument. 

\begin{assumption}[Stability with an instrument]\label{assumption:stability, IV}
Suppose Assumption \ref{assumption:stability} holds replacing $D$ with $(Z,D)$.
\end{assumption}

\begin{assumption}[No direct effects with an instrument]\label{assumption:no direct effects, IV}
Suppose Assumption \ref{assumption:observational}(ii) holds replacing $D$ with $(Z,D)$.
\end{assumption}

\noindent Assumption \ref{assumption:stability, IV} remains the main assumption of our framework: $S\indep R \mid Z,D,Y$. In the presence of an instrument, stability now requires that the conditional distribution of the remotely sensed variable $R$ given $(Z, D, Y)$ is stable across the quasi-experimental and observational samples. 
Analogously, Assumption \ref{assumption:no direct effects, IV} becomes $Z,D \indep R \mid Y$. It now requires that the instrument $Z$ and the treatment $D$ only affect the remotely sensed variable $R$ via their effect on the outcome $Y$.
Together Assumption \ref{assumption:stability, IV} and Assumption \ref{assumption:no direct effects, IV} imply that $(S, D, Z) \indep R \mid Y$, which is a testable implication as before. 

These conditions allow us to identify $\theta^{LATE}$ by combining the quasi-experimental and observational samples. 
As notation, let $\alpha(z) = \E(Y \mid S = e, Z = z)$, and $\beta(z) := \E(D \mid S = e, Z = z)$. By standard arguments, under Assumption \ref{assumption: IV assumptions}, $\theta^{LATE} = \frac{\alpha(1) - \alpha(0)}{\beta(1) - \beta(0)}$. 
Of course, $\beta(0)$ and $\beta(1)$ are identified from the quasi-experimental sample since $(Z,D)$ are observed. 
We will therefore identify $\alpha(0)$ and $ \alpha(1)$ by combining the quasi-experimental and observational samples under Assumptions \ref{assumption:stability, IV} and~\ref{assumption:no direct effects, IV}. 

As a stepping stone, we first identify $\alpha(d, z) := \E(Y \mid S = e, D = d, Z = z)$.

\begin{theorem}[Identification with an instrument]\label{theorem: IV identification with RSV}
Suppose Assumptions \ref{assumption: IV assumptions}, \ref{assumption:stability, IV}, and \ref{assumption:no direct effects, IV} hold. Then, for any $z \in \{0, 1\}$ and $d \in \{0, 1\}$ satisfying $\Pr(D = d \mid Z = z, S = e) > 0$, 
$$
\mathbb{E}[\Delta^e(d,z) - \alpha(d,z) \Delta^o|R] = 0 \mbox{ almost surely},
$$
where $\Delta^e(d, z) := \frac{1\{ D = d, Z = z, S = e \}}{\Pr(D = d, Z = z, S = e)} - \frac{1\{ Y = 0, S = o \}}{\Pr(Y = 0, S = o)}$ and $\Delta^o := \frac{1\{ Y = 1, S = o \}}{\Pr(Y = 1, S = o)} - \frac{1\{ Y = 0, S = o \}}{\Pr(Y = 0, S = o)}$.
\end{theorem}

\begin{proof} See Appendix  \ref{proof:theorem: IV identification with RSV}. 
\end{proof} 
\begin{corollary}[Representation with an instrument]\label{corollary: IV representation}
Under Theorem \ref{theorem: IV identification with RSV}'s conditions, for any $z \in \{0, 1\}$, $d \in \{0, 1\}$ satisfying $\Pr(D = d \mid Z = z, S = e) > 0$ and any representation $H(R)$ with $\E\{H(R) \Delta^o\} \neq 0$, 
$$
    \alpha(d, z) = \frac{\E\{H(R) \Delta^{e}(d, z)\}}{\E\{H(R) \Delta^{o}\}}. 
$$
\end{corollary}

Theorem \ref{theorem: IV identification with RSV} and Corollary \ref{corollary: IV representation} immediately imply that $\alpha(0)$ and $\alpha(1)$ are identified; for $z \in \{0, 1\}$, the law of iterated expectations gives $\alpha(z) = \alpha(1, z) \beta(z) + \alpha(0, z) \{1 - \beta(z)\}$. Therefore, $\theta^{LATE}$ is also identified. 
Estimation and inference can then follow by suitably stacking moments and extending the discussion provided in Appendix \ref{sec:est_inf_direct_effects}.

\subsubsection{Two-Period Difference-in-Differences}

Next, we consider a setting with two periods $t\in\{1, 2\}$. 
Treated units ($D = 1$) receive a treatment between period $t = 1$ and $t = 2$, and untreated units ($D = 0$) remain untreated in both periods.
Each unit is characterized by the random vector 
$$\{S, D, Y_1(0), Y_1(1), Y_2(0), Y_2(1), R_1, R_2\}.$$
For units in the quasi-experimental sample $(S = e)$, we observe $(D, R_1, R_2)$,
For units in the observational sample $(S = o)$, we observe $(Y_1, Y_2, R_1, R_2)$. 

We again modify our previous assumption about the experimental sample, this time to accommodate the panel structure.

\begin{assumption}[Difference-in-differences]\label{assumption: DID assumptions}
Suppose the following:
\begin{enumerate}[label=\roman*.]
\item SUTVA: For $t\in\{1, 2\}$, $Y_t = D Y_t(1) + (1 - D) Y_t(0)$.
\item Overlap: $\Pr(D=1 \mid S=e)$ and $\Pr(Y_t=1 \mid S=o), t \in \{1,2\}$ are bounded away from zero and one.
\item Parallel trends: $\E\{Y_{2}(0) - Y_{1}(0) \mid S=e, D = 1\} = \E\{Y_2(0) - Y_1(0) \mid S=e, D = 0\}$.
\item No anticipation: $\Pr\{Y_1(0) = Y_1(1) \mid S=e,D=1\}=1$.
\end{enumerate}
\end{assumption}

Under Assumption \ref{assumption: DID assumptions}, if we were to observe the outcomes in the quasi-experimental sample, then the average treatment effect on the treated $\theta^{ATT} := \E\{Y_2(1) - Y_2(0) \mid S=e, D = 1\}$ could be identified using standard arguments. 
Assumption \ref{assumption: DID assumptions} is akin to a parallel trends-type assumption on the cumulative distribution functions for untreated potential outcomes \citep[][]{RothSantAnna(23)}.

We modify our stability and observational completeness assumptions to accommodate the panel setting. 

\begin{assumption}[Stability for difference-in-differences]\label{assumption:stability, DID}
Suppose Assumption \ref{assumption:stability} holds for each time period $t \in \{1, 2\}$.
\end{assumption}

\begin{assumption}[No direct effects for difference-in-differences]\label{assumption:no direct effects, DID}
Suppose Assumption \ref{assumption:observational}(ii) holds for each time period $t \in \{1, 2\}$.
\end{assumption}

\noindent In other words, we impose the same assumptions as in the main text, period by period. 
The main conditions are $S\indep R_t \mid D,Y_t$ and $D\indep R_t \mid Y_t$.
Together, Assumption \ref{assumption:stability, DID} and Assumption \ref{assumption:no direct effects, DID} imply that, for each $t\in\{1, 2\}$, $(S, D) \indep R_t \mid Y_t$, which is a testable implication as before. 

These conditions allow us to identify $\theta^{ATT}$ by combining the quasi-experimental and observational samples. 
As notation, let $\alpha_t(d) = \E(Y_t \mid S = e, D = d)$. By standard arguments, under Assumption \ref{assumption: DID assumptions}, $\theta^{ATT} = \left\{ \alpha_2(1) - \alpha_1(1) \right\} - \left\{ \alpha_2(0) - \alpha_1(0) \right\}$. 
We will identify $\alpha_t(d)$ for $t \in \{1, 2\}$ and $d \in \{0, 1\}$ by combining the quasi-experimental and observational samples, under Assumption \ref{assumption:stability, DID} and Assumption \ref{assumption:no direct effects, DID}.

\begin{theorem}[Identification for difference-in-differences]\label{theorem: DID identification with RSV}
Suppose Assumptions~\ref{assumption: DID assumptions}, \ref{assumption:stability, DID} and \ref{assumption:no direct effects, DID} hold. 
Then, for any $t \in \{1, 2\}$ and $d \in \{0, 1\}$, 
$$
\E\{\Delta_t^e(d) - \Delta_t^o \alpha_t(d) \mid R_t\} = 0\mbox{ almost surely},
$$
where $\Delta_t^e(d) := \frac{1\{D = d, S = e\}}{\Pr(D = d, S = e)} - \frac{1\{Y_t = 0, S = o\}}{\Pr(Y_t = 0, S = o)}$ and $\Delta_t^o := \frac{1\{Y_t = 1, S = o\}}{\Pr(Y_t = 1, S = o)} - \frac{1\{Y_t = 0, S = o\}}{\Pr(Y_t = 0, S = o)}$.
\end{theorem}

\begin{proof} See Appendix  \ref{proof:theorem: DID identification with RSV}. 
\end{proof} 
\begin{corollary}\label{corollary: DID representation}
Under Theorem \ref{theorem: DID identification with RSV}'s conditions, for any $t \in \{1, 2\}$, $d \in \{0, 1\}$ and representation $H_t(R_t)$ with $\E\{H_t(R_t) \Delta_t^o\} \neq 0$, 
$$
\alpha_t(d) = \frac{\E\{H_t(R_t) \Delta_t^e(d)\}}{\E\{H_t(R_t) \Delta_t^o\}}.
$$
\end{corollary}

Theorem \ref{theorem: DID identification with RSV} and Corollary \ref{corollary: DID representation} imply that $\alpha_t(d)$ are identified. Therefore, $\theta^{ATT}$ is also identified. 
Estimation and inference can again follow by suitably stacking moments and extending the discussion provided in Appendix \ref{sec:est_inf_direct_effects}.

Our two-period difference-in-differences identification result extends to a setting with multiple periods and a common treatment adoption date. For each period, we apply the same argument to identify $\alpha_t(d)=\E(Y_t \mid S=e,D=d)$ from the remotely sensed variable $R_t$, and then form the desired multi-period contrast by differencing these identified means across time periods and groups.
In a staggered-adoption design (e.g., cohorts $G$ with $D_t = 1\{t\ge G\}$), the same period-by-period argument identifies the means $\mathbb E(Y_t\mid S=e,D_t=d)$. We can then substitute these identified means into  staggered difference-in-differences aggregations, such as \citet[][]{dCdH-DiD} and \citet[][]{CallawaySantanna}, built from the corresponding period-by-period contrasts.

\subsection{Some Spillovers}\label{sec:spillover}

\subsubsection{Generalized Potential Outcomes}

With spillover effects, we write the potential outcome for unit $i$ as $Y_i(d_i,\mathbf{d}_{-i})$. The potential outcome is indexed by the unit of interest's treatment assignment $d_i\in\{0,1\}$, as well as the vector of other units' treatment assignments $\mathbf{d}_{-i}\in\{0,1\}^{n-1}$. 

From this potential outcome, we define the direct potential outcome  by taking the expectation over other units' treatment assignments: $Y_i^{\mathrm{dir}}(d_i)=\E_{\mathbf{D}_{-i}}\{Y_i(d_i,\mathbf{D}_{-i}\}| D_i=d_i,S_i=e\}$. 
Note that it remains random due to unobserved heterogeneity.\footnote{Formally, we may denote unobserved heterogeneity by $\eta_i$ and use the nonseparable model notation $Y_i=Y_i(D_i,\mathbf{D}_{-i},\eta_i)$. Then the potential outcome is $Y_i(d_i,\mathbf{d}_{-i},\eta_i)$ and the direct potential outcome is $Y_i^{\mathrm{dir}}(d_i,\eta_i)=\int Y_i(d_i,\mathbf{d}_{-i},\eta_i)\}\mathrm{d}\Pr(\mathbf{d}_{-i}|D_i=d_i,S_i=e,\eta_i)$. \label{footnote: unobs het}}
The direct potential outcome may be viewed as design-based, since the expectation is over the design‑induced distribution of $\mathbf{D}_{-i}$ conditional on $D_i=d$ in the experiment.

From direct potential outcomes, we define the average direct treatment effect. We relax the assumption of identical distribution across observations that is maintained throughout the main text. As such, it is a sample average direct treatment effect, though we omit ``sample'' for brevity.

\begin{definition}[Average direct treatment effect]
    The average direct treatment effect in the experimental sample is
    $
    \theta_n=\frac{1}{n}\sum_{i=1}^n \E\{Y_i^{\mathrm{dir}}(1)-Y_i^{\mathrm{dir}}(0)|S_i=e\}.
    $
\end{definition}

The parameter $\theta_n$ may be viewed as an average direct effect, because it involves counterfactuals for unit $i$ under a ``direct'' intervention on unit $i$'s treatment assignment.

We now generalize Assumption~\ref{assumption:experimental} for the setting with spillovers.  For clarity of exposition, we focus on the setting of Example~\ref{ex:binary_y}.

\begin{assumption}[Spillovers]\label{assumption:experimental_spill}
    Suppose the following:
    \begin{enumerate}[label=\roman*.]
    \item Spillovers: If $D_i=d_i$ and $\mathbf{D}_{-i}=\mathbf{d}_{-i}$ then $Y_i=Y_i(d_i,\mathbf{d}_{-i})$.
\item Randomization: $D_i \indep \left\{ Y_i(d_i,\mathbf{d}_{-i})\right\} \mid   S_i = e$.\footnote{In nonseparable model notation, $D_i\indep \eta_i |S_i=e$.}
\item Overlap: $\Pr(D_i = 1 \mid  S_i = e) $ is bounded away from zero and one almost surely.
    \end{enumerate}
\end{assumption}

\begin{lemma}\label{lemma:diff}
 Under Assumption~\ref{assumption:experimental_spill}, the average direct treatment effect equals a difference of outcome means. Formally,
 $
 \theta_n=\frac{1}{n}\sum_{i=1}^n\{\E(Y_i|D_i=1,S_i=e)-\E(Y_i|D_i=0, S_i=e)\}.
 $
\end{lemma}

\begin{proof}
    See Appendix~\ref{proof:lemma:diff}.
\end{proof}

If the outcomes were perfectly observed in the experimental sample, then $\theta_n$ would be identified by Lemma~\ref{lemma:diff}. 

Our benchmark estimator in the semi-synthetic exercise is the empirical analogue to this expression. It is the benchmark that an economist would obtain if they could fully observe the treatments and outcomes in the experiment. By Lemma~\ref{lemma:diff}, the benchmark is an unbiased estimator of a reasonable estimand, even in the presence of spillovers.

As before, the crux of our problem is that the outcomes are not perfectly observed in the experimental sample. Instead, we have access to an auxiliary, observational sample. Our method recovers $\theta_n$ under appropriate modifications of our identifying assumptions. We now extend Assumptions~\ref{assumption:stability} and~\ref{assumption:observational}(ii) accordingly. 

\begin{assumption}[Stability with spillovers]\label{assumption:stability_spill}
 Suppose Assumption~\ref{assumption:stability} holds for each unit $i\in \{1,...,n\}$.
\end{assumption}

\begin{assumption}[No direct effects with  spillovers]\label{assumption:observational_spill}
    Suppose Assumption~\ref{assumption:observational}(ii) holds for each unit $i\in\{1,...,n\}$.
\end{assumption}

Most importantly, we modify stability (Assumption~\ref{assumption:stability}). In the presence of spillovers, stability requires $S_i\indep R_i\mid (Y_i,D_i,X_i)$: conditional on a unit's \textit{own} outcome, treatment, and covariates, the distribution of the unit's remotely sensed variable $R_i$ would be the same had it belonged to the experimental or observational samples. In particular, the unit's remotely sensed variable distribution depends on its own outcome, but does not depend on other units' outcomes.

In our semi-synthetic exercise, unit $i$ is a village. Treatments are randomized at the mandal level.  A mandal is a sub-district containing villages, roughly comparable to a U.S. county.
Substantively, this modified stability assumption means that the satellite image of a village depends only on that village's poverty level and on the treatment status of the mandal in which the village lies. It does not depend on the poverty levels of other villages, nor on the treatment statuses of other mandals. We think of this as a plausible approximation; intuitively the satellite image of a village should only depend on variables observed in that village.

In summary, our method extends as long as spillovers are in treatment effects, not in the distribution of the remotely sensed variable. We formalize this claim in what follows.

\begin{theorem}[Identification with spillovers]\label{theorem:spill} Suppose Assumptions~\ref{assumption:experimental_spill},~\ref{assumption:stability_spill}, and~\ref{assumption:observational_spill} hold. Then
$$
\theta_n=\frac{1}{n}\sum_{i=1}^n \frac{\E(\Delta^{e}_i|R_i)}{\E(\Delta_i^o|R_i)}\quad \text{almost surely}, 
$$
    where $\Delta^{e}_i= \frac{ 1(D_i = 1, S_i = e)}{ \Pr(D_i = 1, S_i = e) }-\frac{ 1(D_i = 0, S_i = e)}{ \Pr(D_i = 0, S_i = e) }$ and $\Delta^o_i=
\frac{1(Y_i = 1, S_i = o) }{\Pr(Y_i = 1, S_i = o )}-
\frac{1(Y_i = 0, S_i = o) }{\Pr(Y_i = 0, S_i = o )}.$
\end{theorem}

\begin{proof}
    See Appendix~\ref{proof:theorem:spill}.
\end{proof}

\subsubsection{Average Global Treatment Effect}

Theorem~\ref{theorem:spill} identifies the average direct treatment effect, which is a reasonable estimand. We now ask: when does this reasonable estimand coincide with the standard estimand for models with spillovers, namely the average global treatment effect? We describe two plausible and empirically relevant scenarios. In these two scenarios, our method estimates the average global treatment effect in the presence of spillovers.

\begin{definition}[Average global treatment effect]
    The average global treatment effect in the experimental sample is $\tilde{\theta}_n=\frac{1}{n}\sum_{i=1}^n \E\{Y_i(1,\mathbf{1}_{n-1})-Y_i(0,\mathbf{0}_{n-1})|S_i=e\}$, where $\mathbf{1}_{n-1}$ and $\mathbf{0}_{n-1}$ are vectors of repeated entries in $\R^{n-1}$.
\end{definition}

The parameter $\tilde{\theta}_n$ may be viewed as an average global effect, because it involves counterfactuals for unit $i$ under a ``global'' intervention on all units' treatment assignments.

\begin{corollary}[Spillovers within but not across mandals]\label{cor:spill1}
   Suppose the assumptions of Theorem~\ref{theorem:spill} hold. 
   Suppose that each unit $i$ is a village, and that treatment assignment is randomized at the mandal level, as in the real experiment we study. 
   Now, further assume that spillovers only occur within mandals. Then $\tilde{\theta}_n=\theta_n=\frac{1}{n}\sum_{i=1}^n \frac{\E(\Delta^{e}_i|R_i)}{\E(\Delta_i^o|R_i)}$.
\end{corollary}

\begin{proof}
    See Appendix~\ref{proof:cor:spill1}.
\end{proof}

For simplicity, Corollary~\ref{cor:spill1} assumes that there are no spillovers across mandals. In fact, our argument will go through as long as spillovers across mandals are asymptotically negligible.

Next, we consider an alternative  restriction on spillovers: that they only occur within a fixed geographic radius. This alternative restriction is also widely adopted in the empirical literature. For example, \cite{muralidharan2023general} posit a radius of $20$ km. After appropriately subsetting villages, our method once again estimates the average global treatment effect.

\begin{corollary}[Spillovers within a geographic radius]\label{cor:spill2}
Suppose the assumptions of Theorem~\ref{theorem:spill} hold. 
Suppose that each unit $i$ is a village, and that treatment assignment is randomized at the mandal level, as in the real experiment we study. 
Now, further assume that spillovers only occur within a geographic radius of $20$ km. Then $\tilde{\theta}_n$ coincides with $\theta_n$ after dropping from the experimental sample (both in the definition of $\tilde{\theta}_n$ and $\theta_n$) any village that is within $20$ km of another village with the opposite treatment assignment.
\end{corollary}

\begin{proof}
    See Appendix~\ref{proof:cor:spill2}.
\end{proof}

By assuming no spillovers beyond a $20$ km radius, our method estimates the average global treatment effect from a subset of the sample. On the one hand, this assumption is closer to that in the empirical literature. On the other hand, it comes at the cost of a smaller effective sample size.

\subsection{Variables Collected at Different Times}\label{sec:timing}

In the main text, we avoided time indexing and viewed the remotely sensed variable as a post-outcome variable. In real data, such as the Smartcard illustration, the data have time indices. In particular, the remotely sensed variable and the outcome may be measured at different times. Moreover, the treatment variable may be defined as early adoption, raising the question of whether our method applies to such a setting. We now confirm that it does, and clarify the interpretation of our reported estimate.

To ease exposition, we discuss Example~\ref{ex:binary_y}; we consider a binary outcome, omit pre-treatment covariates, and focus on ``incomplete'' observational cases (Assumption~\ref{assumption:observational}(ii)). The generalization to a discrete outcome with discrete covariates, as in Section \ref{sec:identification} of the main text, is straightforward. 

Suppose each unit is independent and identically distributed. Suppose early adoption of the treatment is randomly assigned at time $t$, the outcome is collected at time $t'\geq t$, and the remotely sensed variable is collected at time $t''\geq t'$. In particular, only the treated experimental units receive treatment at time $t$, while all remaining units receive treatment at time $t'$. Overall, each unit is now characterized by the random vector
$$
\{S,D_t,Y_{t'}(0),Y_{t'}(1),R_{t''}\}
$$
where $Y_{t'}(d_t)$ are potential outcomes. For units in the experimental sample $(S=e)$, we observe $(D_t,R_{t''})$. For units in the observational sample $(S=o)$, we observe $(Y_{t'},R_{t''})$. In this setting, $D_{t'}=1$ and $D_{t''}=1$ for all units, so we omit them. 

The Smartcards illustration takes this form. Simplifying some of the details in Appendix~\ref{section: appendix, smartcards data, details}, only treated units in the experimental sample received Smartcards at time $t=2010$. Untreated units in the experimental sample received Smartcards by the time the outcomes were collected in $t'=2013$. All remaining units, i.e. the observational sample, received Smartcards in $t'=2013$ as well. Satellite images were collected at the later time $t''=2019$.

The parameter of interest is defined from the time $t'$ potential outcomes. In the Smartcards illustration, it is the effect of  $2010$  Smartcards adoption on $2013$ poverty levels.

\begin{definition}[Average time-specific treatment effect]
    The average time-specific treatment effect in the experimental sample is $\theta_{t'}=\E\{Y_{t'}(1)-Y_{t'}(0)|S=e\}$.
\end{definition}

We modify our assumption about the experimental sample to reflect the time variation in the measurement of the outcome and remotely sensed variable. 
\begin{assumption}[Experiment with time variation]\label{assumption:experiment_timing}
Suppose the following:
    \begin{enumerate}[label=\roman*.]
\item SUTVA: $Y_{t'} = D_{t} Y_{t'}(1) + (1 - D_t) Y_{t'}(0)$ almost surely.
\item Randomization: $D_t \indep \left\{ Y_{t'}(0), Y_{t'}(1) \right\} \mid S = e$.
\item Overlap: $\Pr(D_t = 1 \mid S = e) $ is bounded away from zero and one almost surely.
\item Ultimate adoption: $D_{t'}=1$ and $D_{t''}=1$ almost surely.
\end{enumerate}
\end{assumption}

Next, we modify our assumptions on stability and on the observational sample to reflect the time variation in the measurement of the outcome and remotely sensed variable. 

\begin{assumption}[Stability with time variation]\label{assumption:stability_timing}
    Suppose Assumption~\ref{assumption:stability} holds replacing $(D,Y,R)$ with $(D_t,Y_{t'},R_{t''})$.
\end{assumption}

Assumption~\ref{assumption:stability_timing} remains the main assumption of our framework: $S\indep R_{t''}|D_t,Y_{t'}$. In an experiment with time variation, stability now requires that the conditional distribution of the remotely sensed variable $R_{t''}$ given $(D_t,Y_{t'})$ is stable across the experimental and observational samples. Importantly, the remotely sensed variable may be collected at a later time than the outcome. Figure~\ref{fig:smartcards_stability_other_outcomes} provides empirical evidence supporting this assumption.

\begin{assumption}[No direct effects with time variation]\label{assumption:observational_timing}
    Suppose Assumption~\ref{assumption:observational}(ii) holds replacing $(D,Y,R)$ with $(D_t,Y_{t'},R_{t''})$.
\end{assumption}

Assumption~\ref{assumption:observational_timing} becomes $D_t\indep R_{t''}|Y_{t'}$. 
Given the time variation in the measurement of the outcome and the remotely sensed variable, this amounts to a Markov property: the effect of the earlier treatment $D_t$ on the later remotely sensed variable $R_{t''}$ must operate only though its effect on the intermediate outcome $Y_{t'}$. Importantly, $R_{t''}$ may depend on later outcomes $Y_{t''}$, but any effect from the earlier treatment $D_{t}$ must be via the intermediate outcome $Y_{t'}$. 

Concretely, consider the following example: 2010 Smartcards may affect 2013 consumption; 2013 consumption may affect 2019 consumption; and 2013 and 2019 consumption may affect 2019 satellite images. However, in this example, 2010 Smartcards cannot affect 2019 consumption in any channel besides 2013 consumption. This is one possible example that satisfies the Markov restriction. Figure~\ref{fig:no_direct_effects_other_outcomes} provides empirical evidence supporting this assumption.

\begin{theorem}[Identification with time variation]\label{thm:timing}
    Suppose Assumptions~\ref{assumption:experiment_timing},~\ref{assumption:stability_timing}, and~\ref{assumption:observational_timing} hold. Then
    $$
\theta_{t'}=\frac{\E(\Delta^e_t|R_{t''})}{\E(\Delta^o_{t'}|R_{t''})}\quad \text{almost surely,}
$$
where 
    $\Delta^{e}_t= \frac{ 1(D_t = 1, S = e)}{ \Pr(D_t = 1, S = e) }-\frac{ 1(D_t = 0, S = e)}{ \Pr(D_t = 0, S = e) }$
    and
    $\Delta^o_{t'}=
\frac{1(Y_{t'} = 1, S = o) }{\Pr(Y_{t'} = 1, S = o )}-
\frac{1(Y_{t'} = 0, S = o) }{\Pr(Y_{t'} = 0, S = o )}.$
\end{theorem}

\begin{proof}
   See Appendix~\ref{proof:thm:timing}.
\end{proof}

In summary, our results directly apply when $D_{t}$ is an early adoption treatment, and when $Y_{t'}$ and $R_{t''}$ are an outcome and a remotely sensed variable collected later on. 
In the main text, we report the effect of $2010$ Smartcards adoption on $2013$ consumption.
With incomplete cases, there is an implicit Markov restriction that researchers must asses. However, with complete cases, this additional restriction may be relaxed.

\subsection{Partial Identification with Approximate  Assumptions}\label{sec:partial_id}

Our point identification results in Section~\ref{sec:identification} rely on two key conditions beyond experimental unconfoundedness: stability of the remotely sensed variable (Assumption~\ref{assumption:stability}) and, with incomplete observational cases, no direct effects of the treatment on the remotely sensed variable (Assumption~\ref{assumption:observational}(ii)).  
Sometimes researchers may believe these assumptions hold only approximately. For example, the treatment may slightly affect the remotely sensed variable through channels other than the outcome, or the sensing mechanism may vary modestly across samples.
In this section, we develop partial identification results that allow researchers to assess how sensitive their conclusions are to small departures from our assumptions.

We proceed in two steps.  
First, in Section~\ref{sec:partial_direct}, we maintain stability but relax the no-direct-effects condition, bounding the magnitude of any direct effect. The direct effect bias enters additively.  
Second, in Section~\ref{sec:partial_id2}, we also relax stability. The stability violation enters multiplicatively.

For simplicity, we focus on the special case of Example~\ref{ex:binary_y}: a binary outcome, no covariates, and incomplete observational cases. 
Our results extend pointwise in $x$ for discrete covariates. 
Recall that the treatment effect of interest is
$$
\tau=\mu(1)-\mu(0),\quad \mu(d)=\E\{Y(d)|S=e\}.
$$
Recall that, in this special case, the experimental and observational variation are
$$
\Delta^e := \frac{1\{D = 1, S = e\}}{\Pr(D = 1, S = e)} - \frac{1\{ D = 0, S = e \} }{\Pr(D = 0, S = e)},\quad  \Delta^o = \frac{1\{ Y = 1, S = o \}}{ \Pr(Y = 1, S = o)} - \frac{1\{Y = 0, S = o\}}{ \Pr(Y = 0, S = o) }.
$$
Moreover, no
treatment variation is observed in the observational sample: $\Pr(D = 0 \mid S = o) = 1$.

\subsubsection{Approximately No Direct Effects}\label{sec:partial_direct}

We first consider violations of Assumption~\ref{assumption:observational}(ii), allowing the treatment $D$ to have a direct effect on the remotely
sensed variable $R$. 
For any measurable function $H$ with $\E\{\Delta^o H(R)\} \neq 0$, define the quantity
$$
  \widetilde{\tau}(H) := \frac{\E\{\Delta^e H(R)\}}{\E\{\Delta^o H(R)\}}
$$
When Assumption~\ref{assumption:observational}(ii) (no direct effects) holds, Section \ref{sec:identification} showed that $\widetilde{\tau}(H) = \tau$. 
When direct effects are present, $\widetilde{\tau}(H)$ remains identified but may differ from $\tau$.  
We bound this discrepancy.

For the chosen representation $H$, define the outcome-specific direct-effect bias as
$$
  b_y(H) := \E\{H(R) \mid S=e,D=1,Y=y\} - \E\{H(R) \mid S=e,D=0,Y=y\}.
$$
This quantity measures how much the treatment shifts the representation 
$H(R)$, while holding the outcome fixed. 
Under no-direct-effects, $b_y(H)=0$. 
Define the overall direct-effect bias as
$$
b(H)=\{1-\mu(1)\}b_0(H)+\mu(1)b_1(H).
$$

\begin{lemma}[Bias due to direct effects]\label{lem:bias_direct}
Suppose Assumptions~\ref{assumption:experimental} and~\ref{assumption:stability} hold, and $\Pr(D=0\mid S=o)=1$.
Fix $H$ with $\E\{\Delta^o H(R)\} \neq 0$. Then,
\[
 \tau = \widetilde{\tau}(H) - \frac{b(H)}{\E\{\Delta^o H(R)\}}.
\]
\end{lemma}

\begin{proof} 
See Appendix \ref{proof:lem:bias_direct}.  
\end{proof} 

Lemma~\ref{lem:bias_direct} shows that bias is additive. The magnitude of the bias depends on the size of direct effect $b(H)$ and the relevance of the remotely sensed variable  $\E\{\Delta^o H(R)\}$. 
If the researcher is willing to bound $b(H)$, then we can derive an identified set for $\tau$.

\begin{theorem}[Partial identification with incomplete observational cases and bounded direct effects]\label{thm:PI_direct}
Suppose the conditions of Lemma~\ref{lem:bias_direct} hold. Further suppose the direct effects are bounded: $|b_y(H)| \leq \bar{b}(H)$ for $y \in \{0,1\}$.
Then 
$$
  \tau \in
  \left[
    \widetilde{\tau}(H) - \frac{\bar{b}(H)}{|\E\{\Delta^o H(R)\}|},
    \widetilde{\tau}(H) + \frac{\bar{b}(H)}{|\E\{\Delta^o H(R)\}|}
  \right]
  \cap [-\mu(0), 1 - \mu(0)].
$$
Moreover, the set is sharp with respect to the reduced-form moments $\E\{\Delta^e H(R)\}$ and $\E\{\Delta^o H(R)\}$.
\end{theorem}

\begin{proof} 
See Appendix \ref{proof:thm:PI_direct}.  
\end{proof} 

There are several aspects of Theorem~\ref{thm:PI_direct} worth emphasizing.
First, the width of the interval scales with the ratio $\frac{\bar{b}(H)}{|\E\{\Delta^o H(R)\}|}$, i.e. the size of the direct effect over  the relevance of the remotely sensed variable. The bound is most informative when the scope for direct effects is limited (small $\bar{b}(H)$) and the remotely sensed variable is highly predictive of the outcome (large $|\E\{\Delta^o H(R)\}|$). 
In the environmental application of Examples~\ref{ex:environmental}, the treatment (e.g., a PES contract) is unlikely to directly affect the  satellite image, so $\bar{b}(H)$ may be plausibly small.

Second, the feasibility constraint $\tau \in [-\mu(0), 1 - \mu(0)]$ is operational because $\mu(0)$ is identified under Assumptions~\ref{assumption:experimental} and~\ref{assumption:stability} alone, without requiring the no-direct-effects condition.  
By the proof of Lemma~\ref{lem:bias_direct}, stability, and $\Pr(D=0|S=o)=1$, 
\begin{align*}
    \E\{H(R) \mid S = e, D = 0\} 
 &=  \mu(0) \E\{H(R) \mid S = e, D=0,Y = 1\}+\{1 - \mu(0)\} \E\{H(R) \mid S = e,D=0, Y = 0\} \\ 
  &= \mu(0) \E\{H(R) \mid S = o,D=0, Y = 1\}+\{1 - \mu(0)\} \E\{H(R) \mid S = o,D=0, Y = 0\} \\
 &= \mu(0) \E\{H(R) \mid S = o, Y = 1\}+\{1 - \mu(0)\} \E\{H(R) \mid S = o, Y = 0\} 
\end{align*}
so $\mu(0)$ is identified.

Third, because the bound $\bar{b}(H)$ and the relevance $|\E\{\Delta^o H(R)\}|$ both depend on the choice of representation $H$, the researcher has a meaningful degree of freedom: they can select $H$ to load on
features of $R$ for which direct effects are plausibly small while maintaining relevance.
For example, in satellite imagery application, the researcher might
construct $H$ from spectral bands that are specific to the outcome of
interest---such as near-infrared reflectance for vegetation cover, or thermal signatures for crop burning---while avoiding those that could be affected by the treatment through other channels.  

\begin{remark}[Breakpoint analysis.]
Researchers can use Theorem~\ref{thm:PI_direct} to conduct a breakpoint analysis: for each $H$, report the smallest value $\bar{b}(H)$ at which the identified set includes zero (or, more generally, at which a conclusion of interest is overturned) \citep[][]{MastenPoirier(20)}.
This value $\bar{b}(H)^{\mathrm{bp}} = |\widetilde{\tau}(H)| \cdot |\E\{\Delta^o H(R)\}|$ is the smallest direct-effect (measured
on the scale of the representation) required to explain away the estimated treatment effect.
Large breakpoints indicate that the conclusion is robust to substantial
departures from the no-direct-effects assumption.
\end{remark}

\begin{remark}[Multiple moments.]
If we have a class $\mathcal H$ and, for each $H\in\mathcal H$, a bound $\bar{b}(H)$ such that
$|b_y(H)|\le \bar{b}(H)$ for $y\in\{0,1\}$, then intersecting the single-$H$ bounds tightens identification:
\[
\tau \in
\left(\bigcap_{H\in\mathcal H:\E\{\Delta^o H(R)\}\neq 0}
\left[
\widetilde{\tau}(H) -\frac{\bar{b}(H)}{|\E\{\Delta^o H(R)\}|}  ,
\widetilde{\tau}(H) +\frac{\bar{b}(H)}{|\E\{\Delta^o H(R)\}|}
\right]\right)
\cap [-\mu(0),1-\mu(0)].
\]
This intersection can also be viewed  as a test for the stability restriction: if it is empty, then no DGP can satisfy the maintained assumptions.
\end{remark} 

\subsubsection{Approximately No Direct Effects and Approximate Stability}\label{sec:partial_id2}

We now relax both identifying assumptions simultaneously. We allow  direct effects of $D$ on $R$ (relaxing Assumption~\ref{assumption:observational}(ii)) and instability across samples (relaxing Assumption~\ref{assumption:stability}).

For the chosen representation $H$, we introduce the outcome-specific stability violation as
$$
  s_y(H) := \E\{H(R) \mid S=e,D=0,Y=y\} - \E\{H(R) \mid S=o,D=0,Y=y\}.
$$
This quantity measures how much the distribution of $H(R)$ differs across the experimental and observational samples, conditional upon $D=0$ and $Y=y$. 
Under stability (Assumption~\ref{assumption:stability}), $s_y(H) = 0$. 
Define the overall stability violation
$$
s(H)=s_1(H)-s_0(H).
$$

\begin{lemma}[Bias due to direct effects and stability violations]\label{lem:bias_direct_instability}
Suppose Assumption~\ref{assumption:experimental} holds and $\Pr(D=0\mid S=o)=1$.
Fix $H$ with $\E\{\Delta^o H(R)\}+s(H)\neq 0$. 
Then
\[
\tau = \frac{\E\{\Delta^e H(R)\}-b(H)}{\E\{\Delta^o H(R)\}+s(H)}.
\]
\end{lemma}
\begin{proof}
See Appendix \ref{proof:lem:bias_partialid_both}.
\end{proof}

Lemma~\ref{lem:bias_direct_instability} clarifies how the two violations interact. 
The direct-effect bias enters additively, just as in  Lemma~\ref{lem:bias_direct}.  The stability violation enters multiplicatively, distorting the relevance of the remotely sensed variable from $\E\{\Delta^o H(R)\}$ to $\E\{\Delta^o H(R)\} + s(H)$ for some $s(H)$. 
If the researcher is willing to bound both the direct effect and the stability violation, then Lemma~\ref{lem:bias_direct_instability} delivers an identified set for $\tau$.

\begin{theorem}[Partial identification with incomplete observational cases, bounded direct effects, and bounded 
stability violations]\label{thm:PI_direct_instability}
Suppose the conditions of Lemma~\ref{lem:bias_direct_instability} hold. 
Further suppose the direct effects and stability violations are bounded: $|b_y(H)| \leq \bar{b}(H)$ and $|s_y(H)| \leq \bar{s}(H)$ for $y \in \{0,1\}$. 
Then
$$
  \tau \in
  \left\{
    \frac{\E\{\Delta^e H(R)\} - b}{\E\{\Delta^o H(R)\} + s}
    :\;
    |b|\leq \bar{b}(H),\;
    |s|\leq 2\bar{s}(H),\;
    \E\{\Delta^o H(R)\}+s\neq 0
  \right\}.
$$
Moreover, if $|\E\{\Delta^o H(R)\}| > 2\bar{s}(H)$, then the  identified set
 is contained within the interval
$$
  \tau \in
  \left[
    \widetilde{\tau}(H)
    -
    \kappa(H),
    \widetilde{\tau}(H)
    +
    \kappa(H)
  \right],\quad \kappa(H):=\frac{\bar{b}(H) + 2\bar{s}(H) |\widetilde{\tau}(H)|}
         {|\E\{\Delta^o H(R)\}| - 2\bar{s}(H)}.
$$
\end{theorem}

\begin{proof} 
See Appendix~\ref{proof:thm:PI_direct_instability}.  
\end{proof} 

As before, the choice of representation $H$ determines the width of these bounds. Stability is most plausible for representations that depend on the sensing technology rather than on sample-specific factors, e.g spectral indices derived from a common satellite platform.  Choosing $H$ to load on technologically stable features can simultaneously reduce the plausible value of $\bar{s}(H)$ and maintain a strong relevance $|\E\{\Delta^o H(R)\}|$, tightening the bounds in Theorem~\ref{thm:PI_direct_instability}.
\subsection{Continuous Outcomes}\label{sec:cts_outcomes_appendix}

We extend our results to outcomes that are continuous and bounded. 
Our practical suggestion is to discretize the support of a continuous outcome into bins, then to apply our method for discrete outcomes. 
Under a continuity condition on how the remotely sensed variable captures the underlying outcome, we bound the bias incurred from this discretization. 
Finally, we clarify how our relevance condition generalizes to a standard ``completeness'' condition when outcomes are continuous.

For clarity of exposition, we set aside covariates and focus on incomplete observational cases (Assumption~\ref{assumption:observational}(ii)). 
Our results directly extend to settings with covariates and with complete observational cases (Assumption~\ref{assumption:observational}(i)).  
Recall that $f_W(\cdot \mid \ldots)$ is our symbol for the Radon-Nikodym derivative.  

\subsubsection{Discretization}

To begin, we place weak regularity conditions upon $f_{Y(d)}(y \mid S=e)$ (i.e., the conditional density of the potential outcome   $Y(d)\in\mathbb{R}$ given the sample indicator $S=e$).

\begin{assumption}[Bounded outcomes]\label{assumption:cont_outcome}
Suppose that 
the support $\mathcal{Y}$ is uniformly bounded above and below.
\end{assumption}

Assumption~\ref{assumption:cont_outcome} states that the outcomes are uniformly bounded, so their support can be discretized into a finite number of bins $K<\infty$. 

The general technique is to transform a continuous outcome $Y \in\mathcal{Y}$ into a discrete approximation $Y_{\varepsilon} \in \mathcal{Y}_\varepsilon$. Specifically, we discretize the continuous support $\mathcal{Y}$ into a grid $\mathcal{Y}_\varepsilon = \{y_1,\cdots, y_{K}\}$, where each grid value is the center of a bin of radius  $\varepsilon$. 
As notation, let $B_{\varepsilon}(y)$ define an $\ell_{\infty}$-ball of radius $\varepsilon > 0$ around the value $y$, i.e. $B_{\varepsilon}(y)=\{y'\in \mathcal{Y}: \|y-y'\|_{\infty}\leq \varepsilon\}$. 
Define the discretized parameter $\theta_{\varepsilon} \in\mathbb{R}^{K-1}$. The $k$th element of this vector is the average effect of the treatment on a grid value: $\Pr\{Y(1) \in B_\varepsilon(y_k)|S = e\} - \Pr\{Y(0) \in B_\varepsilon(y_k)|S = e\}$.

We will show that the bias from discretizing the outcome is small as long as the remotely sensed variable smoothly reflects the outcome of interest. In other words, the remote sensing is continuous.

\begin{assumption}[Continuous sensing]\label{ass:RSV_continuous} 
Suppose that for all $y, y^\prime \in \mathcal{Y} $, 
$$
\sup_{r \in \mathcal{R}} \, | f_R(r \mid S = o, Y = y) - f_R(r \mid S = o, Y = y^\prime)| \leq \omega(|y - y^\prime|)
$$
for a continuous, non-negative function $\omega(t)$ with $\omega(0) = 0$. Let $\inf_{r \in\mathcal{R}} f_R(r) \ge \underline{\ell} > 0$. 
\end{assumption}

Assumption~\ref{ass:RSV_continuous} states that if the outcome values $y$ and $y^\prime$ are close, then the densities of the remotely sensed variable that they induce are close. Under this plausible assumption, we first characterize the moment restriction then analyze the bias of the treatment effect.

\begin{proposition}[Error from discretization]\label{prop:continuous1} 
Suppose Assumptions~\ref{assumption:experimental}, \ref{assumption:stability}, and \ref{assumption:observational}(ii) hold. Suppose that $X=\varnothing$, and that $Y$ has a continuous support $\mathcal Y \subset\mathbb R$. 
In addition, suppose Assumptions~\ref{assumption:cont_outcome} and \ref{ass:RSV_continuous} hold. Then, 
$$
\left|\E\{\Delta^e - (\Delta_\varepsilon^{o})^{\top} \theta_\varepsilon | R\} \right| \le \frac{2\omega(2\varepsilon)}{\underline{\ell}} \text{ almost surely, where }
$$
$$
\Delta^e = \frac{1\{D = 1, S = e\}}{\Pr(D = 1, S = e)} - \frac{1\{D = 0, S = e\}}{\Pr(D = 0, S = e)},\quad 
\Delta_{\varepsilon,k}^{o}  =\frac{1\{Y \in B_\varepsilon(y_k), S = o\}}{\Pr\{Y \in B_\varepsilon(y_k), S = o\} } - \frac{1\{Y \in B_\varepsilon(y_K), S = o\}}{\Pr\{Y \in B_\varepsilon(y_K), S = o\} }.
$$

\begin{proof} 
See Appendix \ref{proof:prop:continuous1}. 
\end{proof}
\end{proposition}

Proposition~\ref{prop:continuous1} extends our conditional moment restriction from discrete outcomes to continuous outcomes via a discretization technique. The  discretized parameter $\theta_{\varepsilon}$ approximately satisfies the conditional moment restriction, up to a small error that depends on the number of bins $K$ and the smoothness of the sensing mechanism $\omega(\cdot)$. 

\begin{corollary}[Discretization bias]\label{cor:continuous1}
Suppose the conditions of Proposition \ref{prop:continuous1} hold. 
Consider any measurable function $H \colon \mathcal{R} \rightarrow \mathbb{R}^{J}$ with $J = K-1$, $\E\{\|H(R)\|\} < \infty$, and $\E\{H(R) (\Delta_{\varepsilon}^{o})^{\top}\}$ invertible. 
Define the discretized average treatment effect 
$$
\widetilde{\tau}_{\varepsilon} = \lambda^\top_{\varepsilon} \E[\{H(R) (\Delta^{o}_{\varepsilon})^{\top}\}]^{-1} \E\{H(R) \Delta^e\}
$$ 
where $\lambda_{\varepsilon} = \left( y_{1} - y_K, \hdots, y_{K-1} - y_K \right)^\top$ for the grid $\mathcal{Y}_\varepsilon = \{y_1,\cdots, y_{K}\}$.
Then, 
$$
|\widetilde{\tau}_{\varepsilon} - \tau |\leq 2 \varepsilon + \frac{2\omega(2\varepsilon)}{\underline{\ell}} \cdot  \|\lambda_\varepsilon\| \cdot  \| [\E\{H(R) (\Delta_{\varepsilon}^{o})^{\top}\}]^{-1} \|_{\mathrm{op}} \E\{\|H(R)\|\}. 
$$

\begin{proof}
See Appendix \ref{proof:cor:continuous1}.
\end{proof}
\end{corollary}

Corollary \ref{cor:continuous1} further shows that we can bound the bias of the discretized average treatment effect $\widetilde{\tau}_{\varepsilon}$ implied by the average effect of the treatment on grid values $\theta_{\varepsilon}$.

In summary, under a plausible auxiliary assumption, we can bound the bias incurred by applying our estimation algorithm (Algorithm \ref{algorithm:discrete_covar_details}) to discretized outcomes. When the bin size $\varepsilon$ is fixed, one can easily extend our inference guarantees to accommodate this bias, e.g. with bias-aware confidence intervals.

A further extension is to allow the bin size to vanish: $\varepsilon\rightarrow 0$. This choice alters the estimator and asymptotics. Intuitively, it is like using local smoothing with a vanishing bandwidth. On the one hand, the first term in the bound converges to zero. On the other hand, the second term in the bound diverges, due to the factor $\| [\E\{H(R) (\Delta_{\varepsilon}^{o})^{\top}\}]^{-1} \|_{\mathrm{op}}$. Choosing an $\varepsilon$ that balances the two terms leads to an overall slower convergence rate for the treatment effect. More subtly, this choice changes the meaning of the relevance condition, which we now make explicit.

\subsubsection{Generalizing the Relevance Condition}

With binary outcomes, the relevance condition was $\E\{H(R)\Delta^o\}>0$. With discrete outcomes, it was invertibility of $\E\{H(R) (\Delta^o)^\top\}$. With discretized outcomes, it became invertibility of  $\E\{H(R) (\Delta_{\varepsilon}^{o})^{\top}\}$. We now articulate the appropriate generalization of this relevance condition when outcomes are viewed as continuous. It is a familiar condition called ``completeness'' from the literature on nonparametric inverse problems.

In Appendix~\ref{sec:identification_proof}, we defined the treatment weights $\pi_d(R)$ and the outcome weights $\gamma_y(R)$:
$$
\pi_d(R)= \frac{ \Pr\left( D = d, S = e \mid R \right) }{\Pr\left( D = d, S = e \right) },\quad \gamma_{y}(R) = \frac{ f_{Y,S}(y,o|R) }{ f_{Y,S}(y,o) }.
$$
The treatment weights are identified from the experimental sample, while the outcome weights are identified from the observational sample.

As before, we combine the treatment weights and outcome weights to identify the causal parameter. With continuous outcomes, we obtain a conditional moment equation that exactly generalizes our conditional moment equation with discrete outcomes. Specifically, we will verify that Theorem~\ref{theorem:disc_outcomes_incomplete} generalizes to 
$$
\pi_d(R) = \int_{y \in \mathcal{Y}} \gamma_{y}(R)  f_{Y(d)}(y|S=e)  \mathrm{d}y.
$$

To connect this equation to the literature on nonparametric inverse problems, it helps to define the operator $T$ which acts on the counterfactual density $f_{Y(d)}(y|S=e)$ and returns the treatment weights $\pi_d(R)$. The operator may be viewed as a way of encoding the outcome weights. In this notation, our conditional moment equation can be expressed succinctly: 
$$
\pi_d=Tf_{Y(d)},\quad  (Tf)(R)=\int \gamma_y(R) f(y)\mathrm{d}y.
$$
It is a classic inverse problem: to isolate $f_{Y(d)}(y|S=e)$, we must invert $T$. For such inverse problems, regularity conditions are well known \citep{kress1989linear}. 

A solution to the equation exists if $\pi_d$ is in the range $T$. 
For a solution to exist, the treatment weights cannot be too correlated with variation in $R$ that is only weakly generated by the outcome weights. This standard assumption is known as Picard's criterion \citep{kress1989linear}. 

\begin{assumption}[Picard's criterion]\label{assumption:existence}
    Suppose that $T$ is compact and has a singular value decomposition with left singular functions $\{\psi_j(r)\}_{j\geq 1}$, positive singular values $(\sigma_j)_{j\geq 1}$, and right singular functions $\{\phi_j(y)\}_{j\geq 1}$. Suppose that $\pi_d$ is in the range $T$:
    $
    \sum_{j\geq 1} \frac{[\E\{\pi_d(R)\psi_j(R)\}]^2}{\sigma_j^2}<\infty
    $
    and $\pi_d\in \operatorname{null}(T^*)^{\perp}$ (i.e., the orthogonal complement to the null space of the adjoint of $T$).
\end{assumption}

A solution to the equation is unique if $T$ is a one-to-one mapping. In other words, its null space is simply the zero function. Intuitively, no perturbation of the counterfactual density can leave the distribution of the remotely sensed variable unchanged.
This standard assumption is called ``completeness'' \citep{newey2003instrumental}. 
While it is often untestable \citep{canay2013testability}, examples that satisfy this condition are well known  \citep{andrews2017examples}.

\begin{assumption}[Completeness]\label{assumption:uniqueness}
    If a function $h$ satisfies $\int \gamma_y(R)h(y)\mathrm{d}y=0$ almost surely, then $h(y)=0$ almost everywhere.
\end{assumption}

When the outcome has finite support, completeness reduces to linear independence of the outcome weights for different outcome values. After projecting the outcome weights onto a sufficiently rich finite representation $H(R)$, we recover our relevance condition from the main text: invertibility of $\E\{H(R) (\Delta^o)^\top\}$. 

Using this generalized relevance condition for the remotely sensed variable, we identify the counterfactual density.

\begin{proposition}[Identification without discretization]\label{prop:continuous_deconvolution}
Suppose Assumptions~\ref{assumption:experimental}, \ref{assumption:stability},~\ref{assumption:observational}(ii),~\ref{assumption:existence}, and~\ref{assumption:uniqueness} hold.  Suppose that $X=\varnothing$, and that $Y$ has a continuous support $\mathcal Y \subset\mathbb R$.  
Then, for each $d\in\{0,1\}$, the counterfactual density $f_{Y(d)}(y|S=e)$ is identified as the unique solution to the conditional moment equation
$$
\pi_d(R) = \int_{y \in \mathcal{Y}} \gamma_{y}(R)  f_{Y(d)}(y|S=e)  \mathrm{d}y.$$
\end{proposition}

\begin{proof}
    See Appendix~\ref{proof:prop:continuous_deconvolution}.
\end{proof}

With continuous outcomes, the remotely sensed variable estimation problem is closely related to nonparametric instrumental variable regression and deconvolution, for which several nonparametric estimators are available in the literature; see e.g. \cite{carrasco2007linear} for a review. The rate of estimation for $f_{Y(d)}(y\mid S=e)$ will depend on the rates of estimation for $\pi_d(R)$ and $T$, as throttled by the ill-posedness of inverting the operator $T$ \cite[e.g.][]{chen2011rate}. Articulating the technical details of such an estimation strategy is possible and left to future work.

\subsection{Estimation with Covariates}\label{sec:covariates_general}

In this section, we study estimation with discrete or continuous covariates. 
 
\subsubsection{Discrete Covariates and Incomplete Observational Cases}\label{sec:est_inf_disc_x}
 
To begin, suppose we have incomplete observational cases (Assumption \ref{assumption:observational}(ii)), and that the covariates $X$ are discrete with finite support.
Estimation and inference are straightforward: we apply Algorithm \ref{alg:discrete_nocovar_overview} within each covariate stratum and then average over the covariates.

Even in the presence of covariates, results in \citet[][]{chamberlain1987asymptotic} and \citet[][]{newey1993efficient} imply that representation that achieves efficiency, within a class of models satisfying Theorem \ref{theorem:disc_outcomes_incomplete}, is
$$
H^*(X,R)=\frac{\E\{\Delta^o(X) \mid X,R\}}{\sigma^2(\theta,X,R)},\quad \sigma^2(\theta,X,R) = \E[\{\Delta^e(X) - \Delta^o(X)^\top\theta(X)\}^2 \mid X,R].
$$
Here $H^*(X,R)\in\R^{K-1}$, $\E\{\Delta^o(X) \mid X,R\}\in \R^{K-1}$, and $\sigma^2(\theta,X,R)\in\R$. 
Its $k$-th component is the scalar $
H_k^*(X,R)=\frac{\E\{\Delta_k^o(X) \mid X,R\}}{\sigma^2(\theta,X,R)},
$ where $\Delta_k^o(X)=\frac{1\{Y=y_k, S=o\}}{\Pr(Y=y_k, S=o \mid X)} - \frac{1\{Y=y_K, S=o\}}{\Pr(Y=y_K, S=o \mid X)}$.

Like in the main text, the algorithm has three steps: divide the sample into $\tr$ and $\te$ folds; learn the representation on $\tr$; and apply the learned representation to estimate the ATE on $\te$.
Algorithm \ref{algorithm:discrete_covar_details} provides details with sample splitting, though our analysis below allows for cross-fitting with any fixed number of folds.
To analyze the properties of Algorithm \ref{algorithm:discrete_covar_details}, we require only that the predictions, and hence the representation estimator, have some probability limit; they may be misspecified.

\begin{assumption}[Limit]\label{ass:limit_discrete_covar}
For each $x \in \mathcal{X}$, the learned representation has some mean square limit: $\E_R\left\{ \| \widehat{H}(x,R) - \widetilde{H}(x,R)\|^2 \mid X = x \right\} = o_p(1)$ for some limit $\widetilde{H}(x,R) \in \mathbb{R}^{K-1}$ with $\E\{\| \widetilde{H}(x,R) \|^2 \mid X = x\} < \infty$. 
The limit is correlated with outcome variation: $\E\{\widetilde{H}(x, R) \Delta^o(x)^\top \mid X = x\}$ is non-singular for each $x \in \mathcal{X}$, and its smallest singular value is bounded away from zero.
\end{assumption}

The following inference guarantees generalize our results in Section \ref{sec:estimation} to include discrete covariates. 
Let $n_{x} = \sum_{i=1}^{n} 1_{X_i=x}$ denote the number of observations with cell $X_i = x$. We refer to $\Pr(X = x \mid S = e)$ as the marginal cell probabilities. We refer to $\Pr(D = d, S = e \mid X = x)$ and $\Pr(Y = y_k, S = o \mid X = x)$ as the conditional cell probabilities.  

\begin{proposition}[Inference with known conditional cell probabilities]\label{lemma:disc_outcomes_incomplete_known_cell_specific}
Suppose Theorem~\ref{theorem:disc_outcomes_incomplete}'s conditions and Assumption~\ref{ass:limit_discrete_covar} hold. 
Suppose the conditional cell probabilities are known and bounded away from zero.
Then for each $x \in \mathcal{X}$, and for $\lambda=(y_1-y_K,...,y_{K-1}-y_K)^{\top}$, 
$$(n_x)^{1/2}\{\widehat{\tau}(x) - \tau(x)\} \rightsquigarrow \mathcal{N}\{0, \lambda^\top A(x) B(x) A(x)^\top\lambda\}, \text{ where }$$
$$
A(x) = \E\{ \widetilde{H}(x, R) \Delta^o(x)^\top \mid X = x\}^{-1}
\text{ and } B(x) = \E[\{\Delta^e(x) - \Delta^o(x)^\top \theta(x)\}^2 \widetilde{H}(x, R) \widetilde{H}(x, R)^\top \mid X = x].
$$
Moreover, if $\widetilde{H}(x, R) = H^*(x, R)$, then $\widehat{\tau}(x)=\lambda^{\top}\widehat{\theta}(x)$ is semiparametrically efficient for $\tau(x)=\lambda^{\top}\theta(x)$ satisfying the conditional moment restriction $\E[\Delta^e(x)-\{\Delta^o(x)\}^{\top}\theta(x)|X=x,R]=0$ with known conditional cell probabilities.
\end{proposition}

\begin{proof}
    See Appendix~\ref{proof:lemma:disc_outcomes_incomplete_known_cell_specific}.
\end{proof}

\begin{proposition}[Inference with estimated conditional cell probabilities]\label{prop:discrete_incomplete_unknown_covar}
Suppose Theorem~\ref{theorem:disc_outcomes_incomplete}'s conditions and Assumption~\ref{ass:limit_discrete_covar} hold.
If the conditional cell probabilities and their estimators are bounded away from zero, then 
$$(n_x)^{1/2}\{\widehat{\tau}(x) - \tau(x)\} \rightsquigarrow \mathcal{N}\{0, \lambda^\top A(x) V(x) A(x)^\top\lambda\},$$
where $V(x)$ is the conditional generalization of $V$ in Lemma~\ref{lemma:num2_unknown}. 
\end{proposition}

\begin{proof} 
See Appendix \ref{proof:prop:discrete_incomplete_unknown_covar}. 
\end{proof} 

These $n_x^{-1/2}$ inference guarantees for the conditional average treatment effect (CATE) imply $n^{-1/2}$ inference guarantees for ATE by averaging over cells.

\subsubsection{Discrete Covariates and Complete Observational Cases}\label{sec:est_inf_direct_effects}

Next, suppose we have complete observational cases (Assumption \ref{assumption:observational}(i)), and that the covariates $X$ are discrete with finite support. 
In this setting, the observational sample contains $(X,D,Y,R)$, and we allow direct effects of treatment on the remotely sensed variable.

For each treatment arm $d\in \{0,1\}$ and covariate cell $x\in\mathcal X$, define the counterfactual outcome probabilities $\mu_k(d,x)= \Pr\{Y(d)=y_k \mid S=e, X=x\}$. Collect them into the vector $\mu(d,x)= \{\mu_1(d,x),\dots,\mu_{K-1}(d,x)\}^\top \in \mathbb R^{K-1}$. Finally define  $\mu_K(d,x)= 1- \sum_{k=1}^{K-1} \mu_k(d,x)$. 
The CATE is then $\tau(x)= \sum_{k=1}^{K} y_k \{\mu_k(1,x)-\mu_k(0,x)\}$. The ATE  is $\tau= \mathbb \E\{\tau(X)\mid S=e\}$.

Once again, results in \cite{chamberlain1987asymptotic} and \cite{newey1993efficient} pin down the efficient representation. Within the class of models satisfying Theorem~\ref{theorem:disc_outcomes_complete}, the representation to use for $\mu(d,x)$ is 
$$
H_d^*(X, R) = \frac{\E\{\widetilde{\Delta}^o(d, X) \mid X, R\}}{\sigma^2_d(\mu, X, R)},\quad \sigma^2_d(\mu, X, R) = \E[\{\widetilde{\Delta}^e(d, X) - \widetilde{\Delta}^o(d, X)^\top \mu(d, X) \}^2 \mid X, R].
$$
Here $H_d^*(X, R) \in\mathbb{R}^{K-1}$, $\E\{\widetilde{\Delta}^o(d, X) \mid X, R\}\in\mathbb{R}^{K-1}$, and $\sigma^2_d(\mu, X, R)\in\mathbb{R}$.

So far, we have derived an optimal representation for each treatment arm $d$. It turns out that stacking the representations is the efficient choice for CATE. So see why, note that $1_{D=1}1_{D=0}=0$, so mechanically the multiplication of $\widetilde{\Delta}^e(1, X)$ and $\widetilde{\Delta}^o(1, X)$ with $\widetilde{\Delta}^e(0, X)$ and $\widetilde{\Delta}^o(0, X)$ yields zero diagonals.

Having derived the efficient representation, our algorithms and guarantees for inference directly extend from the setting with incomplete observational cases to the setting with complete observational cases.

\subsubsection{Continuous Covariates}\label{sec:est_inf_cts_x}

As a final extension, Algorithm \ref{algorithm:cts_covar_details} modifies our inferential procedure to accommodate continuous covariates.  We focus on low-dimensional covariates with $X \in \mathcal{X} \subseteq \mathbb{R}^{d_x}$. 
For simplicity, we focus on incomplete observational cases (Assumption~\ref{assumption:observational}(ii)). 

Our identification results already apply to continuous covariates as written. The conditional moment equation from Theorem~\ref{theorem:disc_outcomes_incomplete} is
$$
\E\{\Delta^e(X) - \Delta^o(X)^{\top}\theta(X) \mid X, R\} = 0, 
$$
where $\theta_k(x) = \Pr\{Y(1)=y_k\mid S=e, X=x\}-\Pr\{Y(0)=y_k\mid S=e, X=x\}$. 
Corollary~\ref{corr:disc_outcomes_incomplete_representation} gives 
$$
\theta(X) = \E\{H(X, R) \Delta^o(X)^\top \mid X \}^{-1} \E\{H(X, R) \Delta^e(X) \mid X \}.
$$
ATE is then identified as $\tau = \E\{\lambda^\top \theta(X)\}$ for $\lambda = (y_1 - y_K, \hdots, y_{K-1} - y_K)^{\top}$.

With continuous covariates, the key modification to the algorithm is that the conditional expectations 
$\E\{H(X, R) \Delta^o(X)^\top \mid X\}$ and $\E\{H(X, R) \Delta^e(X) \mid X\}$ should be estimated as smooth functions of $X$ using nonparametric methods, e.g., kernel smoothing, local polynomials, or series. With this modification, estimation proceeds in the same manner: divide the sample into $\tr$ and $\te$ folds, learn the representation on $\tr$, then use the learned representation to estimate the ATE on $\te$.

Valid inference would require standard smoothness conditions on these conditional expectations as functions of $X$, as well as undersmoothing or bias correction. The overall analysis with continuous covariates would largely follow our analysis with discrete covariates, combined with standard arguments for $Z$-estimation with nonparametric first stages \citep[e.g.,][]{Newey(97), ai2003efficient, Chen(07)}.  

As in the main text, no rate condition would be required on the estimated representation due to the infinite order Neyman orthogonality of the conditional moment restriction and sample splitting. 
The key technical insight---that the moment condition permits misspecified representation learning---continues to hold. 

\paragraph{Series estimator.} As a concrete example, we describe a series estimator. 

Consider any representation $H$. Write $H_i=H(X_i,R_i)\in \mathbb{R}^{K-1}$. Construct the vector $C_i(H)\in\mathbb{R}^{K(K+2)}$ as the transpose of 
{\footnotesize
$$
 \begin{pmatrix}
  1_{ D_i = 1, S_i = e }, \
   1_{ D_i = 1, S_i = e }H_i^{\top}, \
    1_{ D_i = 0, S_i = e }, \
    1_{ D_i = 0, S_i = e }H_i^{\top}, \
    1_{ Y_i=y_1, S_i = o }, \
    1_{ Y_i=y_1, S_i = o }H_i^{\top}, \
    \cdots, \
     1_{ Y_i = y_K, S_i = o }, \
      1_{ Y_i = y_K, S_i = o }H_i^{\top} 
   \end{pmatrix}^{\top}
$$
}
where the odd entries are scalars and the even entries are vectors in $\mathbb{R}^{K-1}$.
The expectations of these quantities, conditional upon covariates, pin down $\E\{H(X, R) \Delta^o(X)^\top \mid X\}$ and $\E\{H(X, R) \Delta^e(X) \mid X\}$. Therefore, the conditional expectations pin down the CATE and ATE.

Consider a basis $b$ and denote $B_i=b(X_i)\in\mathbb{R}^J$. 
Define the series  estimator as the vector-valued regression
$$
\widehat g_{\tr}(x;H)=b(x)^\top\widehat\beta_{\tr}(H),\quad \widehat\beta_{\tr}(H)=\arg\min_{\beta } \sum_{i\in\mathcal \tr}\|C_i(H)-B_i^\top\beta\|^2.
$$
This vector valued regression is the essential ingredient to estimate $\E\{H(X, R) \Delta^o(X)^\top \mid X\}$ and $\E\{H(X, R) \Delta^e(X) \mid X\}$. An undersmoothed series is one where $J$ grows more quickly than the optimal rate for estimation.

\paragraph{Properties.}  The series estimator will be valid under standard series regularity conditions. First, each component of the vector-valued regression should be H\"older smooth over a compact support, so the approximation error vanishes as $J$ increases. Second, the covariance matrix  $\E_{\tr}(BB^{\top})$ converges in probability to $\E(BB^{\top})$, e.g. when $J^2\ll n$, so that the sampling error vanishes as $J$ increases. Finally, the objects $\E(BB^{\top})$ and $\E\{\widetilde{H}(X, R) \Delta^o(X)^\top \mid X\}$ are well conditioned with singular values bounded away from zero. 

Under such conditions, standard arguments yield pointwise asymptotic normality of $\widehat{\theta}(X)$ after either undersmoothing (i.e., choosing $J$ large enough that the series bias is negligible) or bias correction \citep[e.g.,][]{Newey(97), Chen(07)}. Since $\tau$ is smooth functional of $\theta(X)$, $\widehat{\tau}$ will be $n^{-1/2}$  asymptotically normal, provided that the series dimension $J$ satisfies restrictions ensuring small bias.

\section{Proofs for Additional Theoretical Results}
\subsection{Bias of Alternative Approaches: Extensions}

\subsubsection{Proof of Proposition \ref{proposition:common_practice_complete}} \label{proof:proposition:common_practice_complete}

We prove each result.

\begin{enumerate}
    \item We prove (i) with possibly discrete outcomes. Fix $d \in \{0, 1\}$. Let 
    $$m_d(R) = \E(Y \mid R, D=d, S = o) = \int y f_{Y}(y \mid R, D=d, S = o) \mathrm{d}y.$$
By Bayes' rule,
\begin{align*}
    f_{Y}(Y \mid R, D=d,S = o) &= \frac{f_{R}(R \mid Y, D=d,S = o) f_{Y }(Y \mid D=d,S = o)}{f_{R }(R\mid D=d,S = o)} \\
    f_{R }(R\mid Y, D = d, S = e) &= \frac{f_{Y }(Y \mid R, D = d, S = e) f_{R }(R \mid D = d, S = e)}{f_{Y }(Y \mid D = d, S =e )}.
\end{align*}
By stability,
$$
 f_{R }(R\mid Y, D=d,S = o) = f_{R }(R \mid Y, D = d, S = e).
$$

We combine these expressions to rewrite 
\begin{align*}
    f_{Y}(Y \mid R, D=d,S = o) &= f_{Y }(Y \mid R, D = d, S = e) v_d(R,Y)\\
    v_d(R,Y)& = \frac{f_{Y }(Y \mid D=d,S = o)}{f_{Y }(Y \mid D = d, S =e )} \frac{f_{R }(R\mid D = d, S = e)}{f_{R }(R\mid D=d,S = o)}.
\end{align*}
Consequently, 
$$
m_d(R) = \int y f_{Y }(y \mid R, D = d, S = e) v_d(R,y)  \mathrm{d}y = \E\{Y v_d(R,Y) \mid R, D = d, S = e\}.
$$
Since $\widetilde{\mu}^{\prime}(d) = \E\{m_d(R) \mid D = d, S = e\}$, the result follows by iterated expectations. 
    \item We prove (ii) for binary outcomes, where $\mu(d) = \Pr(Y = 1 \mid D = d, S = e)$. Fix $d\in\{0,1\}$. By iterated expectations and stability, 
\begin{align*}
     f_{R }(R\mid D = d, S = e) 
     &= \mu(d) f_{R }(R\mid Y = 1,D=d, S = e) + \{1 -  \mu(d)\} f_{R }(R \mid Y = 0, D=d,S = e)\\
     &= \mu(d) f_{R }(R\mid Y = 1, D=d,S = o) + \{1 -  \mu(d)\} f_{R }(R \mid Y = 0, D=d,S = o).
\end{align*}
As a consequence, 
\begin{align*}
     \widetilde{\mu}^{\prime}(d) 
     &= \E\{m_d(R) \mid D = d, S = e\} \\
     &= \mu(d) \E\{m_d(R) \mid Y = 1, D=d,S = o\} + \{1 - \mu(d)\} \E\{m_d(R) \mid Y = 0, D=d,S = o\} \\
     &=\mu(d)\kappa_d+\E\{m_d(R) \mid Y = 0, D=d,S = o\},
\end{align*}
where within the first term,
\begin{align*}
    \kappa_d&
    :=\E\{m_d(R) \mid Y = 1, D=d,S = o\}-\E\{m_d(R) \mid Y = 0, D=d,S = o\} \\
    &=\frac{\cov\{m_d(R), Y \mid D=d,S = o\}}{\var(Y \mid D=d,S = o)} \\
    &=\frac{\var\{m_d(R) \mid D=d,S = o\}}{\var(Y \mid D=d,S = o)}.
\end{align*}
By the law of total variance,
$$
\var(Y \mid D=d, S = o)=\var\{m_d(R)\mid D=d, S=o\}+\E\{\var(Y\mid R,D=d,S=o)\mid D=d,S=o\}, 
$$
so $\kappa_d\in[0,1]$. Moreover, $\kappa_d = 1$ if and only if $\E\{\var(Y \mid R, D=d,S = o) \mid D=d,S = o\}= 0$, which is equivalent to $\var(Y \mid R, D=d,S = o) = 0$ almost surely. 

What remains is to show 
$$\E\{m_d(R) \mid Y = 0, D=d,S = o\}=(1-\kappa_d)\E(Y|D=d,S=o).$$
By iterated expectations,
\begin{align*}
    p_d
    &:=\E(Y|D=d,S=o) \\
    &=\E\{\E(Y|R,D=d,S=o)|D=d,S=o\} \\
    &=\E\{m_d(R)|D=d,S=o\} \\
    &=\E\{m_d(R)|Y=1,D=d,S=o\}p_d + \E\{m_d(R)|Y=0,D=d,S=o\}(1-p_d)\\
    &=[\kappa_d+\E\{m_d(R)|Y=0,D=d,S=o\}]p_d + \E\{m_d(R)|Y=0,D=d,S=o\}(1-p_d) \\
    &=\kappa_d p_d+\E\{m_d(R)|Y=0,D=d,S=o\}.
\end{align*}
\end{enumerate}

\qed

\subsubsection{Proof of Proposition \ref{prop:bias_common_practice_random}} \label{proof:prop:bias_common_practice_random}

We prove each result.

\begin{enumerate}
\item Recall that Proposition \ref{prop:bias_common_practice}(i) generally established
$$
\widetilde{\mu}(d) =  \mu(d)+\E[ Y \left\{ w_d(R,Y) - 1 \right\} \mid D = d, S = e],\quad w_d(r,y) := \frac{f_{Y }(y\mid S = o)}{f_{Y }(y\mid D = d, S = e)} \frac{f_{R}(r\mid D = d, S = e)}{f_{R }(r\mid S = o)}.
$$ 

Because $S \indep \{Y(0), Y(1)\}$ and we are in the setting with incomplete observational cases,
$$f_{Y}(y \mid S = o) = f_{Y(0) }(y\mid S = o) = f_{Y(0) }(y\mid S = e)=f_{Y}(y\mid D=0, S = e).$$

Next, we prove that $f_{R}(r \mid S = o) = f_{R }(r\mid D = 0, S = e)$.
First, we write 
\begin{align*}
    f_R(R|S=o)&=\int_y f_R(R|Y=y,S=o)f_Y(y|S=o) \mathrm{d}y\\
    f_R(R|D=0,S=e)&=\int_y f_R(R|Y=y,D=0,S=e)f_Y(y|D=0,S=e)\mathrm{d}y.
\end{align*}
By Assumptions \ref{assumption:stability} and~\ref{assumption:observational}(ii), $(S, D) \indep R \mid Y$, hence
$
f_R(R|Y,S=o)=f_R(R|Y,D=0,S=e).
$ Above, we proved that $f_{Y}(y \mid S = o)=f_{Y}(y\mid D=0, S = e)$.

In summary, we have shown
\begin{align*}
    w_0(r,y)&=\frac{f_{Y }(y\mid S = o)}{f_{Y }(y\mid D = 0, S = e)} \frac{f_{R}(r\mid D = 0, S = e)}{f_{R }(r\mid S = o)}
    =1\\
    w_1(r,y)&=\frac{f_{Y }(y\mid S = o)}{f_{Y }(y\mid D = 1, S = e)} \frac{f_{R}(r\mid D = 1, S = e)}{f_{R }(r\mid S = o)}
    =\frac{f_{Y}(y\mid D=0, S = e)}{f_{Y }(y\mid D = 1, S = e)} \frac{f_{R}(r\mid D = 1, S = e)}{f_{R }(r\mid D = 0, S = e)}.
\end{align*}

    \item Recall that Proposition \ref{prop:bias_common_practice}(ii) established that $\widetilde{\tau} = \kappa \tau$ for $\kappa \in [0, 1)$. We have shown $\widetilde{\mu}(0) = \mu(0)$. Therefore
    $$
    \widetilde{\mu}(1)-\mu(0)=\widetilde{\mu}(1)-\widetilde{\mu}(0)=\widetilde{\tau} = \kappa \tau=\kappa\{\mu(1)-\mu(0)\}.
    $$
\end{enumerate}

\qed 

\subsection{Quasi-Experiments with Remotely Sensed Outcomes}

\subsubsection{Proof of Theorem \ref{theorem: IV identification with RSV}} \label{proof:theorem: IV identification with RSV}

\begin{enumerate}
\item By the law of total probability,
\begin{align*}
    \delta_{d,z}^{e}(R) &:=f_R( R \mid S = e, D = d, Z=z) \\
    &= \int f_{R,Y}( R,y \mid S = e, D = d,Z=z) \mathrm{d}y \\
    &=\int f_R( R\mid S = e, D = d, Z=z, Y=y) f_Y(y \mid S=e,D=d,Z=z) \mathrm{d}y.
\end{align*}
Next, notice that 
\begin{align*}
    f_R( R\mid S = e, D = d, Z=z,Y=y) &= f_R( R \mid Y=y) \\
    &= f_R( R \mid S = o, Y=y)\\
    &=:\delta_y^o(R),
\end{align*}
where the equalities apply the contraction implied by Assumptions \ref{assumption:stability, IV} and~\ref{assumption:no direct effects, IV}: $(S, D, Z) \indep R \mid Y$. 

Combining the previous displays, we arrive at the general result 
$$
    \delta_{d,z}^{e}(R)=\int \delta_y^o(R) f_Y(y \mid S=e,D=d,Z=z) \mathrm{d}y.
$$
When $Y$ is binary,
\begin{align*}
    f_Y(1 \mid S=e,D=d,Z=z)&=\E(Y \mid S=e,D=d,Z=z)=\alpha(d,z) \\
    f_Y(0 \mid S=e,D=d,Z=z)&=1-\E(Y \mid S=e,D=d,Z=z)=1-\alpha(d,z).
\end{align*}
Therefore, the general result specializes to 
$$
\delta_{d,z}^{e}(R)= \delta_1^o(R) \alpha(d,z)+\delta_0^o(R)\{1-\alpha(d,z)\}=\delta_0^o(R)+\{ \delta_1^o(R) -\delta_0^o(R) \}\alpha(d,z).
$$
        
\item We apply Bayes' rule to rewrite
\begin{align*}
\delta_{y}^o(R) &= f_R(R \mid  S = o, Y = y) = \frac{\Pr(Y = y, S = o \mid R) f_R(R )}{\Pr(Y = y, S = o )}, \\
\delta_{d,z}^{e}(R) &= f_R\left( R \mid S = e,  D = d, Z = z \right) = \frac{ \Pr(D = d, Z=z, S = e \mid R) f_R(R ) }{ \Pr(D = d, Z=z, S = e ) }.
\end{align*}
Substituting these expressions into the previous step and canceling $f_R(R)$ yields $$ \E\{\Delta^{e}(d, z) - \Delta^o \alpha(d, z) \mid R\} = 0. $$
\end{enumerate}
 
\qed

\subsubsection{Proof of Theorem \ref{theorem: DID identification with RSV} } \label{proof:theorem: DID identification with RSV}

       \begin{enumerate}
           \item By the law of total probability,
\begin{align*}
    \delta_{d,t}^{e}(R_t) &:=f_{R_t}( R_t \mid S = e, D = d) \\
    &= \int f_{R_t,Y_t}( R_t,y \mid S = e, D = d) \mathrm{d}y \\
    &=\int f_{R_t}( R_t\mid S = e, D = d, Y_t=y) f_{Y_t}(y \mid S=e,D=d) \mathrm{d}y.
\end{align*}
Next, notice that 
\begin{align*}
    f_{R_t}( R_t\mid S = e, D = d, Y_t=y) &= f_{R_t}( R_t \mid  Y_t=y) \\
    &= f_{R_t}( R_t \mid S = o, Y_t=y)\\
    &=:\delta_{y,t}^o(R_t),
\end{align*}
where the equalities apply the contraction implied by Assumptions \ref{assumption:stability, DID} and \ref{assumption:no direct effects, DID}: $(S, D) \indep R_t \mid Y_t$. 

Combining the previous displays, we arrive at the general result 
$$
    \delta_{d,t}^{e}(R_t)=\int \delta_{y,t}^o(R_t) f_{Y_t}(y \mid S=e,D=d) \mathrm{d}y.
$$
When $Y$ is binary,
\begin{align*}
    f_{Y_t}(1 \mid S=e,D=d)&=\E(Y_t \mid S=e,D=d)=\alpha_t(d) \\
    f_{Y_t}(0 \mid S=e,D=d)&=1-\E(Y_t \mid S=e,D=d)=1-\alpha_t(d).
\end{align*}
Therefore, the general result specializes to 
$$
\delta_{d,t}^{e}(R_t)= \delta_{1,t}^o(R_t) \alpha_t(d)+\delta_{0,t}^o(R_t)\{1-\alpha_t(d)\}=\delta_{0,t}^o(R_t)+\{\delta_{1,t}^o(R_t)-\delta_{0,t}^o(R_t)\}\alpha_t(d).
$$
           \item We apply Bayes' rule to rewrite
\begin{align*}
\delta_{y,t}^o(R_t) &= f_{R_t}(R_t \mid Y_t = y, S = o) = \frac{\Pr(Y_t = y, S = o \mid R_t) f_{R_t}(R_t)}{\Pr(Y_t = y, S = o)}, \\
\delta_{d,t}^{e}(R_t) &= f_{R_t}\left( R_t \mid D = d, S = e \right) = \frac{ \Pr(D = d, S = e \mid R_t ) f_{R_t}(R_t) }{ \Pr(D = d, S = e) }.
\end{align*}
Substituting these expressions into the previous step and canceling $f_{R_t}(R_t)$ yields $$ \E\{\Delta_t^e(d) - \Delta_t^o \alpha_{t}(d) \mid R_t\} = 0.$$
       \end{enumerate}

\qed

\subsection{Some Spillovers}

\subsubsection{Proof of Lemma~\ref{lemma:diff}}\label{proof:lemma:diff}

    Using nonseparable model notation with unobserved heterogeneity $\eta_i$, as defined in Footnote~\ref{footnote: unobs het},
\begin{align*}
    \E(Y_i|D_i=d, S_i=e)
    &=\E\{Y_i(D_i,\mathbf{D}_{-i},\eta_i)|D_i=d, S_i=e\}\\
    &=\E[\E\{Y_i(D_i,\mathbf{D}_{-i},\eta_i)|D_i=d, S_i=e,\eta_i)\}|D_i=d, S_i=e] \\
    &=\E\{Y_i^{\mathrm{dir}}(d,\eta_i)|D_i=d, S_i=e\} \\
    &=\E\{Y_i^{\mathrm{dir}}(d,\eta_i)|S_i=e\}
\end{align*}
where the final line appeals to randomization.

\subsubsection{Proof of Theorem~\ref{theorem:spill}}\label{proof:theorem:spill}

    The argument mirrors the proof of Lemma~\ref{lemma:discrete_mixture_incomplete} and Theorem~\ref{theorem:disc_outcomes_incomplete}.
    \begin{enumerate}
        \item By the law of total probability,
\begin{align*}
    \delta_{d,i}^{e}(R_i) &:=f_{R_i}( R_i \mid S_i = e, D_i = d) \\
    &= \int f_{R_i,Y_i}( R_i,y \mid S_i = e, D_i = d) \mathrm{d}y \\
    &=\int f_{R_i}( R_i\mid S_i = e, D_i = d, Y_i=y) f_{Y_i}(y \mid S_i=e,D_i=d) \mathrm{d}y.
\end{align*}

Next, notice that 
\begin{align*}
    f_{R_i}( R_i\mid S_i = e, D_i = d, Y_i=y) &= f_{R_i}( R_i \mid Y_i=y) \\
    &= f_{R_i}( R_i \mid S_i = o, Y_i=y)\\
    &=:\delta_{y,i}^o(R_i),
\end{align*}
where the equalities apply the contraction implied by Assumptions \ref{assumption:stability_spill} and~\ref{assumption:observational_spill}: $(S_i,D_i)\indep R_i \mid Y_i$. 

Combining the previous displays, we arrive at the general result 
$$
    \delta_{d,i}^{e}(R_i)=\int \delta_{y,i}^o(R_i) f_{Y_i}(y \mid S_i=e,D_i=d) \mathrm{d}y.
$$
When $Y$ is binary,
\begin{align*}
    f_{Y_i}(1 \mid S_i=e,D_i=d)&=\E(Y_i \mid S_i=e,D_i=d)=:\mu_i(d) \\
    f_{Y_i}(0 \mid S_i=e,D_i=d)&=1-\E(Y_i \mid S_i=e,D_i=d)=1-\mu_i(d).
\end{align*}
By the proof of Lemma~\ref{lemma:diff}, $\mu_i(d)=\E\{Y_i^{\mathrm{dir}}(d)|S_i=e\}$.
Therefore, the general result specializes to 
$$
\delta_{d,i}^{e}(R_i)= \delta_{1,i}^o(R_i) \mu_i(d)+\delta_{0,i}^o(R_i)\{1-\mu_i(d)\}=\delta_{0,i}^o(R_i)+\{ \delta_{1,i}^o(R_i) -\delta_{0,i}^o(R_i) \}\E\{Y_i^{\mathrm{dir}}(d)|S_i=e\}.
$$

        \item We apply Bayes' rule to rewrite
\begin{align*}
\delta_{y,i}^o(R_i) &= f_{R_i}(R_i \mid  S_i = o, Y_i = y) = \frac{\Pr(Y_i = y, S_i = o \mid R_i ) f_{R_i}(R_i )}{\Pr(Y_i = y, S_i = o )}, \\
\delta_{d,i}^{e}(R_i) &= f_{R_i}\left( R_i \mid S_i = e,  D_i = d \right) = \frac{ \Pr(D_i = d, S_i = e \mid R_i ) f_{R_i}(R_i ) }{ \Pr(D_i = d, S_i = e ) }.
\end{align*}
Substituting these expressions into the previous step and canceling $f_{R_i}(R_i)$ gives
$$ \E(\Delta^{e}_i|R_i)= \E(\Delta_i^o|R_i) \theta_i,\quad \theta_i=\E\{Y_i^{\mathrm{dir}}(1)-Y_i^{\mathrm{dir}}(0)|S_i=e\}.$$
    \end{enumerate}
    We conclude that
    $
    \theta_n=\frac{1}{n}\sum_{i=1}^n \theta_i=\frac{1}{n}\sum_{i=1}^n \frac{\E(\Delta^{e}_i|R_i)}{\E(\Delta_i^o|R_i)}.
    $

\subsubsection{Proof of Corollary~\ref{cor:spill1}}\label{proof:cor:spill1}

    Because spillovers only occur within mandals, the potential outcome of village $i$ only depends on the treatment assignments of other villages within its mandal; it does not depend on treatment assignments of other villages in other mandals. 
    Because treatment assignment is at the mandal level, the treatment of villages $i$ matches the treatment of all other villages in its mandal.
These statements imply that $\E\{Y_i^{\mathrm{dir}}(1)|S_i=e\}=\E\{Y_i(1,\mathbf{1}_{n-1})|S_i=e\}$. The same is true for $d_i=0$. Therefore, $\theta_n=\tilde{\theta}_n$ and we are done.

We formally prove this statement for village $i=1$ and $d_1=1$. Suppose that, among villages $\{1,...,n\}$, the initial $m$ belong to the same mandal. As argued in the proof of Lemma~\ref{lemma:diff}, with nonseparable model notation defined in Footnote~\ref{footnote: unobs het}, 
\begin{align*}
    \E\{Y_1^{\mathrm{dir}}(1,\eta_1)|S_1=e\}
    &=\E\{Y_1(1,D_2,...,D_n,\eta_1)|D_1=1, S_1=e\} \\
    &=\E\{Y_1(1,D_2,...,D_m,\eta_1)|D_1=1, S_1=e\} \\
    &=\E\{Y_1(1,\mathbf{1}_{m-1},\eta_1)|D_1=1, S_1=e\} \\
    &=\E\{Y_1(1,\mathbf{1}_{n-1},\eta_1)|D_1=1, S_1=e\} \\
    &=\E\{Y_1(1,\mathbf{1}_{n-1},\eta_1)| S_1=e\}
\end{align*}
    by spillovers only within mandals, mandal-level treatment assignment, spillovers only within mandals, and randomization.

\subsubsection{Proof of Corollary~\ref{cor:spill2}}\label{proof:cor:spill2}

  The argument is similar to the proof of Corollary~\ref{cor:spill1}. We prove the key equality.

  As before, consider village $i=1$ and $d_1=1$. Suppose that, among villages $\{1,...,n\}$, the initial $m$ are within a $20$ km radius. Each of these $m$ villages must have been treated due to the subetting rule. 
  
  As argued in the proof of Lemma~\ref{lemma:diff}, with nonseparable model notation defined in Footnote~\ref{footnote: unobs het}, 
\begin{align*}
    \E\{Y_1^{\mathrm{dir}}(1,\eta_1)|S_1=e\}
    &=\E\{Y_1(1,D_2,...,D_n,\eta_1)|D_1=1, S_1=e\} \\
    &=\E\{Y_1(1,D_2,...,D_m,\eta_1)|D_1=1, S_1=e\} \\
    &=\E\{Y_1(1,\mathbf{1}_{m-1},\eta_1)|D_1=1, S_1=e\} \\
    &=\E\{Y_1(1,\mathbf{1}_{n-1},\eta_1)|D_1=1, S_1=e\} \\
    &=\E\{Y_1(1,\mathbf{1}_{n-1},\eta_1)| S_1=e\}
\end{align*}
    by spillovers only within the radius, the subsetting rule, spillovers only within the radius, and randomization.

\subsection{Variables Collected at Different Times}

\subsubsection{Proof of Theorem~\ref{thm:timing}}\label{proof:thm:timing}

 The argument is identical to the proof of Lemma~\ref{lemma:discrete_mixture_incomplete} and Theorem~\ref{theorem:disc_outcomes_incomplete}, replacing $(D,Y,R)$ with $(D_t,Y_{t'},R_{t''})$.

\subsection{Partial Identification with Approximate Assumptions}

\subsubsection{Proof of Lemma \ref{lem:bias_direct}} \label{proof:lem:bias_direct}

We proceed in steps.

\begin{enumerate}
\item First, we reduce the numerator $\E\{\Delta^e H(R)\}$ and denominator $\E\{\Delta^o H(R)\}$ to differences in conditional means.
By the definition of $\Delta^e$,
  \[
  \E\{\Delta^e H(R)\}
  = \E\{H(R)\mid S=e,D=1\}-\E\{H(R)\mid S=e,D=0\}.
  \]
By the definition of $\Delta^o$ and $\Pr(D=0\mid S=o)=1$,
  \[
  \E\{\Delta^o H(R)\}
  = \E\{H(R)\mid S=o,D=0,Y=1\}-\E\{H(R)\mid S=o,D=0,Y=0\}.
  \]

\item Next, we analyze the numerator. 
  For each $d\in\{0,1\}$,
  \begin{align*}
      &\E\{H(R)\mid S=e,D=d\} \\
      &=  \E\{H(R)\mid S=e,D=d,Y=1\}\Pr(Y=1|S=e,D=d)+\E\{H(R)\mid S=e,D=d,Y=0\}\Pr(Y=0|S=e,D=d)\\
  &=\mu(d)\E\{H(R)\mid S=e,D=d,Y=1\}+\{1-\mu(d)\}\E\{H(R)\mid S=e,D=d,Y=0\}\\
  &=\mu(d)m_{d,1}+\{1-\mu(d)\}m_{d,0}.
  \end{align*}
by the law of iterated expectations, Assumption~\ref{assumption:experimental}, and the notation $m_{d,y}=\E\{H(R)\mid S=e,D=d,Y=y\}$. 

Taking differences across treatment arms and using $m_{1,y}=m_{0,y}+b_y(H)$,
  \begin{align*}
  \E\{\Delta^e H(R)\}
  &=\E\{H(R)\mid S=e,D=1\}-\E\{H(R)\mid S=e,D=0\}\\
  &=\mu(1)m_{1,1}+\{1-\mu(1)\}m_{1,0}-\mu(0)m_{0,1}-\{1-\mu(0)\}m_{0,0} \\
  &= \mu(1)\{m_{0,1}+b_1(H)\}+\{1-\mu(1)\}\{m_{0,0}+b_0(H)\}-\mu(0)m_{0,1}-\{1-\mu(0)\}m_{0,0} \\
  &= \{\mu(1)-\mu(0)\}(m_{0,1}-m_{0,0}) +\mu(1)b_1(H)+\{1-\mu(1)\}b_0(H)\\
  &=\tau (m_{0,1}-m_{0,0})+b(H).
  \end{align*}

\item Finally, we analyze the denominator. 
By Assumption~\ref{assumption:stability},
  \[
  m_{0,y}=\E\{H(R)\mid S=e,D=0,Y=y\}=\E\{H(R)\mid S=o,D=0,Y=y\}.
  \]
Therefore
\begin{align*}
    \E\{\Delta^o H(R)\}
  &= \E\{H(R)\mid S=o,D=0,Y=1\}-\E\{H(R)\mid S=o,D=0,Y=0\}
  = m_{0,1}- m_{0,0}.
\end{align*}
Collecting results, 
  \[
  \E\{\Delta^e H(R)\}
  = \tau\E\{\Delta^o H(R)\} + b(H).
  \]
  Dividing by $\E\{\Delta^o H(R)\}$ gives the desired conclusion.
\end{enumerate}

\subsubsection{Proof of Theorem \ref{thm:PI_direct}} \label{proof:thm:PI_direct}

We prove each result. 
\begin{enumerate}
    \item First, we prove the set is a valid outer bound. By Lemma~\ref{lem:bias_direct},
\[
\widetilde{\tau}(H)-\tau
=
\frac{b(H)}{\E\{\Delta^o H(R)\}}.
\]
By construction, $b(H)$ is a convex combination of $b_1(H)$ and $b_0(H)$, so the assumed bound implies $|b(H)|\leq \bar{b}(H)$. Therefore, 
$$
|\widetilde{\tau}(H)-\tau|\le \frac{\bar{b}(H)}{|\E\{\Delta^o H(R)\}|}.
$$
This gives the main interval. 

The second interval is due to the binary outcomes. Since $\tau=\mu(1)-\mu(0)$ and $\mu(1)\in[0,1]$, we must have $-\mu(0)\leq \tau\leq 1-\mu(0)$.
    \item Next, we prove the set is sharp with respect to the reduced form moments.  For clarity, let $\E_{\mathrm{obs}}(\cdot)$ denote the moments under the observed data distribution and $\E_{\star}(\cdot)$ the moments under a constructed DGP. 

For any candidate $\tau_{\star}$ in the set, define the required aggregate bias
$$
b_\star := \E_{\mathrm{obs}}\{\Delta^e H(R)\}-\tau_\star\E_{\mathrm{obs}}\{\Delta^o H(R)\}.
$$
Because $\tau_\star$ lies in the interval, $|b_\star|\le \bar{b}(H)$. Now, construct reduced-form moments satisfying
$$
\E_{\star}\{\Delta^o H(R)\}=\E_{\mathrm{obs}}\{\Delta^o H(R)\},\quad \tau=\tau_\star,\quad b(H)=b_\star
$$
Then Lemma~\ref{lem:bias_direct} implies
\[
\E_\star\{\Delta^e H(R)\}
=
\tau_\star
\E_\star\{\Delta^o H(R)\}
+
b_\star.
\]
Substituting the definitions gives
\begin{align*}
\E_\star\{\Delta^e H(R)\}
&=
\tau_\star
\E_{\mathrm{obs}}\{\Delta^o H(R)\}
+
\left[
\E_{\mathrm{obs}}\{\Delta^e H(R)\}
-
\tau_\star
\E_{\mathrm{obs}}\{\Delta^o H(R)\}
\right] \\
&=
\E_{\mathrm{obs}}\{\Delta^e H(R)\}.
\end{align*}
Thus the same observed reduced-form moments are reproduced while the treatment effect equals $\tau_\star$. Since $\tau_\star$ was arbitrary, the bound is sharp with respect to the reduced-form moments.
\end{enumerate}

\subsubsection{Proof of Lemma \ref{lem:bias_direct_instability}}\label{proof:lem:bias_partialid_both}

We proceed in steps. 
\begin{enumerate}
    \item As in the proof of Lemma~\ref{lem:bias_direct}, we rewrite the numerator and denominator as 
    \begin{align*}
         \E\{\Delta^e H(R)\}
  &= \E\{H(R)\mid S=e,D=1\}-\E\{H(R)\mid S=e,D=0\} \\
   \E\{\Delta^o H(R)\}
  &= \E\{H(R)\mid S=o,D=0,Y=1\}-\E\{H(R)\mid S=o,D=0,Y=0\}.
    \end{align*}
    \item Next, we analyze the numerator. As in the proof of Lemma~\ref{lem:bias_direct},
    $$
    \E\{\Delta^e H(R)\}
  =\tau (m_{0,1}-m_{0,0})+b(H),
    $$
    where 
    $
    m_{d,y}=\E\{H(R)\mid S=e,D=d,Y=y\}.
    $
    \item Finally, we analyze the denominator. Using 
    $$m_{0,y}=\E\{H(R)\mid S=o,D=0,Y=y\}+s_y(H),$$
    we have that 
    \begin{align*}
        m_{0,1}-m_{0,0}
        &=\E\{H(R)\mid S=o,D=0,Y=1\}+s_1(H)-\E\{H(R)\mid S=o,D=0,Y=0\}-s_0(H) \\
        &=\E\{\Delta^o H(R)\} + s(H).
    \end{align*}
    Collecting results, 
  \[
  \E\{\Delta^e H(R)\}
  = \tau[\E\{\Delta^o H(R)\}+s(H)] + b(H).
  \]
Rearranging gives the desired conclusion.
\end{enumerate}

\subsubsection{Proof of Theorem \ref{thm:PI_direct_instability}} \label{proof:thm:PI_direct_instability}

For the first claim, Lemma \ref{lem:bias_direct_instability} gives
$$
\tau = \frac{\E\{\Delta^e H(R)\}-b(H)}{\E\{\Delta^o H(R)\}+s(H)}.
$$
Since $b(H)$ is a convex combination of $b_0(H)$ and $b_1(H)$,   $|b(H)|\le \bar{b}(H)$. By the triangle inequality,  $|s(H)|\le 2\bar{s}(H)$.  

For the second claim, suppose $|\E\{\Delta^o H(R)\}|>2\bar{s}(H)$. The reverse triangle inequality gives
\begin{align*}
    |\E\{\Delta^o H(R)\}+s(H)|
    &\ge |\E\{\Delta^o H(R)\}|-|s(H)| \\
    &\ge |\E\{\Delta^o H(R)\}|-2\bar{s}(H)\\
    &>0.
\end{align*}
Substituting $\E\{\Delta^e H(R)\}=\widetilde{\tau}(H)\E\{\Delta^o H(R)\}$ into the expression for $\tau$, we obtain
\[
\tau-\widetilde{\tau}(H)
=
-\frac{b(H)+\widetilde{\tau}(H) s(H)}{\E\{\Delta^o H(R)\}+s(H)}.
\]
Therefore, by the previous bound,
\[
|\tau-\widetilde{\tau}(H)|
\le
\frac{|b(H)|+|\widetilde{\tau}(H)||s(H)|}{|\E\{\Delta^o H(R)\}+s(H)|}
\le
\frac{\bar{b}(H)+2\bar{s}(H)|\widetilde{\tau}(H)|}{|\E\{\Delta^o H(R)\}|-2\bar{s}(H)}=\kappa(H).
\]

\subsection{Continuous Outcomes}

\subsubsection{Proof of Proposition \ref{prop:continuous1}} \label{proof:prop:continuous1} 

Abbreviate each bin as $B_k:=B_\varepsilon(y_k)$. We proceed in steps.

\begin{enumerate}
    \item By the proof of Lemma~\ref{lemma:discrete_mixture_incomplete}, 
    $$
    f_R(r|S=e,D=d)=\int_{y\in\mathcal{Y}} f_{R}(r|S=o,Y=y) f_{Y(d)}(y|S=e)\mathrm{d}y.
    $$
 Partitioning $\mathcal Y$ into $\{B_k\}_{k=1}^K$ gives
$$
f_R(r\mid S=e,D=d)
=
\sum_{k=1}^K \int_{y\in B_k} f_R(r\mid S=o,Y=y)f_{Y(d)}(y|S=e)\mathrm{d}y.
$$
\item Fix $k$ and $y\in B_k$.  
Assumptions~\ref{assumption:observational}(ii)  and~\ref{ass:RSV_continuous} imply
\begin{align*}
    &\sup_{r\in\mathcal R}\big|f_R(r\mid S=o,Y=y)-f_R(r\mid S=o,Y \in B_k)\big| 
\le \omega(2\varepsilon).
\end{align*}

Therefore, adding and subtracting then applying the triangle inequality  gives
\begin{align*}
   &f_R(r\mid S=e,D=d) \\
&=
\sum_{k=1}^K \int_{y\in B_k} \{f_R(r\mid S=o,Y=y)-f_R(r\mid S=o,Y \in B_k)+f_R(r\mid S=o,Y \in B_k)\}f_{Y(d)}(y|S=e)\mathrm{d}y \\
&= \mathrm{err}_{\varepsilon}(r,d)+\sum_{k=1}^K \int_{y\in B_k} f_R(r\mid S=o,Y \in B_k) f_{Y(d)}(y|S=e)\mathrm{d}y\\
&= \mathrm{err}_{\varepsilon}(r,d)+\sum_{k=1}^K f_R(r\mid S=o,Y \in B_k) \Pr\{Y(d)\in B_k |S=e\},
\end{align*}
where, uniformly over $d\in\{0,1\}$ and $r\in\mathcal{R}$,
$$
|\mathrm{err}_{\varepsilon}(r,d)|\leq \omega(2\varepsilon).
$$

\item Subtracting this result across $d\in\{1,0\}$ and appealing to the definition of $\theta_{\varepsilon}$ gives 
\begin{align*}
    &f_R(r\mid S=e,D=1)-f_R(r\mid S=e,D=0) \\
    &=\mathrm{err}_{\varepsilon}(r,1)-\mathrm{err}_{\varepsilon}(r,0)+\sum_{k=1}^K f_R(r\mid S=o,Y \in B_k) [\Pr\{Y(1)\in B_k |S=e\}-\Pr\{Y(0)\in B_k |S=e\}] \\
    &=\mathrm{err}_{\varepsilon}(r)+\sum_{k=1}^K f_R(r\mid S=o,Y \in B_k) \theta_{\varepsilon,k},
\end{align*}
where, uniformly over $r\in\mathcal{R}$, 
$$
|\mathrm{err}_{\varepsilon}(r)|\leq |\mathrm{err}_{\varepsilon}(r,1)|+|\mathrm{err}_{\varepsilon}(r,0)|\leq 2\omega(2\varepsilon).
$$
    \item By Bayes' rule,
\begin{align*}
    f_R(r\mid S=e,D=d)&=\E\left[\frac{1\{D=d,S=e\}}{\Pr(D=d,S=e)}\Bigm|R=r\right]f_R(r) \\
    f_R(r\mid S=o,Y\in B_k)&=\E\left[\frac{1\{Y\in B_k,S=o\}}{\Pr(Y\in B_k,S=o)}\Bigm|R=r\right]f_R(r).
\end{align*}
By the proof of Theorem~\ref{theorem:disc_outcomes_incomplete}, we conclude that
$$
\E\left\{\Delta^e-(\Delta_\varepsilon^{o})^\top\theta_{\varepsilon}\mid R=r\right\}
=
\frac{\mathrm{err}_{\varepsilon}(r)}{f_R(r)}, 
$$
proving the desired result since Assumption \ref{ass:RSV_continuous} implies $|f_R(r)| \ge \underline{\ell}$. 
\end{enumerate}

\subsubsection{Proof of Corollary \ref{cor:continuous1}}\label{proof:cor:continuous1}

To prove this result, it is useful to introduce notation for the discretized outcome and discretized potential outcomes. Define $Y_{\varepsilon} = \sum_{k=1}^{K} y_k 1\{Y \in B_{\varepsilon}(y_k)\}$.
Analogously define $Y_{\varepsilon}(d) = \sum_{k=1}^{K} y_k 1\{Y(d) \in B_{\varepsilon}(y_k)\}$ for $d \in \{0, 1\}$.
By construction, $|Y - Y_{\varepsilon}| \leq \varepsilon$ almost surely and $|Y(d) - Y_{\varepsilon}(d)| \leq \varepsilon$ almost surely. 

We proceed in steps.

\begin{enumerate}
    \item For any $d \in \{0, 1\}$, observe
$$
| \E\{Y(d) \mid S = e\} - \E\{Y_{\varepsilon}(d) \mid S = e\}| \leq \E\{|Y(d) - Y_{\varepsilon}(d)| \mid S = e\} \leq \varepsilon.
$$
Defining $\tau_\varepsilon = \E\{Y_{\varepsilon}(1) - Y_{\varepsilon}(0) \mid S = e\}$, it follows that 
$$
|\tau - \tau_\varepsilon|= | [\E\{Y(1) \mid S = e\} - \E\{Y_{\varepsilon}(1) \mid S = e\}] -  [\E\{Y(0) \mid S = e\} - \E\{Y_{\varepsilon}(0) \mid S = e\}] | \leq 2 \varepsilon.
$$
    \item Defining $U_{\varepsilon} = \Delta^e - (\Delta_{\varepsilon}^{o})^{\top} \theta_{\varepsilon}$, Proposition \ref{prop:continuous1} establishes 
    $$|\E(U_{\varepsilon} \mid R)| \leq \frac{2\omega(2\varepsilon)}{\underline{\ell}}.$$ 
Therefore by the law of iterated expectations,
\begin{align*}
    \| \E\{H(R) U_{\varepsilon}\} \| 
    &=\| \E\{H(R) \E(U_{\varepsilon}|R)\} \| \\
    &\leq  \E\{\|H(R) \E(U_{\varepsilon}|R) \|\} \\
    &\leq  \E\{|\E(U_{\varepsilon}|R)|\cdot \|H(R) \|\} \\
    &\leq \frac{2\omega(2\varepsilon)}{\underline{\ell}}\E\{\|H(R) \|\}.
\end{align*}

    \item 
As notation, let $\widetilde{\theta}_\varepsilon = [\E\{H(R) (\Delta^{o}_{\varepsilon})^{\top}\}]^{-1} \E\{H(R) \Delta^e\}$. 
By definition of $\widetilde{\theta}_\varepsilon$,
$$
\E\{H(R) \Delta^e\}=\E\{H(R) (\Delta^{o}_{\varepsilon})^{\top}\} \widetilde{\theta}_\varepsilon.
$$
By definition of $U_{\varepsilon}$, 
$$
\E\{H(R) \Delta^e\}= \E\{H(R) (\Delta_{\varepsilon}^{o})^{\top}\} \theta_{\varepsilon} + \E\{H(R) U_{\varepsilon}\}.
$$ 
Combining these expressions,
$$
\E\{H(R) (\Delta^{o}_{\varepsilon})^{\top}\} \widetilde{\theta}_\varepsilon = \E\{H(R) (\Delta_{\varepsilon}^{o})^{\top}\} \theta_{\varepsilon} + \E\{H(R) U_{\varepsilon}\}.
$$
Since $\E\{H(R) (\Delta_{\varepsilon}^{o})^{\top}\}$ is invertible, it therefore follows that 
$$
\widetilde{\theta}_\varepsilon - \theta_{\varepsilon} = [\E\{H(R) (\Delta_{\varepsilon}^{o})^{\top}\}]^{-1} \E\{H(R) U_{\varepsilon}\}. 
$$
We conclude that
\begin{align*}
    \| \widetilde{\theta}_\varepsilon - \theta_{\varepsilon} \| 
    &\leq \|[\E\{H(R) (\Delta_{\varepsilon}^{o})^{\top}\}]^{-1}\|_{\mathrm{op}} \|\E\{H(R) U_{\varepsilon}\} \| \\
    &\leq \|[\E\{H(R) (\Delta_{\varepsilon}^{o})^{\top}\}]^{-1}\|_{\mathrm{op}} \frac{2\omega(2\varepsilon)}{\underline{\ell}}\E\{\|H(R) \|\}.
\end{align*}

\item We translate this bound for the discretized parameter into a bound for the discretized average treatment effect. By the Cauchy-Schwarz inequality,
\begin{align*}
| \widetilde{\tau}_{\varepsilon} - \tau_{\varepsilon} |
    &= | \lambda_{\varepsilon}^\top (\widetilde{\theta}_\varepsilon - \theta_{\varepsilon}) | \\
    &\leq \| \lambda_{\varepsilon} \| \cdot \|\widetilde{\theta}_\varepsilon - \theta_{\varepsilon}\| \\
    &\leq \| \lambda_{\varepsilon} \| \cdot  \|[\E\{H(R) (\Delta_{\varepsilon}^{o})^{\top}\}]^{-1}\|_{\mathrm{op}} \cdot \frac{2\omega(2\varepsilon)}{\underline{\ell}}\E\{\|H(R) \|\}
\end{align*}
\item Finally, we appeal to the triangle inequality: 
$
| \widetilde{\tau}_{\varepsilon} - \tau | \leq | \widetilde{\tau}_{\varepsilon} - \tau_{\varepsilon} | + |\tau_{\varepsilon} - \tau|.
$ \qed
\end{enumerate}

\subsubsection{Proof of Proposition~\ref{prop:continuous_deconvolution}}\label{proof:prop:continuous_deconvolution}

The derivation of the conditional moment equation is identical to the proof of Lemma~\ref{lemma:discrete_mixture_incomplete} and step one in the proof of Theorem~\ref{theorem:disc_outcomes_incomplete}, using the generalized outcome weights instead of the previously defined outcome weights.

By standard arguments in \cite{kress1989linear}, Assumption~\ref{assumption:existence} guarantees existence of a solution to this equation. Assumption~\ref{assumption:uniqueness} guarantees uniqueness of the solution to this equation. \qed 

\subsection{Estimation with Covariates}

\subsubsection{Proof of Proposition~\ref{lemma:disc_outcomes_incomplete_known_cell_specific}}\label{proof:lemma:disc_outcomes_incomplete_known_cell_specific}

The proof follows the argument of Proposition \ref{prop:known} within each covariate cell $x \in \mathcal{X}$. \qed

\subsubsection{Proof of Proposition \ref{prop:discrete_incomplete_unknown_covar}} \label{proof:prop:discrete_incomplete_unknown_covar}

The proof follows the argument of Proposition \ref{prop:unknown} within each covariate cell $x \in \mathcal{X}$. \qed


\section{Additional Estimation Algorithms}
\subsection{Estimation with Covariates}
\vspace{-1em}
\begin{algorithm}[H]
\scriptsize
\caption{Estimation with Discrete Outcomes and Discrete Covariates}\label{algorithm:discrete_covar_details}
\begin{algorithmic}[1]
\Require Data $\{S_i,X_i,1_{S_i=e}D_i,1_{S_i=o}Y_i,R_i\}_{i=1}^n$ with
$Y\in\mathcal Y=\{y_1,\ldots,y_K\}$ and $X\in\mathcal X=\{x_1,\ldots,x_{|\mathcal X|}\}$.
\Ensure $\widehat\tau$ and a confidence interval.
\State Randomly split indices into a $\tr$ fold and a $\te$ fold.

\ForAll{$x\in\mathcal X$}

\Statex \textbf{Step 1: Learn a representation on $\tr$.}

    \State Compute $\E_{\tr}\!\left(1_{D_i=d,S_i=e}\mid X_i=x\right)$ for $d\in\{0,1\}$ and
    $\E_{\tr}\!\left(1_{Y_i=y_k,S_i=o}\mid X_i=x\right)$ for $k=1,\ldots,K$.

    \State Train $\textsc{pred}_Y(x,\cdot):\mathcal R\to[0,1]^K$ on $\{(R_i,Y_i): i\in\tr,X_i=x,S_i=o\}$ to estimate
    $\Pr(Y=y_k\mid S=o,X=x,R)$.
    \State Train $\textsc{pred}_D(x,\cdot):\mathcal R\to[0,1]$ on $\{(R_i,D_i): i\in\tr,X_i=x,S_i=e\}$ to estimate
    $\Pr(D=1\mid S=e,X=x,R)$.
    \State Train $\textsc{pred}_S(x,\cdot):\mathcal R\to[0,1]$ on $\{(R_i,S_i): i\in\tr,X_i=x\}$ to estimate
    $\Pr(S=e\mid X=x,R)$.

    \State Define, for generic $(x,R)$,
    \[
    \widehat{\E}\{\Delta^e(x)\mid x,R\}
    =
    \left\{
    \frac{\textsc{pred}_D(x,R)}{\E_{\tr}(1_{D=1,S=e}\mid X=x)}
    -
    \frac{1-\textsc{pred}_D(x,R)}{\E_{\tr}(1_{D=0,S=e}\mid X=x)}
    \right\}\textsc{pred}_S(x,R),
    \]
    and for each $k=1,\ldots,K-1$,
    \[
    \widehat{\E}\{\Delta^o_k(x)\mid x,R\}
    =
    \left\{
    \frac{\textsc{pred}_{Y,k}(x,R)}{\E_{\tr}(1_{Y=y_k,S=o}\mid X=x)}
    -
    \frac{\textsc{pred}_{Y,K}(x,R)}{\E_{\tr}(1_{Y=y_K,S=o}\mid X=x)}
    \right\}\{1-\textsc{pred}_S(x,R)\}.
    \]
    \State Stack $\widehat{\E}\{\Delta^o(x)\mid x,R\}=\big(\widehat{\E}\{\Delta^o_1(x)\mid x,R\},\ldots,\widehat{\E}\{\Delta^o_{K-1}(x)\mid x,R\}\big)^\top$.
    \State Compute
    \[
    \widehat{\theta}_{\textsc{init}}(x)
    =
    \arg\min_{\theta\in\mathbb R^{K-1}}
    \sum_{i\in\tr:X_i=x}
    \left[
      \widehat{\E}\{\Delta^e(x)\mid x,R_i\}
      -
      \widehat{\E}\{\Delta^o(x)\mid x,R_i\}^\top\theta
    \right]^2.
    \]
    \State Compute $\widehat{\sigma}^2(\widehat{\theta}_{\textsc{init}},x,R)$ using Lemma~\ref{lemma:discrete_indicator} by substituting in the predictors above.
    \State Set
    $
    \widehat H(x,R)
    =
    \frac{\widehat{\E}\{\Delta^o(x)\mid x,R\}}{\widehat{\sigma}^2(\widehat{\theta}_{\textsc{init}},x,R)}
    \in\mathbb R^{K-1}.
    $
    
\Statex \textbf{Step 2: Estimate $\widehat\theta$ on $\te$.}

    \State Compute $\E_{\te}\!\left(1_{D_i=d,S_i=e}\mid X_i=x\right)$ for $d\in\{0,1\}$ and
    $\E_{\te}\!\left(1_{Y_i=y_k,S_i=o}\mid X_i=x\right)$ for $k=1,\ldots,K$.

    \ForAll{$i\in\te$ with $X_i=x$}
        \State Set
        \[
        \widehat{\Delta}^e_i(x)
        =
        \frac{1_{D_i=1,S_i=e}}{\E_{\te}(1_{D=1,S=e}\mid X=x)}
        -
        \frac{1_{D_i=0,S_i=e}}{\E_{\te}(1_{D=0,S=e}\mid X=x)}.
        \]
        \For{$k=1,\ldots,K-1$}
            \State Set
            \[
            \widehat{\Delta}^o_{ik}(x)
            =
            \frac{1_{Y_i=y_k,S_i=o}}{\E_{\te}(1_{Y=y_k,S=o}\mid X=x)}
            -
            \frac{1_{Y_i=y_K,S_i=o}}{\E_{\te}(1_{Y=y_K,S=o}\mid X=x)}.
            \]
        \EndFor
        \State Stack $\widehat{\Delta}^o_i(x)=\big\{\widehat{\Delta}^o_{i1}(x),\ldots,\widehat{\Delta}^o_{i(K-1)}(x)\big\}^\top$.
    \EndFor

    \State Compute
    \[
    \widehat{\theta}(x)
    =
    \left\{\sum_{i\in\te:X_i=x} \widehat H(x,R_i)\widehat{\Delta}^o_i(x)^\top\right\}^{-1}
    \left\{\sum_{i\in\te:X_i=x} \widehat H(x,R_i)\widehat{\Delta}^e_i(x)\right\}.
    \]
\EndFor

\Statex \textbf{Step 3: Compute the ATE and standard error.}

\State Compute $\widehat{\theta}=\E_{\te}\{\widehat{\theta}(X)\mid S=e\}$, i.e the empirical mean over $i\in\te$ with $S_i=e$.
\State Compute $\widehat{\tau}=\sum_{k=1}^{K-1} (y_k - y_K) \widehat{\theta}_k$.

\State Bootstrap the standard error $\widehat v$ of $\widehat\tau$, holding $\widehat H$ fixed.
\State \Return the confidence interval $\widehat\tau \pm z_{1-\alpha/2} \ \widehat v$, where $z_{1-\alpha/2}$ is the $(1-\alpha/2)$ quantile of a standard normal distribution.
\end{algorithmic}
\end{algorithm}

\begin{algorithm}[t]
\scriptsize
\caption{Estimation with Discrete Outcomes and Continuous Covariates}\label{algorithm:cts_covar_details}
\begin{algorithmic}[1]
\Require Data $\{S_i,X_i,1_{S_i=e}D_i,1_{S_i=o}Y_i,R_i\}_{i=1}^n$ with
$Y\in\mathcal Y=\{y_1,\ldots,y_K\}$ and $X\in\mathcal X\subseteq\mathbb R^{d_x}$.
\Require Basis $b:\mathcal X\to\mathbb R^J$, an initial representation $H_{\textsc{init}}(X,R)\in\mathbb R^{K-1}$, and the link function $\psi(\cdot)$ such that $\theta(x)=\psi\{g(x)\}$.

\Statex \textbf{Notation.} Consider any representation $H$ and basis $b$. Write $H_i=H(X_i,R_i)\in \mathbb{R}^{K-1}$ and $B_i=b(X_i)\in\mathbb{R}^J$. Construct the vector $C_i(H)\in\mathbb{R}^{K(K+2)}$ as the transpose of 
$$
 \begin{pmatrix}
  1_{ D_i = 1, S_i = e } &
   1_{ D_i = 1, S_i = e }H_i^{\top} &
    1_{ D_i = 0, S_i = e }   &
    1_{ D_i = 0, S_i = e }H_i^{\top}   &
    1_{ Y_i=y_1, S_i = o }  &
    1_{ Y_i=y_1, S_i = o }H_i^{\top}   &
    \cdots   &
     1_{ Y_i = y_K, S_i = o }    &
      1_{ Y_i = y_K, S_i = o }H_i^{\top} 
   \end{pmatrix}
$$
where the odd entries are scalars and the even entries are vectors in $\mathbb{R}^{K-1}$.
Define the series  estimator
$$
\widehat g_{\tr}(x;H)=b(x)^\top\widehat\beta_{\tr}(H),\quad \widehat\beta_{\tr}(H)=\arg\min_{\beta } \sum_{i\in\mathcal \tr}\|C_i(H)-B_i^\top\beta\|^2,
$$
and likewise define $\widehat g_{\te}(x;H)$

\State Randomly split indices into a $\tr$ fold and a $\te$ fold.

\Statex \textbf{Step 1: Learn a representation on $\tr$.}

\State Train $\textsc{pred}_Y(\cdot,\cdot):\mathcal{X}\times \mathcal{R}\rightarrow[0,1]^K$ on $\{(X_i,R_i,Y_i):i\in\tr,S_i=o\}$ to estimate $\Pr(Y=y_k\mid S=o,X,R)$.

\State Train $\textsc{pred}_D(\cdot,\cdot):\mathcal{X}\times \mathcal{R}\rightarrow[0,1]$ on $\{(X_i,R_i,D_i):i\in\tr,S_i=e\}$ to estimate $\Pr(D=1\mid S=e,X,R)$.

\State Train $\textsc{pred}_S(\cdot,\cdot):\mathcal{X}\times \mathcal{R}\rightarrow[0,1]$ on $\{(X_i,R_i,S_i):i\in\tr \}$ to estimate $\Pr(S=e\mid X,R)$.

\State Compute $\widehat g_{\tr}(x)=\widehat g_{\tr}(x;H_{\textsc{init}})$.

\State Let $\widehat{\E}_{\tr}(1_{D=d,S=e}\mid X)$ and $\widehat{\E}_{\tr}(1_{Y=y_k,S=o}\mid X)$ denote the corresponding
components of $\widehat g_{\tr}(X)$.

 \State Define, for generic $(X,R)$,
    \[
    \widehat{\E}\{\Delta^e(X)\mid X,R\}
    =
    \left\{
    \frac{\textsc{pred}_D(X,R)}{\widehat{\E}_{\tr}(1_{D=1,S=e}\mid X)}
    -
    \frac{1-\textsc{pred}_D(X,R)}{\widehat{\E}_{\tr}(1_{D=0,S=e}\mid X)}
    \right\}\textsc{pred}_S(X,R),
    \]
    and for each $k=1,\ldots,K-1$,
    \[
    \widehat{\E}\{\Delta^o_k(x)\mid X,R\}
    =
    \left\{
    \frac{\textsc{pred}_{Y,k}(X,R)}{\widehat{\E}_{\tr}(1_{Y=y_k,S=o}\mid X)}
    -
    \frac{\textsc{pred}_{Y,K}(X,R)}{\widehat{\E}_{\tr}(1_{Y=y_K,S=o}\mid X)}
    \right\}\{1-\textsc{pred}_S(X,R)\}.
    \]
    \State Stack $\widehat{\E}\{\Delta^o(X)\mid X,R\}=\big(\widehat{\E}\{\Delta^o_1(X)\mid X,R\},\ldots,\widehat{\E}\{\Delta^o_{K-1}(X)\mid X,R\}\big)^\top$.

\State Set $\widehat\theta_{\textsc{init}}(x)=\psi\{\widehat g_{\tr}(x)\}$.

    \State Compute $\widehat{\sigma}^2(\widehat{\theta}_{\textsc{init}},X,R)$ using Lemma~\ref{lemma:discrete_indicator}
    by substituting in the predictors above.

    \State Set 
    $
    \widehat H(X,R)
    =
    \frac{\widehat{\E}\{\Delta^o(x)\mid X,R\}}{\widehat{\sigma}^2(\widehat{\theta}_{\textsc{init}},X,R)}
    \in\mathbb R^{K-1}.
    $

\Statex \textbf{Step 2: Estimate $\widehat\theta$ on $\te$.}

\State Compute 
$\widehat g_{\te}(x)=\widehat g_{\te}(x;\widehat H)$.
\State Set $\widehat\theta(x)=\psi\{\widehat g_{\te}(x)\}$.

\Statex \textbf{Step 3: Compute the ATE and standard error.}
\State Compute $\widehat\tau(x)=
\sum_{k=1}^{K-1} (y_k - y_K) \widehat{\theta}_k(x)$.
\State Compute $\widehat\tau=\E_{\te}\big\{\widehat\tau(X)\mid S=e\big\}$, i.e. the empirical mean over $i\in\te$ with $S_i=e$.
\State Bootstrap the standard error of $\widehat\tau$ on $\te$, holding $\widehat H$ fixed.
\State \Return the confidence interval $\widehat\tau \pm z_{1-\alpha/2} \ \widehat v$ for $z_{1-\alpha/2}$ the $(1-\alpha/2)$ quantile of a standard normal distribution.
\end{algorithmic}
\end{algorithm}

\FloatBarrier
\subsection{An Algebraic Simplification}

The following lemma provides algebra that simplifies the denominator of the optimal representation.

\begin{lemma}[An algebraic simplification]\label{lemma:discrete_indicator}
By construction,
\begin{align*}
    &\{ \Delta^e(X) - \Delta^o(X)^\top \theta(X)\}^2 =  \frac{1_{ D = 1, S = e}}{\Pr(D = 1, S = e \mid X)^2} + \frac{1_{ D = 0, S = e }}{\Pr(D = 0, S = e \mid X)^2} \\
    &\quad +\sum_{k=1}^{K-1} \theta_k^2(X) \left\{ \frac{1_{Y = y_k, S = o }}{\Pr(Y = y_k, S = o\mid X)^2} + \frac{1_{Y = y_K, S = o} }{ \Pr(Y = y_K, S = o\mid X)^2 } \right\} \\
    &\quad +\sum_{j \neq k} \theta_j(X) \theta_k(X) \frac{1_{Y = y_K, S = o} }{ \Pr(Y = y_K, S = o\mid X)^2 }.
\end{align*}

\begin{proof}
To lighten notation, we suppress the arguments of $\Delta^e$, $\Delta^o$, and $\theta_j$. Observe that
\begin{align*}
  \{\Delta^e - (\Delta^o)^\top \theta\}^2 &= (\Delta^e)^2 - 2\{ (\Delta^o)^\top \theta \} \Delta^e + \{ (\Delta^o)^\top \theta\}^2 
  = (\Delta^e)^2 + \{ (\Delta^o)^\top \theta \}^2,
\end{align*}
since the events in $\Delta^e$ are exclusive of the events in $\Delta^o$. Specifically, $1_{S=e}1_{S=o}=0$.

Consider the former term. Since $1_{D=1}1_{D=0}=0$, we have by a similar logic that 
$$
(\Delta^e)^2 = \frac{1_{ D = 1, S = e}}{\Pr(D = 1, S = e \mid X)^2} + \frac{1_{ D = 0, S = e }}{\Pr(D = 0, S = e \mid X)^2}.
$$

Consider the latter term. Developing the square,
$$
\{ (\Delta^o)^\top \theta \}^2=\left(\sum_{k=1}^{K-1} \Delta^o_k \theta_k \right)^2=\sum_{j,k }  \theta_j  \theta_k \Delta^o_j \Delta^o_k.
$$
When $j=k<K$, the term becomes $\theta_k^2 (\Delta^o_k)^2$. Since $1_{Y=y_k}1_{Y=y_K}=0$, we have
$$
(\Delta^o_k)^2=\frac{1_{Y = y_k, S = o }}{\Pr(Y = y_k, S = o \mid X)^2} + \frac{1_{Y = y_K, S = o} }{ \Pr(Y = y_K, S = o \mid X)^2 }.
$$
When $j\neq k$ and $j,k<K$, the term becomes $\theta_j  \theta_k \Delta^o_j \Delta^o_k$. Since $1_{Y=y_j}1_{Y=y_k}=0$,  $1_{Y=y_j}1_{Y=y_K}=0$, and $1_{Y=y_k}1_{Y=y_K}=0$, we have 
$$
\Delta^o_j \Delta^o_k=\frac{1_{Y = y_K, S = o} }{ \Pr(Y = y_K, S = o \mid X)^2 }.
$$
\end{proof}
\end{lemma}

\clearpage
\section{Additional Simulation Details}\label{section: additional empirical details}
\subsection{Forest Cover Details}\label{section: appendix, forest cover, details} 

We utilize forest cover measurements from \citet[][]{hansen2013high} for Uganda, which are derived from Landsat satellite imagery at roughly 30-meter resolution. 
In the \cite{hansen2013high} product, tree cover is defined as the fraction of area within a pixel containing vegetation taller than 5 meters. 
We aggregate these pixel-level measurements to a common spatial grid and then construct the outcome used in our simulations.

The unit of analysis is a $0.01^\circ \times 0.01^\circ$ grid cell (approximately $1$ km $\times$ $1$ km) covering Uganda. All variables entering the simulations are aggregated to, or spatially aligned with, this grid. 
For each grid cell, we compute an average of the underlying 30-meter tree-cover pixels to obtain a grid cell-level tree-cover share. 
We then define a binary outcome $Y \in \{0,1\}$ indicating whether at least $80\%$ of the grid cell is forested. 
As noted in the main text, we use this high-forest threshold to align with conservation policies and program rules that are commonly expressed in terms of forest-cover retention requirements, such as mandates under Brazil's Forest Code \citep[][]{AzevedoEtAl(17)} and eligibility criteria in payments for ecosystem services programs \citep[][]{WunderEtAl(08)}.

\begin{figure}[htbp!]
\centering
\includegraphics[width=0.75\textwidth]{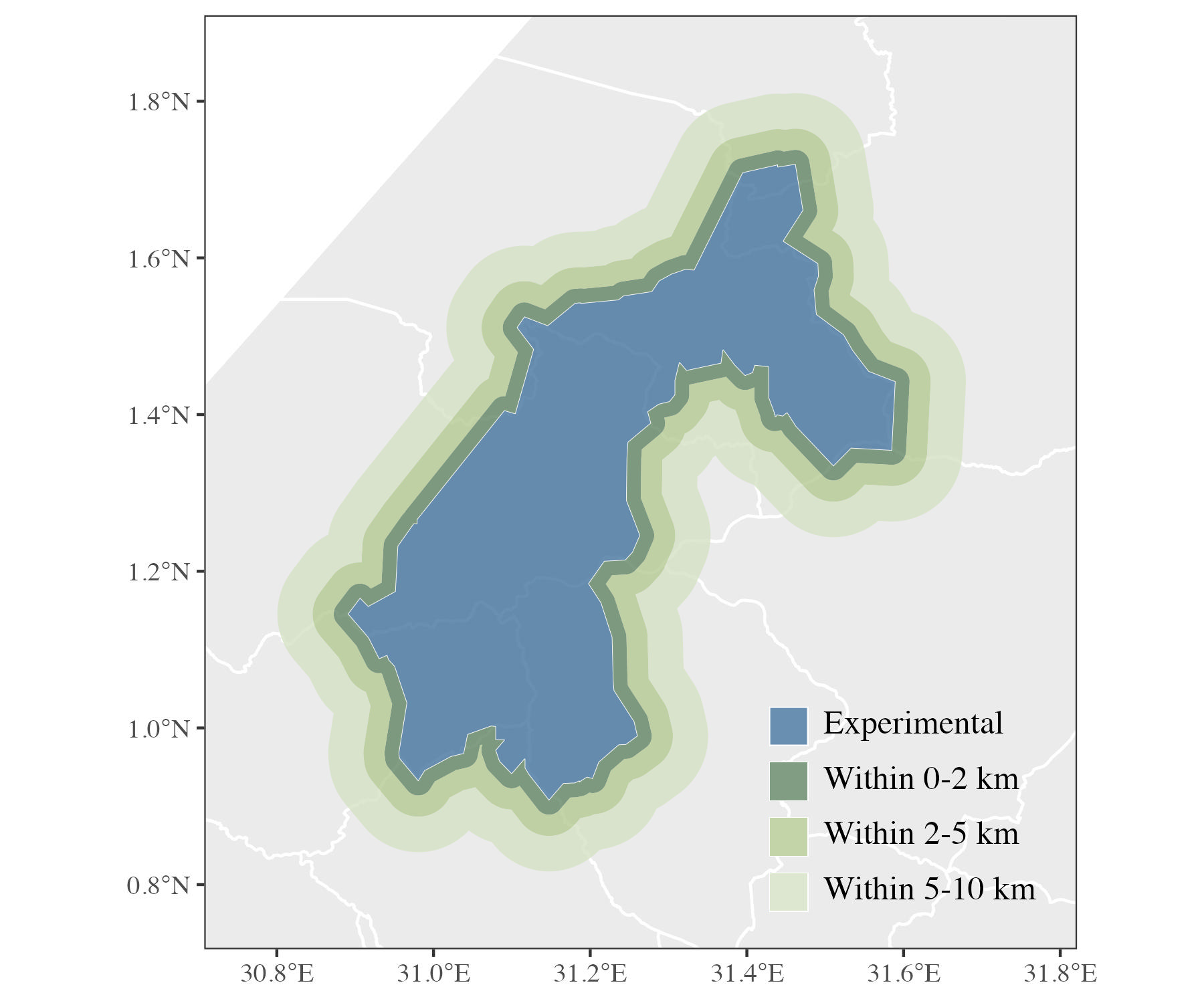}
\caption{We illustrate the experimental and observational samples in the Uganda forest cover data. 
The experimental sample consists of the grid cells covered by the payments for ecosystem services experiment \citep{jayachandran2017cash}. 
The observational sample consists of grid cells in surrounding geographic bands. In the main text, we focus on the 0--2 kilometer geographic band.} 
\label{fig:map_uganda_forestcover}
\end{figure}

The geographic samples used in our simulations are partitioned using the randomized study conducted by \citet[][]{jayachandran2017cash} in Uganda. 
We define the experimental sample as the set of grid cells falling within the geographic boundaries of the original study area. 
We define observational samples using surrounding geographic bands around the experimental region (0--2 km, 0--5 km, and 0--10 km). Appendix Figure~\ref{fig:map_uganda_forestcover} provides a map of these regions. 
In the main text, we focus on the 0--2 km band.

For each $0.01^\circ$ grid cell, we obtain a $4{,}000$-dimensional vector of MOSAIKS satellite embeddings \citep[][]{rolf2021generalizable}, which are derived from publicly available satellite imagery and are distinct from the Landsat imagery underlying the \cite{hansen2013high} forest-cover measurements. 
We merge these embeddings to the grid cell-level forest-cover outcome by geographic coordinates.

We then train a random forest to predict $Y$ from the MOSAIKS embeddings using grid cells from the rest of Uganda, excluding both the experimental region and the surrounding observational bands. 
The resulting predictor achieves high predictive accuracy, with an area under the curve (AUC) of $0.967$ on held-out grid cells. 
We define the remotely sensed variable $R$ as the scalar probability output of this trained random forest.

\subsection{Smartcards Details}\label{section: appendix, smartcards data, details} 

Randomization took place at the level of the mandal, i.e. subdistrict, of Andhra Pradesh, India. 
\cite{muralidharan2023general} partition 396 mandals into the following subgroups:
\begin{itemize}
    \item treated mandals (111), randomly assigned to receive Smartcards in 2010;
    \item buffer mandals (136), randomly assigned to receive Smartcards in 2011;
    \item untreated mandals (44), randomly assigned to receive Smartcards in 2012;
    \item non-study mandals (105), which were excluded from the experiment.
\end{itemize}

We study villages within mandals as the units of analysis, where villages are defined by \cite{asher2021development}. 
For each village, our treatment $D$ indicates whether the village received Smartcards in 2010. 
We interpret villages within the treated mandals as treated experimental units; villages within the buffer and untreated mandals as untreated experimental units; and villages within the non-study mandals as observational units with missing treatment. 
Finally, we drop villages with fewer than $100$ individuals. 
This removes about $2\%$ of villages. 

\begin{figure}
\centering
\includegraphics[width=0.6\textwidth]{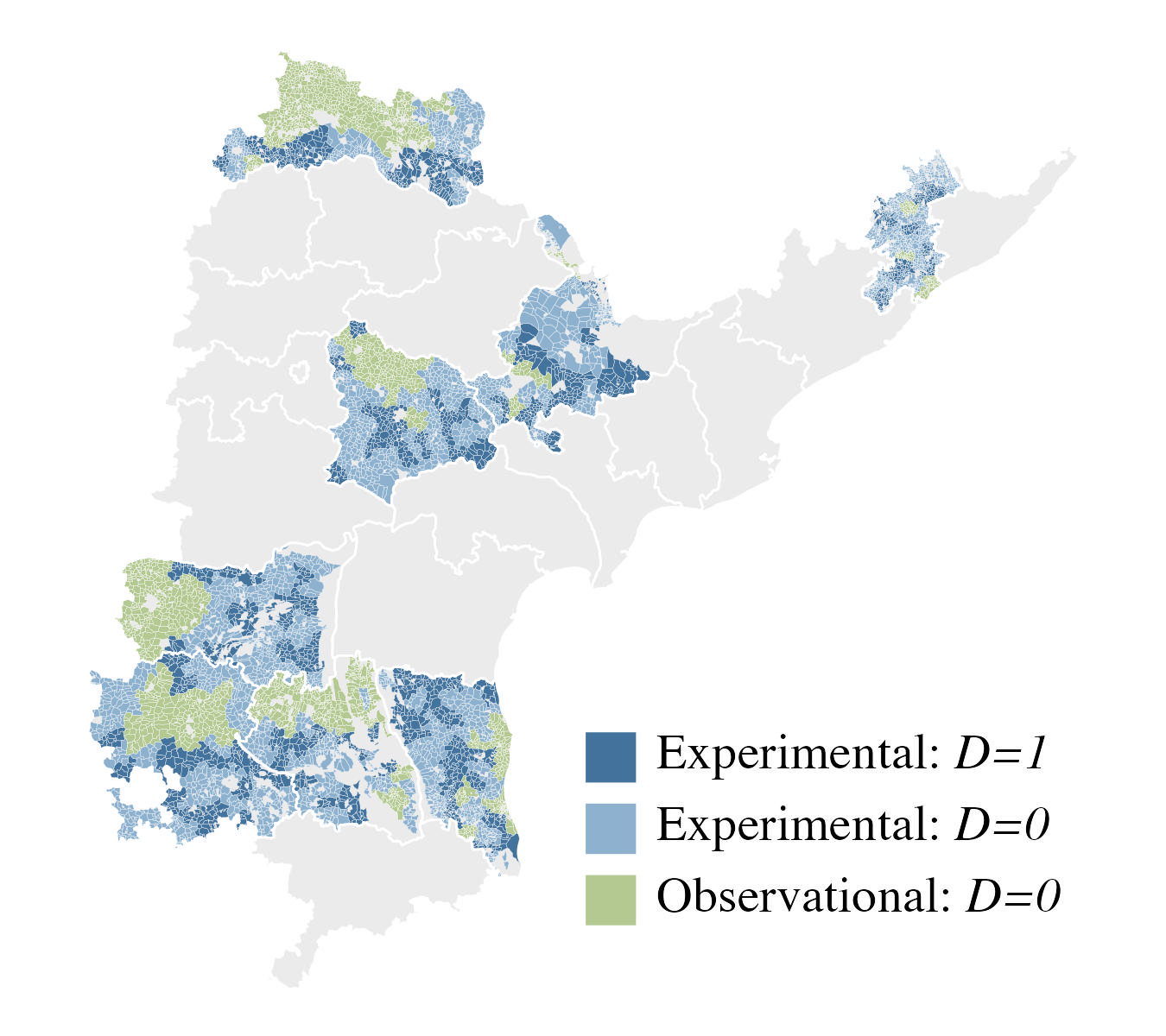}
\caption{We illustrate the experimental and observational samples in the Smartcards simulation \citep{muralidharan2023general}.} 
\label{fig:smartcards_sample_map}
\end{figure}

Appendix Figure~\ref{fig:smartcards_sample_map} illustrates our village classification. 
Appendix Table~\ref{tab:summary_stats_smartcards} summarizes village characteristics.
The causal parameter is the effect of early adoption (2010 Smartcards) for villages in the experiment. 

\begin{table}[htbp!]
\centering 
\caption{Summary statistics for the Smartcards experiment.}
 \label{tab:summary_stats_smartcards}
\scalebox{0.7}{
\begin{tabular}{lcccc}
\toprule
Sample & Non-Study Mandals & Untreated Mandals & Buffer Mandals & Treated Mandals \\
Smartcards Rollout & N/A & 2012 & 2011 & 2010 \\
\midrule 
Number of villages & 2257 & 852 & 2929 & 2274\\
Average population & 1985 & 2255 & 2105 & 2116\\
Average fraction female & 0.491 & 0.492 & 0.496 & 0.497\\
\bottomrule
\end{tabular}

}
\end{table} 

For each village, we collect poverty measurements to serve as the outcome. 
Following \cite{asher2020rural}, the data sources are the 2012-2013 Socio-Economic and Caste Census (SECC), and the 2013 Indian Economic Census. 
We consider three outcomes which we construct directly from these sources. 
The consumption outcome variable indicates whether a village's per capita consumption is in the bottom quartile. 
We consider two additional outcome variables based on income: does a village have only low income households, i.e. no earner making above 5,000 rupees; and does a village only low and middle income households, i.e. no earner making above 10,000 rupees. 
The definitions of low and middle income households are from the SECC.

For each village, we extract satellite images to serve as the remotely sensed variable. First, we extract coordinates for the perimeter of the village \citep{asher2021development}. 
Then, we extract luminosity measures, a vector in $\mathbb{R}^{50}$, which includes the minimum, maximum, mean, and sum of night light within each polygon, along with the total number of pixels in the polygon, from 2012 to 2020 \citep{asher2021development}. 
Finally, we extract satellite images from 2019, summarized as a high-dimensional, pre-trained embedding vector in $\mathbb{R}^{4000}$ \citep{rolf2021generalizable}. 
The concatenation of these objects is our remotely sensed variable $R$.

\subsection{Plausibility of Identifying Assumptions}\label{sec:main_simulation_stability}

We assess whether stability (Assumption~\ref{assumption:stability}) is plausible in each empirical setting. 
Here, stability requires that four conditional distributions of the remotely sensed variable should be the same across samples: $f_R(r \mid S=e, D=d, Y=y) = f_R(r \mid S=o, D=d, Y=y)$, for each $(d,y)$ stratum (or in the absence of direct effects, only for the $(d=0, y=0), (d = 0, y = 1)$ stratum).  
Using ``oracle'' data access, we can visually assess this condition. 
Specifically, we consider the outcomes of untreated units in the experimental and observational samples.\footnote{In the forest cover data, neither sample involves a treatment. In the Smartcards data, we observe outcomes for untreated experimental villages and for observational villages (which are all untreated).}
This allows us to directly evaluate the conditions for $d=0$ and $y\in\{0,1\}$ by subsetting and visualizing densities. 

\begin{figure}[htbp!]
\centering
\captionsetup[subfigure]{justification=Centering}
\begin{subfigure}{\textwidth}
    \centering
    \caption*{Panel A: Forest Cover in Uganda} 
    \begin{subfigure}{0.45\textwidth}
        \centering
        \includegraphics[width=\textwidth]{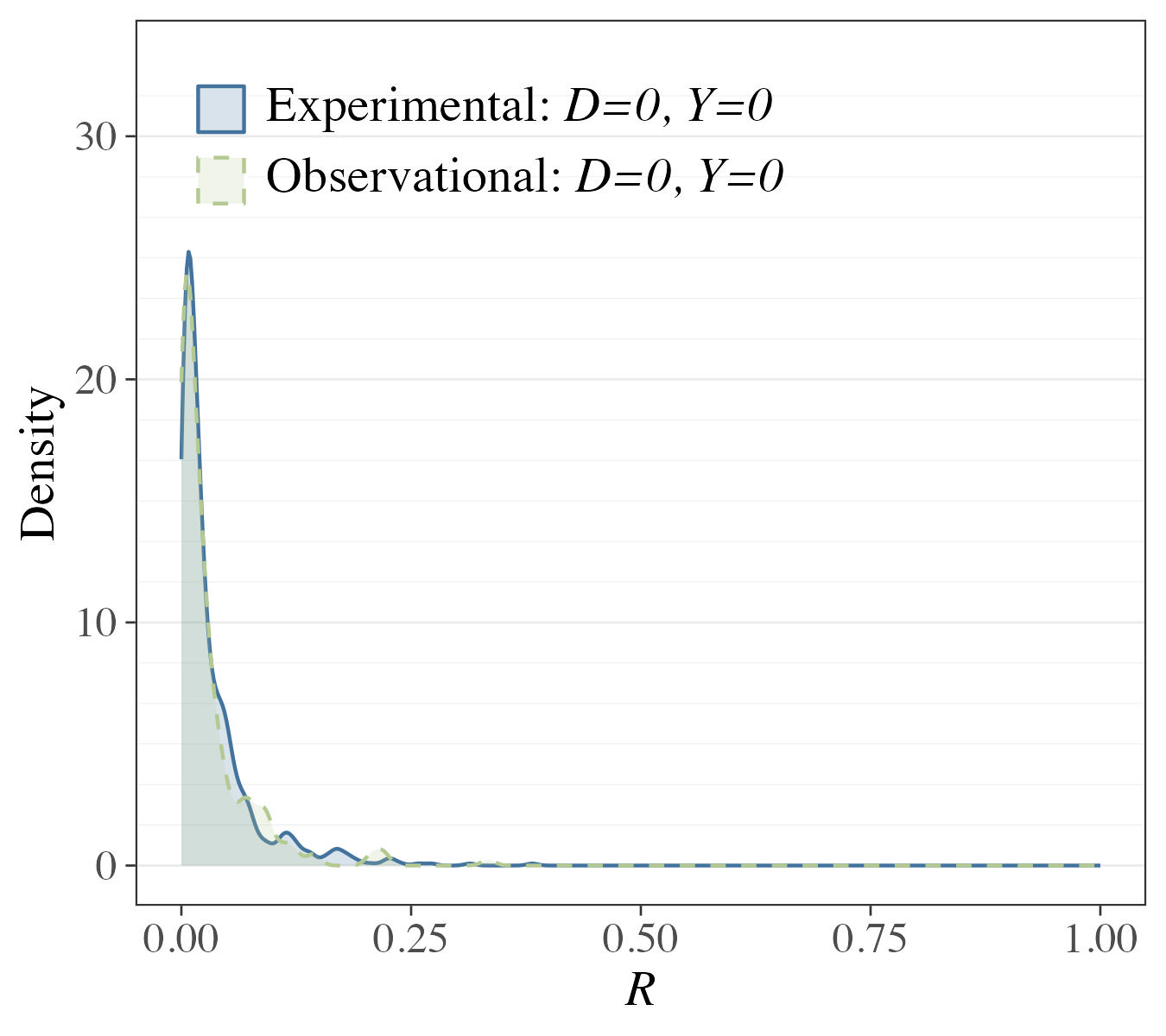}
        \caption{Densities of $R \mid S, D = 0, Y = 0$.}
    \end{subfigure}
    \hfill
    \begin{subfigure}{0.45\textwidth}
        \centering
        \includegraphics[width=\textwidth]{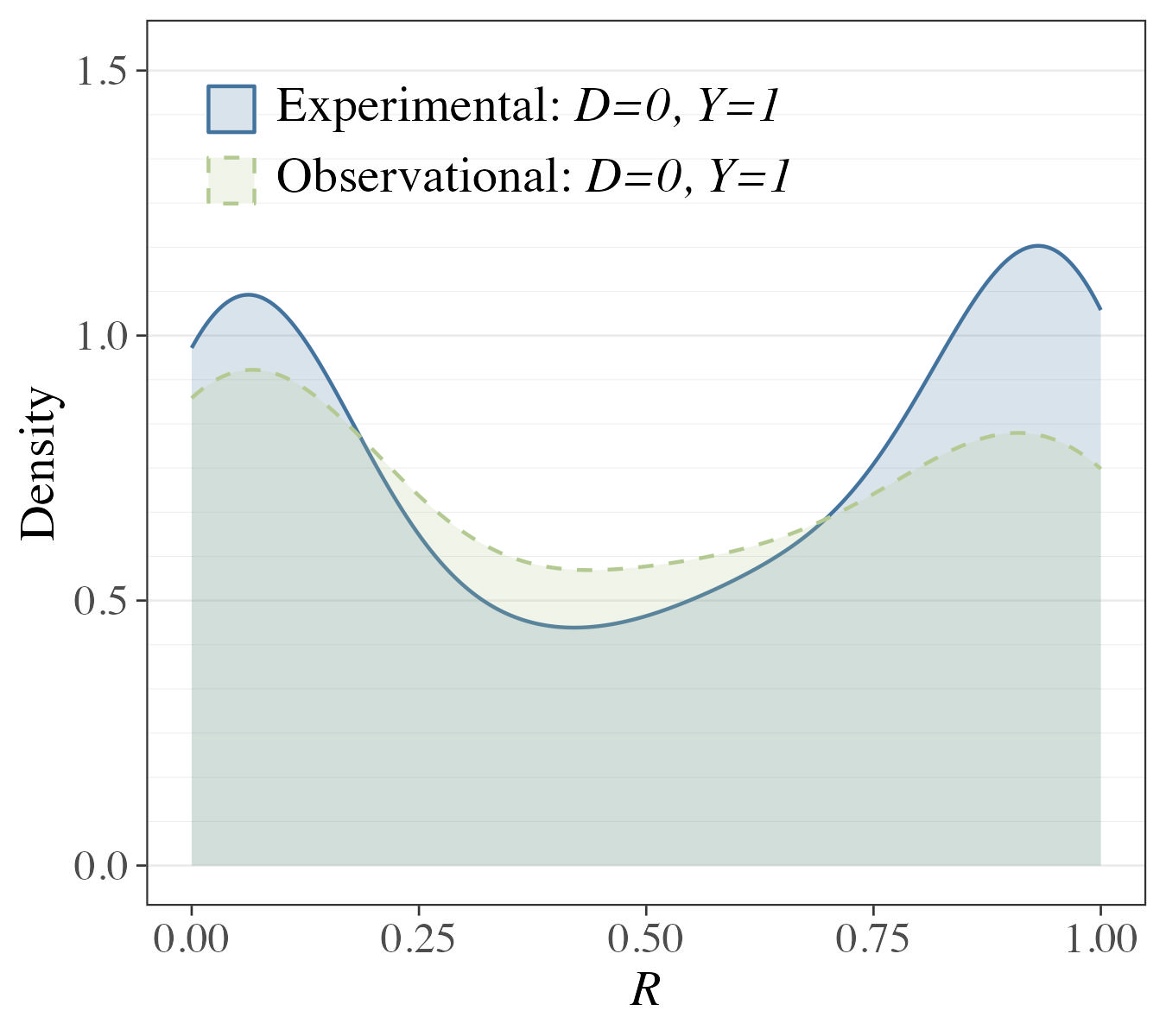}
        \caption{Densities of $R \mid S, D = 0, Y = 1$.}
    \end{subfigure}
\end{subfigure}
\vspace{1cm} 
\begin{subfigure}{\textwidth}
    \centering
    \caption*{Panel B: Smartcards Experiment in India}
    \begin{subfigure}{0.45\textwidth}
        \centering
        \includegraphics[width=\textwidth]{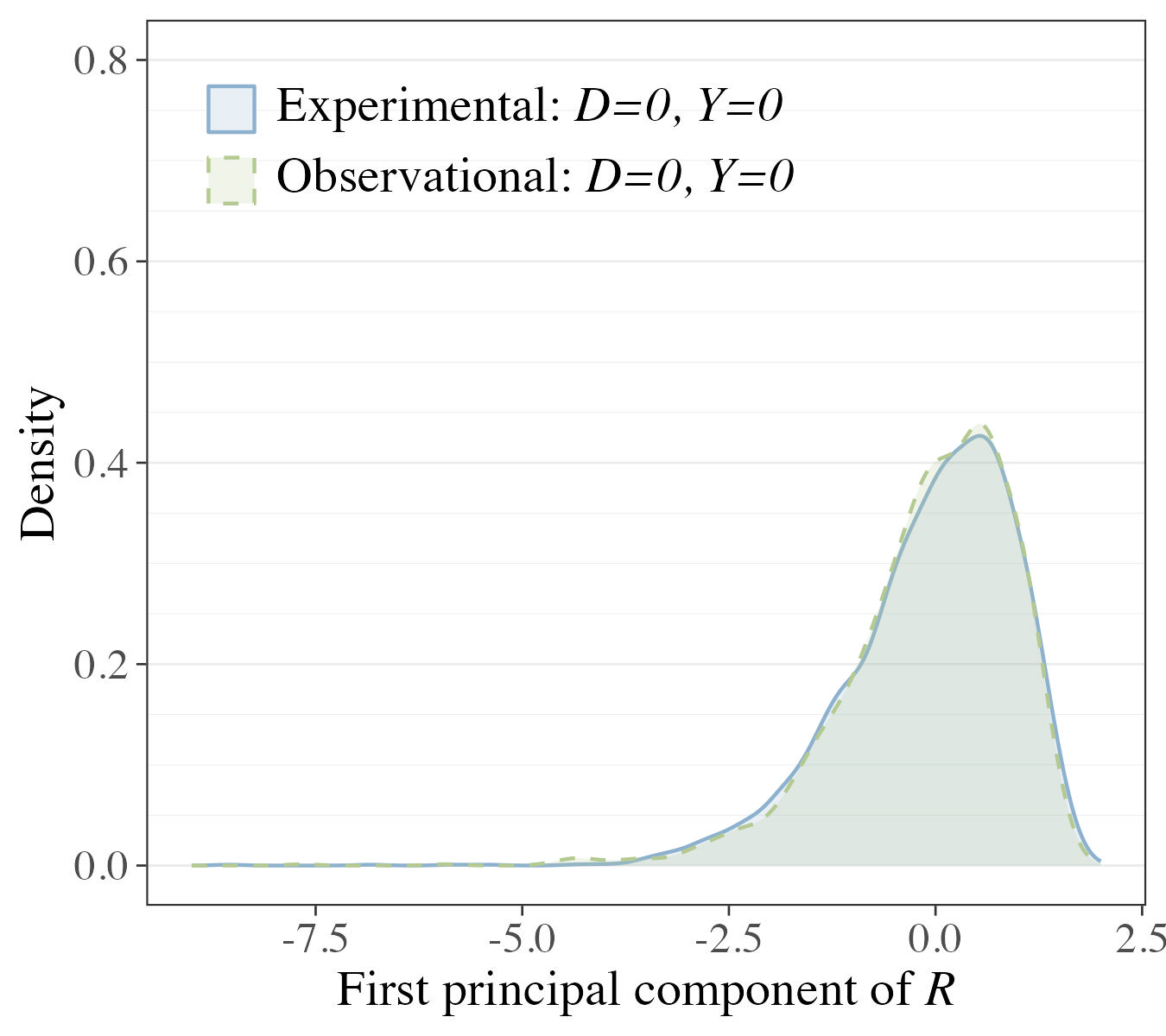}
        \caption{Densities of $R \mid S, D = 0, Y = 0$.}
    \end{subfigure}
    \hfill
    \begin{subfigure}{0.45\textwidth}
        \centering
        \includegraphics[width=\textwidth]{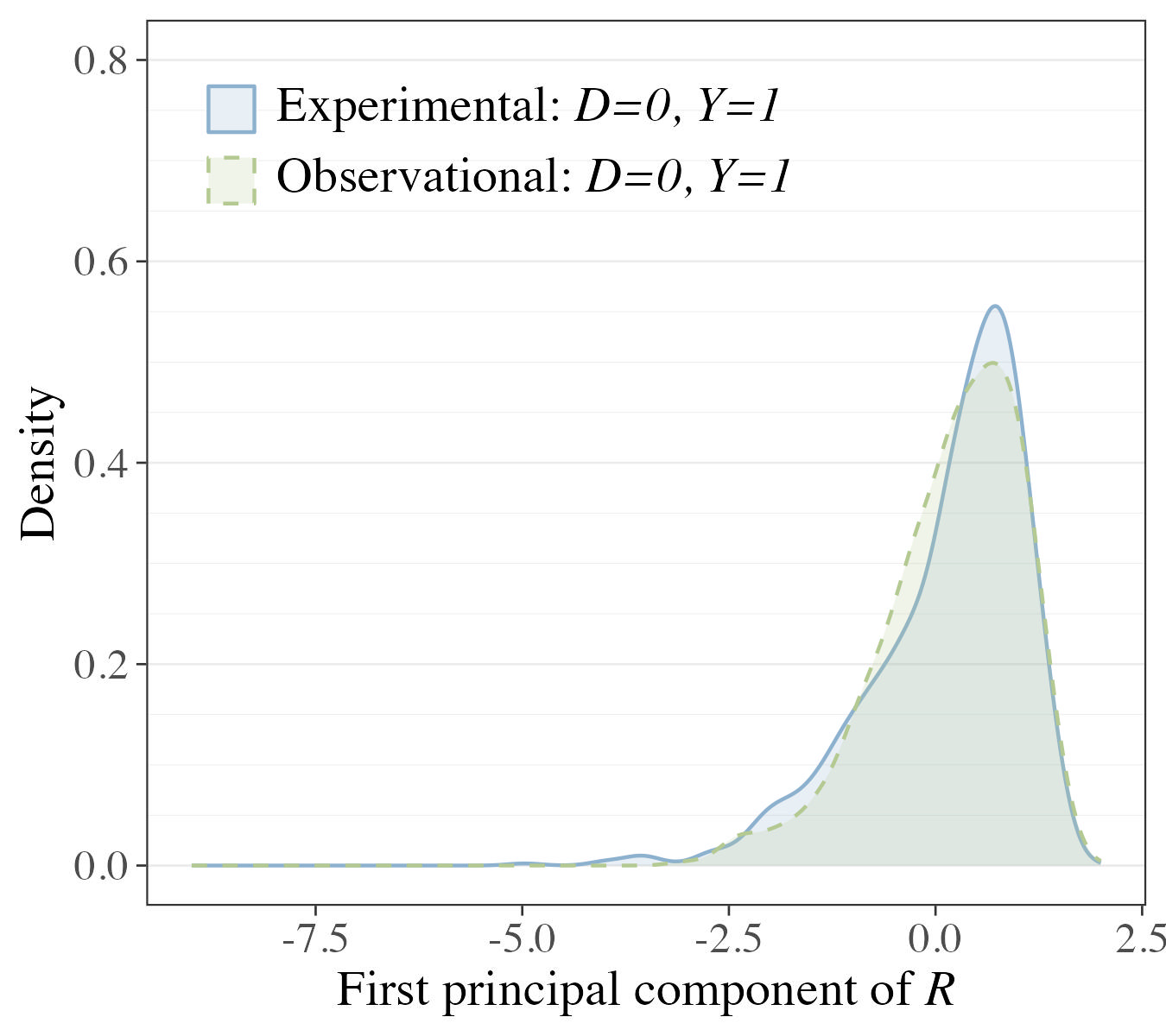}
        \caption{Densities of $R \mid S, D = 0, Y = 1$.}
    \end{subfigure}
\end{subfigure}
\caption{Our stability assumption (Assumption~\ref{assumption:stability}(i)) is plausible in two empirical settings.
We compare $f_R(R \mid S=e,D=d,Y=y)$ with $f_R(R \mid S=o,D=d,Y=y)$ for $d=0$ and $y\in\{0,1\}$.
Because the remotely sensed variable is high-dimensional in the Smartcards experiment, we visualize the density of its standardized first principal component.} 
\label{fig:simulation_stability}
\end{figure}

Figure~\ref{fig:simulation_stability} presents the visual evidence. 
In the forest cover setting (Panel~A), we compare the conditional distributions of $R$ for the experimental sample (grid cells in the region studied by \citet{jayachandran2017cash}) and for the observational sample (the two kilometer geographic band surrounding it). 
The conditional distributions reasonably align. 
In the Smartcards setting (Panel~B), $R$ is high-dimensional, so we plot the conditional distributions of its first principal component, comparing experimental and observational villages. 
The conditional distributions almost coincide.

To complement the visual evidence, we conduct Kolmogorov-Smirnov (KS) tests for equality of the conditional distributions across samples. 
In all comparisons, we cannot reject the null of equal distributions at the $5\%$ level.
Taken together, visual and statistical evidence supports the plausibility of our main identifying assumption in these two empirical settings.

\clearpage
\section{Additional Simulation Results}\label{section: additional monte carlo results}
In Section \ref{sec:simulations}, we reported calibrated simulations for the Uganda forest cover data and the Smartcards experiment in the two sample setting (i.e., experimental and observational samples only). 
In this section, we provide more details, then extend our calibration simulations to incorporate a validation sample (Section~\ref{sec:limited_experimental_outcomes}). 
The extension facilitates comparison with prediction powered inference (PPI).

\subsection{Experimental and Observational Samples Only}\label{section:additional_sims_without_outcomes}

\paragraph{Simulation design.}  
We observe $(D,R)$ when $S=e$ and $(R,Y)$ when $S=o$. Details are given in Section~\ref{sec:main_simulation_design}.

\paragraph{Implementation details.} 

We compare two methods. 

The two-step method trains a predictor of $Y$ from $R$ in the observational sample, then regresses predicted outcomes on $D$ in the experimental sample. Its estimand is $\widetilde{\tau}\neq \tau$ (Section~\ref{sec:common_practice_main_text}).
In the forest cover design, we predict $Y$ from $R$ using logistic regression. 
In the Smartcards design, we predict $Y$ from $R$ using probability random forests. Specifically, we use the \texttt{R} package \texttt{ranger} with $100$ trees and the package default options for all other tuning parameters.

For our method, we use cross-fitting with two folds.  
Its estimand is $\tau$. 
In the forest cover design, we estimate the efficient representation by combining logistic regressions of $\widehat{\Pr}(Y=1 \mid S=o,R)$, $\widehat{\Pr}(D=1\mid S=e,R)$, and $\widehat{\Pr}(S=e\mid R)$.
In the Smartcards design, we do so by combining probability random forests. Specifically, we use the \texttt{R} package \texttt{ranger} with $100$ trees and the package default options.

For computational tractability, we truncate the remotely sensed in the Smartcards design from $\mathbb{R}^{4050}$ to $\mathbb{R}^{1050}$.  
In particular, we use the initial $1,000$ features of the pre-trained MOSAIKS embedding and the nighttime luminosity measures. 

PPI methods are infeasible in this setting because we do not have joint observations of $(D=0,Y,R)$ and $(D=1,Y,R)$.

\paragraph{Additional results.}
We further compare our method and the two-step method in simulations,  varying the baseline outcome probabilities and the sample sizes in the simulation designs.  For these variations on the simulation designs, we find the same results as in the main text.

\begin{figure}[H]
\centering
\captionsetup[subfigure]{justification=Centering}
\vspace{1cm} 
\begin{subfigure}{\textwidth}
    \centering
    \caption*{Panel A: 0--5 Kilometer Geographic Band}
    \begin{subfigure}{0.30\textwidth}
        \centering
        \includegraphics[width=\textwidth]{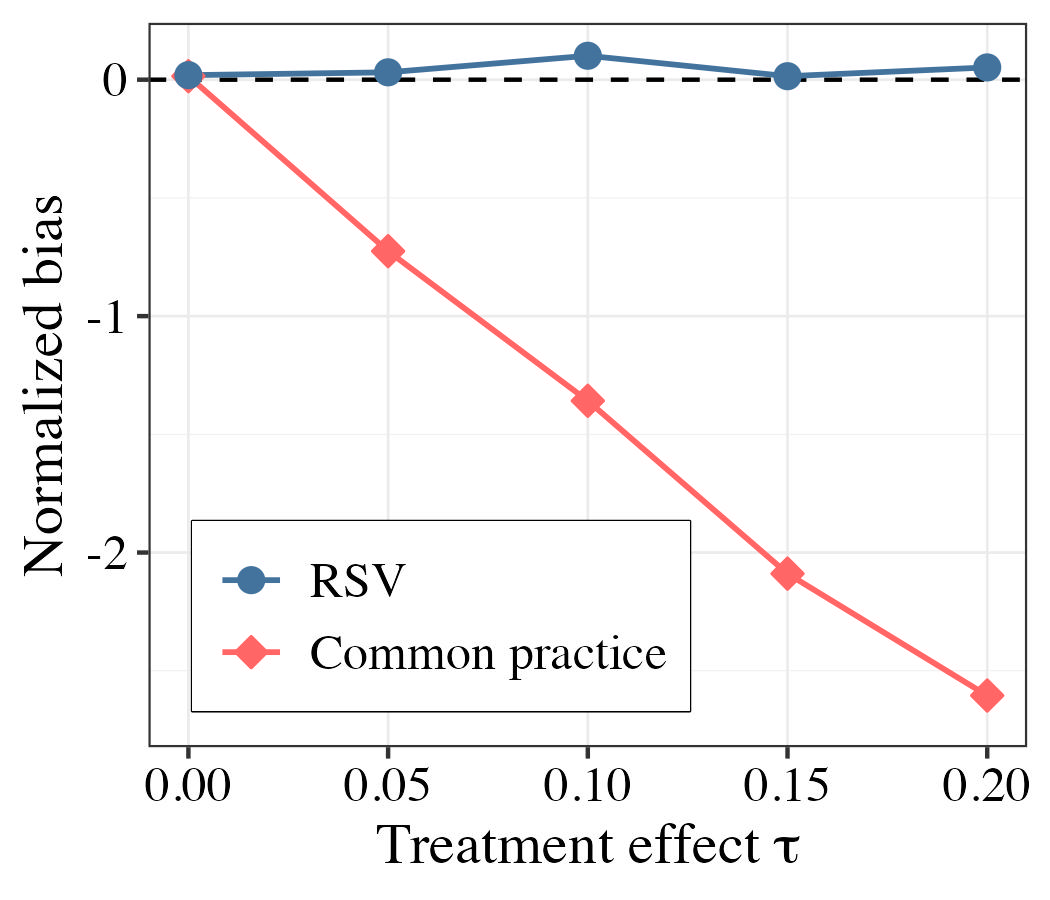}
        \caption{$n_o = 500$.}
    \end{subfigure}
    \hfill
    \begin{subfigure}{0.30\textwidth}
        \centering
        \includegraphics[width=\textwidth]{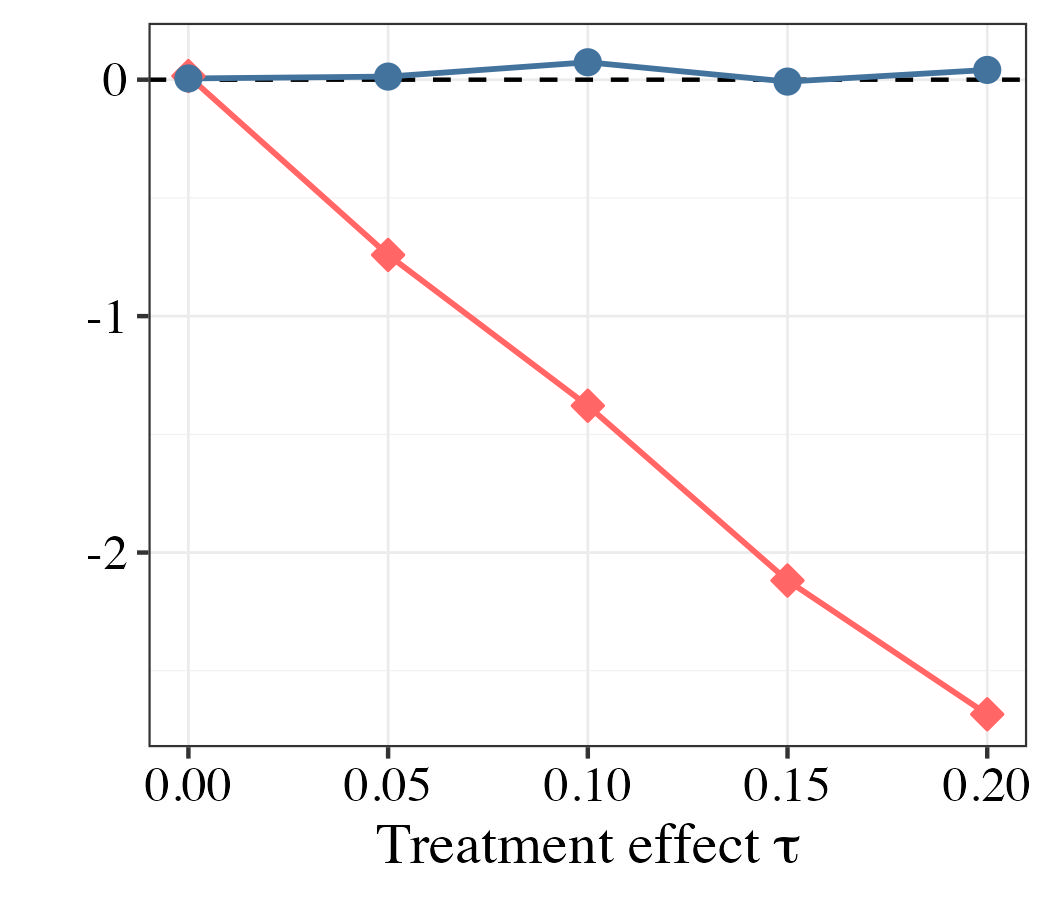}
        \caption{$n_o = 1000$.}
    \end{subfigure}
    \hfill 
    \begin{subfigure}{0.30\textwidth}
        \centering
        \includegraphics[width=\textwidth]{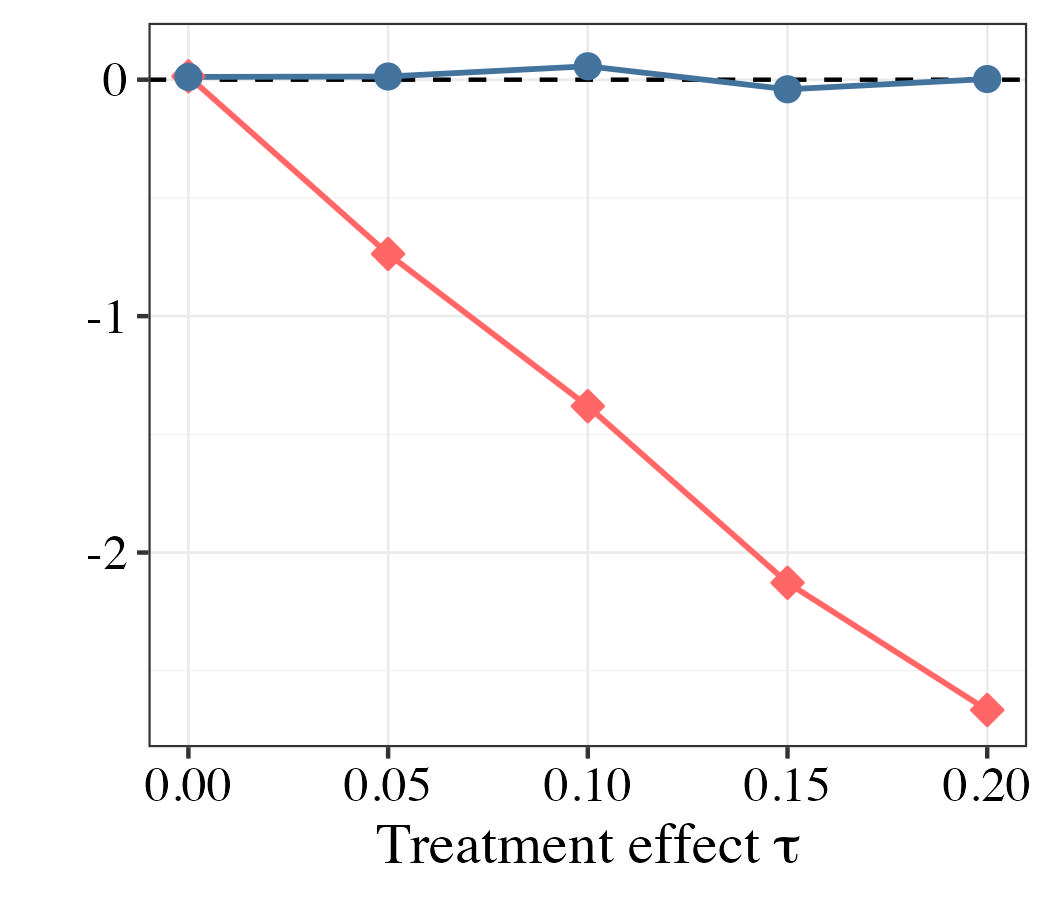}
        \caption{$n_o = 2000$.}
    \end{subfigure}
\end{subfigure}
\vspace{1cm} 
\begin{subfigure}{\textwidth}
    \centering
    \caption*{Panel B: 0--10 Kilometer Geographic Band}
    \begin{subfigure}{0.30\textwidth}
        \centering
        \includegraphics[width=\textwidth]{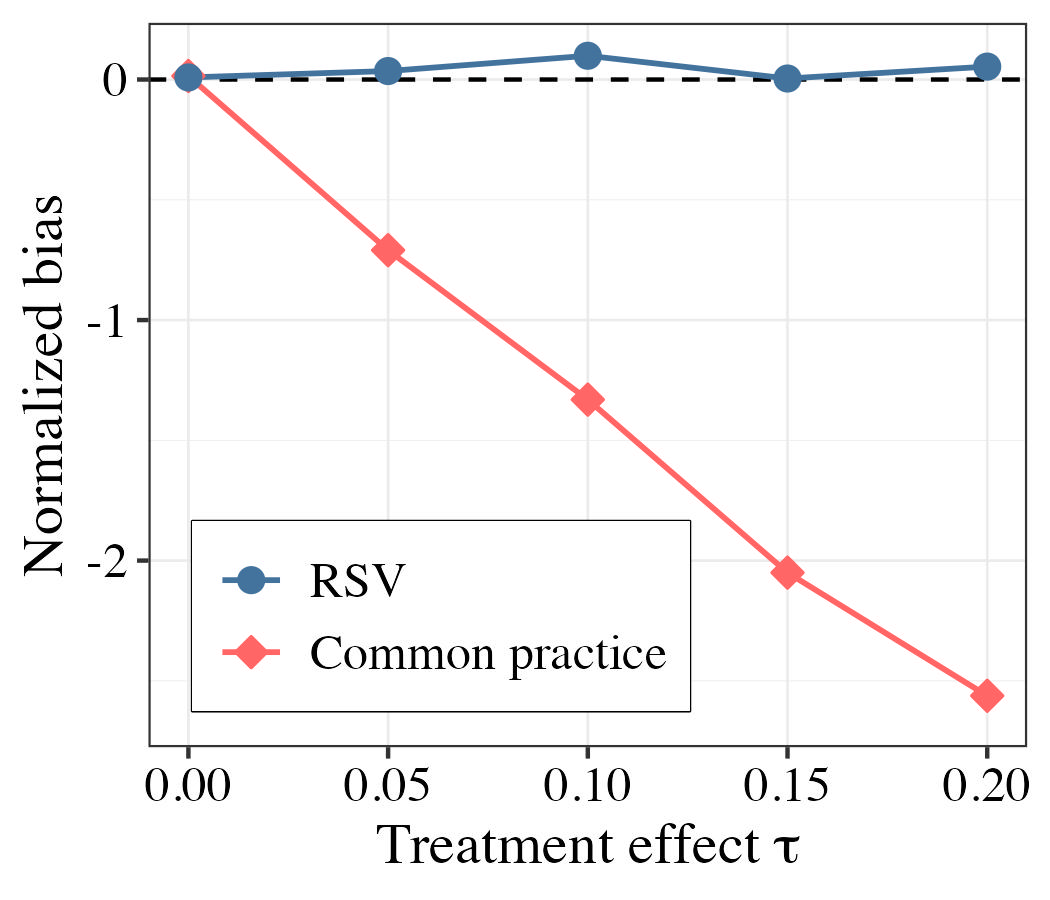}
        \caption{$n_o = 500$.}
    \end{subfigure}
    \hfill
    \begin{subfigure}{0.30\textwidth}
        \centering
        \includegraphics[width=\textwidth]{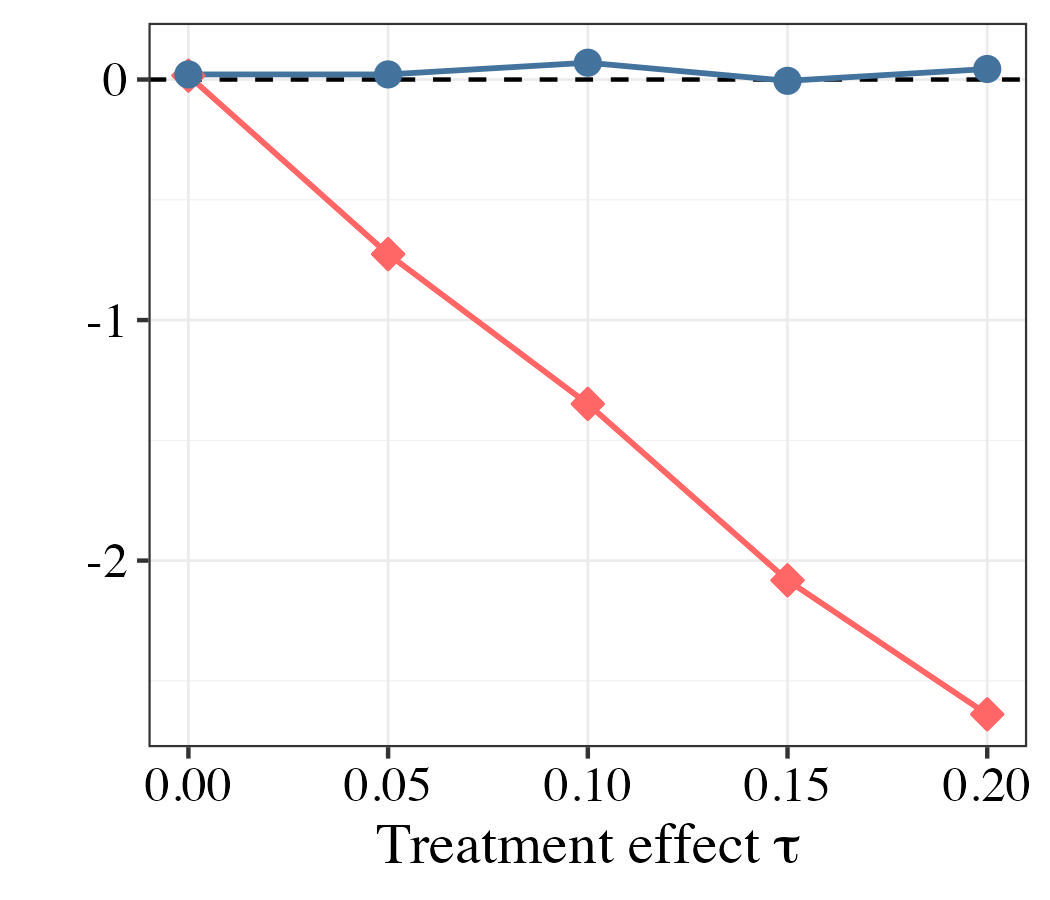}
        \caption{$n_o = 1000$.}
    \end{subfigure}
    \hfill 
    \begin{subfigure}{0.30\textwidth}
        \centering
        \includegraphics[width=\textwidth]{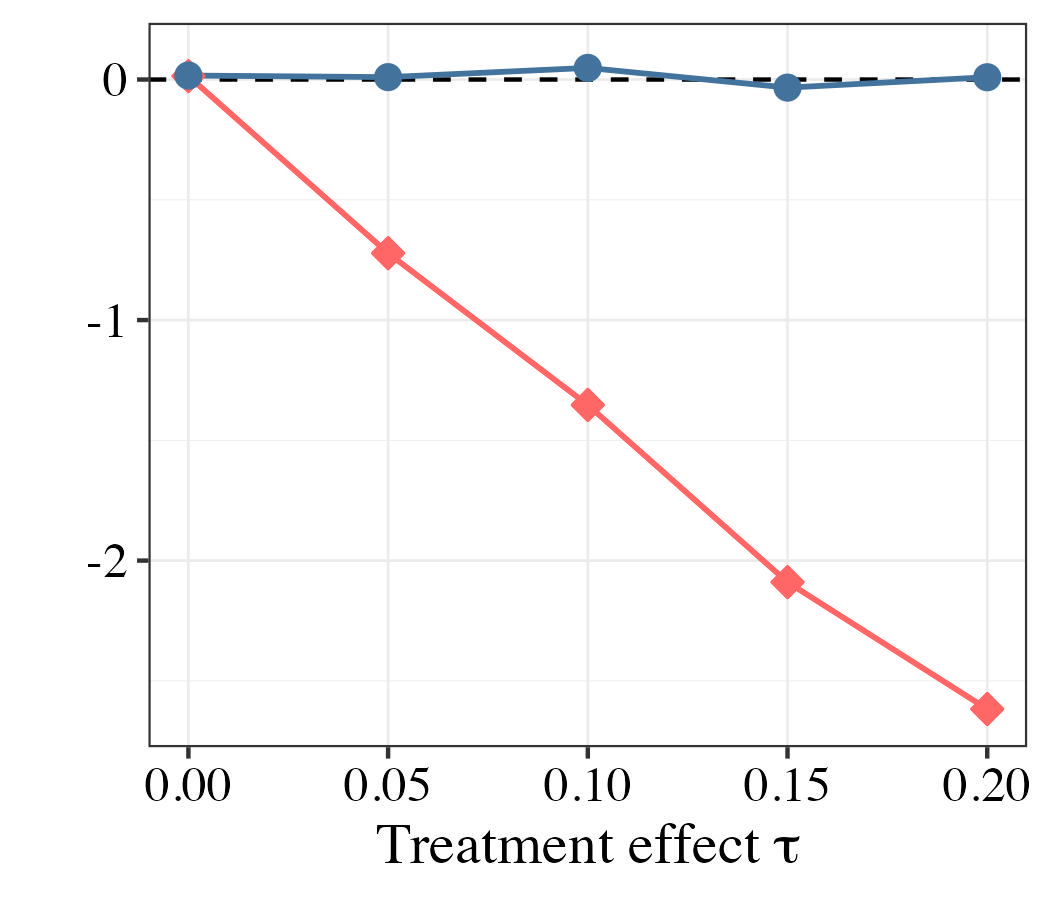}
        \caption{$n_o = 2000$.}
    \end{subfigure}
\end{subfigure}
\caption{In the forest cover design, our method outperforms the two-step method in terms of normalized average bias. 
The different panels vary $\alpha_o$. The different plots vary $n_o$. 
For each value of the synthetic treatment effect $\tau$, we conduct 500 simulations.} 
\label{fig:nooutcomes_uganda_distributionshift_and_samplesize_normalizedbias}
\end{figure}

\begin{figure}[H]
\centering
\captionsetup[subfigure]{justification=Centering}
\vspace{1cm} 
\begin{subfigure}{\textwidth}
    \centering
    \caption*{Panel A: 0--5 Kilometer Geographic Band}
    \begin{subfigure}{0.30\textwidth}
        \centering
        \includegraphics[width=\textwidth]{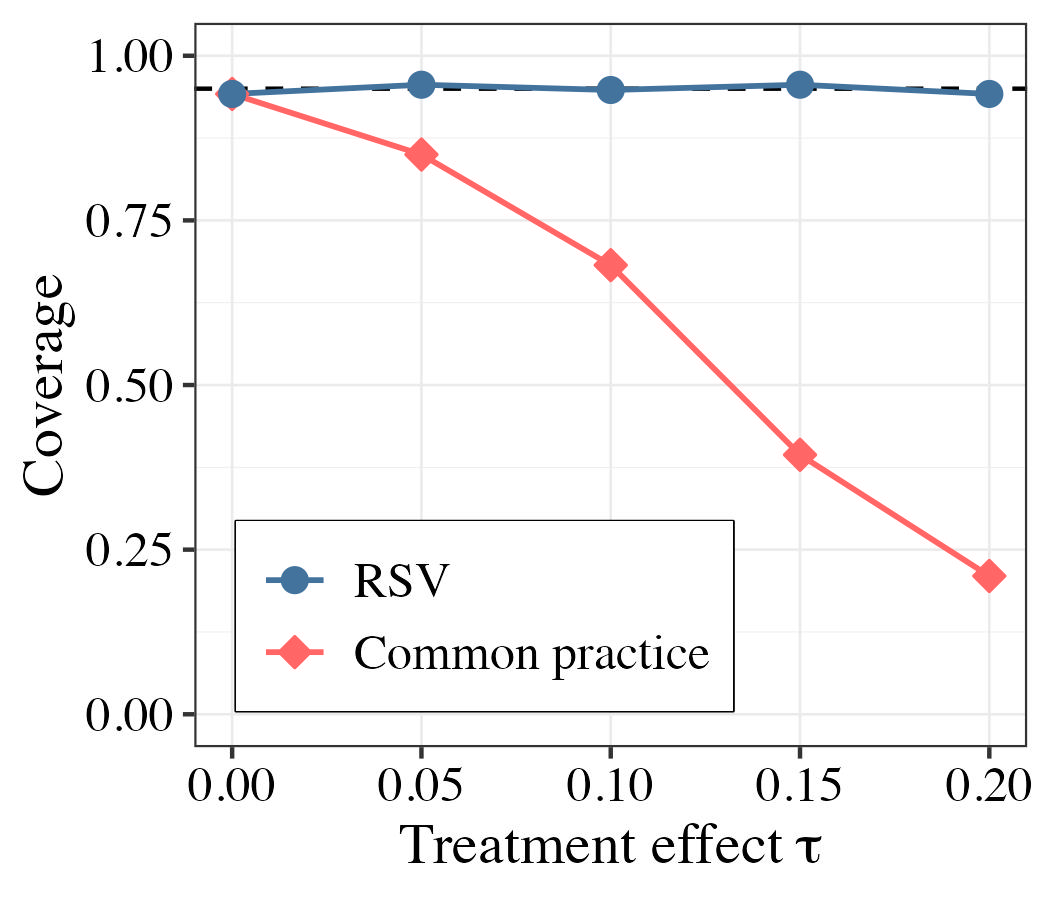}
        \caption{$n_o = 500$.}
    \end{subfigure}
    \hfill
    \begin{subfigure}{0.30\textwidth}
        \centering
        \includegraphics[width=\textwidth]{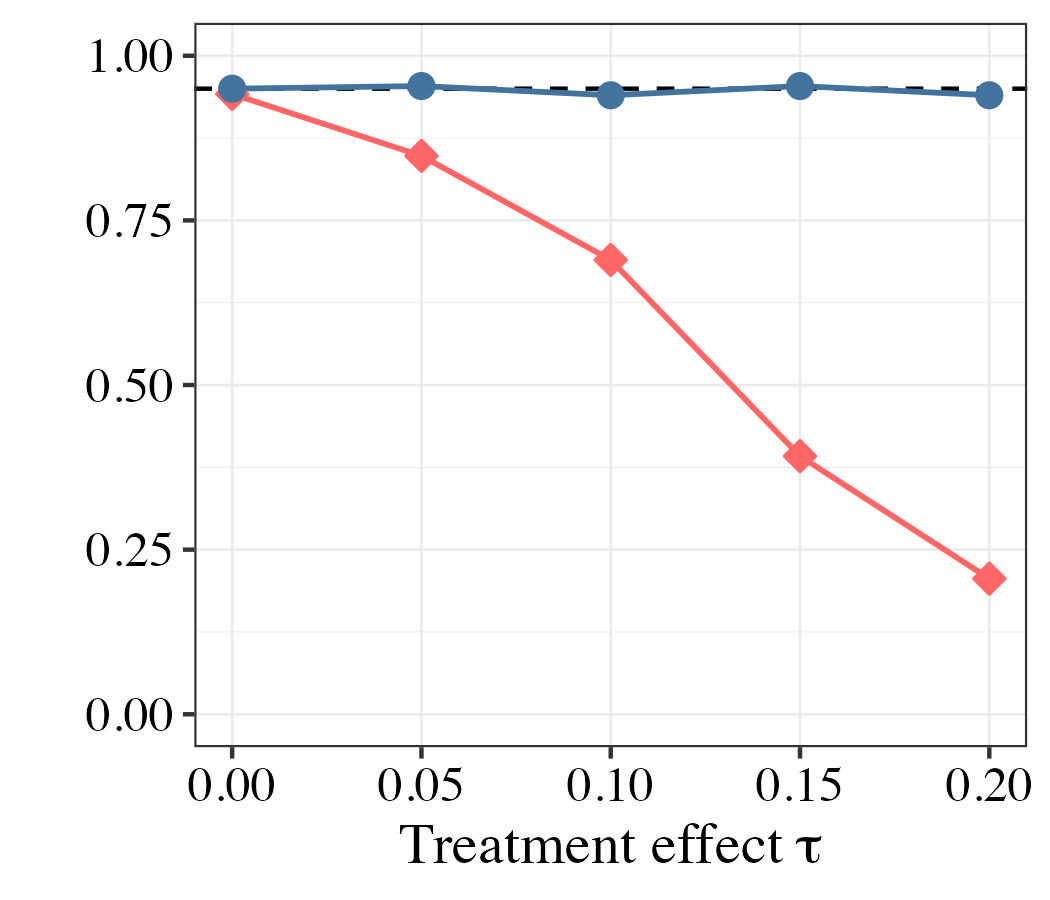}
        \caption{$n_o = 1000$.}
    \end{subfigure}
    \hfill 
    \begin{subfigure}{0.30\textwidth}
        \centering
        \includegraphics[width=\textwidth]{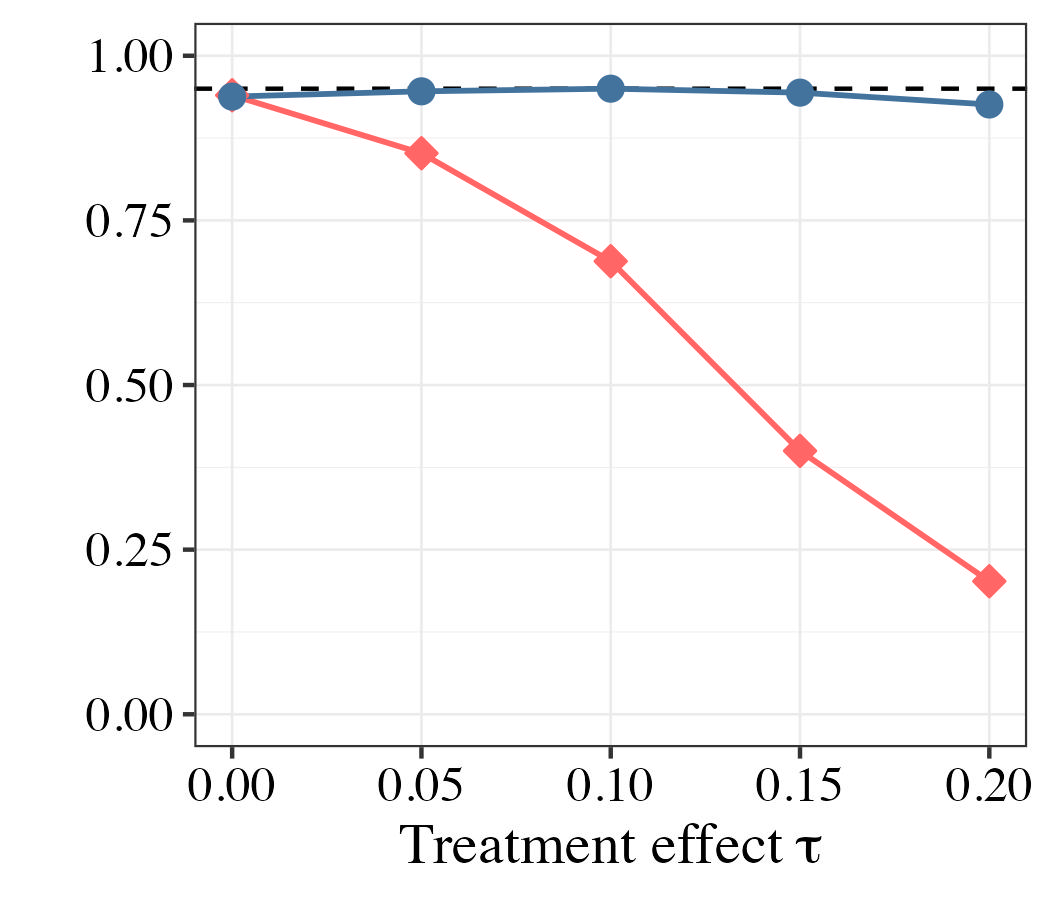}
        \caption{$n_o = 2000$.}
    \end{subfigure}
\end{subfigure}
\vspace{1cm} 
\begin{subfigure}{\textwidth}
    \centering
    \caption*{Panel B: 0--10 Kilometer Geographic Band}
    \begin{subfigure}{0.30\textwidth}
        \centering
        \includegraphics[width=\textwidth]{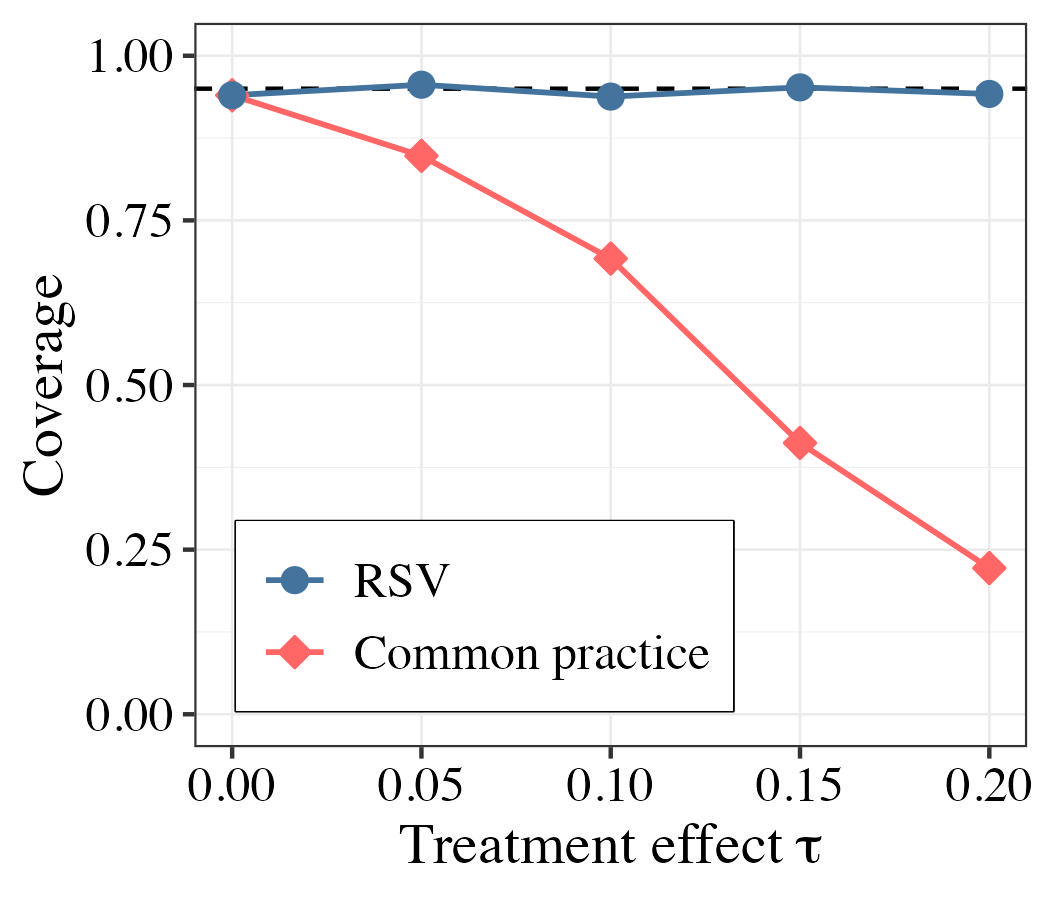}
        \caption{$n_o = 500$.}
    \end{subfigure}
    \hfill
    \begin{subfigure}{0.30\textwidth}
        \centering
        \includegraphics[width=\textwidth]{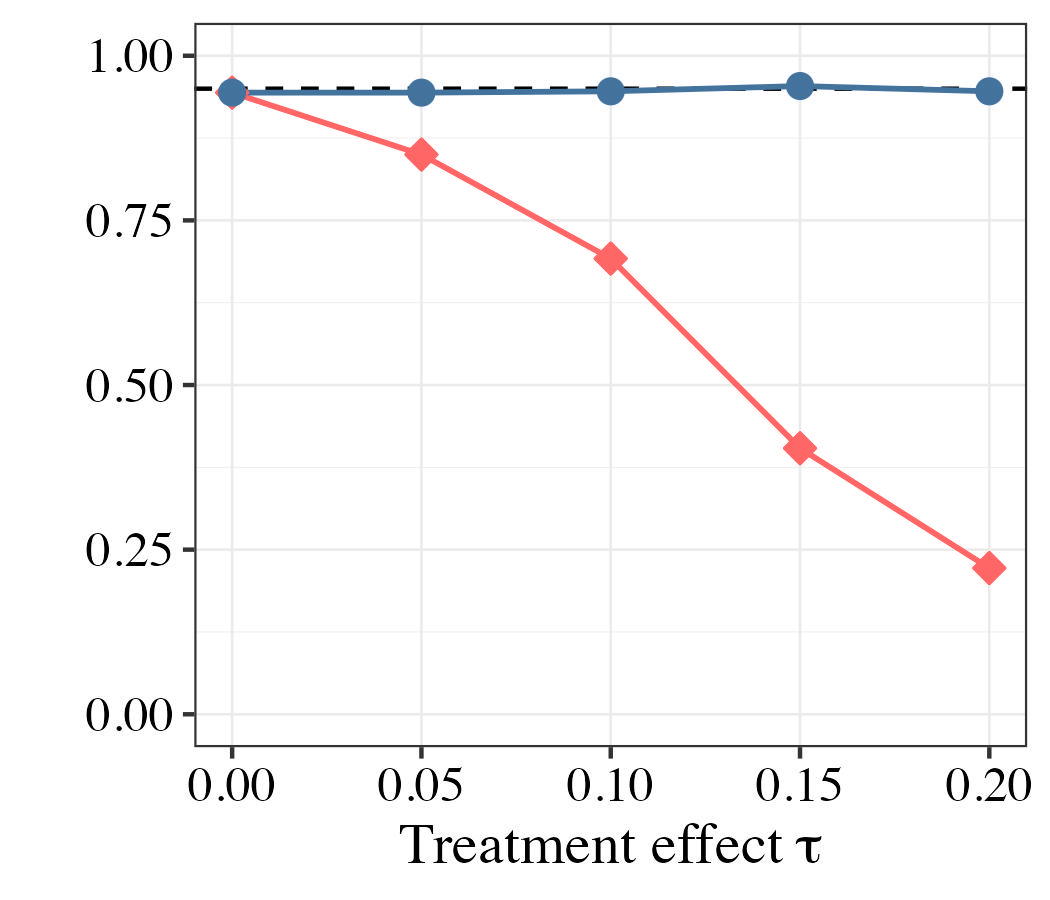}
        \caption{$n_o = 1000$.}
    \end{subfigure}
    \hfill 
    \begin{subfigure}{0.30\textwidth}
        \centering
        \includegraphics[width=\textwidth]{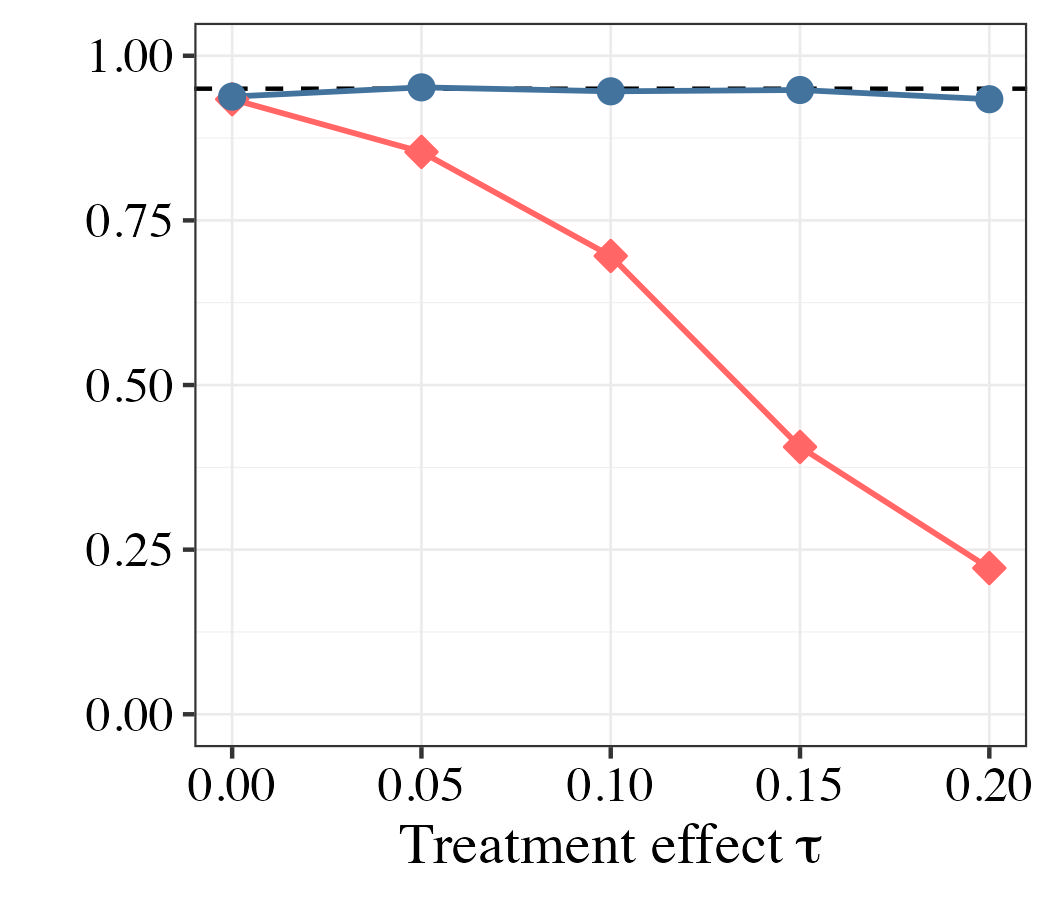}
        \caption{$n_o = 2000$.}
    \end{subfigure}
\end{subfigure}
\caption{In the forest cover design, our method outperforms the two-step method in terms of coverage. 
The different panels vary $\alpha_o$. The different plots vary $n_o$. 
For each value of the synthetic treatment effect $\tau$, we conduct 500 simulations.} 
\label{fig:nooutcomes_uganda_distributionshift_and_samplesize_coverage}
\end{figure}

For the forest cover design,  Figure~\ref{fig:nooutcomes_uganda_distributionshift_and_samplesize_normalizedbias} reports the normalized bias and  Figure~\ref{fig:nooutcomes_uganda_distributionshift_and_samplesize_coverage} reports the coverage of each method. The different panels vary the choice of $\alpha_o$, by calibrating its value to different geographic bands around the experimental sample. The different plots vary the choice of $n_o$.

\begin{figure}[H]
\centering
\captionsetup[subfigure]{justification=Centering}
\begin{subfigure}{\textwidth}
    \centering
    \begin{subfigure}{0.30\textwidth}
        \centering
        \includegraphics[width=\textwidth]{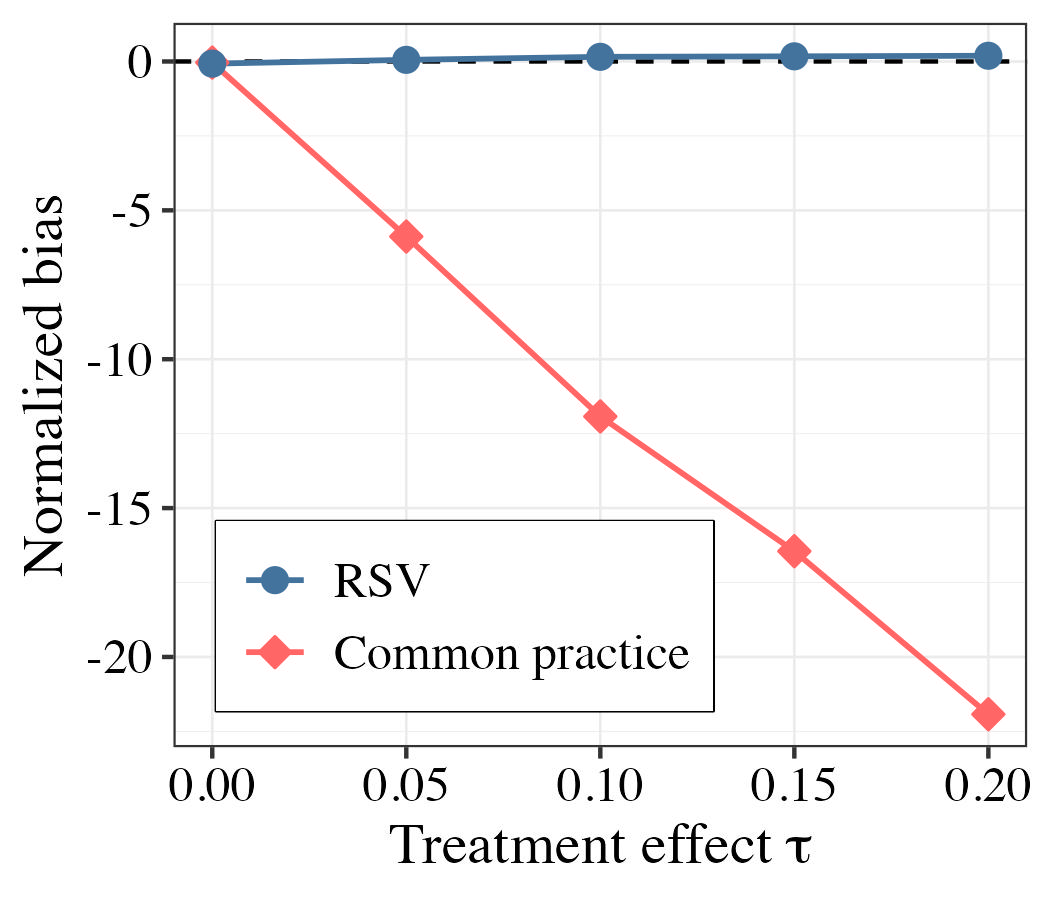}
        \caption{$n_o = 500$.}
    \end{subfigure}
    \hfill
    \begin{subfigure}{0.30\textwidth}
        \centering
        \includegraphics[width=\textwidth]{figures/smartcards/sims_noexpoutcomes/Ymidinc/bias_no1000.jpeg}
        \caption{$n_o = 1000$.}
    \end{subfigure}
    \hfill 
    \begin{subfigure}{0.30\textwidth}
        \centering
        \includegraphics[width=\textwidth]{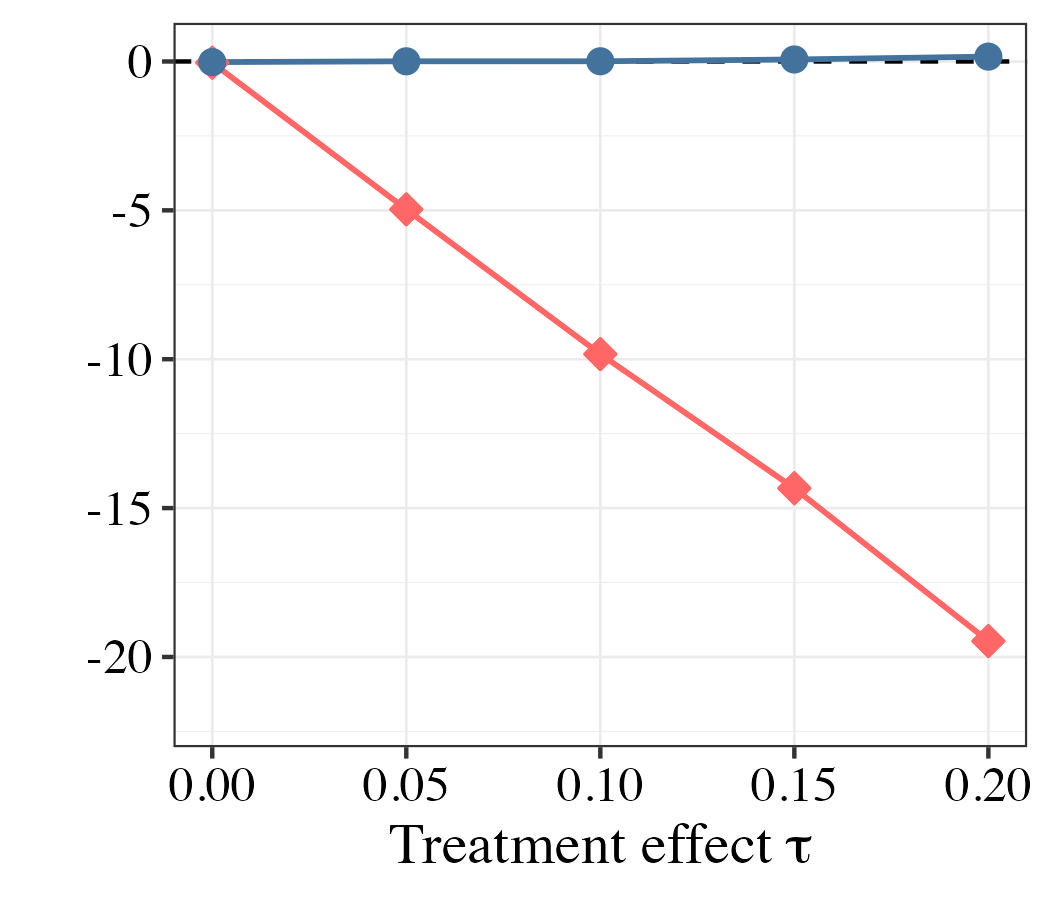}
        \caption{$n_o = 1500$.}
    \end{subfigure}
\end{subfigure}
\caption{In the Smartcards design, our method outperforms the two-step method in terms of normalized average bias. 
The different plots vary $n_o$. 
For each value of the synthetic treatment effect $\tau$, we conduct 500 simulations.} 
\label{fig:nooutcomes_smartcards_samplesize_bias}
\end{figure}

\begin{figure}[H]
\centering
\captionsetup[subfigure]{justification=Centering}

\begin{subfigure}{\textwidth}
    \centering
    \begin{subfigure}{0.30\textwidth}
        \centering
        \includegraphics[width=\textwidth]{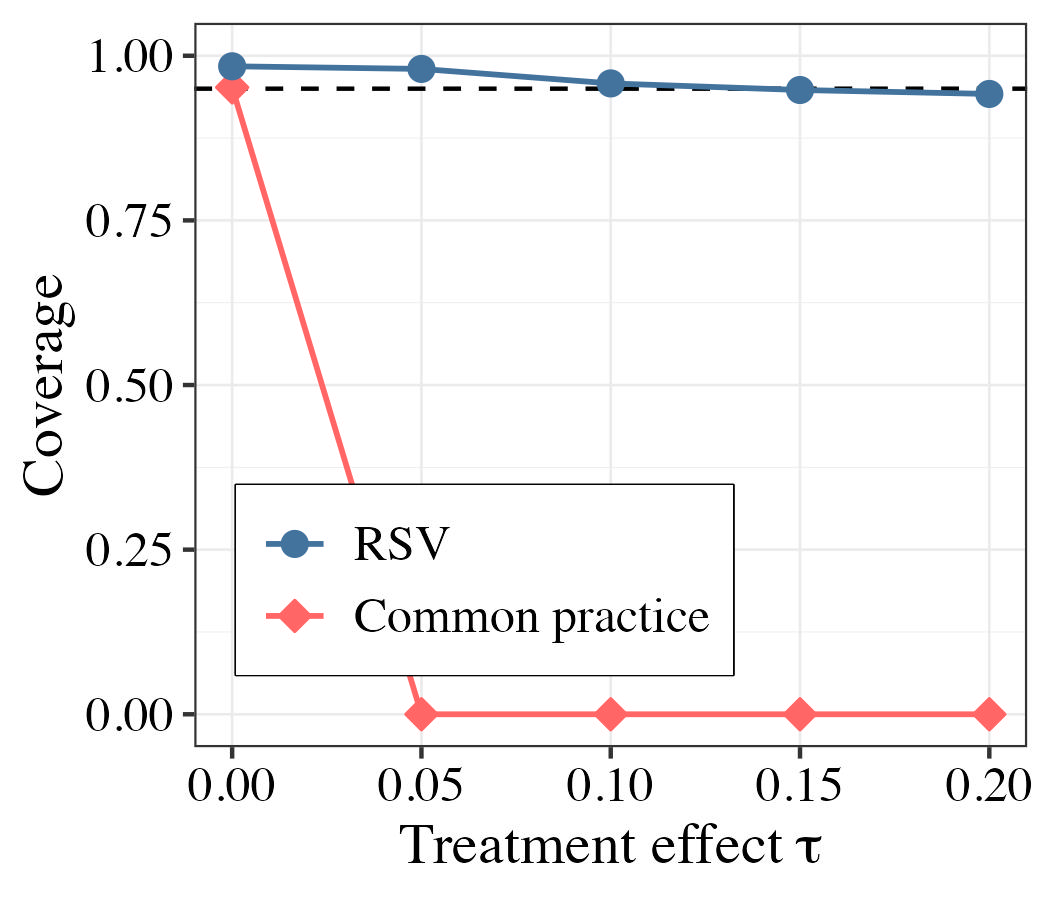}
        \caption{$n_o = 500$.}
    \end{subfigure}
    \hfill
    \begin{subfigure}{0.30\textwidth}
        \centering
        \includegraphics[width=\textwidth]{figures/smartcards/sims_noexpoutcomes/Ymidinc/coverage_no1000.jpeg}
        \caption{$n_o = 1000$.}
    \end{subfigure}
    \hfill 
    \begin{subfigure}{0.30\textwidth}
        \centering
        \includegraphics[width=\textwidth]{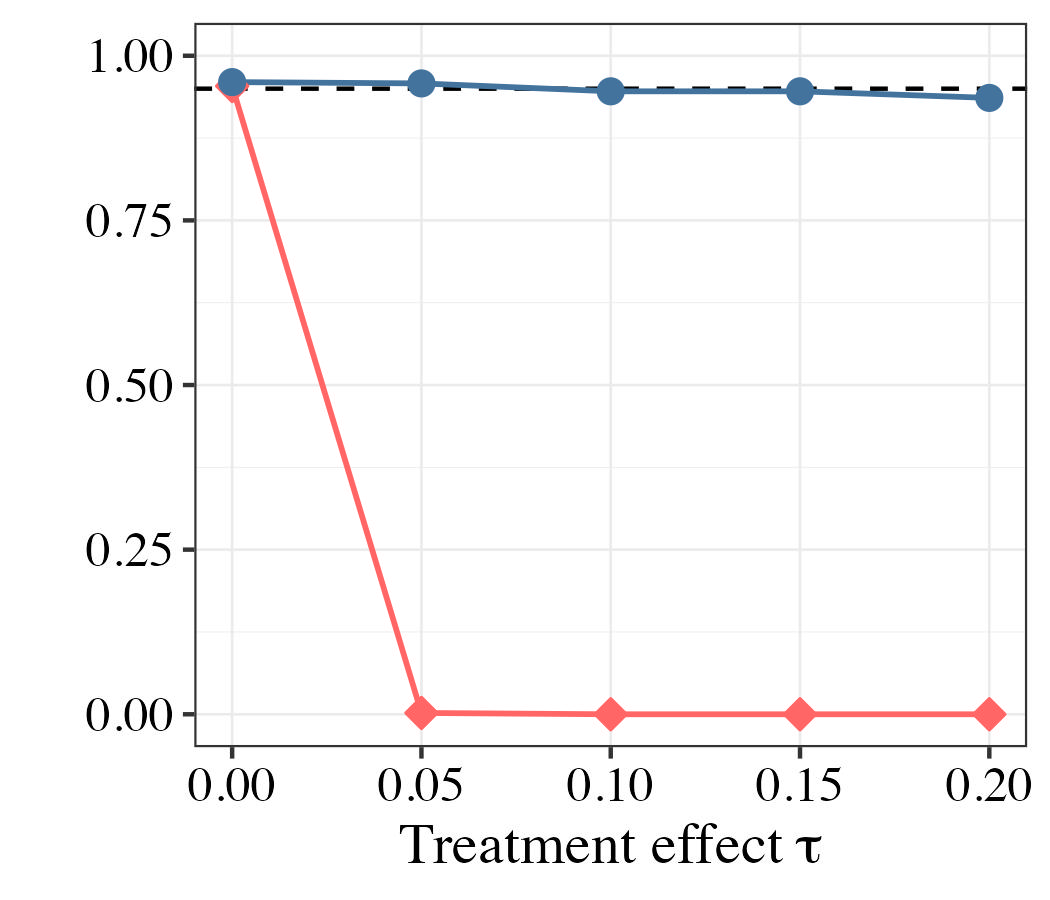}
        \caption{$n_o = 1500$.}
    \end{subfigure}
\end{subfigure}
\caption{In the Smartcards design, our method outperforms the two-step method in terms of coverage. 
The different plots vary $n_o$. 
For each value of the synthetic treatment effect $\tau$, we conduct 500 simulations.} 
\label{fig:nooutcomes_smartcards_samplesize_cov}
\end{figure}

For the Smartcards design, Figure~\ref{fig:nooutcomes_smartcards_samplesize_bias} reports the normalized bias and  Figure~\ref{fig:nooutcomes_smartcards_samplesize_cov} reports the coverage of each method. The different plots vary the choice of $n_o$.

\clearpage 
\subsection{Experimental, Observational, and Validation Samples}\label{section:additional_sims_with_limitedoutcomes}

To compare our method with PPI methods, we create a new simulation design with a randomly selected validation sample, as discussed in Section \ref{sec:limited_experimental_outcomes}. 
PPI methods correct the errors of generic machine learning outputs, so we focus on the forest cover design in which the remotely sensed variable $R$ is the scalar output of the pretrained random forest. 

\paragraph{Simulation design.} 
We observe $(D,R)$ when $\widetilde{S}=e$, $(R,Y)$ when $\widetilde{S}=o$, and now $(D,Y,R)$ when $\widetilde{S}=v$.
The simulation design is as described in Section~\ref{sec:main_simulation_design}, except we now retain outcomes for a random subset of experimental units. 
Specifically, we fix the total number of experimental units at $n_e+n_v=1000$ and the observational sample size at $n_o = 1000$. 
We vary the number of validation units: $n_v \in \{100, 250, 500\}$.
These choices correspond to scenarios in which outcomes are randomly collected for 10\%, 25\%, or 50\% of experimental units while relying on the remotely sensed outcome for the remainder. 
For the $n_v$ randomly selected validation units, we have $(D, Y, R)$; for the remaining $n_e=1000 - n_v$ experimental units, we have only $(D, R)$ as before.

\paragraph{Implementation details.} We compare four methods. 

Our method uses all three samples:  $\widetilde{S}\in\{e,v,o\}$. Its estimand is $\tau$. 
We estimate the efficient representation by combining logistic regressions of 
$\widehat{\Pr}(Y=1 \mid \widetilde{S}\in\{o,v\},R)$, $\widehat{\Pr}(D=1\mid \widetilde{S}\in \{e,v\},R)$, and $\widehat{\Pr}(\widetilde{S}\in \{e,v\}\mid R)$.
As before, we use cross-fitting with two folds. Our approach is unbiased and efficient.

One variation of PPI uses all three samples: $\widetilde{S}\in\{e,v,o\}$. We call it PPIO. Its estimand is
\begin{align*}
 \widetilde{\tau}^{\mathrm{PPIO}} 
 & = \E(R \mid  D = 1,\widetilde{S} = e) - \E(R \mid  D = 0,\widetilde{S} = e)
  + \E(Y - R \mid  D=1, \widetilde{S}=v)-\E(Y - R \mid D=0, \widetilde{S} \in\{v,o\}),
\end{align*}
which does not equal $\tau$ because the rectifier for the untreated arm uses $S=o$. Therefore, PPIO is biased.

Another variation uses only the $\widetilde{S}=e$ and $\widetilde{S}=v$ samples. We call it PPIV. Its estimand is
\begin{align*}
\widetilde{\tau}^{\mathrm{PPIV}} 
 &=  \E(R \mid  D = 1,\widetilde{S} = e) - \E(R \mid  D = 0,\widetilde{S} = e)
 + \E(Y - R \mid  D=1,\widetilde{S}=v)-\E(Y - R \mid D=0,\widetilde{S}=v),
\end{align*}
which equals $\tau$ because $\widetilde{S}=v$ is randomly selected in this simulation design. PPIV is unbiased but inefficient because it discards $\widetilde{S}=o$.

Finally, we consider a benchmark regression using only the $\widetilde{S}=v$ sample. Its estimand is $\tau$. Using this small subset of units, it is a least squares regression of $Y$ on $D$, ignoring $R$. It is unbiased but inefficient because it discards $\widetilde{S}=e$ and $\widetilde{S}=o$.

\paragraph{Additional results.}
Two findings emerge, pertaining to bias and variance. 

\begin{figure}[H]
\centering
\captionsetup[subfigure]{justification=Centering}
\begin{subfigure}{\textwidth}
    \centering
    \caption*{Panel A: 0--2 Kilometer Geographic Band}
    \begin{subfigure}{0.30\textwidth}
        \centering
        \includegraphics[width=\textwidth]{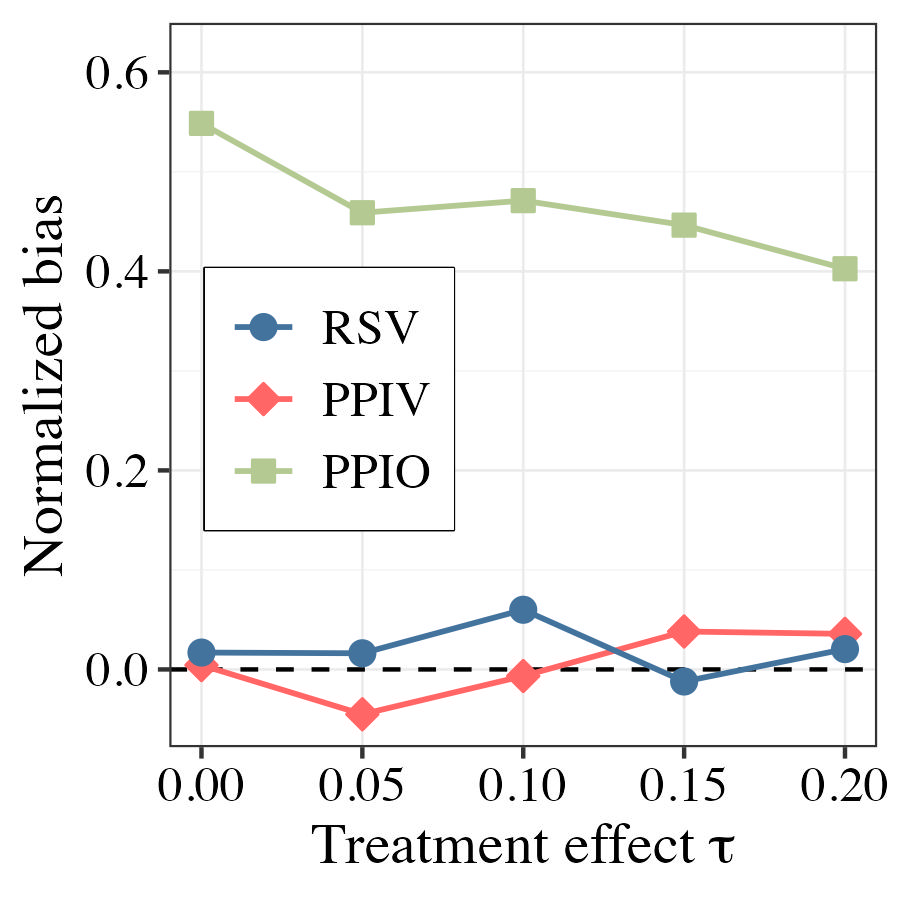}
        \caption{$n_v = 100$.}
    \end{subfigure}
    \hfill
    \begin{subfigure}{0.30\textwidth}
        \centering
        \includegraphics[width=\textwidth]{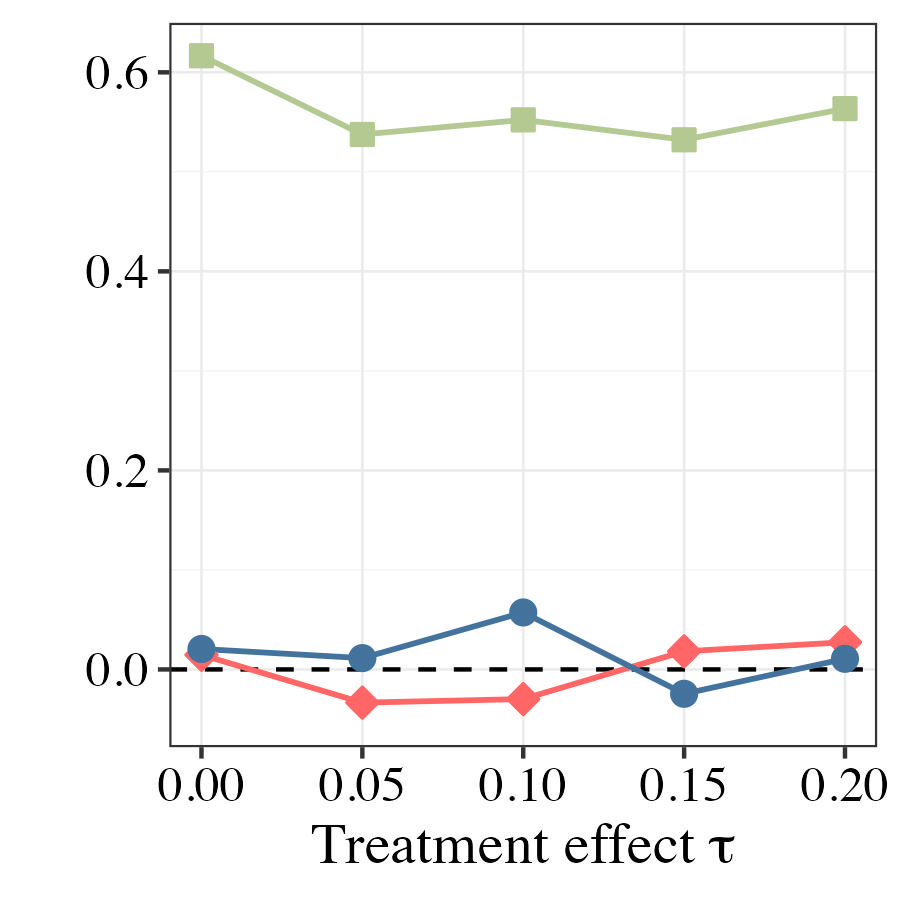}
        \caption{$n_v = 250$.}
    \end{subfigure}
    \hfill 
    \begin{subfigure}{0.30\textwidth}
        \centering
        \includegraphics[width=\textwidth]{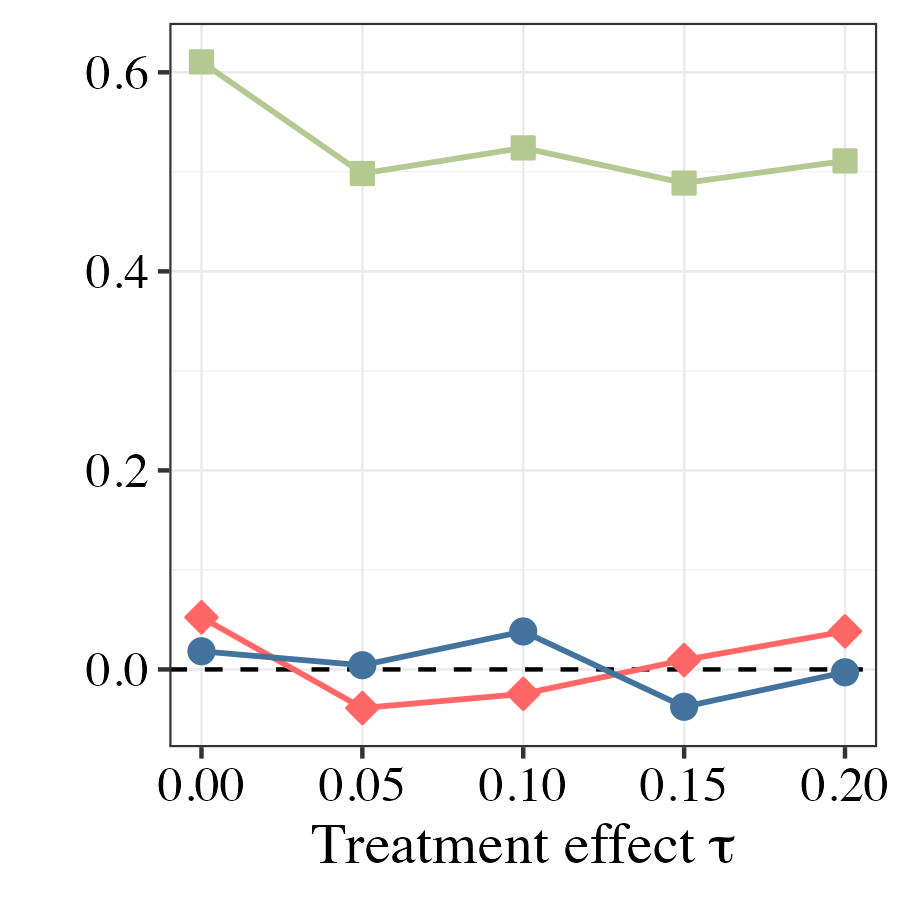}
        \caption{$n_v = 500$.}
    \end{subfigure}
\end{subfigure}
\begin{subfigure}{\textwidth}
    \centering
    \caption*{Panel B: 0--5 Kilometer Geographic Band}
    \begin{subfigure}{0.30\textwidth}
        \centering
        \includegraphics[width=\textwidth]{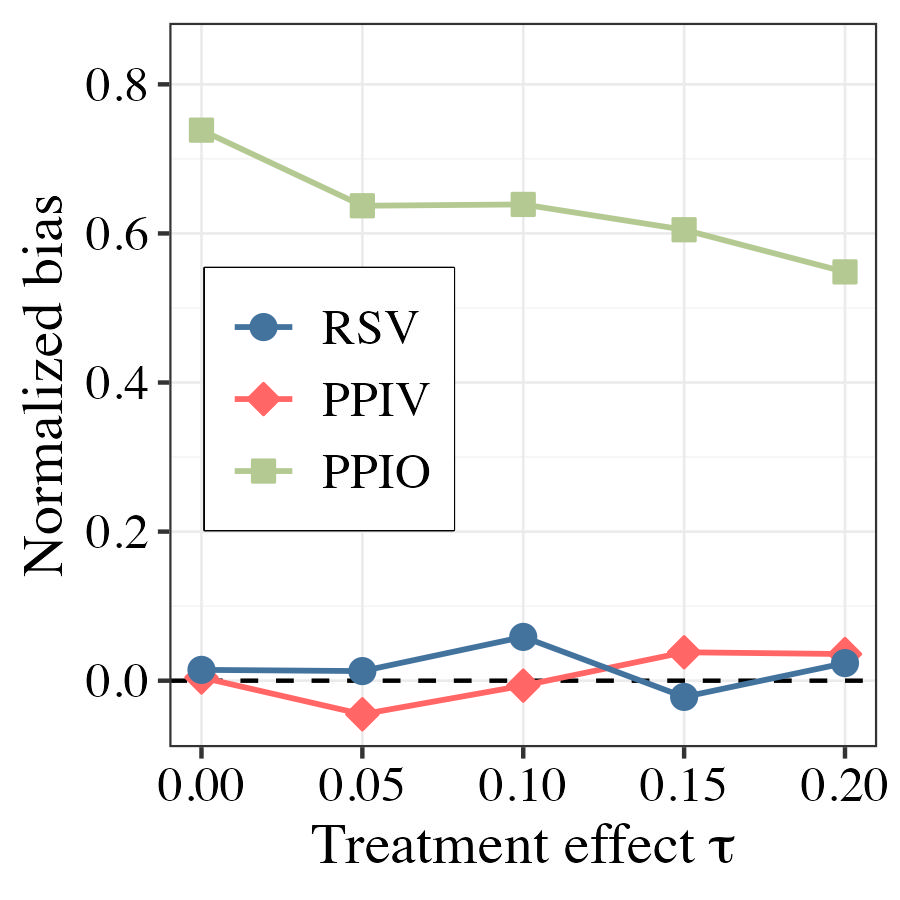}
        \caption{$n_v = 100$.}
    \end{subfigure}
    \hfill
    \begin{subfigure}{0.30\textwidth}
        \centering
        \includegraphics[width=\textwidth]{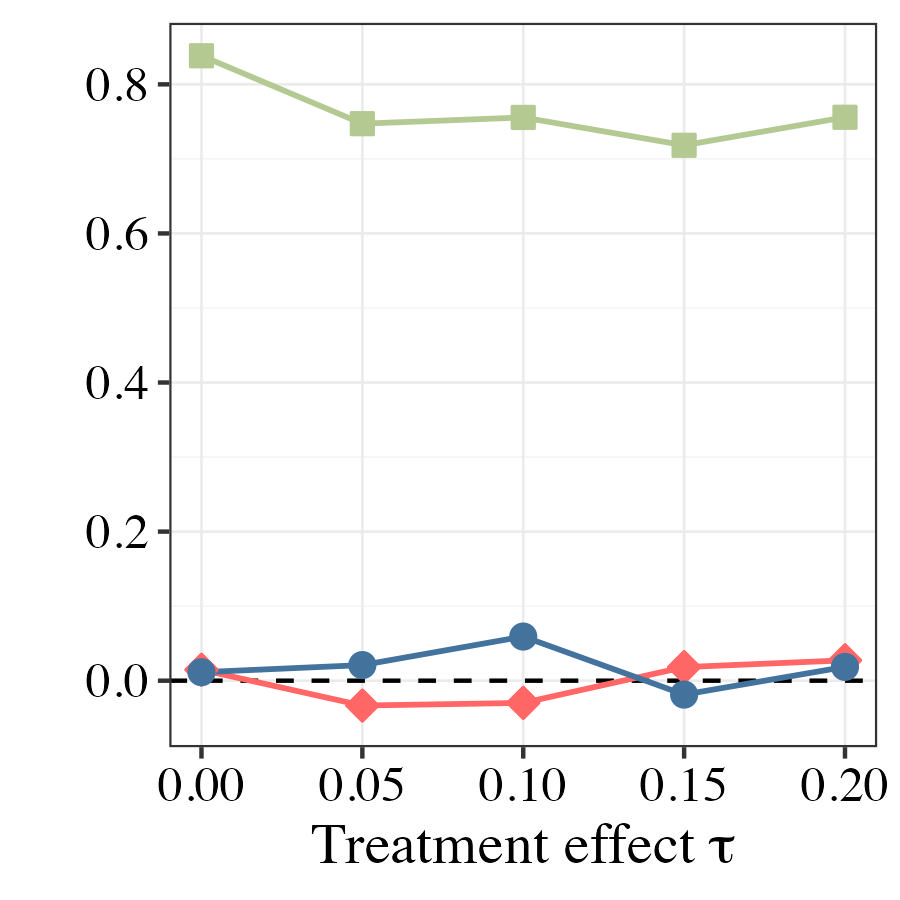}
        \caption{$n_v = 250$.}
    \end{subfigure}
    \hfill 
    \begin{subfigure}{0.30\textwidth}
        \centering
        \includegraphics[width=\textwidth]{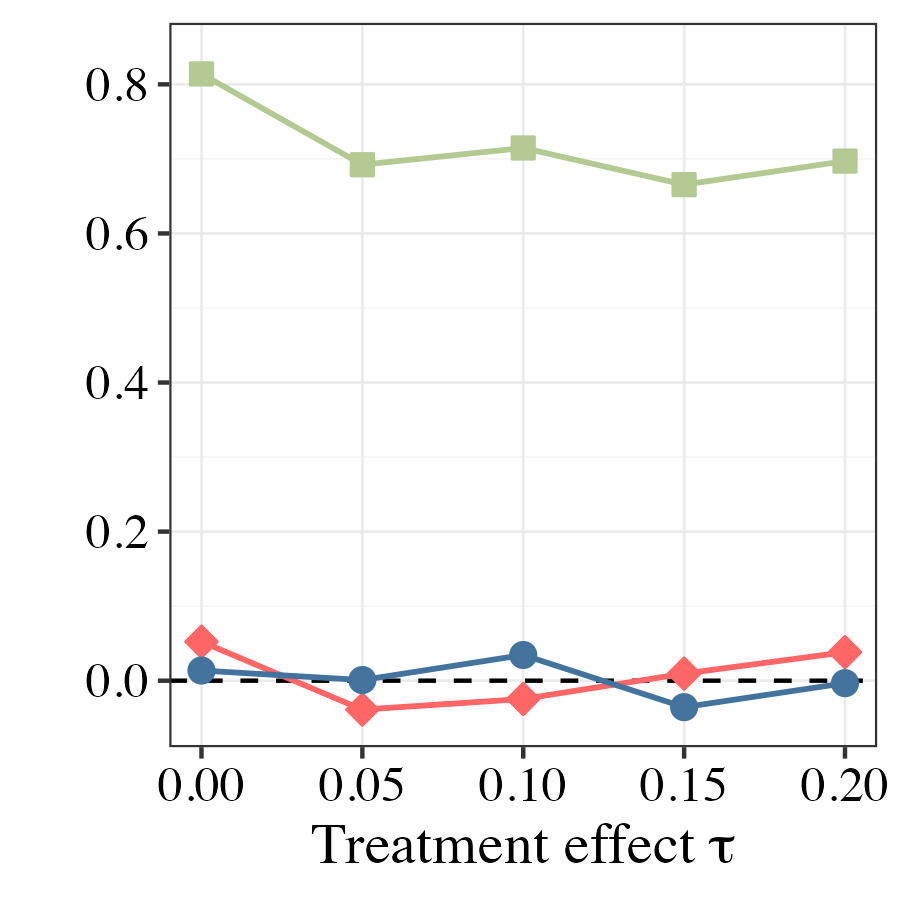}
        \caption{$n_v = 500$.}
    \end{subfigure}
\end{subfigure}
\begin{subfigure}{\textwidth}
    \centering
    \caption*{Panel C: 0--10 Kilometer Geographic Band}
    \begin{subfigure}{0.30\textwidth}
        \centering
        \includegraphics[width=\textwidth]{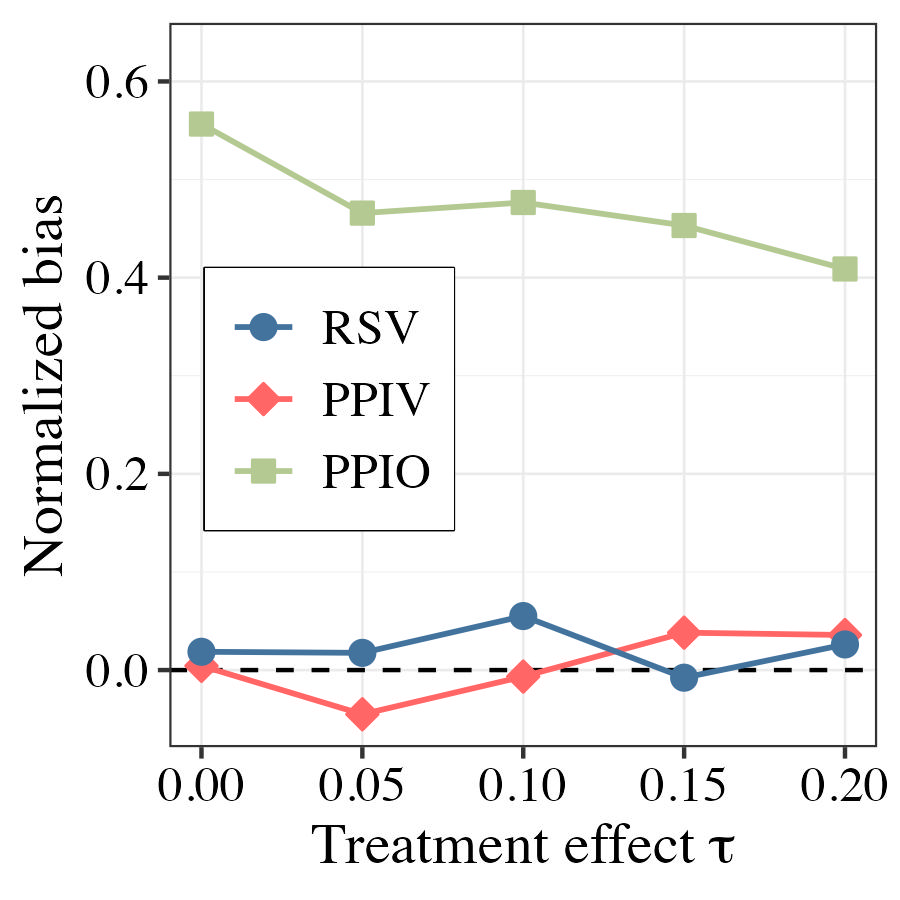}
        \caption{$n_v = 100$.}
    \end{subfigure}
    \hfill
    \begin{subfigure}{0.30\textwidth}
        \centering
        \includegraphics[width=\textwidth]{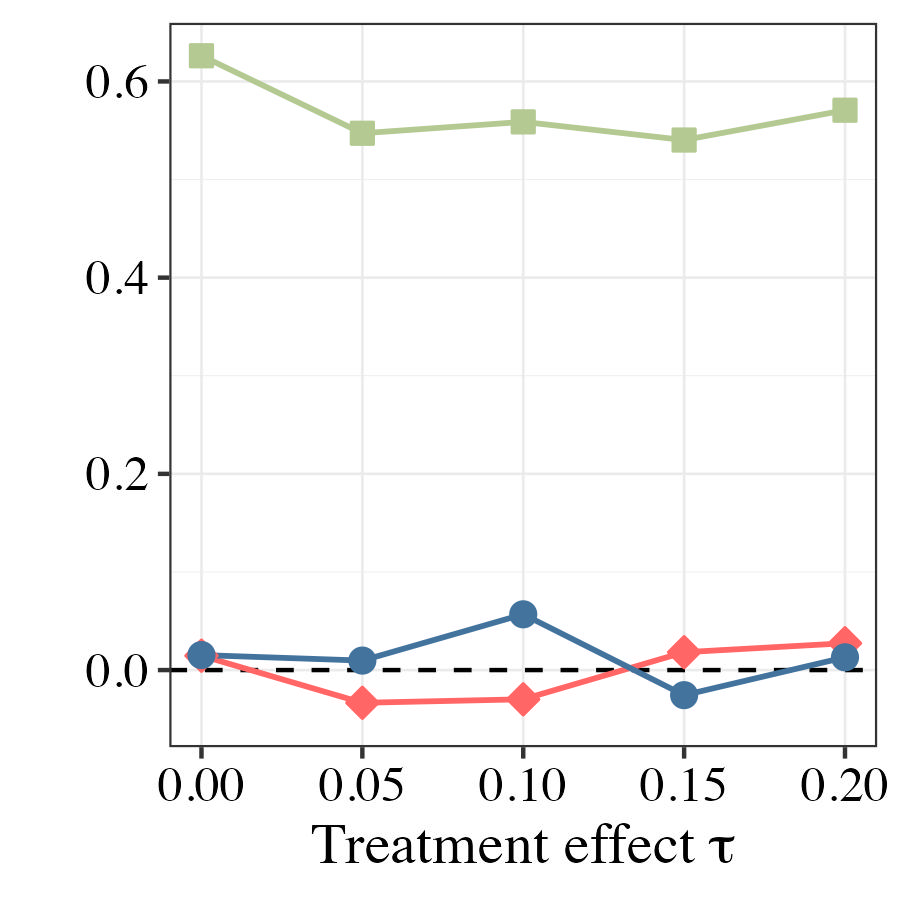}
        \caption{$n_v = 250$.}
    \end{subfigure}
    \hfill 
    \begin{subfigure}{0.30\textwidth}
        \centering
        \includegraphics[width=\textwidth]{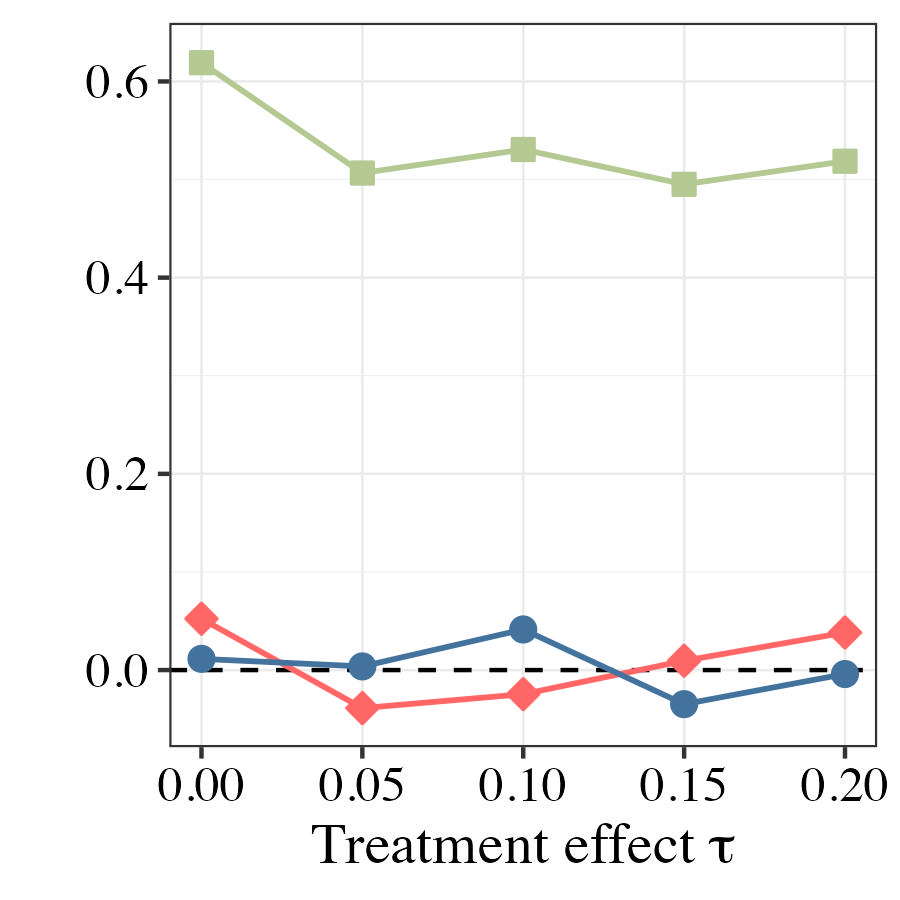}
        \caption{$n_v = 500$.}
    \end{subfigure}
\end{subfigure}
\caption{In the forest cover design with three samples, we compare the normalized average bias of four methods: our proposal using $\widetilde{S}\in\{e,v,o\}$, PPIO  using $\widetilde{S}\in\{e,v,o\}$, PPIV using $\widetilde{S}\in\{e,v\}$, and a benchmark regression using $\widetilde{S}=v$. PPIO is the only method that exhibits substantial bias. The different panels vary $\alpha_o$. The different plots vary $n_v$. 
For each value of the synthetic treatment effect $\tau$, we conduct 500 simulations.
}
\label{fig:normalizedbias_limitedoutcomes}
\end{figure}

Our method, PPIV, and the benchmark regression are unbiased, but PPIO is biased. Figure~\ref{fig:normalizedbias_limitedoutcomes} visualizes the normalized bias of the four methods in simulations. As before, the different panels vary the choice of $\alpha_o$, by calibrating its value to different geographic bands around the experimental sample. The different plots vary the choice of $n_v$.
Since $\alpha_e \neq \alpha_o$, PPIO displays substantial bias, confirming   Proposition~\ref{prop:bias_PPI}. 

\begin{figure}[H]
\centering
\captionsetup[subfigure]{justification=Centering}
\begin{subfigure}{\textwidth}
    \centering
    \caption*{Panel A: 0--2 Kilometer Geographic Band}
    \begin{subfigure}{0.30\textwidth}
        \centering
        \includegraphics[width=\textwidth]{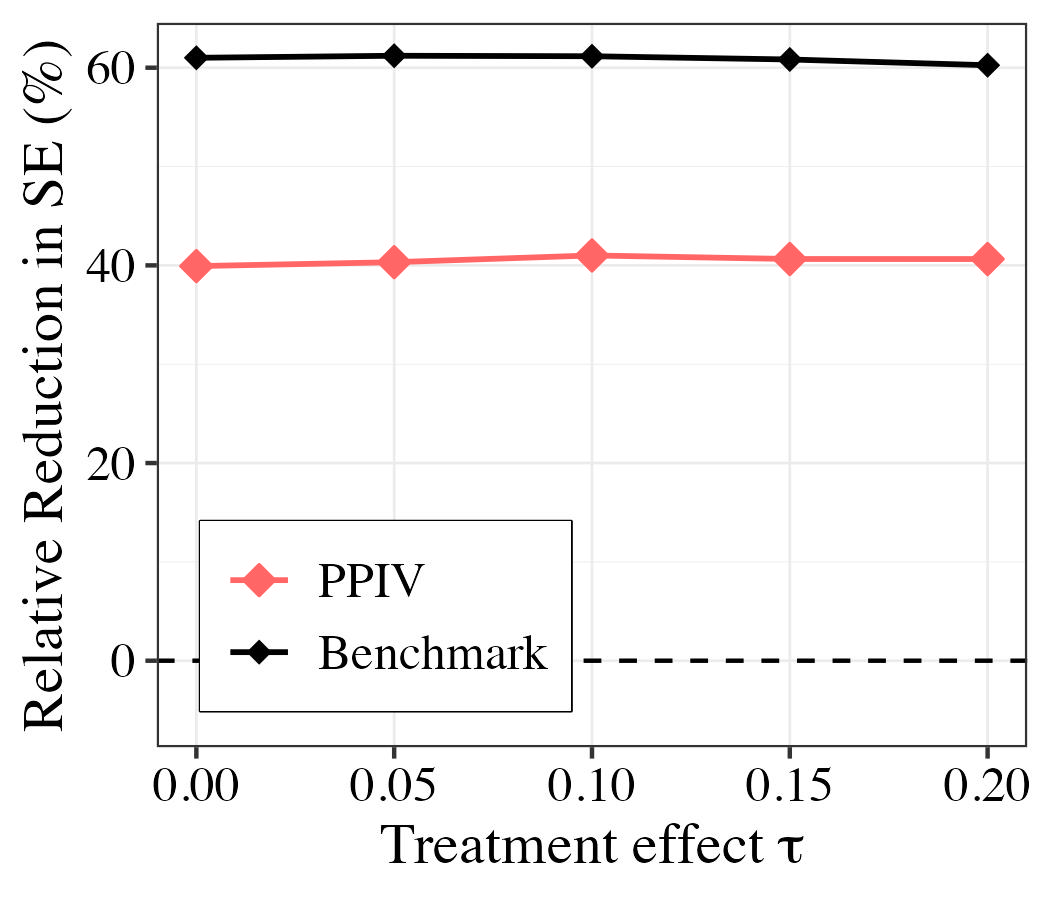}
        \caption{$n_v = 100$.}
    \end{subfigure}
    \hfill
    \begin{subfigure}{0.30\textwidth}
        \centering
        \includegraphics[width=\textwidth]{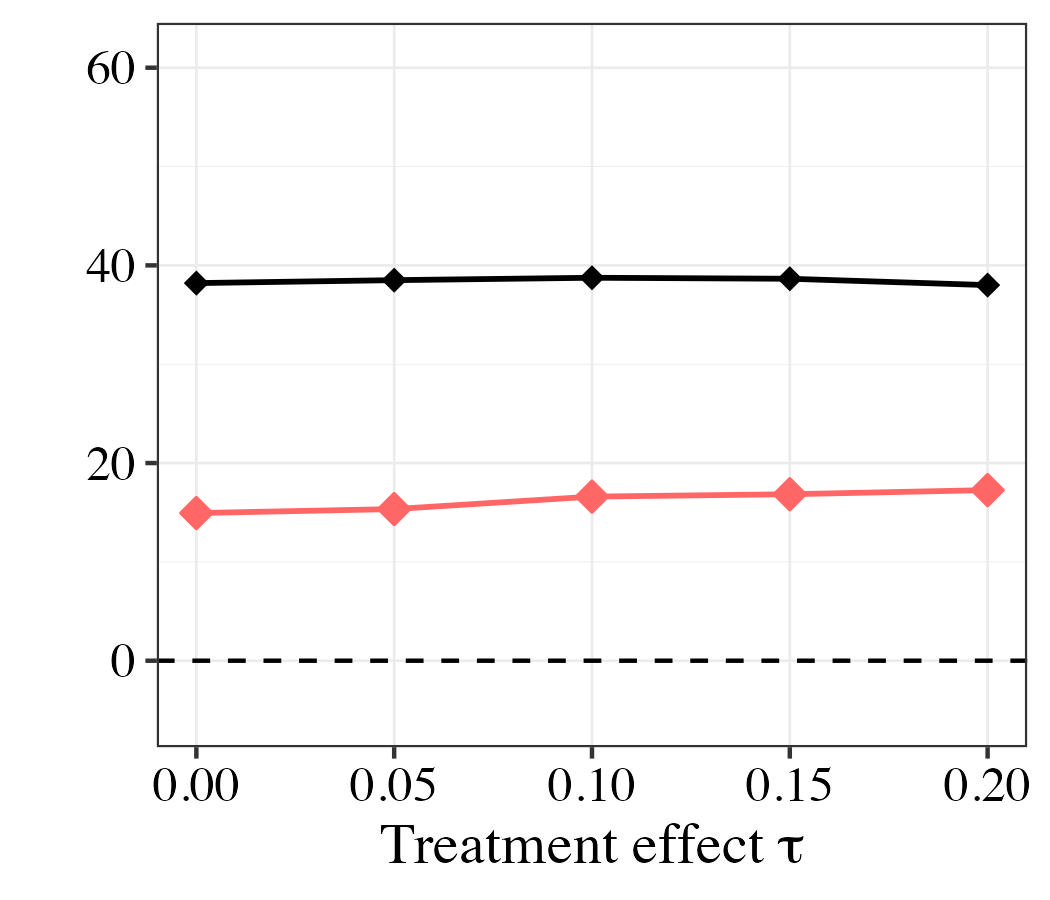}
        \caption{$n_v = 250$.}
    \end{subfigure}
    \hfill 
    \begin{subfigure}{0.30\textwidth}
        \centering
        \includegraphics[width=\textwidth]{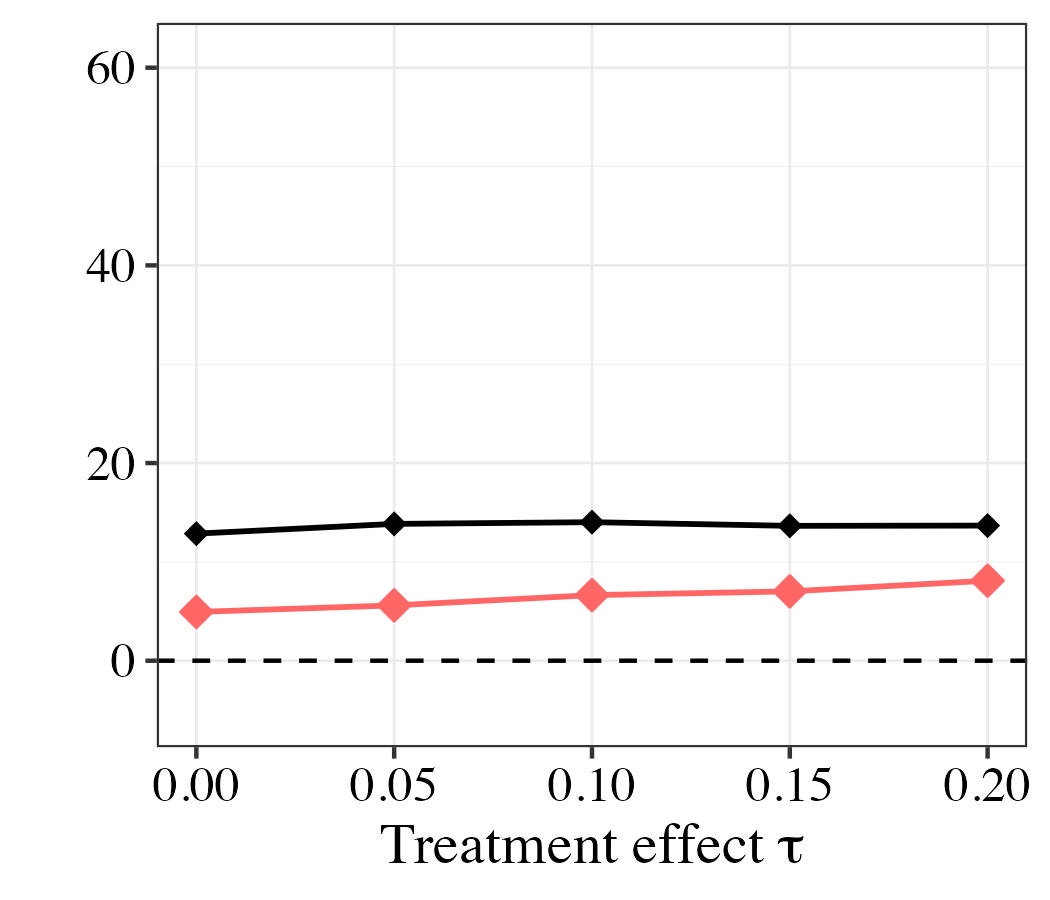}
        \caption{$n_v = 500$.}
    \end{subfigure}
\end{subfigure}
\begin{subfigure}{\textwidth}
    \centering
    \caption*{Panel B: 0--5 Kilometer Geographic Band}
    \begin{subfigure}{0.30\textwidth}
        \centering
        \includegraphics[width=\textwidth]{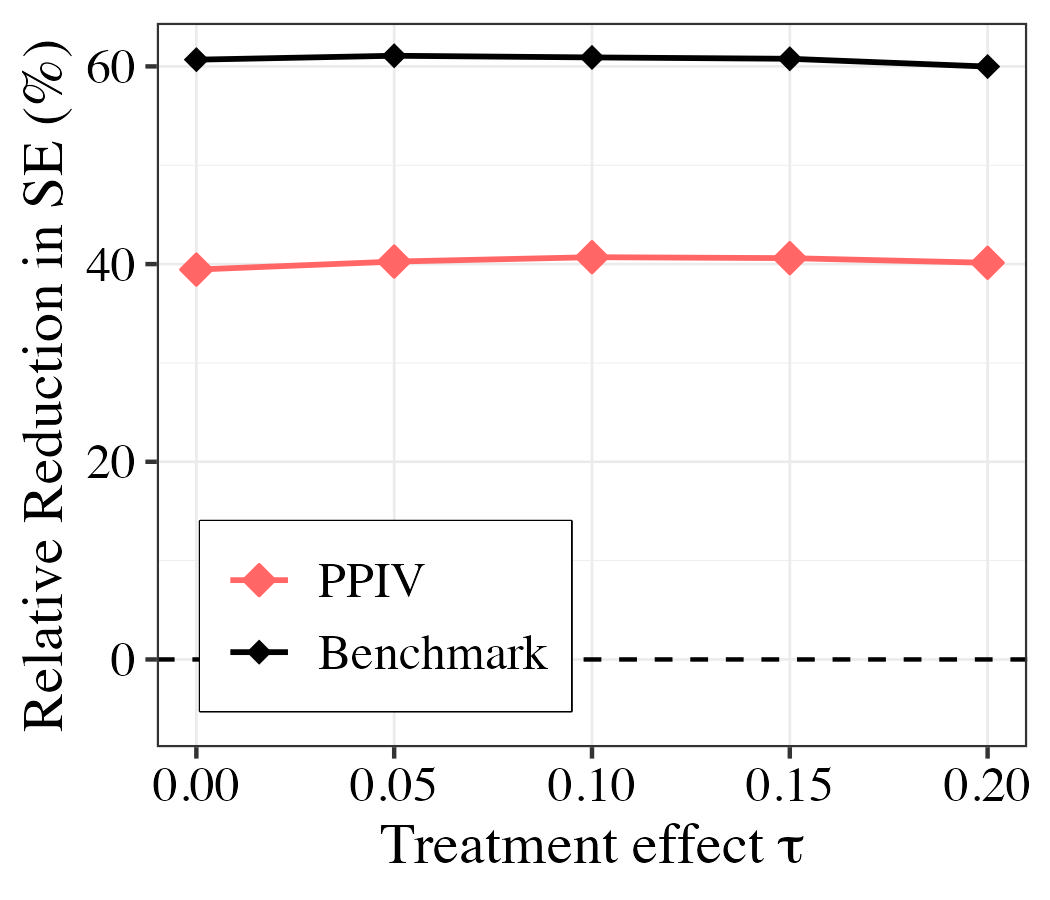}
        \caption{$n_v = 100$.}
    \end{subfigure}
    \hfill
    \begin{subfigure}{0.30\textwidth}
        \centering
        \includegraphics[width=\textwidth]{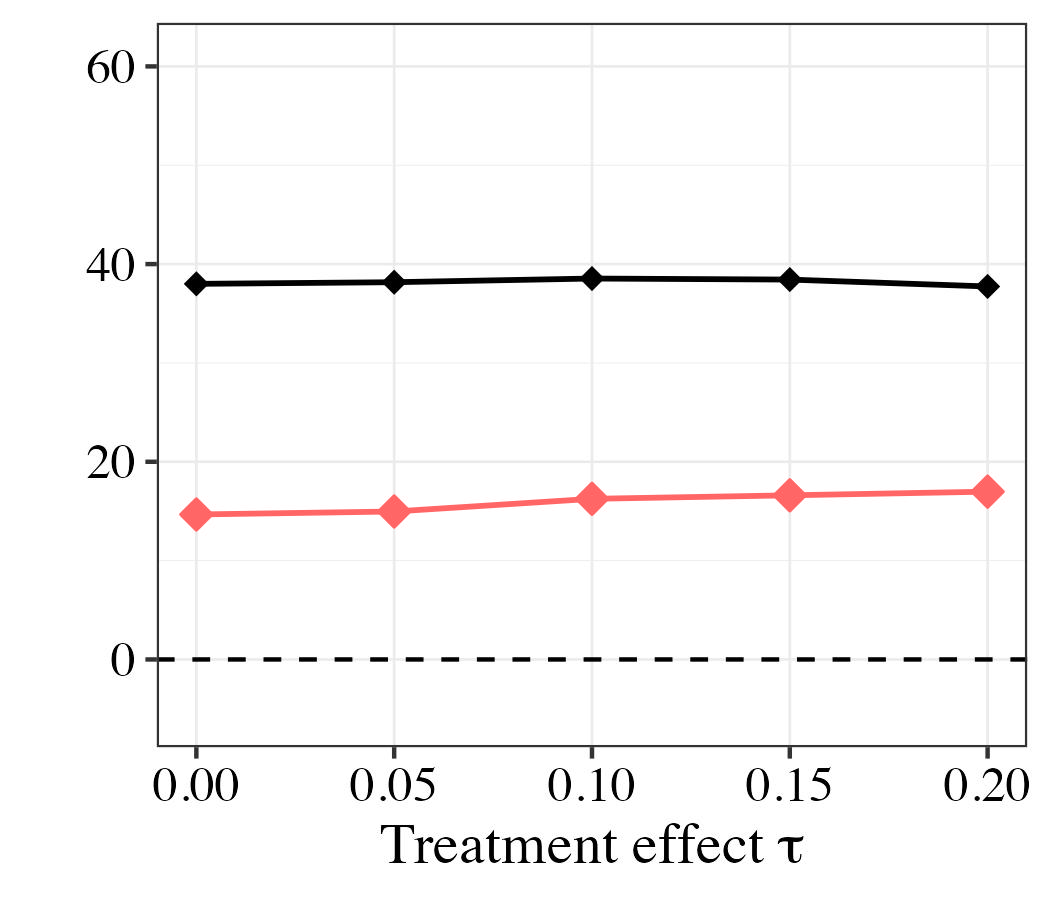}
        \caption{$n_v = 250$.}
    \end{subfigure}
    \hfill 
    \begin{subfigure}{0.30\textwidth}
        \centering
        \includegraphics[width=\textwidth]{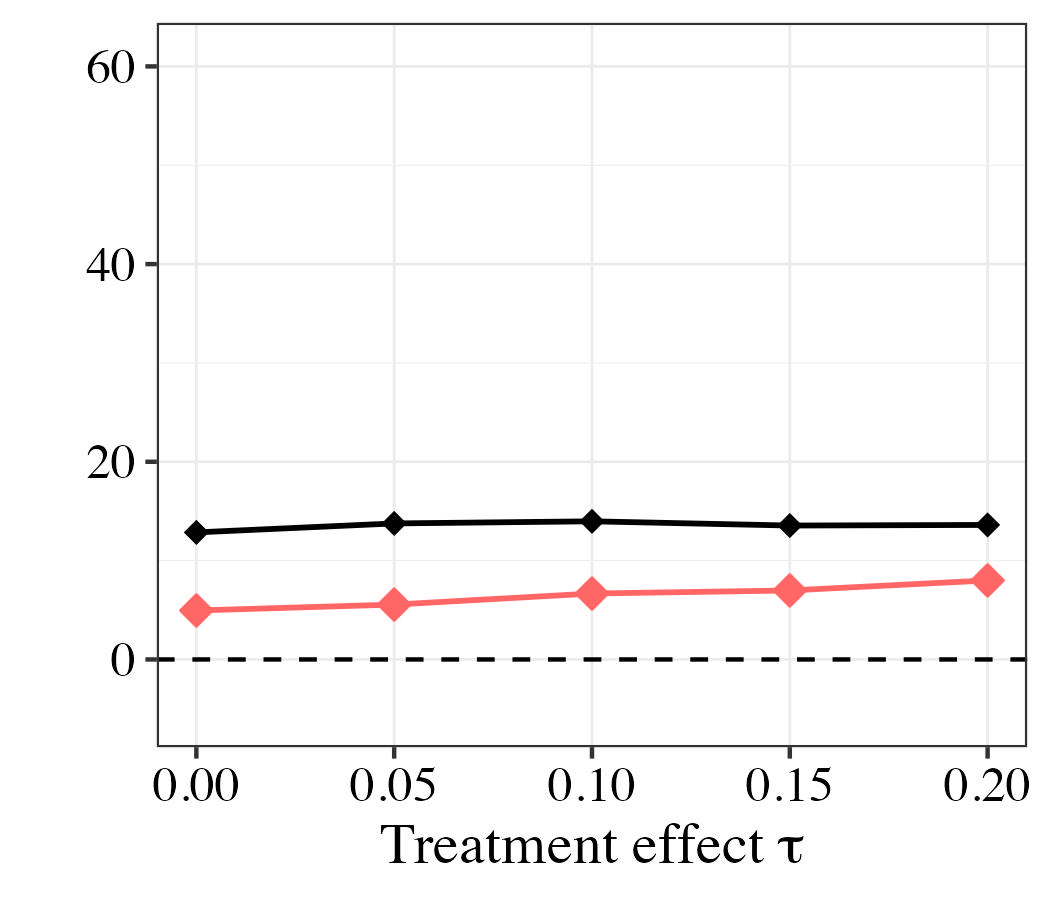}
        \caption{$n_v = 500$.}
    \end{subfigure}
\end{subfigure}
\begin{subfigure}{\textwidth}
    \centering
    \caption*{Panel C: 0--10 Kilometer Geographic Band}
    \begin{subfigure}{0.30\textwidth}
        \centering
        \includegraphics[width=\textwidth]{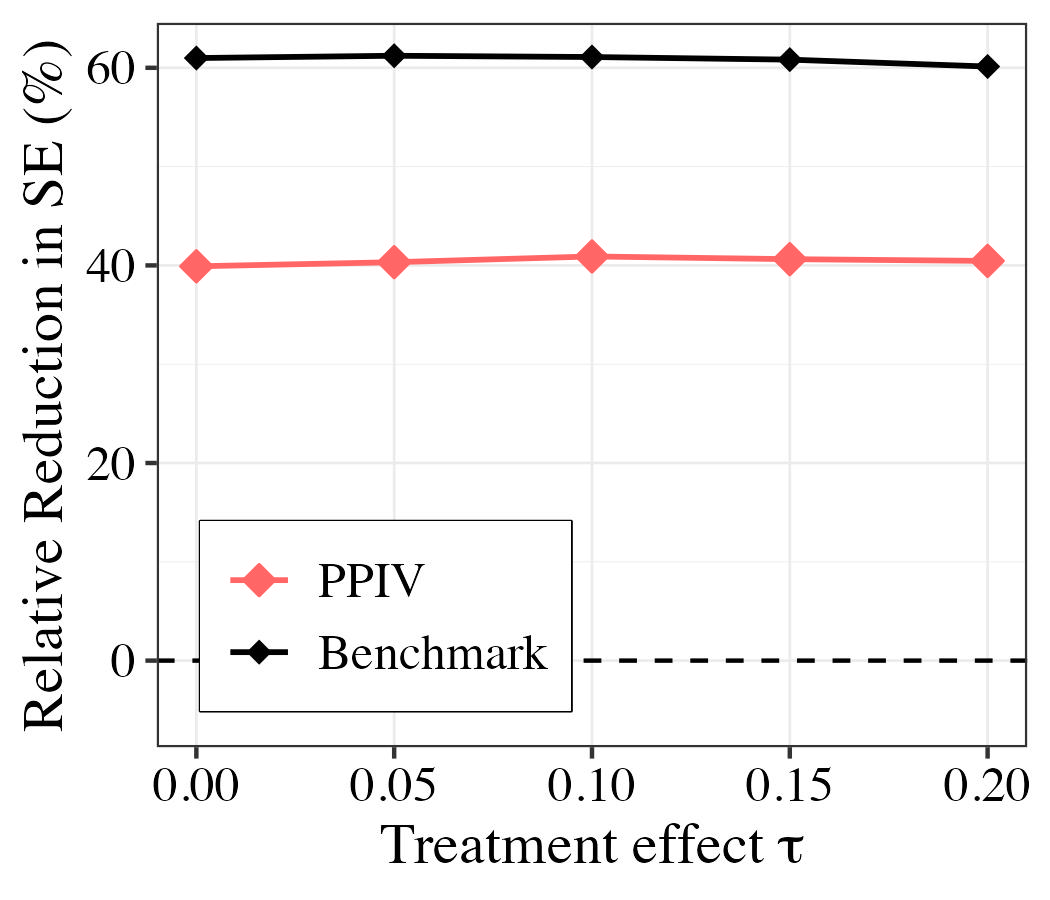}
        \caption{$n_v = 100$.}
    \end{subfigure}
    \hfill
    \begin{subfigure}{0.30\textwidth}
        \centering
        \includegraphics[width=\textwidth]{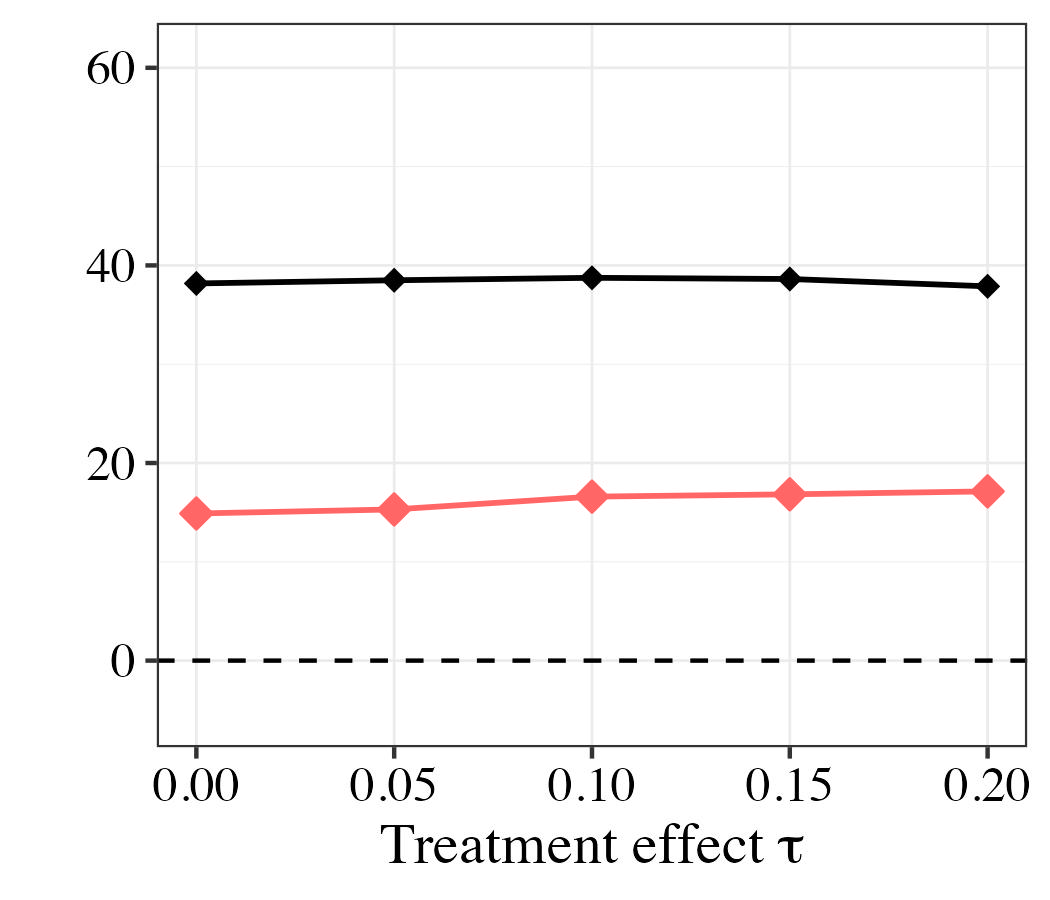}
        \caption{$n_v = 250$.}
    \end{subfigure}
    \hfill 
    \begin{subfigure}{0.30\textwidth}
        \centering
        \includegraphics[width=\textwidth]{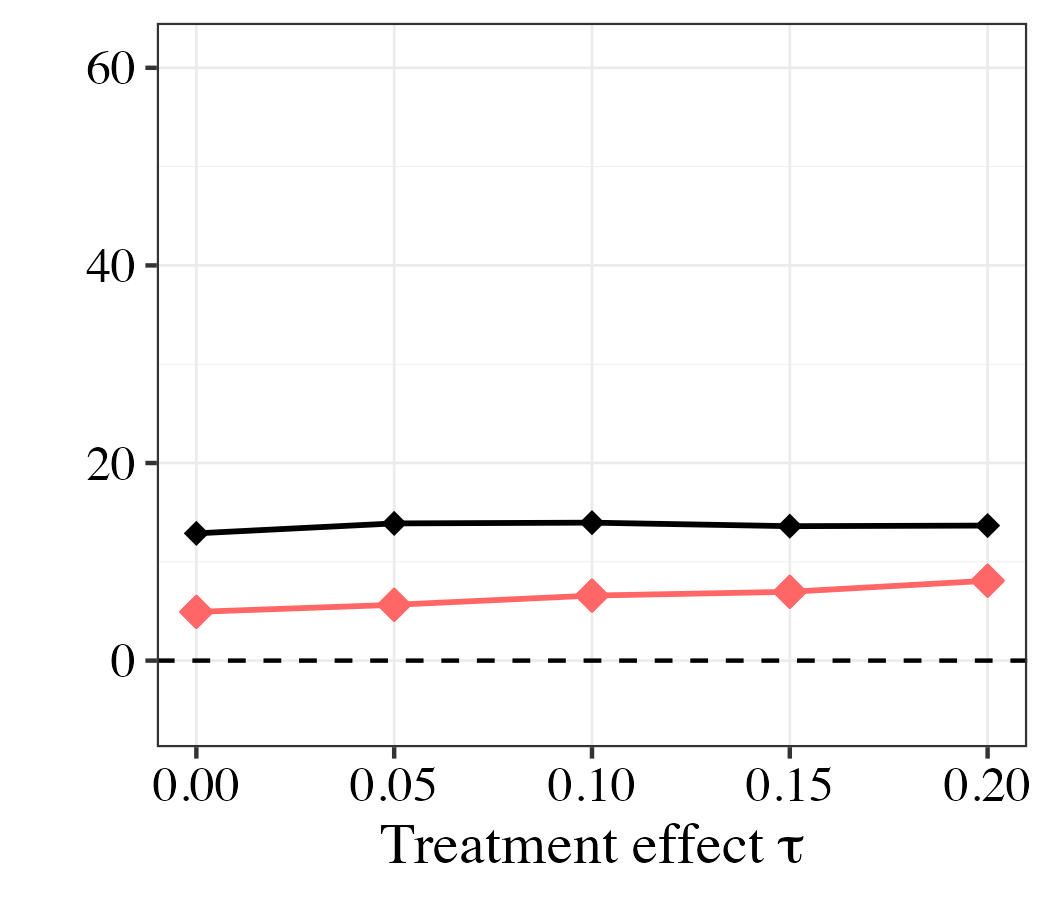}
        \caption{$n_v = 500$.}
    \end{subfigure}
\end{subfigure}
\caption{In the forest cover design with three samples, we compare the standard errors of the three unbiased methods: our proposal using $\widetilde{S}\in\{e,v,o\}$, PPIV using $\widetilde{S}\in\{e,v\}$, and a benchmark regression using $\widetilde{S}=v$. Specifically, we report the percent reduction in standard errors of our method relative to each unbiased alternative. 
Our method is the most efficient, since it uses all three samples.
The different panels vary $\alpha_o$. The different plots vary $n_v$. 
For each value of the synthetic treatment effect $\tau$, we conduct 500 simulations.}
\label{fig:main_relse_limitedoutcomes}
\end{figure}

Focusing on the unbiased methods, we find that our method is more efficient than PPIV, which is more efficient than the benchmark regression. Intuitively, our method uses all three samples, PPIV uses two samples, and the benchmark regression uses one sample. 
Figure~\ref{fig:main_relse_limitedoutcomes} visualizes the percent reduction of standard errors. As before, the different panels vary the choice of $\alpha_o$, by calibrating its value to different geographic bands around the experimental sample. The different plots vary the choice of $n_v$.
Our method yields standard errors that can be 40\% smaller than PPIV, and 60\% smaller than the benchmark.

In summary, in the presence of a randomly selected validation sample $\widetilde{S}=v$, 
researchers may turn to our method to improve efficiency. By exploiting the post-outcome structure of the remotely sensed variable, our method efficiently incorporates information from the observational sample $\widetilde{S}=o$, while some existing methods do not.

\clearpage
\section{Additional Empirical Results}\label{section: additional empirical results}
\subsection{Smartcards and Village-Level Poverty}\label{section:smartcards_empirical_additionaldetails}

\begin{figure}
\centering
\includegraphics[width=0.6\textwidth]{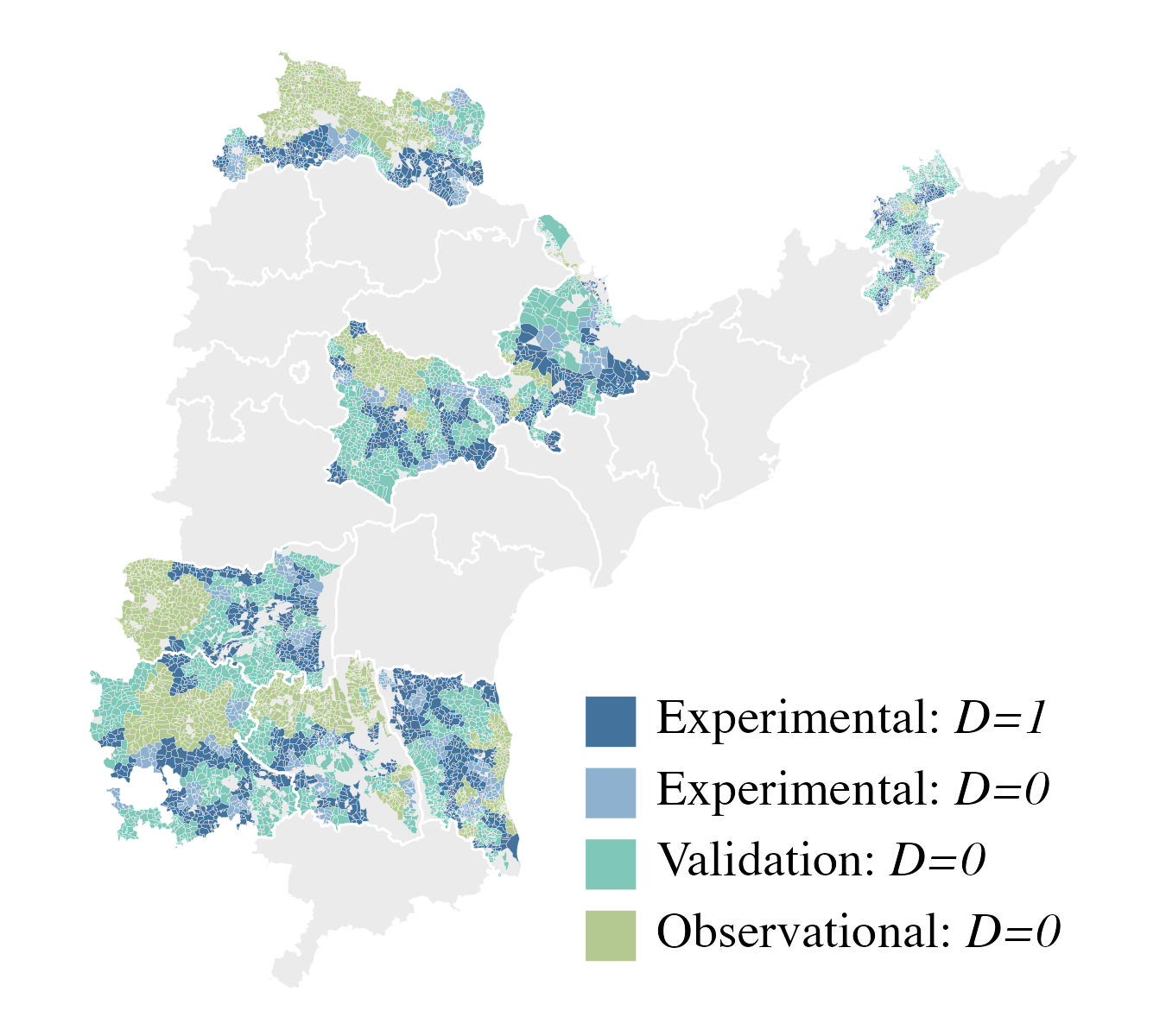}
\caption{We illustrate the experimental, validation, and observational samples in the Smartcards application \citep{muralidharan2023general}.}
\label{fig:smartcards_sample_map2}
\end{figure}

In this section, we provide additional details and results for our re-analysis of the Smartcards experiment of \citet[][]{muralidharan2016building, muralidharan2023general} in Section~\ref{section:smartcards_empirical_maintext}. Appendix Figure~\ref{fig:smartcards_sample_map2} illustrates our village classification. 

We implement our method using probability random forests to estimate the nuisance functions that enter the representation. Specifically, we use the \texttt{R} package \texttt{ranger} with $1,000$ trees  \texttt{min.node.size = 20}, and \texttt{max.depth = 10}. 
Random forests satisfy stability conditions, which allow us to eliminate cross-fitting; the argument is a straightforward extension of Proposition~\ref{prop:unknown}, using stability in place of independence to handle the stochastic equicontinuity terms. 
See e.g. \citet[Theorem 2]{chernozhukov2020adversarial}. Alternatively, we could use the limited complexity of random forests, along the lines of \citet[Theorem 3]{chernozhukov2020adversarial}.

Appendix Table~\ref{table:smartcards_main_results} reports the point estimates and standard errors associated with Figure~\ref{fig:smartcards_main_results} in the main text. We also report the point estimates and standard errors for the differences between the unbiased benchmark and our method. The differences are statistically insignificant.

\begin{table}
\centering
\begin{threeparttable}
\begin{tabular}{l | ccc}
\toprule
& \multicolumn{1}{c}{Consumption} & \multicolumn{1}{c}{Low income} & \multicolumn{1}{c}{Middle income} \\
\midrule
  RSV & -0.0525 & -0.0261 & -0.0450 \\
           & (0.0436) & (0.0220) & (0.0360) \\
\midrule
\addlinespace[0.3em]
  Benchmark & -0.0746 & -0.0235 & -0.0397 \\
           & (0.0445) & (0.0148) & (0.0274) \\
\midrule
  Difference & -0.0222 & 0.0026 & 0.0053 \\
           & (0.0271) & (0.0142) & (0.0246) \\
\bottomrule
\end{tabular}
\end{threeparttable}

\caption{Our method's point estimates approximately recovers the unbiased benchmark estimates.
For each poverty outcome, we report point estimates and standard errors for the benchmark and for our method. 
We also report the difference between the benchmark and our method as well as its standard error.
Bootstrap standard errors, based on $1000$ replications, are clustered at the sub-district level.
}
\label{table:smartcards_main_results}
\end{table}

\begin{figure}[htbp!]
\centering
\captionsetup[subfigure]{justification=Centering}
\begin{subfigure}{\textwidth}
    \centering
    \caption*{Panel A: Consumption} 
    \begin{subfigure}{0.45\textwidth}
        \centering
        \includegraphics[width=0.8\textwidth]{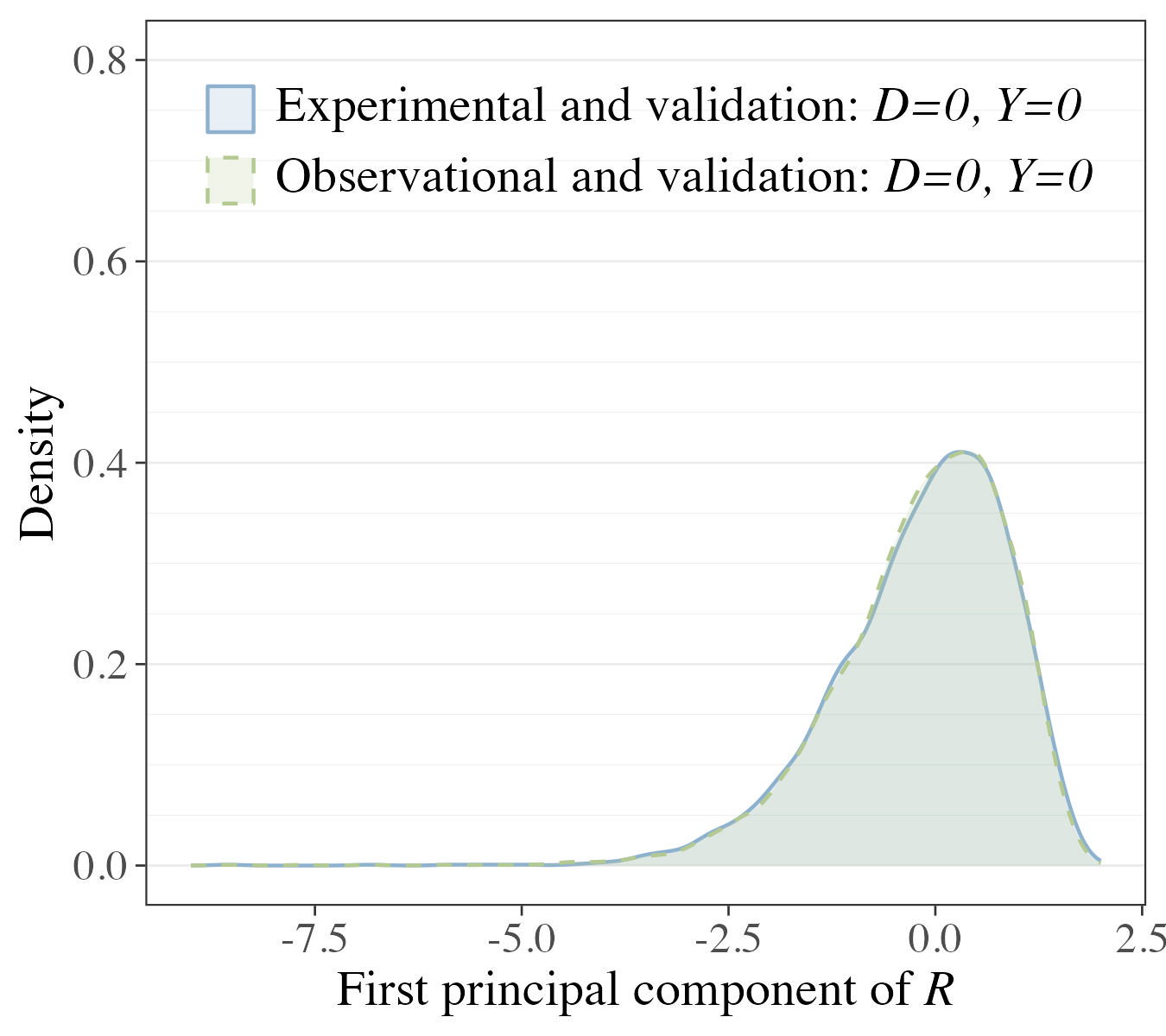}
        \caption{Densities of $R \mid \widetilde{S}, D = 0, Y = 0$.}
    \end{subfigure}
    \hfill
    \begin{subfigure}{0.45\textwidth}
        \centering
        \includegraphics[width=0.8\textwidth]{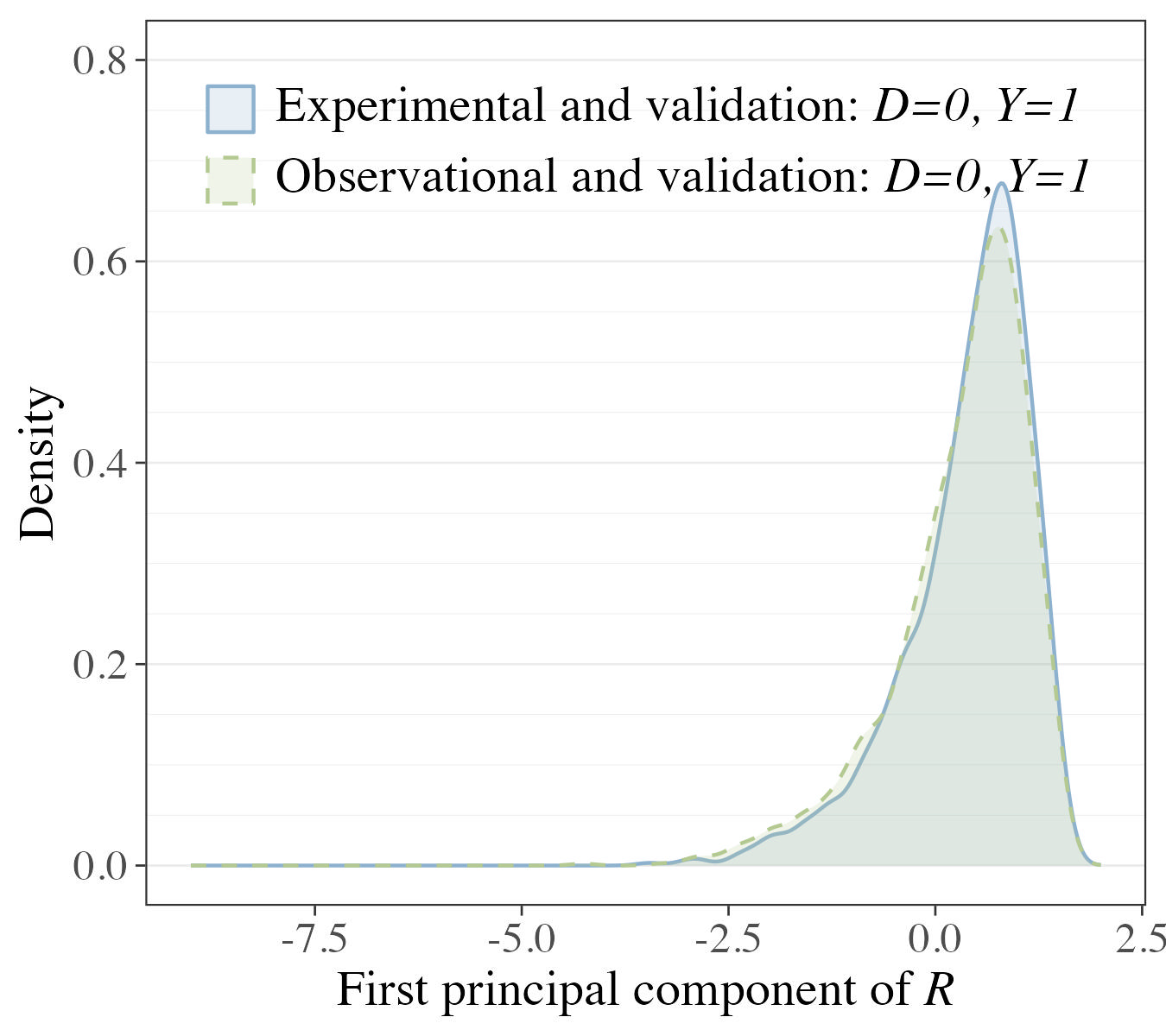}
        \caption{Densities of $R \mid \widetilde{S}, D = 0, Y = 1$.}
    \end{subfigure}
\end{subfigure}
\begin{subfigure}{\textwidth}
    \centering
    \caption*{Panel B: Low Income}
    \begin{subfigure}{0.45\textwidth}
        \centering
        \includegraphics[width=0.8\textwidth]{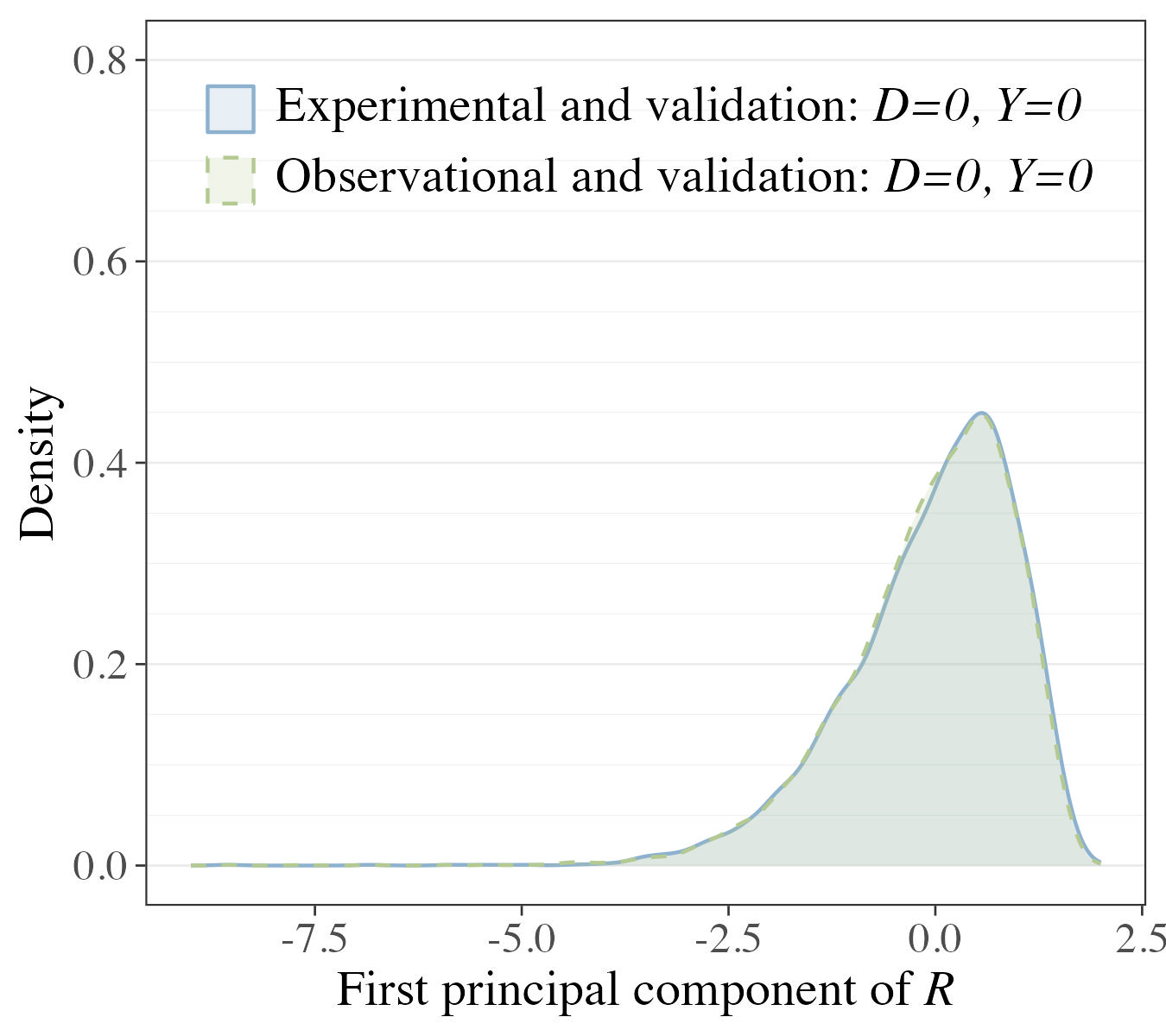}
        \caption{Densities of $R \mid \widetilde{S}, D = 0, Y = 0$.}
    \end{subfigure}
    \hfill
    \begin{subfigure}{0.45\textwidth}
        \centering
        \includegraphics[width=0.8\textwidth]{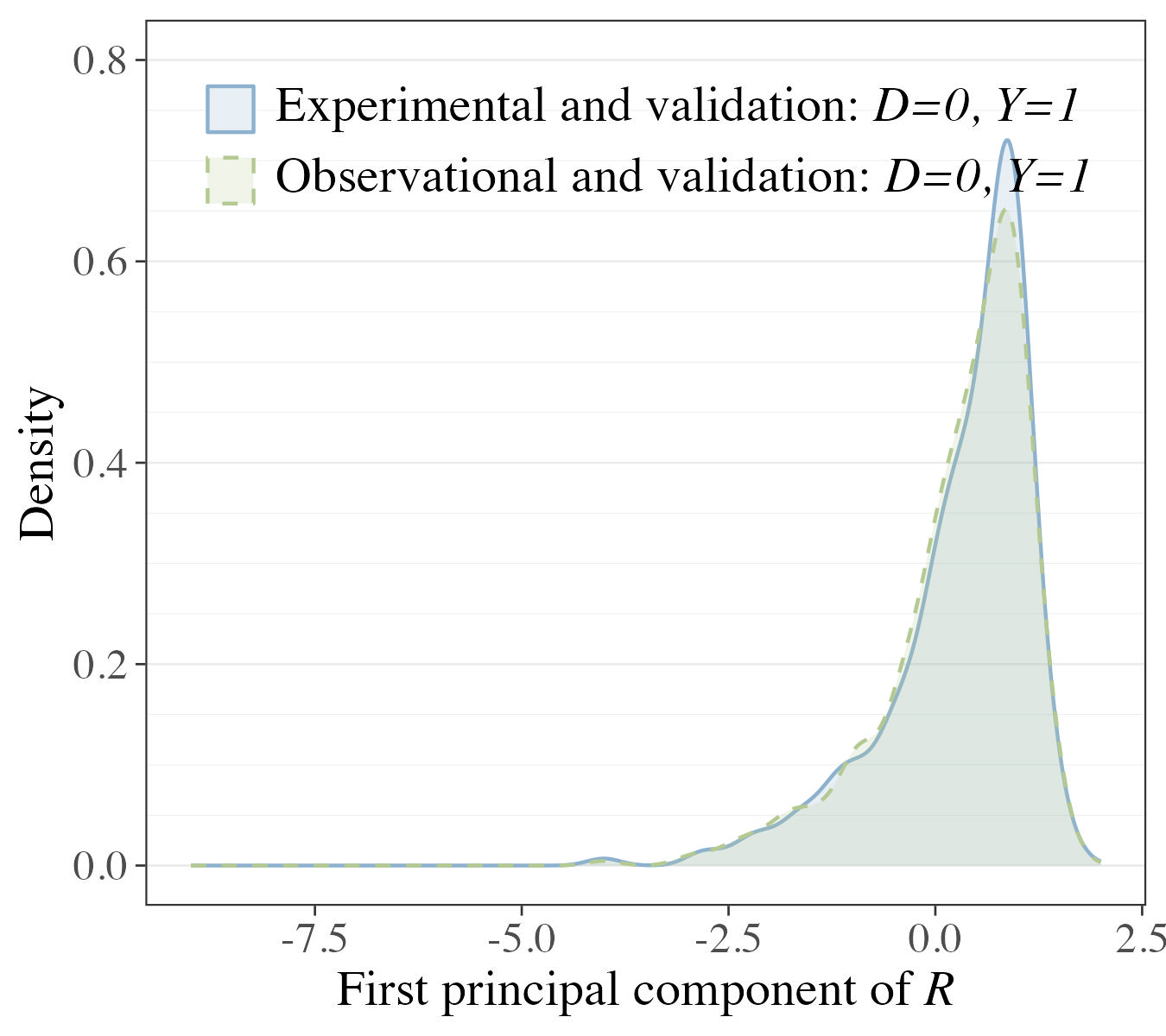}
        \caption{Densities of $R \mid \widetilde{S}, D = 0, Y = 1$.}
    \end{subfigure}
\end{subfigure}
\begin{subfigure}{\textwidth}
    \centering
    \caption*{Panel C: Middle Income}
    \begin{subfigure}{0.45\textwidth}
        \centering
        \includegraphics[width=0.8\textwidth]{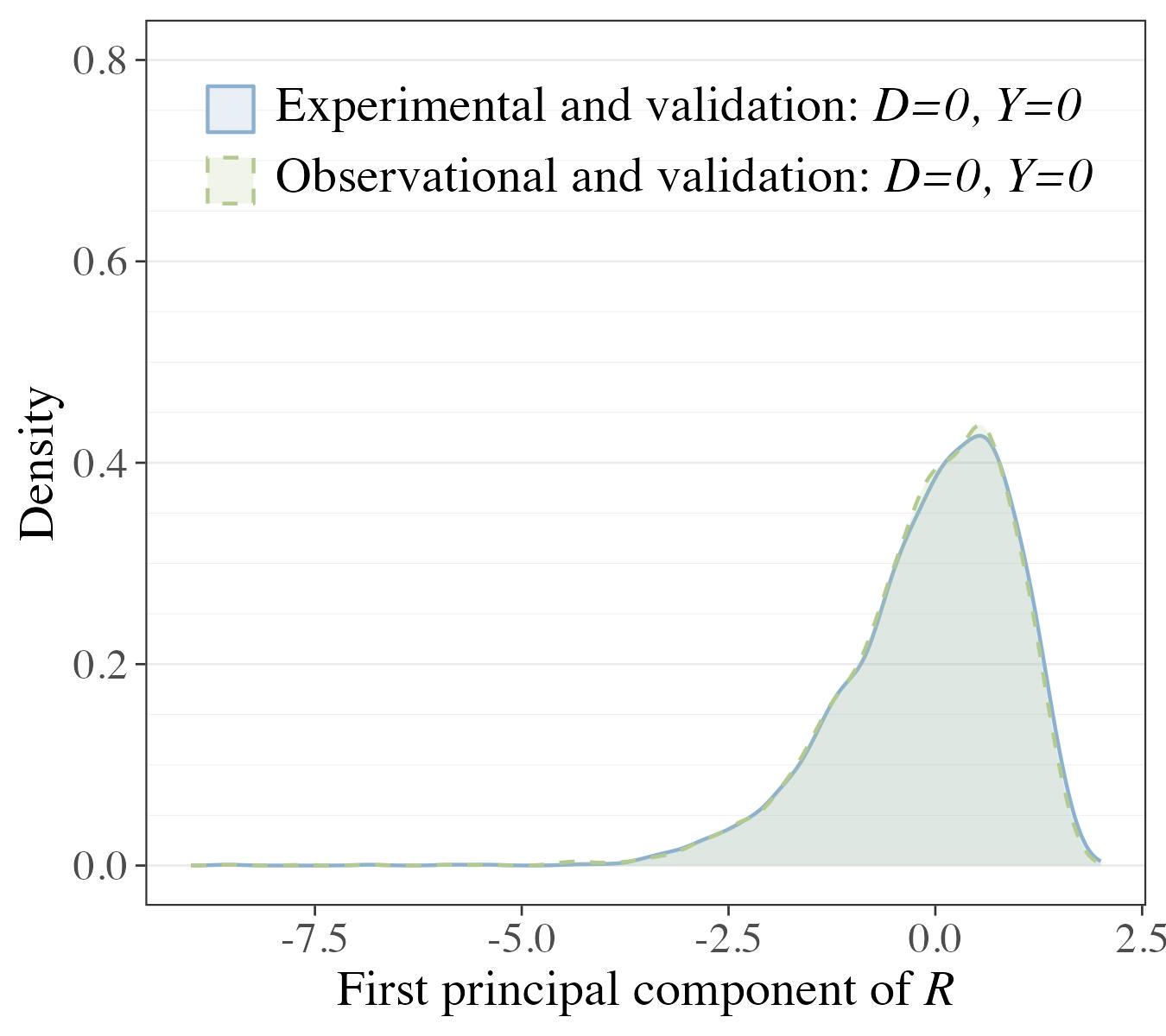}
        \caption{Densities of $R \mid \widetilde{S}, D = 0, Y = 0$.}
    \end{subfigure}
    \hfill
    \begin{subfigure}{0.45\textwidth}
        \centering
        \includegraphics[width=0.8\textwidth]{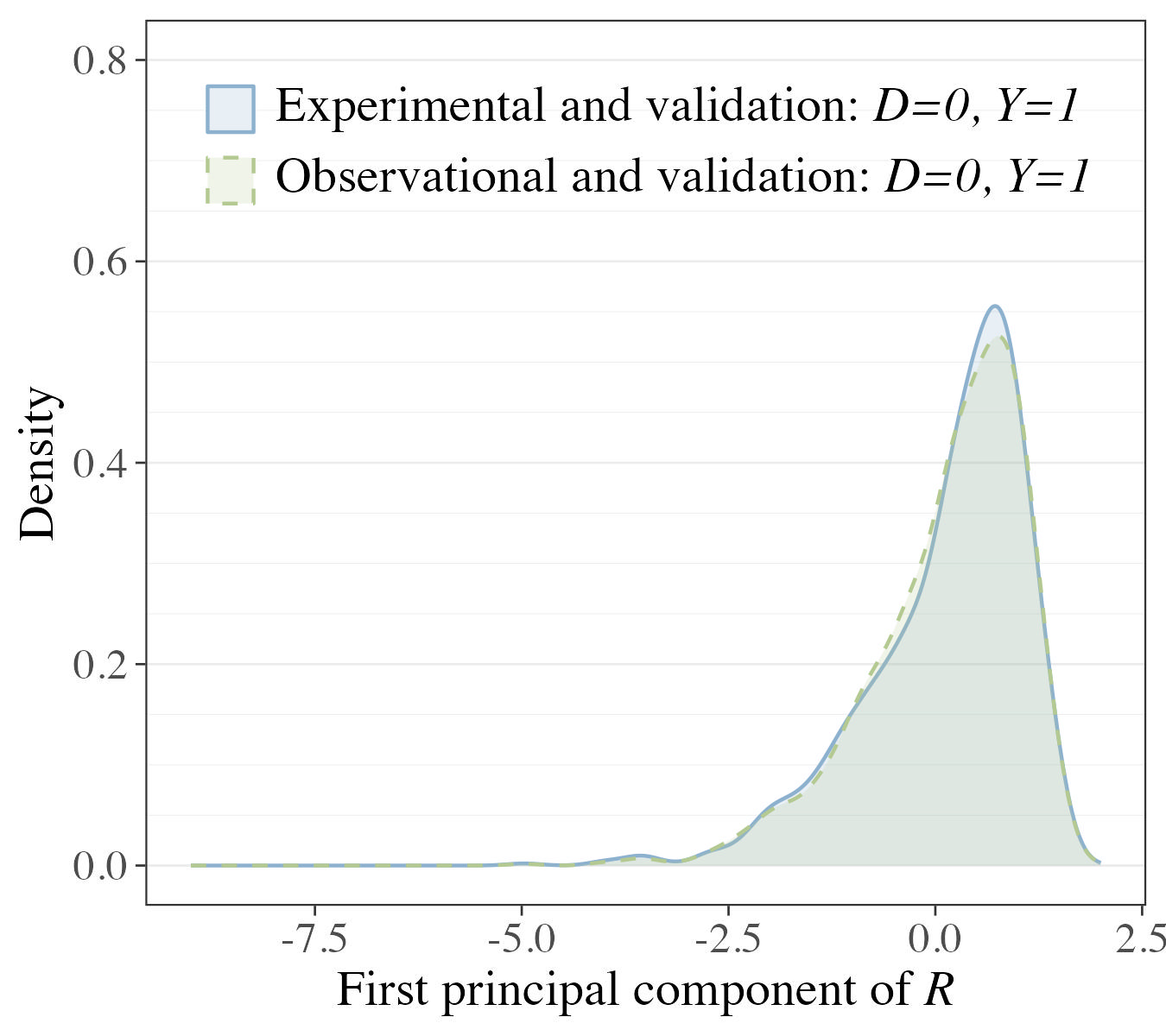}
        \caption{Densities of $R \mid \widetilde{S}, D = 0, Y = 1$.}
    \end{subfigure}
\end{subfigure}
\caption{Our stability assumption (Assumption~\ref{assumption:stability}(i)) is plausible for each poverty outcome in the Smartcards experiment.
We compare $f_R(R \mid \widetilde{S}\in\{e,v\},D=d,Y=y)$ with $f_R(R \mid \widetilde{S}\in\{o,v\},D=d,Y=y)$ for $d=0$ and $y\in\{0,1\}$.
Because $R$ is high-dimensional, we visualize the density of its standardized first principal component.
} 
\label{fig:smartcards_stability_other_outcomes}
\end{figure}

\begin{figure}[htbp!]
\captionsetup[subfigure]{justification=Centering}
\captionsetup[subfigure]{justification=Centering}
\begin{subfigure}{\textwidth}
    \centering
    \caption*{Panel A: Consumption} 
    \begin{subfigure}{0.45\textwidth}
        \centering
        \includegraphics[width=0.8\textwidth]{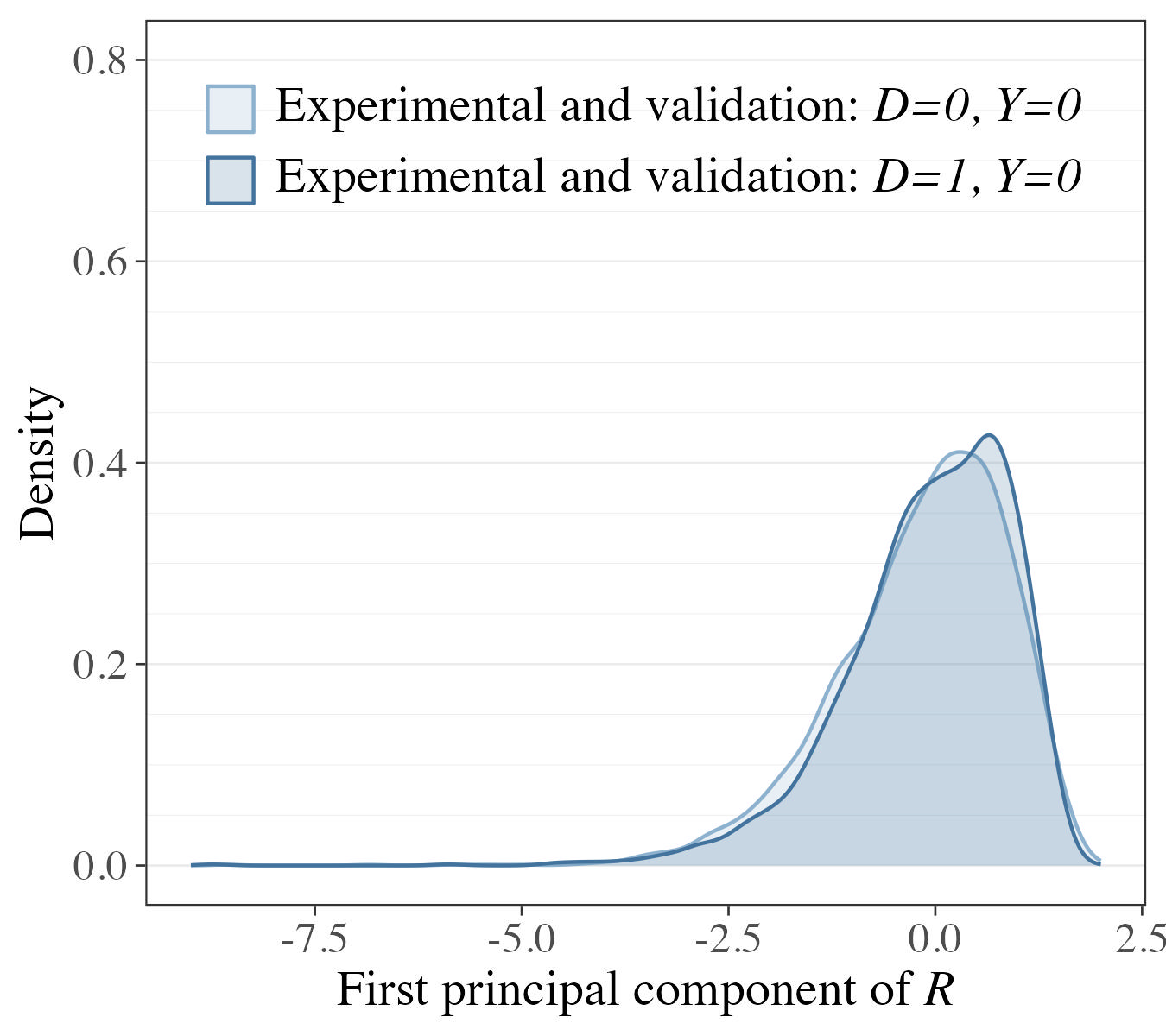}
        \caption{Densities of $R \mid \widetilde{S}\in\{e,v\}$, $D$, $Y=0$.}
    \end{subfigure}
    \hfill
    \begin{subfigure}{0.45\textwidth}
        \centering
        \includegraphics[width=0.8\textwidth]{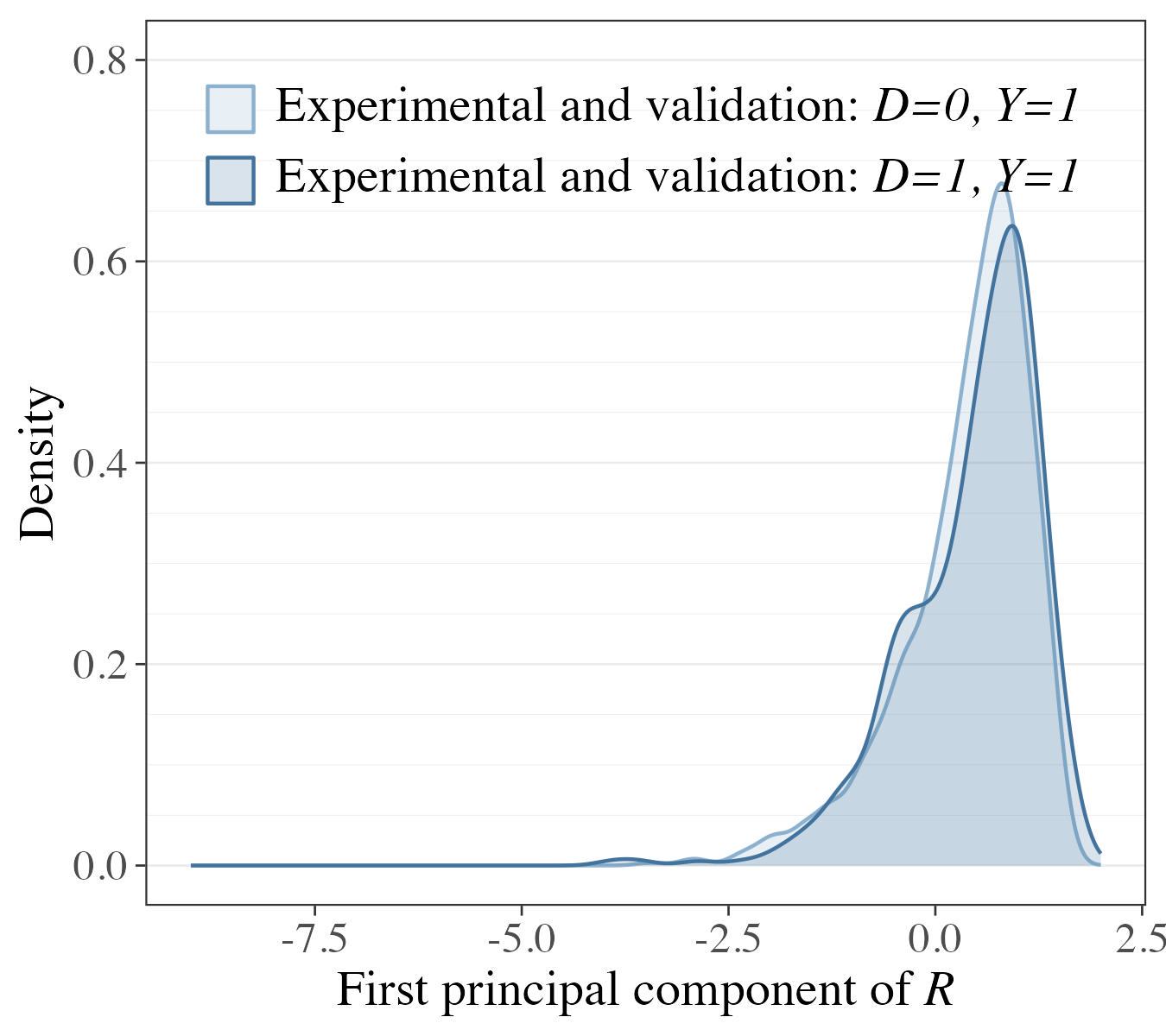}
        \caption{Densities of $R \mid \widetilde{S}\in\{e,v\}$, $D$, $Y=1$.}
    \end{subfigure}%
\end{subfigure}%
\vspace{0.25cm} 
\begin{subfigure}{\textwidth}
    \centering
    \caption*{Panel B: Low Income}
    \begin{subfigure}{0.45\textwidth}
        \centering
        \includegraphics[width=0.8\textwidth]{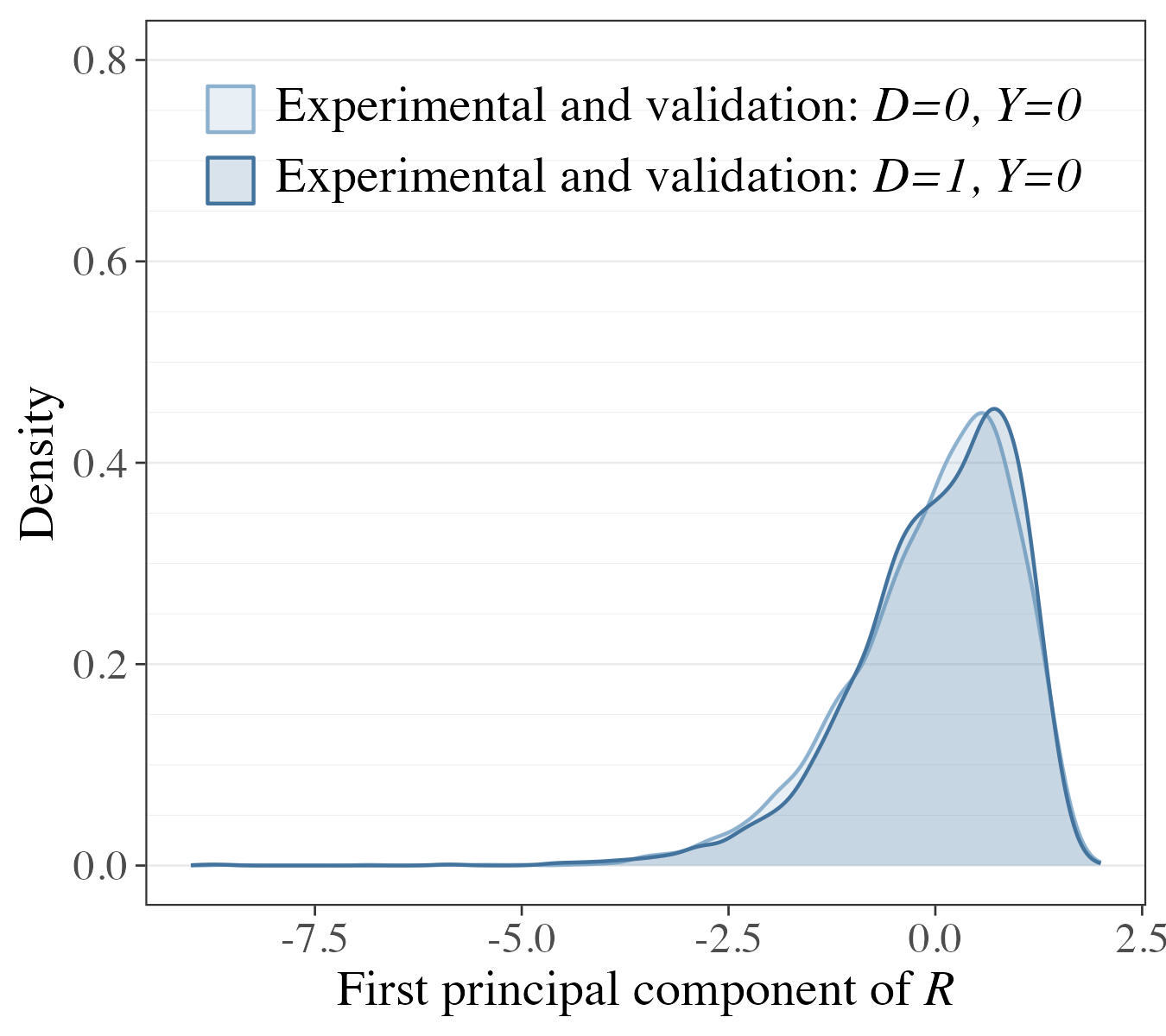}
        \caption{Densities of $R \mid \widetilde{S}\in\{e,v\}$, $D$, $Y=0$.}
    \end{subfigure}
    \hfill
    \begin{subfigure}{0.45\textwidth}
        \centering
        \includegraphics[width=0.8\textwidth]{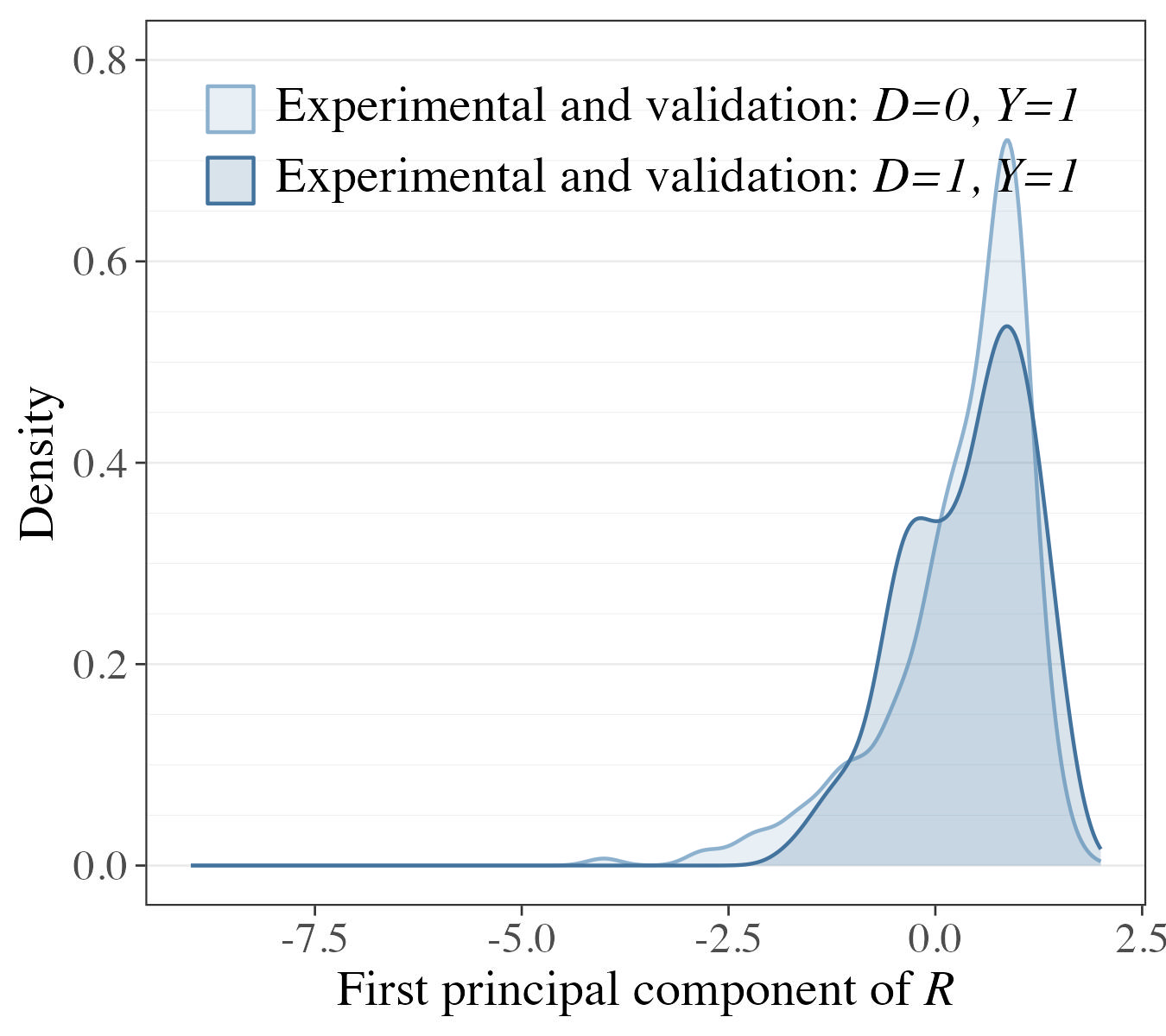}
        \caption{Densities of $R \mid \widetilde{S}\in\{e,v\}$, $D$, $Y=1$.}
    \end{subfigure}%
\end{subfigure}%
\vspace{0.25cm} 
\begin{subfigure}{\textwidth}
    \centering
    \caption*{Panel C: Middle Income}
    \begin{subfigure}{0.45\textwidth}
        \centering
        \includegraphics[width=0.8\textwidth]{figures/smartcards/no_direct_effect/Ymidinc/y0.jpeg}
        \caption{Densities of $R \mid \widetilde{S}\in\{e,v\}$, $D$, $Y=0$.}
    \end{subfigure}
    \hfill
    \begin{subfigure}{0.45\textwidth}
        \centering
        \includegraphics[width=0.8\textwidth]{figures/smartcards/no_direct_effect/Ymidinc/y1.jpeg}
        \caption{Densities of $R \mid \widetilde{S}\in\{e,v\}$, $D$, $Y=1$.}
    \end{subfigure}%
\end{subfigure}
\caption{
No direct effects (Assumption~\ref{assumption:observational}(ii)) is plausible in the Smartcards experiment, for each poverty outcome.
We compare $f_R(R \mid \widetilde{S}\in \{e,v\}, D = 0, Y = y)$ with $f_{R}(R \mid \widetilde{S}\in \{e,v\}, D = 1, Y = y)$ for $y\in\{0,1\}$.
Because $R$ is high-dimensional, we visualize the density of its standardized first principal component. 
} 
\label{fig:no_direct_effects_other_outcomes}
\end{figure}

\paragraph{Plausibility of identifying assumptions.}
Appendix Figures~\ref{fig:smartcards_stability_other_outcomes} and~\ref{fig:no_direct_effects_other_outcomes} demonstrate that the stability and no-direct-effect assumptions are plausible for each poverty outcome.

First, stability requires that four conditional distributions of the remotely sensed variable should be the same across samples: $f_R(r \mid \widetilde{S}\in\{e,v\}, D=d, Y=y) = f_R(r \mid \widetilde{S}\in\{o,v\}, D=d, Y=y)$, for each $(d,y)$ stratum.\footnote{This condition follows from the proof of Lemma~\ref{lemma:discrete_mixture_incomplete}.}  
Appendix Figure~\ref{fig:smartcards_stability_other_outcomes} evaluates the conditions for $d=0$ and $y\in\{0,1\}$ by subsetting and visualizing densities. The conditional distributions almost coincide.

To complement the visual evidence, we conduct Kolmogorov-Smirnov tests for equality of the conditional distributions across samples. 
For each poverty outcome, we cannot reject the null of equal distributions at the $5\%$ level.

Second, with incomplete observational cases, we require no direct effects (Assumption~\ref{assumption:observational}(ii)):  the treatment should affect the remotely sensed variable only through the outcome. 
Using ``oracle'' data access, we can visually assess this condition; the densities $f_R(r \mid \widetilde{S}\in \{e,v\}, D = 0, Y = y)$ and $f_R(r \mid \widetilde{S}\in \{e,v\}, D = 1, Y = y)$ should coincide for $y\in\{0,1\}$.
In Figure~\ref{fig:no_direct_effects_other_outcomes}, the conditional densities closely align for each poverty outcomes

Again, we conduct Kolmogorov-Smirnov tests. For the ``consumption'' outcome, we can marginally reject the null of equal distributions at the $5\%$ level for the $Y = 0$ case ($p = 0.048)$ and the $Y = 1$ case ($p = 0.033$). For the ``low income'' and ``middle income'' outcomes, we cannot reject the null of equality at the 5\% level. 

\begin{figure}[htbp!]
\begin{subfigure}[t]{0.32\textwidth}
  \centering
  \includegraphics[width=\textwidth]{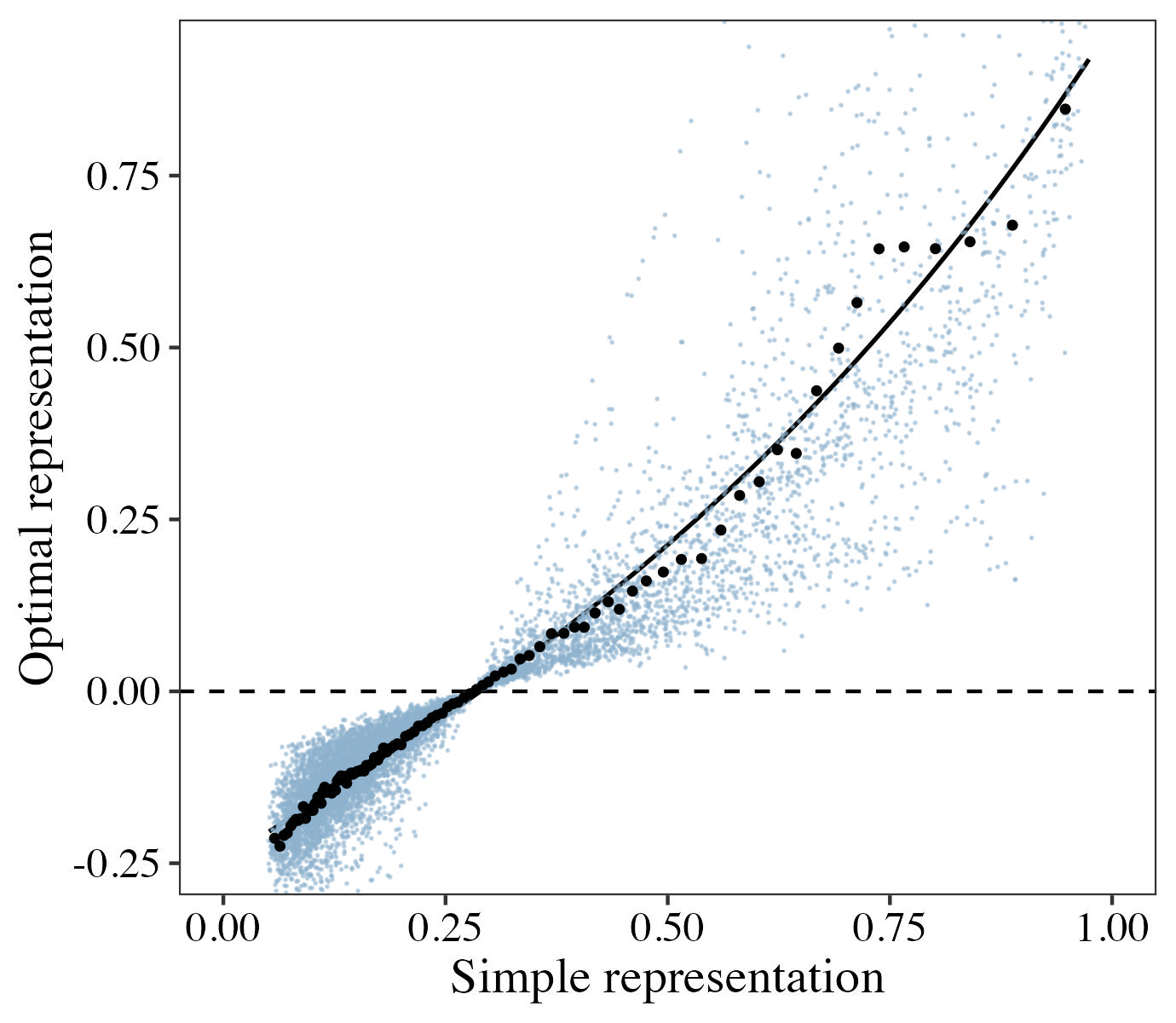}
  \caption{\footnotesize Consumption.}
\end{subfigure}\hfill
\begin{subfigure}[t]{0.32\textwidth}
  \centering
  \includegraphics[width=\textwidth]{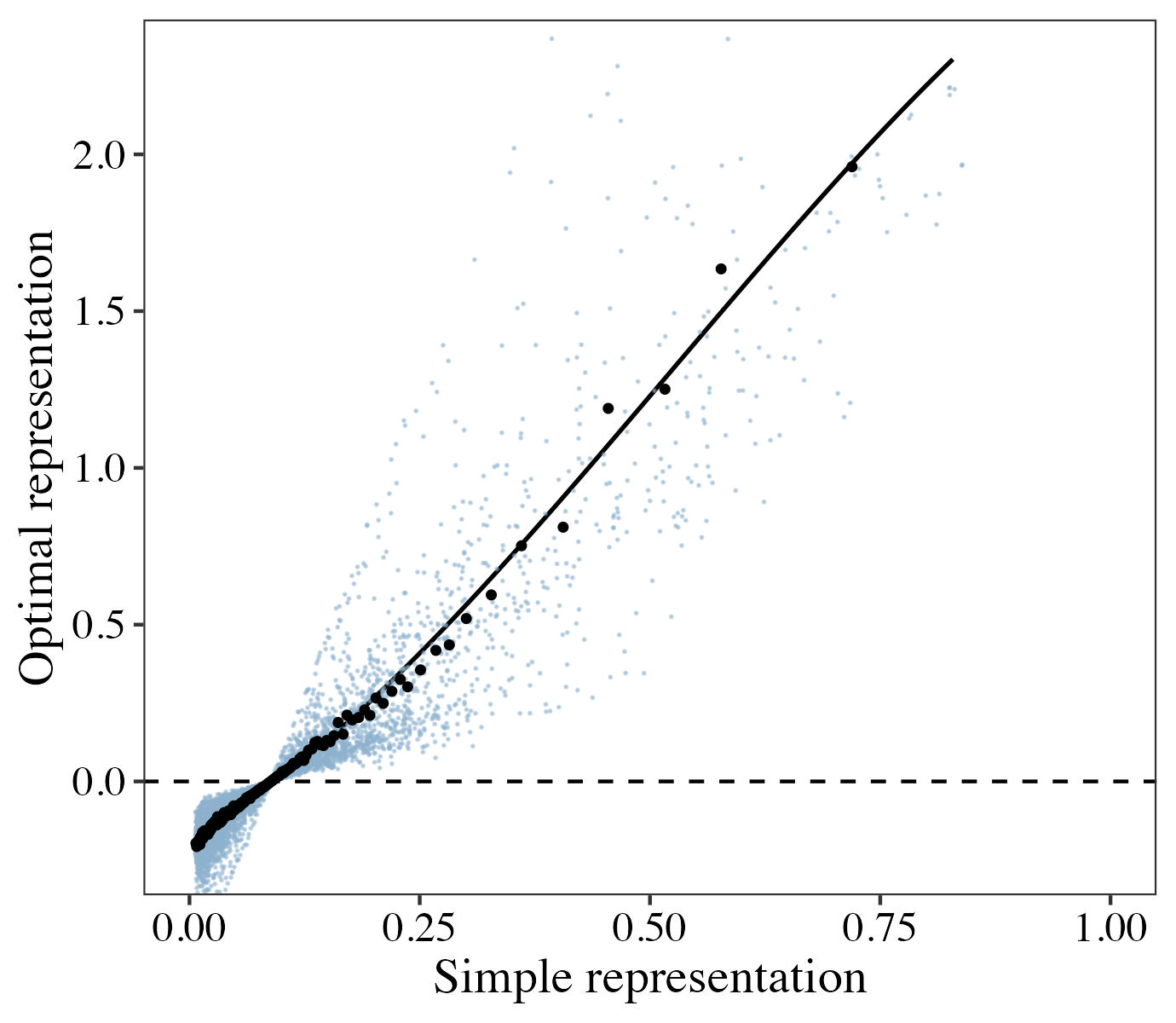}
  \caption{\footnotesize Low Income.}
\end{subfigure}\hfill
\begin{subfigure}[t]{0.32\textwidth}
  \centering
  \includegraphics[width=\textwidth]{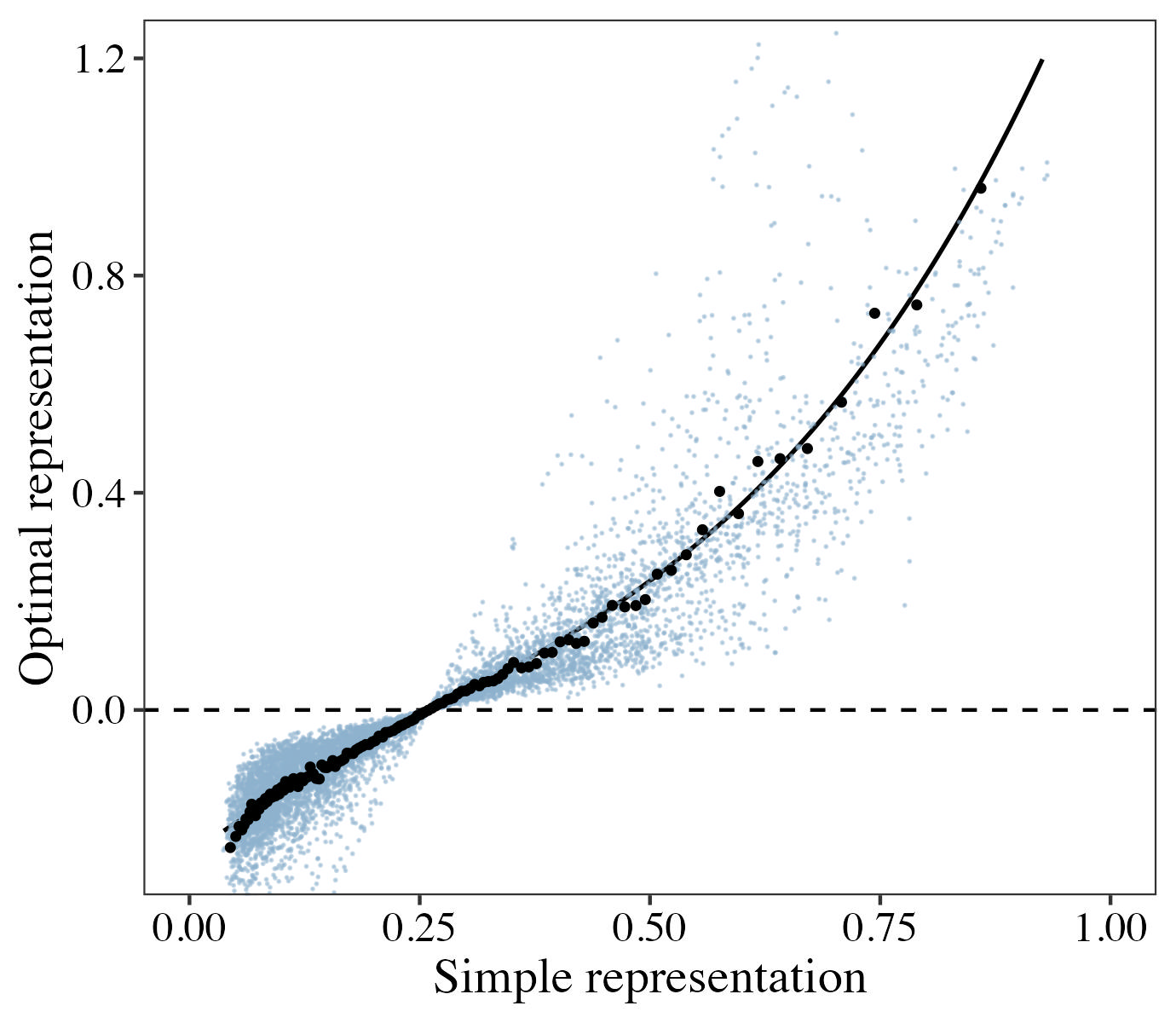}
  \caption{\footnotesize Middle Income.}
\end{subfigure}
\caption{We contrast optimal versus simple representations of the remotely sensed variable. The optimal representation is $H^*(R)$ from Section~\ref{sec:optimal}, combining three predictions. The simple representation is one prediction: $\Pr(Y = 1 \mid R, \widetilde{S} \in \{o,v\})$.} 
\label{fig:representation_learning}
\end{figure}

\paragraph{Comparing representations.}
Appendix Figure~\ref{fig:representation_learning} visualizes how our optimal representation $H^*(R)$ compares to the simple representation $\Pr(Y = 1 \mid R, \widetilde{S} \in \{o,v\})$. 
Even though the remotely sensed variable is high-dimensional, these representations are scalars. 
Therefore, we can visualize each village as a point in a plot with $H^*(R)$ on the vertical axis and $\Pr(Y = 1 \mid R, \widetilde{S} \in \{o,v\})$ on the horizontal axis. 
For the sake of visualization, we also plot the binscatter and the best fitting curve. 

There are some notable differences. 
The optimal representation $H^*(R)$ often takes negative values, while the simple representation $\Pr(Y = 1 \mid R, \widetilde{S} \in \{o,v\})$ is within the unit interval $(0,1)$. 
It is optimal to extrapolate beyond observed villages. 
Using $\Pr(Y = 1 \mid R, \widetilde{S} \in \{o,v\})$ in place of $H^*(R)$ in Algorithm~\ref{alg:discrete_nocovar_overview} would only interpolate among observed villages. 
The representations $H^*(R)$ and $\Pr(Y = 1 \mid R, \widetilde{S} \in \{o,v\})$ are generally correlated, but their relationship is nonlinear.
As discussed in the main text, these differences matter for the precision of downstream inference.

\paragraph{Spillovers.}
By assuming no spillovers across mandals, our method estimates the average global treatment effect from the full sample.
Therefore, under the conditions of Corollary~\ref{cor:spill1}, Figure~\ref{fig:smartcards_main_results} and Table~\ref{tab:rsv_relevance_smartcards} report average global treatment effect estimates and relevance in the presence of spillovers.

\begin{table}
\centering
\begin{threeparttable}
\begin{tabular}{l | ccc}
\toprule
 & \multicolumn{1}{c}{Consumption} & \multicolumn{1}{c}{Low income} & \multicolumn{1}{c}{Middle income} \\
\midrule
  Relevance, $\E_n\{\widehat{H}(R) \widehat{\Delta}^o\}$ & 0.2349 & 0.5068 & 0.2145 \\
           & (0.0810) & (0.1715) & (0.0708) \\
\bottomrule
\end{tabular}
\end{threeparttable}

\caption{Satellite images are relevant to the poverty outcomes in the presence of spillovers. 
For each poverty outcome, we report $\E_n\{\widehat{H}(R) \widehat{\Delta}^o\}$ and its standard error using our learned representation. 
The estimates correspond to Corollary~\ref{cor:spill2}.
Bootstrap standard errors, based on $1000$ replications, are clustered at the sub-district level.
}
\label{tab:rsv_relevance_smartcards_wo_spillovers}
\end{table}

As a robustness check, we now repeat our empirical analysis after subsetting the villages in the manner described in Corollary~\ref{cor:spill2}. The subsetting step eliminates about $27\%$ of the total number villages. In particular, about $27\%$ of villages in the full sample have at least one village within a $20$ km radius exposed to opposite treatment status. The distance between two villages is measured using their centroids, following \cite{muralidharan2023general}. 

\begin{figure}[ht]
\centering
\begin{subfigure}[t]{0.32\textwidth}
    \centering
    \includegraphics[width=\textwidth]{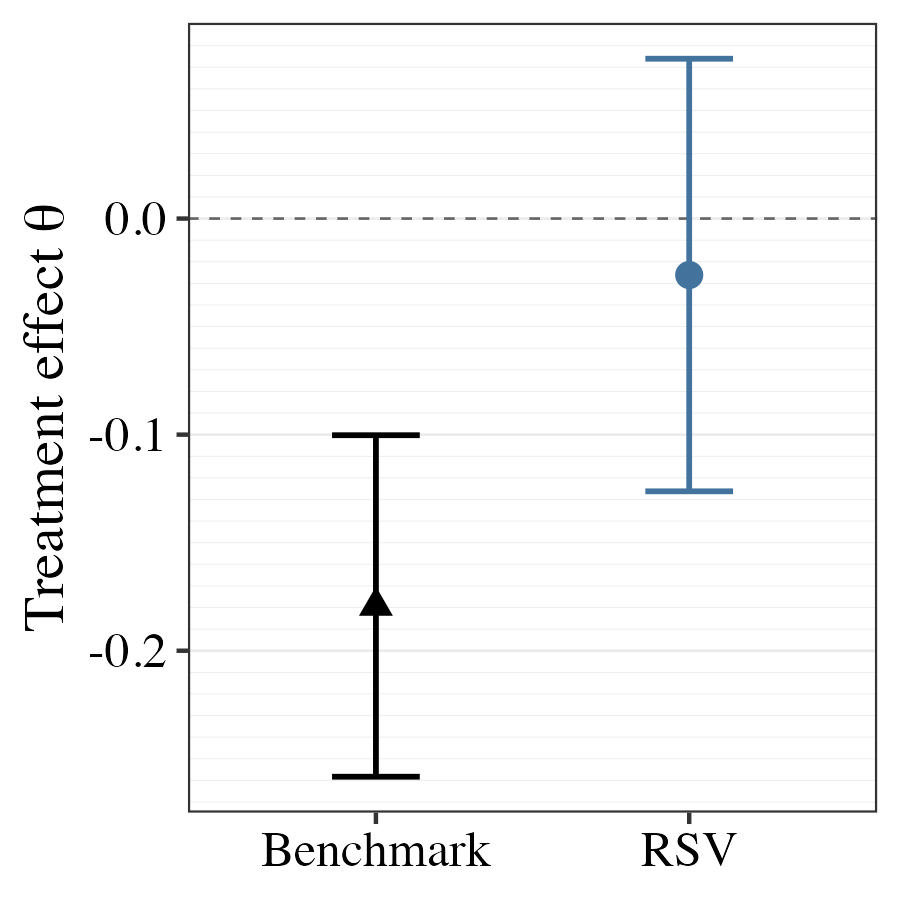}
    \caption{\footnotesize Consumption.}
\end{subfigure}
\hfill 
\begin{subfigure}[t]{0.32\textwidth}
    \centering
    \includegraphics[width=\textwidth]{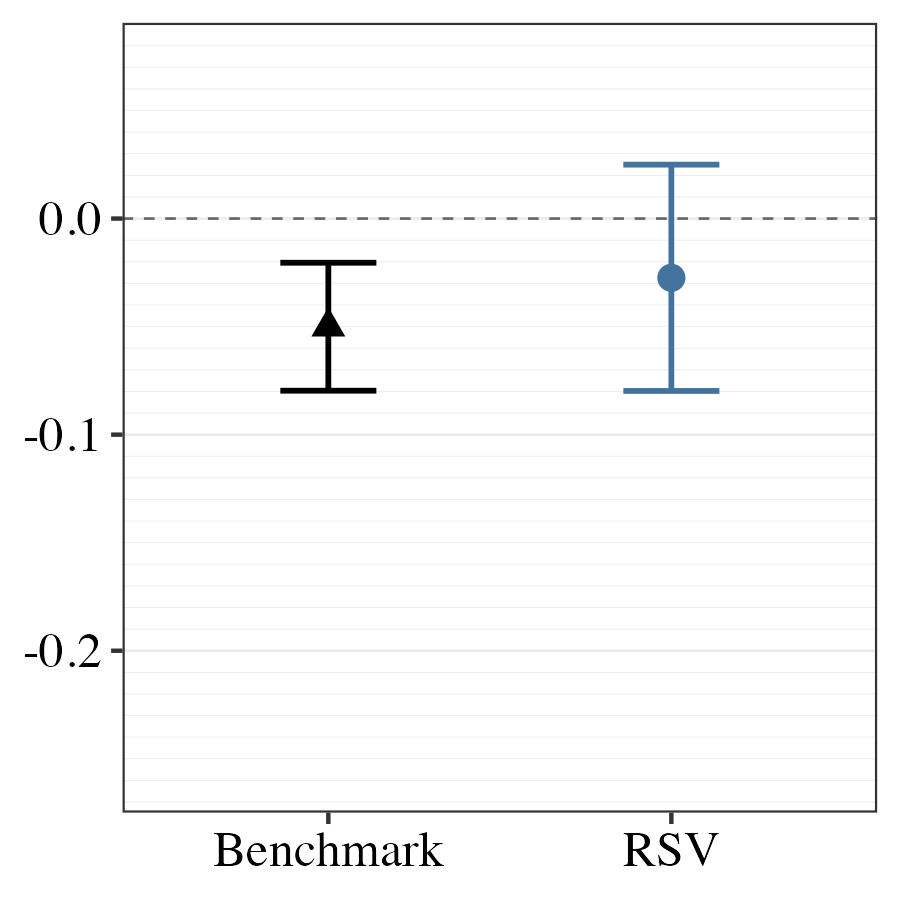}
    \caption{\footnotesize Low income.}
\end{subfigure}
\hfill 
\begin{subfigure}[t]{0.32\textwidth}
    \centering
    \includegraphics[width=\textwidth]{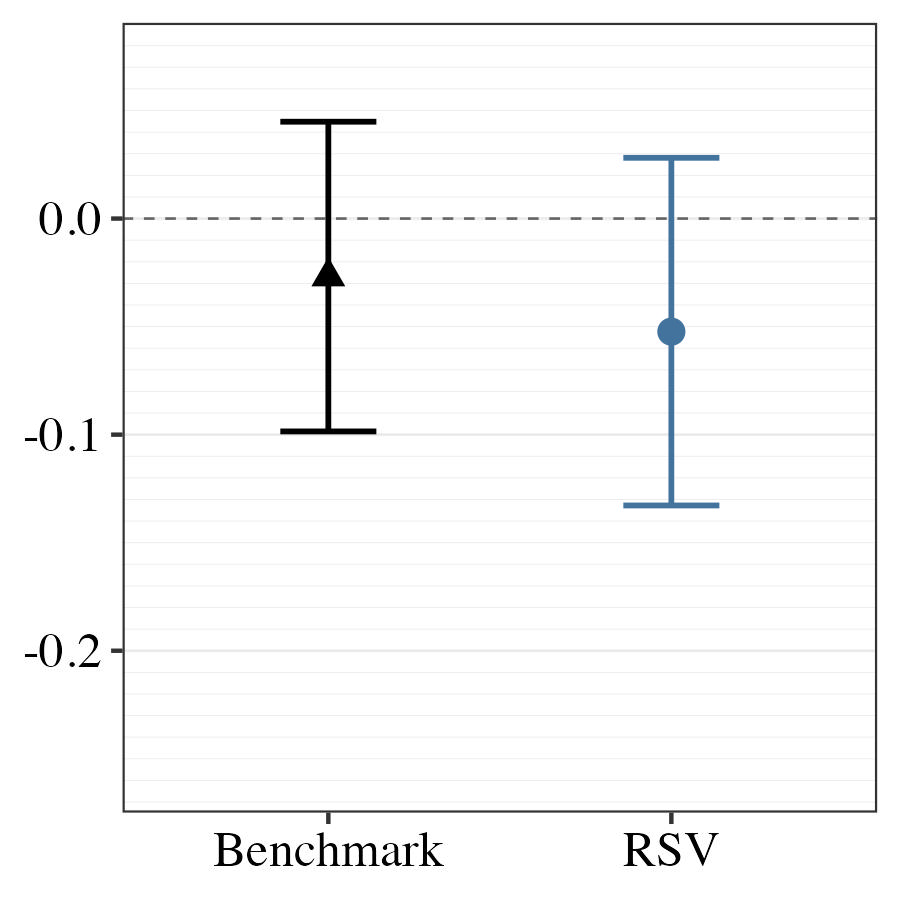}
    \caption{\footnotesize Middle income.}
\end{subfigure}
\caption{
Our method's estimates approximately recover the unbiased benchmark estimates in the presence of spillovers. 
For each poverty outcome, we report 90\% confidence intervals for the benchmark and for our method. 
The estimates correspond to Corollary~\ref{cor:spill2}, i.e. using the subsetted sample. 
Bootstrap standard errors, based on $1000$ replications, are clustered at the sub-district level.
}
\label{figure:smartcards_main_results_wo_spillover}
\end{figure}

Under the conditions of Corollary~\ref{cor:spill2}, Figure~\ref{figure:smartcards_main_results_wo_spillover} and Table~\ref{tab:rsv_relevance_smartcards_wo_spillovers} report global average treatment effect estimates and relevance in the presence of spillovers. We find qualitatively similar results as before. Here, the benchmark uses the same subset of villages as our method.

\paragraph{Time variation.}

We discuss time variation in detail in Appendix~\ref{sec:timing}. Here, we summarize the consequences for the interpretation of results presented in Section~\ref{section:smartcards_empirical_maintext}. 

Under the conditions of Theorem~\ref{thm:timing}, Figure~\ref{fig:smartcards_main_results} reports the effect of  $2010$  Smartcards adoption on $2013$ poverty, for treated, untreated, and buffer mandals.
Our results apply when $D$ is an early adoption treatment deployed in $2010$, when $Y$ is a poverty outcome collected in $2013$, and when $R$ is a remotely sensed variable collected separately in, say, $2019$. 

With incomplete cases, there is an implicit Markov restriction that researchers must assess. Concretely, 2010 Smartcards may affect 2013 poverty; 2013 poverty may affect 2019 poverty; and 2013 and 2019 poverty may affect 2019 satellite images. However, in this example, 2010 Smartcards cannot affect 2019 poverty via any channel besides 2013 poverty. Figure~\ref{fig:no_direct_effects_other_outcomes} provides empirical evidence supporting this assumption.

\subsection{Payments for Ecosystem Services and Crop Burning}\label{section:cropburning_empirical_additionaldetails}

We present additional results for the empirical application in Section~\ref{sec:crop_burning}, based on~\cite{jack2022money}. 
Specifically, Table~\ref{table:crop_other_outcomes} presents additional variations of the two-step method.
In Table~\ref{table:crop}, we used the authors' ``maximum accuracy'' classifier for whether a field has not been burned as the predicted outcome. Now, we also use the authors' ``balanced accuracy'' classifier as the predicted outcome. 
Finally, we use the continuous random forest prediction as the predicted outcome. 
Both variations of the two-step method appears to have economically meaningful attenuation bias (Proposition~\ref{prop:bias_common_practice}).

\begin{table}[ht!]
\centering
\begin{threeparttable}
\begin{tabular}{l | c c c}
\toprule
 & \shortstack{Maximum \\ accuracy} & \shortstack{Balanced \\ accuracy} & \shortstack{Random forest \\ prediction} \\
\midrule
  Common practice & 0.075 & 0.067 & 0.030 \\
           & (0.038) & (0.046) & (0.017) \\
\bottomrule
\end{tabular}
\end{threeparttable}

\caption{Variations of the common practice exhibit attenuation bias. The different variations use different predicted outcomes within the two-step method: the authors' ``maximum accuracy'' classifier, their ``balanced accuracy'' classifier, or continuous random forest predictions. We separately regress each predicted outcome on the treatment. Bootstrap standard errors in parentheses, based on 5000 replications, are clustered at the village level.}
\label{table:crop_other_outcomes}
\end{table}


\end{document}